\documentclass[12pt]{article}
\newlength{\abstractwidth}
\usepackage{amsmath, nccmath}
\usepackage{amsfonts}
\usepackage{amssymb}
\usepackage{latexsym}

\usepackage{color}
\usepackage{graphicx}
\usepackage{tikz}
\usepackage{dsfont}

\usetikzlibrary{arrows,shapes,positioning}
\usetikzlibrary{decorations.markings}
\usepackage[rightcaption]{sidecap}
\tikzstyle arrowstyle=[scale=1]
\tikzstyle directed=[postaction={decorate,decoration={markings,
    mark=at position .65 with {\arrow[arrowstyle]{stealth}}}}]
\tikzstyle reverse directed=[postaction={decorate,decoration={markings,
    mark=at position .65 with {\arrowreversed[arrowstyle]{stealth};}}}]
\usetikzlibrary{positioning}

\usepackage{hyperref}
\definecolor{darkred}{rgb}{0.8,0.1,0.1}
\hypersetup{colorlinks=true, linkcolor=darkred, citecolor=blue, linktoc=page}

\renewcommand{\thefootnote}{\fnsymbol{footnote}}
\renewcommand{\thanks}[1]{\footnote{#1}}
\newcommand{\starttext}{
\setcounter{footnote}{0}
\setcounter{section}{0}
\renewcommand{\thefootnote}{\arabic{footnote}}}

\newcommand{\bea}{\begin{eqnarray}}
\newcommand{\eea}{\end{eqnarray}}
\newcommand{\be}{\begin{eqnarray}}
\newcommand{\ee}{\end{eqnarray}}
\newcommand{\bma}{\begin{matrix}}
\newcommand{\ema}{\cr\end{matrix}}
\newcommand{\<}{\langle}
\renewcommand{\>}{\rangle}

\def\cB{{\cal B}}
\def\cC{{\cal C}}

\def\cF{{\cal F}}
\def\cG{{\cal G}}
\def\cH{{\cal H}}
\def\cI{{\cal I}}
\def\cJ{{\cal J}}

\def\cL{{\cal L}}
\def\cM{{\cal M}}
\def\cN{{\cal N}}
\def\cO{{\cal O}}
\def\cP{{\cal P}}

\def\cS{{\cal S}}

\def\cV{{\cal V}}
\def\cW{{\cal W}}

\def\cY{{\cal Y}}
\def\cZ{{\cal Z}}

\def\mA{\mathfrak{A}}
\def\mB{\mathfrak{B}}

\def\mF{\mathfrak{F}}

\def\mH{\mathfrak{H}}

\def\mJ{\mathfrak{J}}

\def\mL{\mathfrak{L}}
\def\mM{\mathfrak{M}}
\def\mN{\mathfrak{N}}

\def\mP{\mathfrak{P}}

\def\mS{\mathfrak{S}}

\def\me{\mathfrak{e}}

\def\mq{\mathfrak{q}}

\def\mt{\mathfrak{t}}

\def\mz{\mathfrak{z}}

\def\ZZ{{\mathbb Z}}
\def\RR{{\mathbb R}}

\def\CC{{\mathbb C}}

\def\Im{{\rm Im \,}}
\def\tr{{\rm tr}}

\def\det{{\rm det \,}}

\def\half{{1\over 2}}
\def\thalf{{\tfrac{1}{2}}}
\def\p{\partial}

\def\a{\alpha}
\def\b{\beta}

\def\eps{\epsilon}
\def\f{\varphi}
\def\tet{\vartheta}
\def\ep{\varepsilon}
\def\om{\omega}

\def\kap{\mu}

\def\Wt{\tilde{W}}
\def\Dt{\tilde{{\cal D}}}

\def\pbz{\p _{\bar z}}
\def\pbw{\p_{\bar w}}
\def\chiz{\chi_{\bar z} ^+}

\def\ti{\tilde}
\def\no{\nonumber}
\def\sm{\smallskip}

\begin{document}
\starttext
\setcounter{footnote}{0}

\begin{flushright}
2021 August 2 \\
%revised 2021 November 27 \\
revised 2022 July 24 \\
UUITP-35/21 \\
%RNS-paper-42.tex
\end{flushright}

\bigskip

\begin{center}

{\Large \bf Two-loop superstring five-point amplitudes III }

\vskip 0.1in

{ \bf Construction via the RNS formulation: even spin structures   }

\vskip 0.4in

{  \bf Eric D'Hoker$^{(a)}$ and Oliver Schlotterer$^{(b)}$}

\vskip 0.1in

 ${}^{(a)}$ {\sl Mani L. Bhaumik Institute for Theoretical Physics}\\
 { \sl Department of Physics and Astronomy }\\
{\sl University of California, Los Angeles, CA 90095, USA}\\

\vskip 0.1in

 ${}^{(b)}$ { \sl Department of Physics and Astronomy,} \\ {\sl Uppsala University, 75108 Uppsala, Sweden}
 
 \vskip 0.1in
 
{\sl dhoker@physics.ucla.edu},  {\sl oliver.schlotterer@physics.uu.se}

\vskip 0.5in

{\sl Dedicated to Professor Michael Green on the occasion of his 75-th birthday}

\hskip 0.5in

\begin{abstract}
The contribution from even spin structures to the genus-two amplitude for five massless external NS states in Type II and Heterotic superstrings is evaluated from first principles in the RNS formulation. Using chiral splitting with the help of loop momenta this problem reduces to the evaluation of the corresponding chiral amplitude, which is carried out using the same techniques that were used for the genus-two amplitude with four external NS states. The results agree with the parity-even NS components of a construction using chiral splitting and pure spinors given in earlier companion papers arXiv:2006.05270 and arXiv:2008.08687. 
\end{abstract}

\end{center}

\newpage

\setcounter{tocdepth}{2} 
\tableofcontents

\newpage

\baselineskip=16pt
\setcounter{equation}{0}
\setcounter{footnote}{0}

\newpage

%%%%%%%%%%%%%%%%%%%%%%%%%%%%%%%%%%%%%%%%%%%
%%%%%%%%%%%%%%%%%%%%%%%%%%%%%%%%%%%%%%%%%%%
\section{Introduction}
\label{sec:1}
\setcounter{equation}{0}
%%%%%%%%%%%%%%%%%%%%%%%%%%%%%%%%%%%%%%%%%%%
%%%%%%%%%%%%%%%%%%%%%%%%%%%%%%%%%%%%%%%%%%%

The study of perturbative scattering amplitudes in string theory dates back to the founding of the subject and continues to provide deep insights into the dynamics of string theory today. String amplitudes enjoy a rich mathematical structure and have proven fundamental in establishing and verifying predictions from string dualities and space-time supersymmetry. Furthermore, the appearance of super-Yang-Mills and supergravity amplitudes 
in their low-energy limit has become increasingly significant to expose the intimate connections
between gauge theory and gravity.

\sm

Early investigations of superstring amplitudes, including calculations of tree-level and genus-one amplitudes with a small number of external string states, were carried out in a variety of formulations (see for example \cite{GSW} and references therein). More recent result have been obtained primarily using the Ramond-Neveu-Schwarz (RNS) formulation with the Gliozzi-Scherk-Olive (GSO) projection \cite{Friedan:1985ge, RMP, DHoker:2002hof, Witten:2012bh} as well as the manifestly spacetime supersymmetric pure-spinor formulation \cite{Berkovits:2000fe, Berkovits:2004px,Berkovits:2005bt}. 

\sm

In particular, for higher loop amplitudes, the genus-two measure for even spin structures was obtained  in the RNS formulation by direct calculation using the super period matrix  in  \cite{DP1, DP2, DP3, DP4}, and by purely algebraic-geometry methods in \cite{Witten:2013tpa}. The corresponding amplitudes in both Type II and Heterotic strings with up to four external massless NS bosons were constructed in the RNS formulation in \cite{DP5,DP6} and generalized to include external fermions in the pure-spinor formulation in \cite{Berkovits:2005df, Berkovits:2005ng}. For tree-level and genus-one amplitudes with larger numbers of external states, much progress has been made towards the construction of amplitudes with an arbitrary number of external states (see for example 
\cite{Tsuchiya:1988va, Mafra:2011nv, Mafra:2018qqe} and references therein).

\sm

Applications to the predictions of S-duality and space-time supersymmetry were analyzed for successively higher orders in the low energy or $\a'$ expansion and in the genus expansion, starting with \cite{Green:1997tv, Green:1999pu, Green:2005ba,Green:2013bza} at tree level and genus one,  \cite{DP6, DGP, DHoker:2014oxd, Gomez:2015uha} at genus two, and \cite{Gomez:2013sla} at genus three.

\sm

In recent work, the genus-two amplitude for five external massless string states was obtained for both the Type~II and Heterotic strings in \cite{DHoker:2020prr}.  The construction produces a consistent expression for the chiral amplitude by an amalgam of results from chiral splitting in terms of loop momenta and homology shift invariance of the RNS formulation \cite{RMP, DHoker:1989cxq} (see also \cite{Verlinde:1987sd}) and results from zero-mode counting and the BRST cohomology in the pure-spinor formalism  \cite{Berkovits:2005bt}. The physical string amplitudes  are given  in terms of convergent integrals over the genus-two moduli  space of compact Riemann surfaces. The moduli-space integrands in turn are themselves integrals over the vertex points of combinations of Green functions and Abelian differentials~\cite{DHoker:2020prr}. The  overall normalization of the Type~II amplitude was fixed using methods analogous to those used in \cite{DGP,Gomez:2010ad} for the four-point amplitude. The low-energy expansion of the amplitude was obtained in \cite{DHoker:2020tcq} using the methods of higher-genus modular graph functions \cite{DHoker:2014oxd, DHoker:2017pvk, DHoker:2018mys, Basu:2018bde}. As a result, a  variety of S-duality predictions for Type~IIB components with different R-symmetry charges were substantiated, while a detailed study of the tropical limit \cite{Tourkine:2013rda} to the corresponding supergravity amplitude, obtained in \cite{Carrasco:2011mn, Mafra:2015mja}, provides further validation of the chiral amplitude \cite{DHoker:2020tcq}.

\sm

In the present work, we present a first-principles derivation of the even parity part of the chiral genus-two amplitude for five massless NS states. The methods adopted here are those used for the construction of the genus-two four-point massless NS amplitude in \cite{DP5,DP6}. The driving forces are chiral splitting in terms of internal loop momenta combined with the holomorphic projection from supermoduli space onto the moduli space of compact Riemann surfaces using the super period matrix \cite{RMP,DHoker:1989cxq} (see also \cite{Witten:2012ga}). The full physical amplitudes for both the Type~II  and Heterotic strings are obtained by pairing left and right chiral amplitudes of the respective theories at fixed loop momenta and then integrating over the loop momenta. Our results are derived for flat Minkowski space-time $\RR^{1,9}$ but can be readily generalized to a $d$-dimensional  toroidal compactification \cite{Alvarez-Gaume:1986rcs} by replacing the continuous loop momenta of $\RR^d$ by the discrete momenta of the torus $T^d$.

\sm 

As will be explained in the sequel, the five-point computation in this work successfully  overcomes several technical challenges that go considerably beyond those encountered in the four-point case \cite{DP6}. Among other things, we present a variety of new simplified spin structure sums of multiple products of Szeg\"o kernels and streamline the manipulations of the Beltrami differentials in the moduli-space integrand. These new techniques will be, no doubt, crucial for subsequent investigations of multiparticle and higher-genus amplitudes. Our results may also be of use in relating the correlators of the ambi-twistor string \cite{Geyer:2016wjx, Geyer:2018xwu} to those of the   conventional superstring  \cite{Kalyanapuram:2021xow, Kalyanapuram:2021vjt}. This relation was already used in the proposal of \cite{Geyer:2021oox} for the genus-three four-point amplitude as an uplift of the low-energy limit of the amplitude derived in \cite{Gomez:2013sla} to higher orders in $\alpha'$.

\sm

This is a long paper in which many conceptual and technical developments are discussed and used. The final expression for the amplitude is, however, remarkably simple and we shall begin by a summary in the sequel of this introduction.

\newpage

%%%%%%%%%%%%%%%%%%%%%%%%%%%%%%%%%%%%%%%%%%%
%%%%%%%%%%%%%%%%%%%%%%%%%%%%%%%%%%%%%%%%%%%
\subsection{Summary of results}
\label{sec:int.1}

The main result of this work is the construction, from first principles in the RNS formulation, of the even spin-structure contribution to the genus-two five-point amplitude of massless NS-NS-states in Type II and massless NS states in Heterotic superstrings. 

\sm

Using chiral splitting \cite{RMP, DHoker:1989cxq} these physical closed-string amplitudes ${\cal A}_{(5)}^{\rm even}$ may be obtained by pairing left and right chiral amplitudes ${\cal F}$ and $\tilde \cF$ at fixed loop momenta $p^I$ with $I=1,2$. For the Type II strings, the physical amplitude takes the form,\footnote{The external momenta $k_j$ are taken to be complexified in (\ref{newsumm.1}), (\ref{newsumm.1a}) and below,
and the integration domain $ \mathbb R^{20}$ for the loop momenta $p_\mu^I$ results from the standard
Wick-rotation of their timelike components~$p_0^I$.}
\bea
\label{newsumm.1}
{\cal A}_{(5)}^{\rm even} \, \Big|_{\rm Type \, II} &=  \mN_{{\rm II}} \,
\delta (k) \int_{{\cal M}_2} \!\! |d^3 \Omega|^2 \int_{\Sigma^5} \int_{ \mathbb R^{20} } \!\! dp \, 
{\cal F}(z_i,\ep_i, k_i,p^I| \Omega ) \, \overline{  \tilde{ {\cal F}}(z_i,\tilde \ep_i,-k_i^{\ast},-p^I |\Omega) }
\quad 
\eea
while for the Heterotic strings, the amplitude is given by,
\bea
\label{newsumm.1a}
{\cal A}_{(5)}^{\rm even} \, \Big|_{\rm Heterotic} &=  \mN_{\rm H} \,
\delta (k) \int_{{\cal M}_2} \!\! |d^3 \Omega|^2 \int_{\Sigma^5} \int_{ \mathbb R^{20} } \!\! dp \, 
{\cal F}(z_i,\ep_i, k_i,p^I | \Omega ) \, \overline{  \tilde{ {\cal F}}_{\rm bos}(z_i,\tilde \me_i, -k_i^{\ast},-p^I |\Omega) }
\qquad
\eea
The construction of the physical amplitudes is thus reduced to the determination of the chiral amplitudes, $\cF$, $\tilde \cF$ and $\tilde \cF_{{\rm bos}}$. We begin by explaining the ingredients in (\ref{newsumm.1}) and (\ref{newsumm.1a}).

\sm

The external string states are labelled by $i=1,\cdots, 5$ and have lightlike momenta $k_i$; the total momentum $k=\sum_{i=1}^5 k_i$ vanishes  by the momentum conserving $\delta$-function; the periods $\Omega$ are integrated over the moduli space $\cM_2$ of genus-two compact Riemann surfaces $\Sigma$; and the chiral amplitudes $\cF$, $\tilde \cF$ and $\tilde \cF_{\rm bos}$ are $(1,0)$-forms in each vertex point $z_i \in \Sigma$ which is integrated over $\Sigma$. 

\sm

For both the Type II and Heterotic strings, $\ep_i$ are the polarization vectors for the left chiral amplitude $\cF$ of the superstring. For Type II, $\tilde \ep_i$ are the polarization vectors for right chirality and $\tilde \cF$ is again the chiral amplitude of the superstring, and the polarization tensors $\ep_i \otimes \tilde \ep _i$ parametrize all the NS-NS states in the $\cN=2$ supergravity multiplet of Type~II. The sign of the parity odd terms in $\tilde \cF$ is the same as for those in $\cF$ in case of Type~IIB amplitudes and opposite in case of Type~IIA.
For the NS states of the Heterotic string, the assignment $\tilde \me_i$ may correspond either to a polarization vector $\tilde \ep_i$ to form the polarization tensor $\ep_i \otimes \tilde \ep_i$ of the $\cN=1$ supergravity multiplet, or to the $E_8 \times E_8$ or $Spin(32)/\ZZ_2$ gauge algebra data $\tilde t^a_i$ for the polarizations $\ep_i \otimes \tilde t^a_i$ in the $\cN=1$ super-Yang-Mills multiplet. In both cases $\tilde {\cal F}_{\rm bos}$ stands for the chiral amplitude of the (compactified) bosonic string. Its construction for the five-point function was 
outlined in \cite{DHoker:2020prr}, and the corresponding result may be carried over the present work 
without modification. 

The normalization factors $\mN_{{\rm II}}$ and $\mN_{{\rm H}}$ are proportional to $e^{2 \phi}$, where $\phi$ is the expectation value of the dilaton field and, by dimensional analysis, to powers of the 10-dimensional Newton's constant  $\kappa_{10}$. The overall normalization for the Type II case was obtained in \cite{Gomez:2015uha, DHoker:2020tcq} following the pure-spinor calculations of \cite{Gomez:2010ad,Gomez:2013sla}.
In the RNS formulation used here, the calculation of the normalization factors $\mN_{{\rm II}}$ and $\mN_{{\rm H}}$ proceeds via the methods of physical factorization of the amplitude used in \cite{DGP}. One may factorize the genus-two five-point function in either one of two ways. A first way to factorize is onto a massive intermediate NS-NS state into a genus-one three-point function (already normalized in \cite{DGP}) and a genus-one four-point function (which may be obtained by factorizing the genus-one five-point function of \cite{Tsuchiya:1988va} onto a massive pole). A second way to factorize is onto a massless NS-NS state into a massless tree-level three-point function (which does not vanish for complex momenta) and a massless genus-two four-point function normalized in \cite{DGP}. A detailed implementation of these factorization methods
is left for the future and in fact not needed here for the Type II case in view of the normalization factors known from \cite{Gomez:2015uha, DHoker:2020tcq} and the matching between RNS and pure-spinor computations to be detailed below.

\sm

We note that our result for $\cF$ can be readily exported to determine the integrand of Type~I superstrings with respect to the real moduli of an open-string worldsheet with boundaries.

\sm

The final result for the even spin-structure contribution to the chiral amplitudes $\cF$ and $\tilde \cF$ is remarkably simple and compact, especially so in view of the complexity of its derivation obtained in this paper, and is given by,\footnote{Throughout the body of the paper we shall adopt the Einstein summation convention for the indices $I,J=1,2$ that label the homology cycles so that a pair of an upper and a lower identical indices is summed over without exhibiting the summation sign. In the appendices, however, all summation signs are being kept explicitly to ensure maximal clarity.}
\begin{align}
{\cal F} (z_i, \ep_i, k_i, p^I|\Omega ) = 
{ i \, {\cal N}_5 \over 8 \pi^2}  &\bigg\{ {-}  \mt_1  \,  \ep_1 \cdot   \mP^I(z_1) 
\Big ( k_1 \cdot k_2 \, \om_I(4) \Delta (5,1) \Delta (2,3) + {\rm cycl}(1,2,3,4,5) \Big )
\no \\ & \ \
+\half  k_1 \cdot \mP^I(z_1) \Big [ 
(\mt_{12} -2 \mt_1 \, \ep_1 \cdot k_2  ) \om_I(4) \Delta (1,3) \Delta (2,5) 
\no \\ & \ \ \hskip 1in 
+ (\mt_{13} -2 \mt_1 \, \ep_1 \cdot k_3 ) \om_I(4) \Delta (1,2) \Delta (3,5) 
\label{newsumm.2}
\\ &\ \  \hskip 1in 
+ (\mt_{14} -2 \mt_1 \, \ep_1 \cdot k_4  ) \om_I(3) \Delta (1,2) \Delta (4,5)  \Big ] 
 \no \\ & \ \
+  k_1 \cdot k_2 \, \om_I(4) \Delta (5,1) \Delta (2,3)  \sum_{1\leq i < j}^5 \mt_{ij} \, g^I _{i,j}
+ {\rm cycl}(1,2,3,4,5) \bigg\}
\notag
\end{align}
The prefactor ${\cal N}_5$ denotes the ubiquitous chiral Koba-Nielsen factor, which is common to both Type II and Heterotic strings, and is given by, 
\bea
 {\cal N}_5  = \exp \bigg(
 i \pi \, \Omega _{IJ} \, p^I \cdot p^J + 2 \pi i  \sum_{j=1}^5 k_j \cdot p^I \int ^{z_j}_{z_0} \om_I + \sum _{1 \leq i <j}^5
k_i \cdot k_j \ln E(z_i,z_j) \bigg)
\label{newsumm.3}
\eea
The holomorphic Abelian differentials $\om_I$ and the prime form $E(z_i,z_j)$ are reviewed in appendices \ref{sec:A.RS} and \ref{sec:A.Fay}, respectively, while the bi-holomorphic form $\Delta$ is reviewed in appendix \ref{sec:B1} and given in terms of $\om_I(i) = \om_I(z_i)$ by,\footnote{Following the conventions
in \cite{DHoker:2020prr, DHoker:2020tcq}, we will only display the coefficient functions $\omega_I(z)$ of the differentials $dz$ in $(1,0)$ forms $\omega_I(z)dz$, with local complex coordinates $(z,\bar z)$ on the surface $\Sigma$, that is why (\ref{newsumm.2}), (\ref{defofdelta}) and later equations do not involve any antisymmetric wedge products.}
\bea
\Delta(i,j) = \omega_1(i) \omega_2(j) -  \omega_2(i) \omega_1(j)
\label{defofdelta}
\eea
Momentum conservation guarantees that $\cN_5$ is independent of the arbitrary point $z_0$.
The first two lines of (\ref{newsumm.2}) feature the following combinations,
\begin{align}
\label{newsumm.3a}
\mP^I(z_i) &=2\pi i p^I + \sum_{j\neq i}^5 k_j \,  g^I_{i,j}
& \hskip 0.8in 
g^I_{i,j} &= \frac{\partial}{\partial \zeta_I} \ln \vartheta[\nu](\zeta|\Omega) \Big | _{
\zeta_I = \int^{z_i}_{z_j} \omega_I}
\end{align}
of the loop momentum and theta functions reviewed in appendix \ref{sec:A.theta}. Here $\nu$ is a common, but otherwise arbitrary, odd spin structure whose dependence drops out from the chiral amplitude $\cF$. The dependence of the chiral amplitude (\ref{newsumm.2}) on the polarization vectors $\varepsilon_j^\mu$ is governed in part by the permutation-invariant $t_8$-tensor, which may be defined 
in terms of the linearized field strengths $f_i^{\mu \nu} = \varepsilon_i^\mu k_i^\nu-\varepsilon_i^\nu k_i^\mu$ by,\footnote{Throughout, products, traces, and commutators of $f$ are to be understood in the sense of matrix multiplication, so that we have $(f_if_j)^{\mu \nu} = (f_i)^{\mu \rho} (f_j)_\rho {}^\nu$,  $\tr(f_i \cdots f_j)=(f_i \cdots f_j)^\mu {}_\mu$, and $[f_i, f_j] = f_i f_j - f_j  f_i$. }
\bea
t_8(f_i,f_j,f_k,f_\ell) = {\rm tr}(f_i f_j f_k f_\ell ) - \frac{1}{4}  {\rm tr}(f_i f_j)  {\rm tr}(f_k f_\ell ) + {\rm cycl}(j,k,\ell)
\label{newsumm.5}
\eea
More specifically, the kinematic factors $\mt_i$ and $\mt_{ij}$ that enter (\ref{newsumm.2}) are defined by, 
\bea
\mt_1 & = & t_8(f_2,f_3,f_4,f_5) 
\no \\ 
\mt_{12} & = & t_8([f_1,f_2],f_3,f_4,f_5)
\label{newsumm.5z}
\eea
and permutations thereof. The cyclic permutations in the last line of (\ref{newsumm.2}) apply to the external-state labels of both the differential forms and kinematic factors in all the five lines of the expression. An alternative representation for the chiral amplitude with the manifest permutation symmetry is, 
\bea
\label{Falt}
{\cal F} (z_i, \ep_i, k_i, p^I|\Omega ) = \frac{ i {\cal N}_5 }{16\pi^2} \bigg\{ {\cal D}_I \sum_{i=1}^5 \mt_i \varepsilon_i \cdot \mP^I(z_i)
- {\cal D}_I \sum_{1\leq i<j}^5 \mt_{ij} g^I_{i,j} + \sum_{i=1}^5 k_i \cdot \mP^I(z_i) T_{iI}  \bigg\}
\quad
\eea
where we have made use of the following combinations,
\bea
{\cal D}_I &=&
-2 k_1 \cdot k_2 \, \om_I(4) \Delta (5,1) \Delta (2,3) + {\rm cycl}(1,2,3,4,5) \label{dperminv}
 \\
T_{1I} &= &\frac{1}{4}(\mt_{12} -2 \mt_1 \varepsilon_1 \cdot k_2) \big( \omega_I(3) \Delta(1,5) \Delta(2,4) + {\rm cycl}(3,4,5) \big) + {\rm cycl}(2,3,4,5) 
\no
\eea
and cyclic permutations of $T_{1I}$ to produce $T_{iI}$.
The expression for $\cF$ in (\ref{Falt}) differs from the one given in (\ref{newsumm.2}) by exact holomorphic differentials in the vertex points, and therefore yields the same physical amplitude. Full permutation symmetry is the result of the facts that the combination ${\cal D}_I $ is invariant under all permutations of $1,2,3,4,5$ (even though only cyclic permutations are manifestly a symmetry)\footnote{Permutation invariance of ${\cal D}_I $
follows from the relations among five-forms reviewed in appendix \ref{sec:B1}, and the relations among five-point bilinears $k_i \cdot k_j$ due to momentum conservation and $k_j^2 = 0$ for all $j=1,2,3,4,5$. The five-dimensional cyclic basis of bilinears $k_i \cdot k_j$ in (\ref{dperminv}) can be obtained from cyclic permutations of  $k_1\cdot k_3= k_4\cdot k_5-k_1\cdot k_2-k_2\cdot k_3$.} and that $T_{1I}$ is invariant under all permutations of  $2,3,4,5$.

\sm

The expression for the chiral amplitude $\cF$ in (\ref{newsumm.2}) leads to convergent integrals in the full-fledged Type II amplitude of (\ref{newsumm.1}) and its Heterotic counterpart, in the sense of analytic continuation, as was the case for the genus-one four-string amplitude \cite{DHoker:1993hvl,DHoker:1994gnm}. Specifically, at fixed moduli $\Omega$, the integral over the vertex points $z_i$ may be analytically continued in the kinematic variables $s_{ij} =  - \alpha ' (k_i+k_j)^2/4$ to produce poles corresponding to physical intermediate one-particle states. These poles are  governed by the operator product expansion of pairs of colliding vertex operators. The combined integrations over $\Omega$ and the loop momenta $p^I$ are convergent only for purely imaginary values of the $s_{ij}$, but may be analytically continued to produce branch cuts  in the $s_{ij}$ corresponding to physical  multi-particle states.

\sm

The contribution from even spin structures produces the even parity part of the chiral amplitude with external NS states. This even parity part will be shown to agree with the bosonic supermultiplet components of the chiral amplitude in pure-spinor superspace derived in \cite{DHoker:2020prr}. The chiral genus-two five-point amplitude additionally has an odd parity part,\footnote{We are grateful to Alex Edison, Carlos Mafra and Fei Teng for 
 discussions that led to identifying a missing factor of $1/8$ in the 
 parity-odd terms in earlier versions.}
\begin{align}
 {\cal F}_{\rm odd} &=  
% \frac{{\cal N}_5 }{16\pi}
 \frac{{\cal N}_5 }{128\pi} 
  \, {\cal D}_I \, 
\epsilon_{10}(p^I,\varepsilon_1,f_2,f_3,f_4,f_5)
\label{summ.10}
\end{align}
with $\epsilon_{10}$ denoting the ten-dimensional Levi-Civita symbol and the permutation invariant
five-form ${\cal D}_I$ is defined in (\ref{dperminv}). The odd parity part was obtained using chiral splitting and the pure-spinor formalism in \cite{DHoker:2020prr, DHoker:2020tcq}. In the RNS formulation, the odd parity part derives from the contributions of the odd spin structures. A bootstrap approach to its construction is described in section \ref{sec:boots}. The construction of the odd spin structure part is complicated by the zero modes of the RNS worldsheet fermion fields, which in turn complicate the structure of the super-Riemann surfaces and their supermoduli space. Its derivation from first principles in the RNS formulation is relegated to future work.

%%%%%%%%%%%%%%%%%%%%%%%%%%%%%%%%%%%%%%%%%%%
%%%%%%%%%%%%%%%%%%%%%%%%%%%%%%%%%%%%%%%%%%%
\subsection{Comparison with lower genus}
\label{sec:int.2}

The new representation (\ref{newsumm.2}) of the chiral genus-two five-point amplitude may be compared with its counterparts at lower genus. The genus-one amplitude $\cF_{g=1}$ was obtained in \cite{Tsuchiya:1988va},  
\bea
\label{newsum.13}
{\cal F}_{\! g=1}(z_i, \ep_i, k_i, p|\tau) =  \cN_5 \bigg\{ 
\sum_{i=1}^5 \mt_i \, \varepsilon_i \cdot  \mP(z_i)  
-  \sum_{1\leq i < j}^5 \mt_{ij} \, g_{i,j}  
%- \pi i 
- \frac{ \pi i }{8}  \epsilon_{10}(p,\varepsilon_1,f_2,f_3,f_4,f_5)
\bigg\}
\qquad
\eea
The kinematic factors $\mt_i$ and $\mt_{ij}$ are those that occurred in the genus-two amplitude and were defined in (\ref{newsumm.5z}); the unique modulus for genus one is denoted by $\Omega _{11}=\tau$; there is a single loop momentum $p$; the single Abelian differential is constant; the prime form simplifies to $E(z_i,z_j|\tau)= \tet_1(z_i-z_j|\tau)/\tet'_1(0|\tau)$; the chiral Koba-Nielsen factor is obtained from (\ref{newsumm.3}) by these restrictions; and the combinations $ \mP(z_i) $ and $g_{i,j}$ are given by,  
\bea
\mP(z_i) =2\pi i p + \sum_{j\neq i}^5 k_j \, g_{i,j}\,,
\hskip 0.6in
g_{i,j}=\partial_{z_i} \ln \vartheta_1(z_i{-}z_j|\tau)
\label{newsum.14}
\eea
The similarities and differences in the structures of the genus-one and genus-two expressions for the chiral amplitudes are striking. The contributions proportional to $k_i \cdot \mP^I(z_i)$ to the genus-two amplitude in the second to fourth line of (\ref{newsumm.2}) become exact differentials in $z_i$ upon restriction to a single loop, and are absent from the genus-one formula in (\ref{newsum.13}). All other terms in the genus-two chiral amplitude sport additional kinematic factors of $k_i \cdot k_j$ and the bi-holomorphic form $\Delta$ which are absent from the genus-one amplitude. More specifically, they enter through the permutation invariant ${\cal D}_I$ defined in (\ref{dperminv}) which occurs in the first two parity-even terms in (\ref{Falt}) and multiplies the entire parity-odd contribution (\ref{summ.10}).

\sm

As a result, the genus-two contribution is generally softer-behaved at low energies, consistently with predictions from S-duality and space-time supersymmetry. Finally, we note that the supersymmetrization of (\ref{newsum.13}) is known from both the non-minimal pure-spinor formalism \cite{Mafra:2009wi} and the minimal one \cite{Mafra:2018qqe}. 

\sm

For completeness, we quote the tree-level contribution to the chiral five-point amplitudes in Type~II superstrings, which  
can be brought into the following compact form \cite{Mafra:2011nv},
\begin{align}
{\cal F}_{\! g=0}(z_i, \ep_i, k_i) =
\cN_5 \bigg\{
 \frac{ s_{12} s_{34} A_{\rm SYM}^{\rm tree}(1,2,3,4,5) }{z_{12} z_{25} z_{53} z_{34} z_{41} }
+   \frac{ s_{13} s_{24}  A_{\rm SYM}^{\rm tree}(1,3,2,4,5)}{z_{13} z_{35} z_{52} z_{24} z_{41} } \bigg\}
\label{newsum.11}
\end{align}
The expression for $\cN_5$ simplifies for tree-level as there are no loop momenta, and the prime form $E(z_i,z_j)$  reduces to $z_{ij}=z_i-z_j$. Furthermore,  the color-ordered tree-level amplitudes $A_{\rm SYM}^{\rm tree}$ of ten-dimensional super-Yang-Mills are related to the $t_8$-tensors in (\ref{newsumm.5z}) via,
\begin{align}
&(s_{12}s_{34}-s_{34}s_{45}-s_{51} s_{12} ) A_{\rm SYM}^{\rm tree}(1,2,3,4,5)
+ s_{13}s_{24} A_{\rm SYM}^{\rm tree}(1,3,2,4,5) 
\label{newsum.12} \\
&=  %\frac{(\alpha')^2}{(k_{1} {+} k_2)^2}  \Big ( \mt_1 \, (\varepsilon_1 \cdot k_2)  - \mt_2 \, (\varepsilon_2 \cdot k_1)  -\mt_{12} \Big ) + {\rm cycl}(1,2,3,4,5)
\frac{(\alpha')^3}{4 s_{12}}  \Big (    \mt_2 \, (\varepsilon_2 \cdot k_1) - \mt_1 \, (\varepsilon_1 \cdot k_2) +\mt_{12} \Big ) + {\rm cycl}(1,2,3,4,5)
\notag
\end{align}
Earlier work on five-point superstring tree-level amplitudes includes  \cite{Medina:2002nk, Mafra:2009bz} and the  pure-spinor-superspace uplift of the super-Yang-Mills amplitudes can be  found in \cite{Mafra:2010ir, Mafra:2010jq, Mafra:2015vca}.

%\newpage

\subsection*{Organization}

The remainder of the paper is organized as follows. We start by reviewing the RNS prescription for genus-two amplitudes  and spelling out the opening line for the parity-even part of massless five-point amplitudes in section~\ref{sec:2}. The sums over even spin structures needed at five points are carried out in section~\ref{sec:3}, followed  by a discussion of closely related fundamental simplifications and cancellations in section~\ref{sec:4}. The non-vanishing  contributions to the chiral amplitude are identified in section~\ref{sec:5}. All the $(0,1)$-forms and double poles in the chiral amplitude are shown to reduce to exact differentials which cancel out from the physical amplitude  in section~\ref{sec:6}. In section~\ref{sec:7}, the  results of the previous sections are simplified and brought into the manifestly gauge-slice independent form  (\ref{newsumm.2}). This end result of our RNS computation is shown in section~\ref{sec:int.3} to  match the bosonic components of the chiral five-point amplitude in the pure-spinor formalism. Appendix \ref{sec:A} provides a summary of the function theory on Riemann surfaces of arbitrary genus, while appendix \ref{sec:B} collects a number of properties specific to genus two. Appendices \ref{sec:C} to \ref{sec:J} elaborate on various more technical results used in this work.

\subsection*{Acknowledgments}

ED is grateful to Duong H.\ Phong  for earlier collaboration on genus-two amplitudes upon which the present work rests. We also gratefully acknowledge our collaboration with Carlos Mafra and Boris Pioline on
the closely related pure-spinor computation of the amplitude under investigation. 
The research of ED is supported in part by NSF grant PHY-19-14412. 
The research of OS  is supported by the European Research Council under ERC-STG-804286 UNISCAMP.

\newpage

%%%%%%%%%%%%%%%%%%%%%%%%%%%%%%%%%%%%%%%%%%%
%%%%%%%%%%%%%%%%%%%%%%%%%%%%%%%%%%%%%%%%%%%
\section{Structure of the five-point NS string amplitude}
\label{sec:2}
\setcounter{equation}{0}
%%%%%%%%%%%%%%%%%%%%%%%%%%%%%%%%%%%%%%%%%%%
%%%%%%%%%%%%%%%%%%%%%%%%%%%%%%%%%%%%%%%%%%%

In this section, we shall present the general outline of the construction of the even spin structure contribution to the genus-two chiral amplitude for five external massless NS states. We begin by reviewing the conceptual aspects of the construction in terms of the super period matrix \cite{DP5,DP6}, then summarize the results obtained in \cite{DP1, DP2, DP3, DP4} for the chiral measure (see also  \cite{DHoker:2002hof} for a review), to be followed by several subsections in which the  contributions to the chiral amplitude will be made explicit. The construction will closely  parallel the one for four external massless NS states presented in \cite{DP5,DP6}, and we shall follow the same notations and conventions unless specified otherwise.

\subsection{The super period matrix}
\label{sec:SP}

The fundamental tool for the concrete construction  of the genus-two chiral amplitude for even spin structures in the RNS formulation is the holomorphic projection provided by the super period matrix. The supermoduli space $\mM_{2,+}$ of compact super Riemann surfaces with even spin structures is holomorphically projected to the moduli space $\cM_2$ of compact Riemann surfaces. Super moduli space $\mM_{2,+}$ may be parametrized locally by the ordinary period matrix $\Omega$ and a pair of odd Grassmann-valued parameters $\zeta ^\a$ with $\a=1,2$. Denoting the super period matrix by $\hat \Omega$, the holomorphic projection may be represented as follows \cite{DHoker:1989hhv},
\bea
\label{2.proj}
\mM_{2,+} \mapsto \cM_2: (\Omega, \zeta) \mapsto \hat \Omega
\eea
The projection associates to a super Riemann surface $\Sigma$ with super moduli $(\Omega, \zeta)$  an ordinary Riemann surface $\Sigma _{{\rm red}}$ with moduli $\hat \Omega$. The data of this projection are related as follows \cite{RMP},\footnote{Following \cite{RMP} and \cite{DP6}, we use a system of local complex coordinates $(z,\bar z)$ in which the metric takes the form $ds^2 = 2 |dz|^2$. To make contact with the standard Euclidean metric in $\RR^2$, we set $z=(x+iy)/\sqrt{2}$ and $\bar z = (x-iy)/\sqrt{2}$ with $x,y\in \RR$ so that $ds^2 = dx^2+dy^2$ and the volume form is  $d^2 z = i dz \wedge d \bar z = dx \wedge dy$. The (positive) Laplacian is given by $\Delta = - 2 \p_z \pbz$, and the Cauchy-Riemann operator acts by $\pbz (z-w)^{-1} = 2 \pi \delta (z,w)$, where the $\delta$-function is normalized as follows, $\int _\Sigma d^2z \, \delta(z,w) f(z) = f(w)$. We note that these conventions differ by factors of 2 from those used in \cite{DHoker:2020prr, DHoker:2020tcq}.}
\bea
\label{2.SP}
\hat \Omega_{IJ} = \Omega _{IJ} - { i \over 8 \pi} 
\int_\Sigma d^2 z \int_\Sigma d^2 w \, \om_I(z) \chi(z) S_\delta (z,w) \chi(w) \om_J(w)
\eea
where $\om_I$ with $I=1,2$ are the holomorphic $(1,0)$-forms and $S_\delta(z,w)$ is the Szeg\"o kernel for even spin structure $\delta$ which is a $(\half, 0)$ form in both $z$ and $w$ (see appendix \ref{sec:A} for details), and $\chi(z)$ is the worldsheet gravitino field which is a $(-\half, 1)$-form linear in $\zeta ^\a$. Performing a local worldsheet supersymmetry transformation on both $\chi$ and $\Omega$ (caused by the supersymmetry transformation of the worldsheet metric), leaves the super period matrix invariant. 

\sm

To construct the chiral amplitude, we make use of NS vertex operators without ghosts, following \cite{DP5,DP6}. Each operator insertion point, with local coordinates $(z_i,\theta_i)$, is  treated as a marked point, to be integrated over the super Riemann surface $\Sigma$. As a result, the string amplitudes properly reduce to integrals over the supermoduli space $\mM_{2,+}$, to which the holomorphic projection (\ref{2.proj}) may be applied.\footnote{This approach should be contrasted with the one in which vertex operators involve $c$ and $\delta(\gamma)$ ghosts and vertex insertion  points are punctures rather than marked points. In the approach involving ghosts, the group of super diffeomorphisms is reduced to the one preserving the punctures and, as a result, the string amplitudes are reduced to integrals over the super moduli space of punctured super Riemann surfaces. This last super moduli space does not necessarily possess the holomorphic projection onto ordinary moduli space \cite{Donagi:2013dua} that is being used here. However, with Ramond punctures, under certain conditions, there does exist a super period matrix, as was shown in  \cite{Witten:2015hwa, DHoker:2015gwa}.}
To carry out the holomorphic projection in the superstring amplitudes, it will be convenient to change coordinates from $(\Omega, \zeta)$ to $(\hat \Omega, \zeta)$ using (\ref{2.SP}), so that  the projection (\ref{2.SP}) reduces to,\footnote{While the supermoduli space $\mM_{2,+}$ is projected, its Deligne-Mumford compactification $\overline{\mM}_{2,+}$ is not projected  \cite{Witten:2013cia}, due to singular behavior at the compactification divisor. In certain compactifications of space-time, such as on Calabi-Yau orbifolds, the obstruction to holomorphic projection leads to physical effects such as the breaking of space-time supersymmetry  and the non-vanishing of the genus-two contribution to the cosmological constant \cite{Witten:2013cia,DHoker:2013sqy} (see also \cite{Berkovits:2014rpa}). For flat Minkowski space-time as is being considered in this paper, however, the superstring measure vanishes at the relevant separating nodes \cite{DHoker:2002hof} and no contribution to the amplitude arises from the fact that $\overline{\mM}_{2,+}$ is not projected.}
\bea
\label{2.proja}
\mM_{2,+} \mapsto \cM_2: (\hat \Omega, \zeta) \mapsto \hat \Omega
\eea
and may be carried out simply by integration over $\zeta$. 

\sm

This change of variables may be implemented in conformal field theory correlators by using a Beltrami differential $\hat \mu$ paired against an insertion of the stress tensor,  
\bea
\< \cO  \> (\Omega)  \to  \< \cO \> (\hat \Omega) + {1 \over 2 \pi} \int_\Sigma d^2 w \, \hat \mu (w)  \< T(w) \cO \>
\eea
where $\< \cO \>(\Omega) $ and $\< \cO \>(\hat \Omega)$  stand for the expectation values of an arbitrary  operator $\cO$ evaluated at moduli $\Omega$ and $\hat \Omega$, respectively, and $T(z)$ stands for the total stress tensor. A first-order deformation in $\hat \mu$ suffices here because the Beltrami differential $\hat \mu$ may be chosen to be proportional to $\zeta ^1 \zeta ^2$ and is thus nilpotent.  To see this, we recall that a Beltrami differential $\hat \mu(w)$ produces the following first-order variation in the period matrix,
\bea
\delta \Omega _{IJ} = {1 \over 2 \pi} \int_\Sigma d^2 w \, \hat \mu(w) \delta _{ww} \Omega _{IJ} 
\hskip 0.6in 
\delta _{ww} \Omega _{IJ} = 2 \pi i \om_I (w) \om_J(w)
\eea 
so that the Beltrami differential that deforms $\Omega$ to $\hat \Omega$ must satisfy,
\bea
 \int_\Sigma d^2 w \, \hat \mu  (w) \,  \om_I (w) \om_J(w)
 = i (\hat \Omega _{IJ} - \Omega _{IJ})
\eea
and therefore may be chosen to be nilpotent, just as $\hat \Omega - \Omega$ is.

\subsection{Parametrization of super moduli space adapted to projection}
\label{sec:Par}

An equivalent, but  more geometrical, formulation of the projection treats the gravitino slice $\chi $ and the Beltrami differential $\hat \mu$ on a more equal footing. To do so, one starts with a bosonic Riemann surface $\Sigma$ with period matrix $\Omega$ (corresponding to a Riemann surface $\Sigma _{{\rm red}}$ with super period matrix $\hat \Omega$ in the formulation given in subsection \ref{sec:SP}). Next, one parametrizes supermoduli space  by turning on the field $\chi(z)$ linear in the odd moduli $\zeta ^\a$, while at the same time turning on $\hat \mu$ bilinear in $\zeta ^\a$ in precisely such a manner as to keep the period matrix $\Omega$ unchanged. Adding both deformations of the period matrix, namely the bilinear deformation due to $\chi$ of (\ref{2.SP}), and the linear deformation due to $\hat \mu$, we obtain, 
\bea  
\label{SPb}
{ 1 \over 8 \pi}  \int_\Sigma d^2 z \int_\Sigma d^2 w \, \om_I(z) \chi(z) S_\delta (z,w) \chi(w) \om_J(w)
-  \int_\Sigma d^2 w \, \hat \mu  (w) \,  \om_I (w) \om_J(w) =0
\eea
whose interpretation is that the original period matrix $\Omega$ is left invariant under the combined deformation, and indeed plays the role of the super period matrix in the formulation of subsection \ref{sec:SP}.
Equation (\ref{SPb}) determines $\hat \mu$ in terms of $\chi$, up to diffeomorphisms $\hat \mu (w) \to \hat \mu(w) + \pbw v^w(w)$ where $v^w$ is an arbitrary $(-1,0)$-form diffeomorphism vector field. It is this formulation that was used in \cite{DP5,DP6} to construct the chiral amplitude for four massless NS states, and will be used also here for the amplitude with five massless NS states.

\subsection{The chiral measure}

The procedure outlined above leads to a remarkably simple genus-two superstring chiral measure on $\mM_{2,+}$ evaluated in \cite{DP2}. For fixed even spin structure $\delta$, one finds \cite{DP2, DP3,DP4},\footnote{The holomorphic  volume form $d^3 \Omega = d \Omega _{11} \, d \Omega _{12} \, d \Omega _{22}$   on $\cM_2$ is included in the measure $d \mu$, while the volume form on the  odd fiber will be denoted by $d^2 \zeta = d \zeta ^1 d \zeta ^2$. } 
\bea
\label{measure}
\qquad
d\mu [\delta] (\Omega, \zeta) = d\mu_0 [\delta] (\Omega)d^2 \zeta   
+ d\mu_2[\delta](\Omega) \zeta ^1 \zeta ^2 d^2 \zeta
\eea
Following the parametrization and notation  introduced in subsection \ref{sec:Par}, the measure and amplitudes  will be considered on a Riemann surface $\Sigma$ with period matrix $\Omega$ which, in the discussion and notations of subsection \ref{sec:SP}, correspond to $\Sigma _{{\rm red}}$ and $\hat \Omega$, respectively.

\subsubsection{Top component of the chiral measure}

The top component of the chiral measure, for flat Minkowski  space-time $\RR^{1,9}$,  is given by,
\bea
d\mu_2[\delta](\Omega)
= {\tet[\delta](0|\Omega)^4\Xi_6[\delta](\Omega) \over 16\pi^6\Psi_{10}(\Omega)} d^3 \Omega
\eea
It is independent of the choices of slice for $\chi$ and $\hat \mu$, and involves only Siegel modular forms. The
Riemann $\tet$-functions of rank 2 are reviewed in appendix \ref{sec:A.theta} and the weight 10 Igusa cusp form $\Psi_{10}$ is given by,
\bea
\Psi _{10} ( \Omega) = \prod _{\delta \, {\rm even}} \tet [\delta ](0| \Omega )^2
\label{defigusa}
\eea
To define $\Xi_6[\delta] (\Omega)$, we express the even spin structure $\delta$ in terms of a sum of three of the six odd spin structures,  $\delta = \nu _1+ \nu_2 + \nu_3 = \nu_4+\nu_5+\nu_6$, where all $\nu_i$ are distinct from one another. The modular form $\Xi_6 [\delta ](\Omega)$ is then defined by the following sum of products,
\bea
\label{Xi6}
\Xi_6 [\delta ] (\Omega )= \sum _{1 \leq i < j \leq 3} \< \nu_i | \nu_j\> 
\prod _{k=4,5,6} \tet [\nu_i + \nu_j + \nu_k] (0| \Omega )^4
\eea
The symplectic pairing between half-integer characteristics $\nu_i= \nu_i ''+ \Omega \nu_i'$, with the components of $\nu_i', \nu_i'' $ taking the value 0 or $\half$ modulo 1,  is given by, 
\bea
\label{sympair}
\< \nu_i | \nu_j\>  = \exp \{ 4 \pi i ( \nu_i ' \nu_j '' - \nu _i '' \nu '_j) \}
\eea
and takes values in $\{ \pm 1\}$.  The top component of the chiral measure transforms under $Sp(4,\ZZ)$ modular transformations as a modular form of weight $-5$, 
\bea
\label{9f}
d\mu_2 [\tilde \delta ] (\tilde \Omega) = \det (C\Omega+D)^{-5} d\mu_2 [ \delta ] ( \Omega)
\eea
where the transformation laws for $\Omega$ and $\delta$ are given in appendix \ref{sec:A}.
Modular weight $-5$  is the correct value for superstring theory in dimension $d=10$, as the 
modular transformations of the $2\times 10$ internal loop momenta will then make the 
combined integrand of left and right chiralities modular invariant \cite{DHoker:2002hof}.

\subsubsection{Bottom component of the measure}

The bottom component of the measure does depend on the choice of slice for $\chi$, a dependence which will be compensated for by the dependence on $\chi$ of the correlators of the vertex operators in the full superstring amplitude. We set,
\bea
\label{dmu0dmu2}
d\mu_0[\delta](\Omega) = \cZ[\delta] (\Omega) d^3 \Omega
\eea
where $\cZ [\delta](\Omega)$ is the chiral partition function incorporating contributions of the worldsheet matter fields as well as from the worldsheet $(b,c)$ and $(\beta, \gamma)$ ghost systems. To make the dependence on the gauge choices explicit, we shall use local supersymmetry to choose a convenient slice for the worldsheet gravitino field $\chi$ with support on two points $q_1,q_2\in \Sigma$,
\bea
\label{chiqq}
\chi (z) =  \zeta ^1 \delta (z,q_1) + \zeta ^2 \delta (z,q_2)
\eea 
and specialize the points $q_1,q_2$ to be the zeros of a holomorphic $(1,0)$-form $\varpi(z)$,
see (\ref{varpi2}) to (\ref{varpi4}) for further details,
\bea
\varpi(q_\a)=0 \hskip 1in \a=1,2
\eea
This gauge choice is referred to as {\sl unitary gauge} and the relation between the points may alternatively be presented in the equivalent forms,
\bea
\label{qqdel}
q_1 + q_2 - 2 \Delta = 2 \kappa 
\hskip 0.4 in \hbox{i.e.} \hskip 0.4 in
\int ^{q_1} _{z_0} \om _I + \int ^{q_2} _{z_0} \om _I - 2\Delta _I(z_0) = 2 \kappa_I
\eea
where $\Delta_I$ is the Riemann vector (\ref{Riemvec}) and $\kappa = \kappa '' + \Omega \kappa '$ is an arbitrary even or odd half characteristic so that $2 \kappa$ is a full period. In this gauge, $\cZ[\delta](\Omega)$ evaluates as follows,
\bea
\label{zeedelta}
\cZ [\delta ] = \cZ_0 \, E(q_1,q_2) \, e^{4 \pi i \kappa ' \Omega \kappa'} ~
\< \kappa |\delta \> ~ \tet [\delta ](0)^4
\eea
where the prime form $E(q_1,q_2)$ is reviewed in appendix \ref{sec:A.Fay} and $\cZ_0$ is a $\delta$-independent form of weight $(-1,0)$ in both $q_1$ and $q_2$ with non-trivial monodromy and a double pole at $q_1=q_2$, given by,
\bea
\label{zeedelta1}
\cZ _0 = { Z^{12} \over \pi^{12} \Psi _{10} (\Omega) E(q_1, q_2)^2 \sigma (q_1)^2 \sigma (q_2)^2}
\eea
see (\ref{sigmaz}) for Fay's form $\sigma$.
The explicit form of the chiral scalar partition function $Z=Z(\Omega)$ (which is holomorphic in $\Omega$)
can be found in appendix \ref{apponchiralZ} and \cite{DP4}, it can be evaluated via chiral bosonization \cite{Verlinde:1986kw}. The genus-two chiral measure for even spin structures (\ref{measure}) was alternatively obtained by exploiting the conditions of holomorphicity and modular invariance in \cite{Witten:2013tpa}.

\subsection{The chiral amplitude}

The full chiral amplitude is obtained as an integral over odd moduli and sum over spin structures $\delta$ of the product of the chiral measure, discussed in the previous subsection, and correlators constructed from the vertex operators. The correlators take the following form,\footnote{We note that no terms of degree one in $\zeta ^\a$ arise due to worldsheet fermion number conservation of the correlators that produce $\cC[\delta]$.}
\bea
\cC[\delta ](\Omega, \zeta) = \cC_0 [\delta] (\Omega) + \cC_2[\delta](\Omega) \zeta ^1 \zeta ^2
\label{c0andc2}
\eea
and the chiral amplitude is given by,
\bea
\cF(\Omega) d^3 \Omega = \sum _\delta 
\int d^2 \zeta \, d\mu [\delta] (\Omega, \zeta) \, \cC[\delta ] ( \Omega , \zeta)
\label{chiralF}
\eea
The meromorphic expression for $\cF(\Omega)$ to be derived in this work will allow to 
assemble massless five-point amplitudes of both Type II and Heterotic theories from the 
pairings of left and right movers given in (\ref{newsumm.1}) and (\ref{newsumm.1a}). 
In (\ref{chiralF}) and throughout the rest of this work, the sum $\sum_\delta$ is understood to run 
over the ten even spin structures $\delta$.

\sm

Upon carrying out the integration over the odd moduli $\zeta$, the chiral amplitude (\ref{chiralF}) receives contributions from both the top and bottom components of the chiral measure, and may be decomposed as follows,
\bea
\label{splittingF}
\cF (\Omega) & = & \cF^{(c)} (\Omega) + \cF^{(d)} (\Omega)
\eea
The superscripts $(c)$ and $(d)$ refer to the connected and disconnected parts of the correlators, respectively, which in turn are given as follows,
\bea
\cF^{(c)} (\Omega)  & = & \sum _\delta \cZ  [\delta] (\Omega) \, \cC_2[\delta] ( \Omega) 
\no \\
\cF^{(d)} (\Omega)  & = & 
\sum _\delta { \Xi_6[\delta](\Omega)   \, \tet[\delta](0|\Omega)^4  \over 16 \pi^6 \Psi _{10} (\Omega)} \,  \cC_0 [\delta] (\Omega)
\label{splitFcd}
\eea
with $\Xi_6[\delta]$ and $ \cZ  [\delta] $ given by (\ref{Xi6}) and (\ref{zeedelta}), respectively.
The disconnected part contains the contributions from the self-contractions of the total stress tensor and finite-dimensional determinants produced by gauge fixing. The connected part contains all others. It remains to evaluate the components of the correlators, which was carried out in \cite{DP6} for the amplitude with four massless NS bosons. 

\subsubsection{Correlators of chiral vertex operators}

The components $\cC_0[\delta ](\Omega)$ and $\cC_2[\delta ](\Omega)$ are obtained by chiral splitting  from correlators of the vertex operators for the superstring. The vertex operator for the $i$-th massless NS state 
depending on its lightlike external momentum $k_i$ and transverse polarization vector $\varepsilon_i$ 
is given as a sum of three parts,
\bea
\label{V}
{\cal V}(z_i,\varepsilon_i,k_i)= {\cal V}^{(0)}(z_i,\varepsilon_i,k_i) 
+ {\cal V}^{(1)}(z_i,\varepsilon_i,k_i) + {\cal V}^{(2)}(z_i,\varepsilon_i,k_i)
\eea
These parts were obtained in  \cite{DP5} using chiral splitting and are given by,\footnote{Here and throughout this work, we do not distinguish between Lorentz indices $\mu,\nu,\ldots =0,1,\ldots,9$ in superscripts or subscripts and choose their position such as to minimize clashes with other kinds of labels.}
\bea
\label{vertex}
{\cal V}^{(0)}(z_i,\varepsilon_i,k_i) 
& = &
dz_i \Big (\varepsilon_i^\mu\p_{z_i}x_+^\mu (z_i) - \tfrac{i}{2} f_i^{\mu\nu}
\psi_+^\mu\psi_+^\nu (z_i) \Big )\, e^{ik_i\cdot x_+(z_i)} 
\no \\
{\cal V}^{(1)}(z_i,\varepsilon_i,k_i)
&=& -\half d\bar z_i \varepsilon_i^\mu  \chiz \psi_+^\mu(z_i)\,e^{ik_i \cdot x_+(z_i)}
\no \\
{\cal V}^{(2)}(z_i,\varepsilon_i,k_i)
&=&
- d\bar z _i \hat \mu_{\bar z}{}^z  \Big (\varepsilon_i^\mu\p_{z_i}x_+^\mu (z_i) - \tfrac{i}{2} f_i^{\mu\nu}
\psi_+^\mu\psi_+^\nu (z_i) \Big )\, e^{ik_i\cdot x_+(z_i)}
\eea
where the gauge invariant chiral field strength is given by,
\bea
\label{f}
f_i^{\mu\nu}=\varepsilon_i^\mu k_i^\nu-\varepsilon_i^\nu k_i^\mu
\eea
and $x_+^\mu$ and $\psi_+^\mu$ are the worldsheet chiral scalar and fermion fields, respectively (see \cite{DP1, DP2} for the details of their construction). 

\sm

Since the Beltrami differential $\hat \mu$ is bilinear in odd moduli, $\cV^{(n)}$ has degree $n$ in the odd moduli. The part $\cV^{(0)}$ is a $(1,0)$-form in $z$ familiar from the standard RNS treatment. The parts $\cV^{(1)}$ and  $\cV^{(2)}$ are $(0,1)$-forms in $z$  and are required by local supersymmetry invariance. Although such $(0,1)$-forms would naively appear to violate the meromorphicity of the chiral amplitudes, they eventually combine into exact differentials in the vertex points $z_i$ that integrate to zero upon pairing left and right chiral amplitudes in (\ref{newsumm.1}) or (\ref{newsumm.1a}). The role of the $(0,1)$-forms in (\ref{vertex}) for the even spin structure contributions  to genus-two amplitudes was established for an arbitrary number of external massless NS states. As demonstrated in \cite{DP7}, these $(0,1)$-forms guarantee that the chiral $n$-point amplitudes are independent of gauge slice choices, and meromorphic in the vertex points $z_i$, up to the addition of exact differentials.

\sm

The correlators may be computed by Wick contractions of the free fields $x_+^\mu$ and $\psi_+^\mu$ with the help of the effective rules of chiral splitting and the following two-point functions \cite{DP1, DP2}, 
\bea
\label{Wick}
\< x^\mu _+ (z) x^\nu _+(w) \> & = & - \eta ^{\mu \nu} \ln E(z,w)
\no \\
\< \psi _+^\mu (z) \psi ^\nu _+(w) \> & = & - \eta^{\mu \nu} S_\delta (z,w)
\eea
where $E(z,w)$ is the prime form and $S_\delta (z,w)$ is the Szeg\"o kernel (see appendix \ref{sec:A.Fay}). 

\sm

Since the bottom component $\cC_0[\delta] (\Omega)$ of the correlator is independent of odd moduli, it is given by the correlator of the zero-th component of the vertex operators, 
\bea
\cC_0[\delta] (\Omega) = \Big \<Q(p)\prod_{i=1}^5 \cV^{(0)}(z_i,\varepsilon_i,k_i) \Big \>
\eea
Here, it is understood that the correlator is evaluated at spin structure $\delta$, and chiral splitting at fixed loop momenta produces the insertion of the operator $Q(p)$, 
\bea
\label{Q}
Q(p)= \exp \,\bigg \{ ip^I_\mu\oint_{\mB_I}dz\,\p_zx_+^\mu(z) \bigg \}
\eea
through which loop momenta are introduced in the conformal field theory correlators. Our conventions for the homology cycles $\mA_I,\mB_J$ are fixed in  appendix \ref{sec:A.RS}.

\subsection{Contribution from disconnected correlators}

The contribution from the disconnected correlators is then given by,
\bea
\label{Fdiscon}
\cF^{(d)} (\Omega) = \sum _\delta { \Xi_6[\delta](\Omega)   \, \tet[\delta](0|\Omega)^4  \over 16 \pi^6 \Psi _{10} (\Omega)} \Big \<Q(p)\prod_{i=1}^5 \cV^{(0)}(z_i,\varepsilon_i,k_i) \Big \>
\eea
The top component $\cC_2[\delta] (\Omega)$ of the correlator is bilinear in odd moduli, and contains all remaining contributions to be discussed in the next subsection. All contributions to the disconnected correlators
from three or fewer fermion bilinears in the vertex operators will vanish by carrying out the spin structure sums for $I_{17}, I_{18}$ and $I_{19}$ in (\ref{I17}), as is familiar from the evaluation of the four-point NS amplitude in \cite{DP6}.  The remaining disconnected contributions to the chiral five-point amplitude arise 
from the insertions of four and five fermion bilinears, and we shall decompose $\cF^{(d)}$ accordingly as follows, 
\bea
\cF^{(d)} (\Omega) = \cF^{(d4)} (\Omega) + \cF^{(d5)} (\Omega)
\label{simpdisc}
\eea
The four fermion bilinear  contribution is given as follows, 
\bea
\cF^{(d4)} (\Omega) &= & \frac{1}{16}  \sum_{\ell=1}^5 dz_\ell \,\varepsilon_\ell^{\mu_\ell} 
\sum _\delta { \Xi_6[\delta](\Omega)   \, \tet[\delta](0|\Omega)^4  \over 16 \pi^6 \Psi _{10} (\Omega)} 
\no \\ && \qquad 
 \times \Big \<Q(p) \partial_{z_\ell} x_+^{\mu_\ell}(z_\ell) e^{ik_\ell \cdot x_+(z_\ell)} \prod_{i=1 \atop{i \neq \ell}}^5 dz_i \, f^{\mu_i \nu_i}_i  \psi^{\mu_i}_+ \psi^{\nu_i}_+(z_i) e^{ik_i \cdot x_+(z_i)}  \Big \>
\label{d4andd5}
\eea
The spin structure sums of its fermion correlators are precisely those encountered in the case of four external NS states, and correspond to the sums $I_{20}$ and $I_{21}$ in (\ref{I20}). 
The five fermion bilinear contribution is given by,
\bea
\cF^{(d5)} (\Omega) = - \frac{i}{32} \sum _\delta { \Xi_6[\delta](\Omega)   \, \tet[\delta](0|\Omega)^4  \over 16 \pi^6 \Psi _{10} (\Omega)} \Big \<Q(p)\prod_{i=1}^5 dz_i \, f^{\mu_i \nu_i}_i  \psi^{\mu_i}_+ \psi^{\nu_i}_+(z_i) e^{ik_i \cdot x_+(z_i)}  \Big \>
\label{mored4andd5}
\eea
The spin structure sums of its fermion correlators  correspond to the sums $J_1$ and $J_2$ in (\ref{J12}) and will be evaluated in section \ref{sec:3}.

\subsection{Contribution from connected correlators}

The evaluation of the contribution from the connected parts in (\ref{splittingF}) constitutes the most difficult aspect of this project, and we shall divide the task of the evaluation by decomposing the contributions as follows,
\bea
\label{Fc}
\cF ^{(c)} = \sum _{a=1}^5 
 \cF_a
\hskip 1in
 \cF_a =  \sum _\delta  \int d^2 \zeta \, \cZ[\delta] \,  \cY_a[\delta] 
\eea
where $\cZ[\delta]$ was defined in (\ref{dmu0dmu2}) and an explicit formula was given in (\ref{zeedelta}).
The individual contributions to the correlator $\cC_2[\delta] ( \Omega) $ in (\ref{splittingF}) are given by\footnote{Throughout,  the integration over the worldsheet $\Sigma$ will be  abbreviated by $\int _\Sigma d^2 z \to \int$ when no confusion  is expected to arise, and we shall write
$\int \chi S = \int _\Sigma d^2 z \chiz S(z)$ and $\int \hat \mu T =  \int _\Sigma d^2 z \hat \mu _{\bar z} {}^z   T(z)$. Moreover, we will employ the shorthand $\cV^{(n)}_i=\cV^{(n)}(z_i,\varepsilon_i,k_i)$.}
\bea
\label{Ys}
\cY _1[\delta]
& = &
{1 \over 8 \pi ^2} \bigg \< Q(p)\,  \int \! \chi S ~ \int \! \chi S ~
\prod _{i=1}^5 \cV_i ^{(0)} \bigg \> _{(c)}
\no \\
\cY _2 [\delta] 
& = &
{1 \over 2 \pi } \bigg \< Q(p)\, \int \! \hat \mu \Big ( T_x + T_\psi \Big ) ~
\prod _{i=1}^5 \cV_i ^{(0)} \bigg \> _{(c)}
\no \\
\cY _3 [\delta] 
& = &
{1 \over 2 \pi } \sum _{i=1}^5 \bigg \< Q(p)\, \int \! \chi S ~ \cV ^{(1)} _i
~ \prod _{j \not= i}^5  \cV_j ^{(0)}  \bigg \>
\no \\
\cY _4 [\delta] 
& = &
\half \sum _{i \not= j} \bigg \< Q(p) ~
 \cV^{(1)} _i  ~ \cV^{(1)} _j  ~ \prod _{l \not= i,j}^{5}  \cV_l ^{(0)} \bigg \>
\no \\
\cY _5 [\delta ]
& = &
\sum _{i =1}^5 \bigg \< Q(p) ~  \cV ^{(2)} _i ~
\prod _{j \not= i}^{5}  \cV_j ^{(0)} \bigg \>
\eea
Here, $\cY_1[\delta]$ arises from the insertion of two matter supercurrents $S$ paired against the gravitino slice; $\cY_2[\delta]$ arises from the insertion of the stress tensor paired against the Beltrami differential; $\cY_3[\delta]$ arises from a mixed contribution which has one first order vertex operator $\cV^{(1)}_i$ and one supercurrent; while  $\cY_4[\delta]$ and $\cY_5[\delta]$ arise from the first and second order corrections to the vertex operators, respectively.
Since the above correlators must be connected, and the vertex operators are independent of the ghosts, the ghost parts of the supercurrent and stress tensor  do not contribute, and only their $x_+$ and $\psi_+$ dependent parts contribute  to (\ref{Ys}). They take the form produced by the chiral splitting procedure,
\bea
\label{ST}
S= - \half \psi ^\mu _+ \p x_+ ^\mu 
\hskip 0.6in
T_x = - \half \p x_+^\mu \p x_+^\mu 
\hskip 0.6in 
T_\psi = \half \psi_+^\mu \p \psi _+^\mu
\eea
Note that, in view of the special gauge (\ref{chiqq}),  the insertions of $\int \chi S$ and $\cV_i^{(1)}$ in $\cY_1[\delta]$, $\cY_3[\delta]$ and $\cY_4[\delta]$ will have support limited to the points $q_1$ and $q_2$.

\subsubsection{The connected parts $\cY_1$ and $\cY_2$}

It will be understood that in the connected correlators of (\ref{Ys}) both $x_+$ fields in $T_x$ and both $\psi_+$ fields in $T_\psi$ are to be Wick contracted onto the corresponding fields in the vertex operators, and not onto one another.  (The self-contractions of $T_x$ and $T_\psi$ are part of the disconnected correlator contributions.) It will be convenient to decompose $\cY_2$ further,
\bea
\label{Y2x}
\cY_2[\delta] = \cY_{2x} [\delta] + \cY_{2\psi} [\delta]
\eea
where $\cY_{2x} [\delta]$ and $ \cY_{2\psi} [\delta]$ are given by $\cY_2[\delta]$ in (\ref{Ys}) restricted to $T_x$ and $T_\psi$, respectively.

\sm

It will also be useful to decompose $\cY_1[\delta]$ according to the three possible ways the fields in the supercurrents can be Wick contracted with one another, or not, 
\bea
\cY _1[\delta] = \cY _{1xx} [\delta]  + \cY _{1\psi \psi} [\delta]  + \cY _{1n}[\delta] 
\eea
In $\cY _{1xx} [\delta] $ the two $\psi_+$ fields in the supercurrents are Wick contracted with one another, in $\cY _{1\psi \psi} [\delta] $ the two $x_+$ fields are Wick contracted with one another, and in $\cY _{1n} [\delta] $ no fields are contracted between the two supercurrents. Choosing the gauge of (\ref{chiqq}) for the gravitino field $\chi$, these contributions may be made more explicit as follows, 
\bea
\label{Y1xx}
\cY _{1\psi \psi} [\delta]  & = & { \zeta ^1 \zeta ^2 \over 16 \pi^2} \p_{q_1} \p_{q_2} \ln E(q_1,q_2) 
\Big \< Q(p)\,  \psi _+^\mu (q_1) \psi_+^\mu (q_2)  \prod _{i=1}^5 \cV_i ^{(0)} \Big \> _{(c)}
\no \\
\cY _{1xx} [\delta] & = & { \zeta ^1 \zeta ^2 \over 16 \pi^2} S_\delta (q_1, q_2) 
\Big \< Q(p)\,  \p x _+^\mu (q_1) \p x_+ ^\mu (q_2)  \prod _{i=1}^5 \cV_i ^{(0)} \Big \> _{(c)}
\no \\
\cY _{1n}[\delta]  & = & - { \zeta ^1 \zeta ^2 \over 16 \pi^2} 
\Big \< Q(p)\,  \psi ^\mu _+ (q_1) \p x _+^\mu (q_1) \psi ^\nu _+ (q_2) \p x_+ ^\nu (q_2)  \prod _{i=1}^5 \cV_i ^{(0)} \Big \> _{(c)}
\eea
The  prescription $(c)$ requires that none of the fields at the points $q_1$ and $q_2$ are Wick contracted with one another. The functions $\cF_1$ and $\cF_2$ in (\ref{Fc}) are decomposed accordingly.

\subsubsection{Structure of the spin summands}
\label{sec:struct}

In the next section, we shall carry out the summations over even spin structures needed to evaluate $\cF^{(c)}$ and $\cF^{(d)}$. To do so, it will be convenient to  extract the non-trivial spin structure $\delta$-dependence of the Beltrami differential, 
\bea
\label{hatmu}
\hat \mu (w) = S_\delta (q_1,q_2) \mu(w)
\eea
where $\mu(w)$ satisfies,
\bea
\label{muom}
\int \mu \om_I \om_J = { \zeta ^1 \zeta ^2 \over 8 \pi} \Big ( \om_I (q_1) \om_J(q_2) + \om_J (q_1) \om_I(q_2) \Big )
\eea
and may be chosen to be independent of $\delta$. In preparation for computing the spin structure sums in the next section,  we list the various types of summands we shall need in these spin structure sums, which we do below. Wick contractions of the $\psi_+$ fields,
\begin{enumerate}
\itemsep=-0.03in
\item in $\cF^{(d)}$ produce closed loops of Szeg\"o kernels multiplied by $\Xi_6[\delta] \tet[\delta]^4$; 
\item in $\cY_{1xx}[\delta] $, $\cY_{2x}[\delta]$ and $\cY_5[\delta]$ produce closed loops of Szeg\"o kernels multiplied by $S_\delta(q_1,q_2)$;
\item in $\cY_{2\psi } [\delta]$ produce closed loops of Szeg\"o kernels with one stress tensor $T_\psi$ insertion;
\item in $\cY_{1\psi \psi}[\delta]$, $\cY_{1n}[\delta]$, $\cY_3[\delta] $ and $\cY_4[\delta]$ produce open chains of Szeg\"o kernels beginning at $q_1$ and ending at $q_2$, possibly times a closed loop.
\end{enumerate}

The spin structure sums of many of these contributions will in fact cancel, and therefore it will be convenient to perform those spin structure sums before writing out all contributing kinematic arrangements.

\subsection{Preview figure for the simplification process}

In the following sections \ref{sec:3} to \ref{sec:6}, we will simplify the contributions
(\ref{d4andd5}), (\ref{mored4andd5}) and (\ref{Ys}) to the chiral amplitude and identify a 
wealth of cancellations.
Figure \ref{fig:1} below gives an overview of cancellations that rely on the interplay of 
several ${\cal F}_a$ in (\ref{Fc}). Contributions to the final form of the chiral amplitude 
in section \ref{sec:7} will be denoted by $\mF_{a}$ with $a=1,2,\ldots,10$, and 
figure \ref{fig:1} also indicates their origin.

\begin{figure}[h]
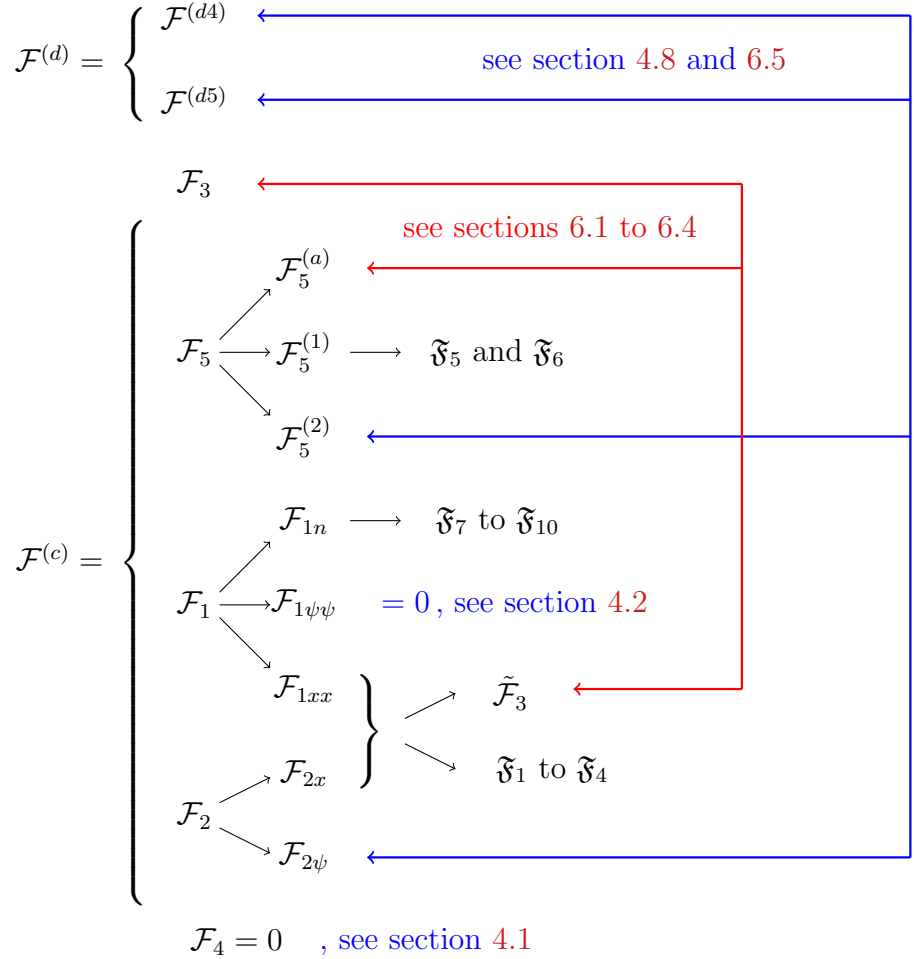

\begin{center}
\tikzpicture[scale=0.56]
\draw(-0.2,0)node{${\cal F}^{(d)} = $}; 
\draw(1.9,-0.15)node{$\left\{ \begin{array}{c}  \\ \\ \\ \end{array} \right.$};
\draw(3,1)node{${\cal F}^{(d4)} $};
\draw(3,-1)node{${\cal F}^{(d5)} $};
\draw(-0.2,-11.85)node{${\cal F}^{(c)} = $}; 
\draw(1.9,-12)node{$\left\{ \begin{array}{c}  \\ \\ \\ \\ \\ \\ \\ \\ \\ \\ \\ \\ \\ 
\\ \\ \\ \\ \\  \end{array} \right.$};
\draw(3,-3)node{${\cal F}_3 $};
\draw(3,-7)node{${\cal F}_5$};
\draw[->](3.6,-6.7) -- (4.8,-5.5);
\draw[->](3.6,-7) -- (4.8,-7);
\draw[->](3.6,-7.3) -- (4.8,-8.5);
\draw(5.6,-5)node{${\cal F}_5^{(a)}$};
\draw(5.6,-7)node{${\cal F}_5^{(1)}$};
\draw(5.6,-9)node{${\cal F}_5^{(2)}$};
\draw[->](6.7,-7) -- (7.9,-7);
\draw(10.2,-7)node{$\mF_5$ and $\mF_{6}$};
\draw(3,-13)node{${\cal F}_1$};
\draw[->](3.6,-12.7) -- (4.8,-11.5);
\draw[->](3.6,-13) -- (4.8,-13);
\draw[->](3.6,-13.3) -- (4.8,-14.5);
\draw(5.6,-11)node{${\cal F}_{1n}$};
\draw(5.6,-13)node{${\cal F}_{1\psi\psi}$};
\draw(5.6,-15)node{${\cal F}_{1xx}$};
\draw[->](6.7,-11) -- (7.9,-11);
\draw(10.2,-11)node{$\mF_7$ to $\mF_{10}$};
\draw[blue] (10.6,-13)node{$=0\, $, see section \ref{sec:4.2}};
\draw(3,-18)node{${\cal F}_2$};
\draw[->](3.6,-17.7) -- (4.8,-17.1);
\draw[->](3.6,-18.3) -- (4.8,-18.9);
\draw(5.6,-17)node{${\cal F}_{2x}$};
\draw(5.6,-19)node{${\cal F}_{2\psi}$};
\draw(6.8,-16)node{$\left. \begin{array}{c}  \\ \\ \\ \end{array} \right\}$};
\draw[->](8,-15.7) -- (9.2,-15.1);
\draw[->](8,-16.3) -- (9.2,-16.9);
\draw(10.5,-15.1)node{$\tilde \cF_3$};
\draw(11.5,-16.9)node{$\mF_1$ to $\mF_4$};
\draw(4,-21)node{${\cal F}_4=0$};
\draw[blue] (8.5,-21)node{, see section \ref{sec:4.1}};
\draw[<-,blue, line width =0.3mm](4.5,1)--(20,1);
\draw[<-,blue, line width =0.3mm](4.5,-1)--(20,-1);
\draw[<-,blue, line width =0.3mm](7.1,-19)--(20,-19);
\draw[<-,blue, line width =0.3mm](7.1,-9)--(20,-9);
\draw[blue, line width =0.3mm](20,-19)--(20,1);
\draw[blue](13.45,0)node{see section \ref{sec:4.8} and \ref{sec:5.6}};
\draw[<-,red, line width =0.3mm](4.5,-3)--(16,-3);
\draw[<-,red, line width =0.3mm](7.1,-5)--(16,-5);
\draw[<-,red, line width =0.3mm](12,-15)--(16,-15);
\draw[red, line width =0.3mm](16,-3)--(16,-15);
\draw[red](11.45,-4)node{see sections \ref{sec:6.1} to \ref{sec:6.4}};
\endtikzpicture
\caption{Overview of the components $\cF_1, \cdots \cF_5$ and ${\cal F}^{(d4)},{\cal F}^{(d5)}$, their partial cancellations, and the origin of the contributions $\mF_1 , \cdots, \mF_{10}$ to the chiral amplitude. The text and arrows in blue and red refer to cancellations discussed in sections \ref{sec:4} and \ref{sec:6}, respectively.} \label{fig:1}
\end{center}
\end{figure}

\subsection{Wick contractions of fermions}

The remaining Wick contractions of the fermions, required to evaluate $\cY_n [\delta]$, fall into three different groups, according to whether they correspond to closed loops of Szeg\"o kernels on vertex points (for items 1 and 2 in the list at the end of subsection \ref{sec:struct}), a closed loop with a fermionic stress tensor inserted (for item 3), or an open chain of Szeg\"o kernels beginning at $q_1$ and ending at $q_2$ (for item 4). These three cases are given respectively by,
\bea
\cW_{1,2, \cdots, n} [\delta ] & = & 
\left ( - { i \over 2} \right )^n \left \< \prod _{i=1}^n f_i ^{\mu_i \nu_i} \psi _+ ^{\mu_i} \psi _+ ^{\nu_i} (z_i) \right \>
\no \\
\cW'_{1,2, \cdots, n} [\delta ] & = & 
{ 1 \over 2 \pi} \left ( - { i \over 2} \right )^n \int d^2 w \, \hat \mu (w) \left \< T_\psi (w) \prod _{i=1}^n f_i ^{\mu_i \nu_i} \psi _+ ^{\mu_i} \psi _+ ^{\nu_i} (z_i) \right \>_{(c)}
\no \\
\cW^{\mu \nu} _{1,2,\cdots, n} [\delta] & = & \left ( - { i \over 2} \right )^n \left \< \psi _+ ^\mu (q_1) \psi _+^\nu (q_2) \prod _{i=1}^n f_i ^{\mu_i \nu_i} \psi _+ ^{\mu_i} \psi _+ ^{\nu_i} (z_i) \right \>_{(c)}
\label{defWs}
\eea
where the subscript $(c)$ excludes self-contractions of $T_\psi$ and of $\psi ^\mu _+(q_1) \psi ^\nu _+(q_2)$ in the second and third lines, respectively.

\subsubsection{Notations}
\label{sec:notations}

Throughout, we shall use the following notations for various permutation sums.
\begin{itemize}
\itemsep=0in
\item The symbol $(i_1, \cdots, i_n)$ denotes the sum over all permutations of $(1,\cdots, n)$;
\item The symbol $(i_1, \cdots, i_n)_r$ denotes the sum over all permutations $(i_1, \cdots, i_n)$ of $(1,\cdots, n)$ modulo reversal $(i_1, \cdots, i_n) \to (i_n, \cdots, i_1)$;
\item The symbol $(i_1, \cdots, i_n)_c$ denotes the sum over all permutations $(i_1, \cdots, i_n)$ of $(1,\cdots, n)$ modulo reversal and cyclic permutations;
\item The symbol $(i_1, \cdots, i_m | i_{m+1} , \cdots, i_n)$ with $1\leq m<n$ denotes the sum over all inequivalent partitions of the set $(1,\cdots, n)$ into sets of cardinality $m$ and $n-m$. Equivalently, $(i_1, \cdots, i_m | i_{m+1} , \cdots, i_n)$ denotes the sum over all permutations $(i_1, \cdots, i_n)$ of $(1,\cdots, n)$ modulo all permutations of $(i_1, \cdots, i_m)$ and all permutations of $(i_{m+1} , \cdots, i_n)$.
\end{itemize}
The Wick contractions of the fermions are carried out using the two-point function given by the Szeg\"o kernel in (\ref{Wick}). Throughout, we shall use the following notations, 
\bea
(f_1 f_2 \cdots f_k) & = & f_1 ^{\mu _1 \mu_2} f_2^{\mu_2 \mu_3} \cdots f_k^{\mu_k \mu_1} 
\no \\
(f_1 f_2 \cdots f_k)^{\mu \nu}  & = & f_1 ^{\mu \, \mu_2} f_2^{\mu_2 \mu_3} \cdots f_k^{\mu_k \, \nu} 
\eea
for closed loops and open chains, respectively, with $(f_1 \cdots f_k)= \tr (f_1 \cdots f_k)$.

\subsubsection{Closed loops of Szeg\"o kernels}
\label{sec:241}

Performing the Wick contractions explicitly for $n \leq 4$, we have
\bea
\cW_{1,2} [\delta ]  & = & \half (f_1 f_2) S_\delta (1,2) S_\delta (2,1)
\no \\
\cW_{1,2,3} [\delta ]  & = & i (f_1 f_2 f_3 ) S_\delta (1,2) S_\delta (2,3) S_\delta (3,1)
\no \\
\cW_{1,2,3,4} [\delta ]  & = &
- (f_1f_2f_3f_4) \, S_\delta (1,2) S_\delta (2,3) S_\delta (3,4) S_\delta (4,1) 
\no \\ &&
- (f_1f_3f_2f_4) \, S_\delta (1,3) S_\delta (3,2) S_\delta (2,4) S_\delta (4,1) 
\no \\ &&
- (f_1f_3f_4f_2) \, S_\delta (1,3) S_\delta (3,4) S_\delta (4,2) S_\delta (2,1) 
\no \\ && 
+  \cW_{1,2}[\delta ]  \cW_{3,4}[\delta ]  + \cW_{1,3}[\delta ]  \cW_{2,4}[\delta ]  + \cW_{1,4}[\delta ]  \cW_{2,3}[\delta ]  
\eea
with $S_\delta(a,b)=S_{\delta}(z_a,z_b)$. The sum for the four-point correlator may be expressed in an equivalent and more succinct manner by using the following notations.
\bea
\cW_{1,2,3,4} [\delta ]  =  
-  \!\! \sum _{(i,j,k)_r} (f_1 f_i f_j f_k ) \, S_\delta (1,i) \, S_\delta (i,j) \,  S_\delta (j,k) \, S_\delta (k,1)
+ \! \! \sum _{(i,j|k,\ell)} \cW_{i,j} [\delta ] \, \cW_{k,\ell } [\delta ] 
\qquad
\eea
The sum on the first line runs over all 3 permutations $(i,j,k)$ of $( 2,3,4)$ modulo reversal $(i,j,k) \to (k,j,i)$, while the sum on the second line runs over 3 inequivalent partitions of $(1,2,3,4)$ into $2+2$, in accord with our general notational conventions of subsection \ref{sec:notations}. 
For $n=5$ we have,
\bea
\cW_{1,2,3,4,5} [\delta ]  & = &  
- i \sum _{(i,j,k,\ell,m)_c} (f_i f_j f_k f_\ell f_m) \, S_\delta (i,j) \, S_\delta (j,k) \, 
S_\delta (k,\ell) \, S_\delta (\ell,m) \, S_\delta (m,i)
\no \\ &&
+ \sum _{(i,j|k,\ell,m)} \cW_{i,j} [\delta ] \, \cW_{k,\ell ,m} [\delta ] 
\eea
Under reversal $(i,j,k,\ell,m) \to (m,\ell,k,j,i)$ on the first line, the product of Szeg\"o kernels reverses sign, and so does the trace of the $f$-matrices since $\tr (f_i f_j f_k f_\ell f_m) = - \tr (f_m f_\ell f_k f_j f_i)$, so that each term is invariant. Therefore, the sum on the first line runs over all 12 permutations $(i,j,k,\ell,m)$ of $(1, 2,3,4,5)$ modulo reversal $(i,j,k,\ell) \to (\ell,k,j,i)$ and modulo cyclic permutations, while the sum on the second line runs over all 10 inequivalent partitions of $(1,2,3,4,5)$ into $2+3$.

\subsubsection{Closed loops of Szeg\"o kernels with a fermion stress tensor}
\label{sec:242}

To organize the Wick contraction of the correlators ${\cal W}'$ in (\ref{defWs}) with a fermionic stress tensor, it will be convenient to use the {\sl deformed Szeg\"o kernel}, defined as the two-point function of $\psi$ with an insertion of the stress tensor  integrated against the Beltrami differential~$\hat \mu$,\footnote{The object $S'_\delta(x,y)$ is defined here with the opposite sign of its definition in \cite{DP6} in order to match the sign of the undeformed Szeg\"o kernel in (\ref{Wick}).} 
\bea
\label{defSprime}
S'_\delta (x,y) \, \eta ^{\mu \nu} = - { 1 \over  2 \pi}  \int d^2 w \, \hat \mu(w)   \, \< T_\psi (w) \psi _+^\mu (x) \psi _+^\nu (y) \>
\eea
We note  that $S'_\delta$ is antisymmetric $S'_\delta (x,y)  = - S'_\delta (y,x)$ just as $S_\delta$ itself is.

\sm

The fermion correlators with the insertion of a single fermion stress tensor are given as follows. For two and three vertex points, we have respectively,
\bea
\cW_{1,2}'  [\delta ]  & = & (f_1 f_2) \, S_\delta ' (1,2) \, S_\delta (2,1)
\no \\
\cW_{1,2,3}' [\delta ]  & = & i \sum_{(i,j,k)_r } (f_i f_j f_k ) S'_\delta(i,j) S_\delta (j,k) S_\delta (k,i)
\eea
The sum is over all 3 permutations $(i,j,k)$ of $(1,2,3)$ modulo reversal $(i,j,k) \to (k,j,i)$, under which both $(f_i f_j f_k)$ and the product of Szeg\"o kernels are odd making the summand invariant. For four points, we have, 
\bea
\label{wprime1234}
\cW_{1,2,3,4} '[\delta ]  & = &  
-  \sum _{(i,j,k,\ell)_r} (f_i f_j f_k f_\ell )  \, S_\delta '(i,j) \, S_\delta (j,k) \, S_\delta (k,\ell) \, S_\delta (\ell,i)
\no \\ &&
+ \sum _{(i,j|k,\ell)} \Big ( \cW_{i,j} '[\delta ]  \, \cW_{k,\ell} [\delta ] + \cW_{i,j} [\delta ] \, \cW_{k,\ell} '[\delta ]  \Big )
\eea
The sum on the first line is over all 12 permutations of $(1,2,3,4)$ modulo the reversal $(i,j,k,\ell) \to (\ell, k, j, i)$, under which both 
$(f_i f_j f_k f_\ell )$ and the product of Szeg\"o kernels are invariant. The sum on the second line is over all 3 inequivalent partitions of $4\to 2+2$. Finally, for five points we have,
\bea
\cW_{1,2,3,4,5} '[\delta ]  & = &  
- i \sum _{(i,j,k,\ell,m)_r} (f_i f_j f_k f_\ell f_m) \,   
S_\delta '(i,j) \, S_\delta (j,k) \, S_\delta (k,\ell) \, S_\delta (\ell,m) \, S_\delta (m,i)
\no \\ &&
+ \sum _{(i,j|k,\ell,m)} \Big ( \cW_{i,j} '[\delta ] \, \cW_{k,\ell ,m} [\delta ] + \cW_{i,j} [\delta ] \, \cW_{k,\ell ,m} '[\delta ]  \Big )
\eea
The sum on the first line is over all 60 permutations of $(1,2,3,4,5)$ modulo the reversal $(i,j,k,\ell,m) \to (m,\ell, k, j, i)$, under which both  $(f_i f_j f_k f_\ell f_m )$ and the product of Szeg\"o kernels are odd so that their product is even.  The sum on the second line is over all 10 inequivalent partitions of $5 \to 3+2$.

\subsubsection{Open chains of Szeg\"o kernels}
\label{sec:243}

For open chains $\cW^{\mu \nu}$ in (\ref{defWs}), we have the following Wick contractions for three vertex points or fewer, 
\bea
\cW_1^{\mu \nu} [\delta] & = & - i f_1^{\mu \nu} S_\delta (q_1,1)S_\delta(1,q_2)
\no\\
\cW_{1,2}^{\mu \nu} [\delta] & = &  \sum_{(i,j)}  (f_if_j)^{\mu \nu} S_\delta (q_1,i)S_\delta(i,j) S_\delta (j,q_2)
\label{wlessmunu} \\
\cW_{1,2,3}^{\mu \nu} [\delta] & = & i \sum_{(i,j,k)} (f_i f_j f_k)^{\mu \nu} S_\delta (q_1,i) S_\delta (i,j) S_\delta (j,k)  S_\delta (k,q_2)
%\no \\ &&
+ \sum _{(i|j,k)} \cW^{\mu \nu} _i [\delta] \, \cW_{j,k} [\delta]
\no
\eea
The sum in $\cW_{1,2}^{\mu \nu}$ is over 2 permutations $(i,j)$ of $(1,2)$. The first sum in $\cW^{\mu \nu}_{1,2,3}$ is over all 6 permutations $(i,j,k)$ of $(1,2,3)$ while the second sum is over all 3 partitions $(i|j,k)$ of $(1,2,3)$ into $1+2$.  For four vertex points, we have, 
\bea
\label{w1234munu}
\cW_{1,2,3,4}^{\mu \nu} [\delta] & = &  
 - \sum_{(i,j,k,\ell)} (f_i f_j f_k f_\ell )^{\mu \nu} S_\delta (q_1,i) S_\delta(i,j) S_\delta(j,k) S_\delta (k,\ell) S_\delta (\ell,q_2)
\\ &&
+ \sum _{(i,j|k,\ell)} \Big ( \cW^{\mu \nu} _{i,j} [\delta] \, \cW_{k,\ell} [\delta]+\cW^{\mu \nu} _{k,\ell} [\delta] \, \cW_{i,j} [\delta] \Big )
+ \sum _{(i|j,k,\ell)} \cW^{\mu \nu} _i [\delta] \, \cW_{j,k,\ell} [\delta]
\no \quad
\eea
The sum on the first line is over all 24 permutations $(i,j,k,\ell)$ of $(1,2,3,4)$. The first sum on the second line is over all 3 inequivalent partitions $(i,j|k,\ell)$ of $(1,2,3,4)$ into $2+2$. The second sum on the second line  is over all 4 inequivalent partitions  $(i|j,k,\ell)$ of $(1,2,3,4)$ into $1+3$.  Finally for five vertex points, we have,
\bea
\label{Wmn5}
\cW_{1,2,3,4,5}^{\mu \nu} [\delta] & = &  -i \sum_{(i,j,k,\ell,m)} (f_i f_j f_k f_\ell f_m )^{\mu \nu} 
S_\delta (q_1,i) S_\delta(i,j) S_\delta(j,k) S_\delta (k,\ell) S_\delta (\ell, m) S_\delta (m,q_2)
\no \\ &&
+ \sum _{(i,j|k,\ell,m)}  \cW^{\mu \nu} _{i,j} [\delta] \, \cW_{k,\ell,m} [\delta] 
+ \sum _{(i|j,k,\ell, m)} \cW^{\mu \nu} _i [\delta] \, \cW_{j,k,\ell,m} [\delta]
\\ &&
- {i \over 2} \sum _{ (i,j,k,\ell,m)/\ZZ_2}  (f_i f_j f_k)^{\mu \nu} (f_\ell f_m) S_\delta (q_1,i) S_\delta (i,j) S_\delta (j,k)  S_\delta (k,q_2) S_\delta(\ell,m)^2 
\no
\eea
The sum on the first line is over all permutations $(i,j,k,\ell,m)$ of $(1,2,3,4,5)$. The first sum on the second line is over all 10 inequivalent partitions  $(i,j|k,\ell,m)$ of $(1,2,3,4,5)$ into $2+3$.  The second sum on the second line is over all 5 inequivalent partitions  $(i|j,k,\ell,m)$ of $(1,2,3,4,5)$ into $1+4$. Finally, the sum on the third line is over  all 60 permutations $(i,j,k,\ell,m)/\ZZ_2$ of $(1,2,3,4,5)$ modulo swapping $\ell$ and $ m$, under which the summand is invariant. Equivalently, this sum may be expressed as half of the sum over {\sl all} permutations~$(i,j,k,\ell,m)$.

\newpage

%%%%%%%%%%%%%%%%%%%%%%%%%%%%%%%%%%%%%%%%%%%
%%%%%%%%%%%%%%%%%%%%%%%%%%%%%%%%%%%%%%%%%%%
\section{Spin structure sums}
\label{sec:3}
\setcounter{equation}{0}
%%%%%%%%%%%%%%%%%%%%%%%%%%%%%%%%%%%%%%%%%%%
%%%%%%%%%%%%%%%%%%%%%%%%%%%%%%%%%%%%%%%%%%%

A crucial ingredient in the construction of the amplitudes in the RNS formulation is the evaluation of the sums over spin structures. The contributions to the genus-two amplitude evaluated here will be the parity conserving part only, restricted to external NS states. Therefore, the spin structure sums will be over even spin structures only. For the five-point string amplitude, some of the sums were required already for the evaluation of the four-point string amplitude in \cite{DP6}, and those results, denoted by $I_1$ to $I_{21}$, are summarized in appendix \ref{sec:D}.  The spin structure sums that first occur in the five-point string amplitude will be discussed in this section, with detailed calculations relegated to appendix \ref{sec:E}. We group the different sums according to their structure.

\subsection{Auxiliary holomorphic forms}

Of crucial importance is the bi-holomorphic form $\Delta$, which was already introduced in the summary of results, and which we recall here for convenience,
\bea
\label{Delta}
\Delta(z,w) = \om_1(z) \om_2(w) - \om_2(z) \om _1(w)
\eea
The holomorphic $(1,0)$-form $\varpi$ may be defined for either $\a=1,2$ by,
\bea
\label{varpi2}
\varpi (z) = c_\a \Delta (q_\a,z) 
\eea
or equivalently in terms of $\tet$-functions by,\footnote{From the different contexts in which they appear, the bi-holomorphic form $\Delta$ should be clearly distinguished from the Riemann vector $\Delta _I(z_0)$ of (\ref{Riemvec}). The half-characteristics $\kappa'$ on the right-hand side of (\ref{varpi2a}) and (\ref{varpi3}) is
defined by (\ref{qqdel}).} 
\bea
\label{varpi2a}
\varpi(z) = (-)^{\a-1}  \om_I (z) \, \p^I \tet (q_\a-\Delta) \, e^{2 \pi i \kappa ' (q_\a-\Delta)}
\eea
The holomorphic $(-1,0)$-forms with non-trivial monodromy $c_\a$ were evaluated in~\cite{DP6},
\bea
\label{varpi3}
c_\a = (-)^{\a-1} Z^3 \sigma (q_\a)^{-1} e^{2 \pi i \kappa '(q_\a-\Delta)}
\eea
in terms of the chiral scalar partition function $Z$ given in appendix \ref{apponchiralZ} and the Fay form $\sigma$ given in appendix \ref{sec:A.Fay}. These forms satisfy the following further relations,
\bea
\label{varpi4}
c_1 \om_I(q_1) - c_2 \om_I(q_2) & = & 0
\no \\
c_1^2 \, \p \varpi(q_1) + c_2^2 \, \p \varpi(q_2) & = & 0
\no \\
\cZ_0 c_1 c_2 \p \varpi(q_1) \p \varpi(q_2) & = & 1
\eea
which were proven in~\cite{DP6}, and will be used extensively throughout.

\subsection{Sums involving $\Xi_6[\delta]$}

The spin structure sums that involve the modular form $\Xi_6[\delta]$ defined in (\ref{Xi6}) are the correlators that contribute to $\cF^{(d)}$ in (\ref{Fdiscon}). Those with three or fewer vertex points $I_{17}, I_{18}$ and $I_{19}$ of (\ref{I17}) vanish, leading to the simplified form (\ref{simpdisc}); those with four vertex points $I_{20}$ and $I_{21}$ of (\ref{I20}) are given in (\ref{I20S}); and those with five vertex points are the sums,\footnote{When no confusion is expected, we shall often abbreviate dependence on a vertex point $z_i$ simply by its index~$i$ so that, for example, $S_\delta(i,j) = S_\delta (z_i,z_j)$, $\Delta (i,j) = \Delta (z_i,z_j)$, $\om_I(i) = \om_I(z_i)$, and $\varpi(i)=\varpi(z_i)$. The dependence on other points, such as $q_1,q_2,w$, will always be kept unabbreviated.}
\bea
\label{J12}
J_1 (z_1, z_2, z_3; z_4, z_5) & = & \sum _\delta \Xi _6 [\delta] \, \tet [\delta](0)^4 \, 
S_\delta (1, 2)S_\delta (2, 3)S_\delta (3, 1) S_\delta (4, 5)^2
\\
J_2 (z_1, z_2, z_3, z_4, z_5) & = & \sum _\delta \Xi _6 [\delta] \, \tet [\delta](0)^4 \, 
S_\delta (1, 2)S_\delta (2, 3)S_\delta (3, 4) S_\delta (4, 5) S_\delta (5,1)
\no
\eea
The key to their calculation is the use of the reduction formulas for the product of Szeg\"o kernels, presented in appendix \ref{sec:C}. To evaluate $J_1$, we apply  formula (\ref{tri1}) to the product of three Szeg\"o kernels. The spin structure sum of the first term on the right side of (\ref{tri1}) vanishes in view of the vanishing of  $I_{18}$ in (\ref{I17}). Similarly, $J_2$ may be evaluated by reducing the product of five Szeg\"o kernels using the results of  (\ref{penta2}). The spin structure sums of the terms in the first line on the right of (\ref{penta2}) cancel in view of the vanishing of $I_{17}$ and $I_{18}$, respectively. The remaining spin structure sums may be evaluated in terms of $I_{20}$ as given in (\ref{I20S}).  The resulting expressions are as follows, 
\bea
J_1 & = & 
- 2 \pi^4 \Psi_{10} \, { x_{12} \, dx_3 \over  x_{23} \, x_{31} } 
\Big ( \Delta (1,4)\Delta (2,5) + \Delta (1,5) \Delta (2,4) \Big ) 
+ \hbox{cycl}(1,2,3)
\no \\
J_2 & = & 
-  \pi^4 \Psi _{10}  \, {   x_{14} \, dx_5 \over x_{45}x_{51}} 
\Big ( \Delta (1,2)\Delta(3,4) - \Delta (1,4)\Delta (2,3) \Big )
+ \hbox{cycl} (1,2,3,4,5)
\label{J2.1}
\eea
The prefactor is expressed in the hyper-elliptic formulation where $z_i=(x_i, s(x_i))$.   The combination $x_{ij} dx_k /(x_{jk} x_{ki})$  is a meromorphic $(1,0)$-form in $z_k=(x_k,s(x_k))$ with simple poles at $z_i, z_j$ and at their images $ \cI (z_i)$, $\cI(z_j) $ under the involution $\cI(z)= (x,-s(x))$. Its simple zeros are the six branch points of $\Sigma$.
It is shown in appendix \ref{sec:C} that this form may be expressed in terms of $\Delta$ as follows,
\bea
\label{Deltaprime}
{ x_{12} \, dx_3 \over  x_{23} \, x_{31} } = { \Delta (z_1,z_2) \Delta'(z_3) \over \Delta (z_2,z_3) \Delta (z_3,z_1)} dz_3
\hskip 0.6in 
\Delta '(z) = \p_w \Delta (w,z) \Big |_{w=z}
\eea
The differential $\Delta'(z) (dz)^3$ is a holomorphic $(3,0)$-form with zeros at the six branch points; it is unique up to an overall constant. Alternatively, the differential may also be expressed in terms of the first-order derivatives of the prime form, which we decompose as follows, 
\bea
\p_{z_i} \ln E(z_i,z_j) = \om_I(z_i) g^I_{i,j} - \p_{z_i} \ln h_\nu (z_i) 
\eea
where $h_\nu$ is the holomorphic $(\half, 0)$-form for odd spin structure $\nu$ given in (\ref{omnu}), while $g_{i,j}^I$ is defined as follows. For a $\tet$-function with odd characteristic $\nu$ we introduce the vector of functions, 
\bea
\label{gI}
g^I_{x,y} =  { \p \over \p \zeta _I} \ln \tet [\nu] (\zeta |\Omega) ~~~ \hbox{evaluated at } ~~~ \zeta _I = \int ^{x} _{y} \om_I
\eea
When no confusion is expected, the functions $g^I_{x,y} $ for $x,y$ evaluated on the vertex points $z_i,z_j$ will be abbreviated as follows,
\bea
g^I_{i,j} = g^I_{z_i,z_j}
\eea
The function $g^I_{z_i,z_j}$ depends on $\nu$ but  this dependence will not be exhibited. It changes sign under swapping the points $z_i$ and $z_j$, and has trivial monodromy under $\mA_I$ cycles but non-trivial monodromy around $\mB_I$ cycles, which is given by, 
\bea
(z_i,z_j) \to (z_i +  \mB_J, z_j) \, : \qquad g_{i,j}^I \to g_{i,j}^I - 2 \pi i \delta ^I_J
\eea
We shall also introduce cyclic sums of $g_{i,j}^I$, all with a common spin structure $\nu$, 
\bea
\label{GI}
G^I_{i_1, i_2, \cdots ,i_n} = g^I _{i_1, i_2} + g^I_{i_2, i_3} + \cdots + g^I_{i_{n-1}, i_n} + g^I_{i_n, i_1}
\eea
Clearly, we have $G^I_{i,j}=0$ while for $n\geq 3$ the combinations $G^I$ are independent of the spin structure, and single-valued in all points. Antisymmetry of $g^I _{i,j}= - g^I _{j,i}$ leads to the reflection
property $G^I_{i_1, i_2, \cdots ,i_n} = - G^I_{i_n,  \cdots ,i_2,i_1}$ such that $G^I_{i,j,k}$ is totally antisymmetric
in $i,j,k$.

\sm

 It is shown in appendix \ref{sec:E} that both $J_1$ and $J_2$ may be simply expressed in terms of the functions $G^I$ and holomorphic Abelian differentials, and the final results are as follows,
\bea
 \label{J12.6}
J_1 & = & 
- 2 \pi^4 \Psi _{10} \, \om_I (1) \Big ( \Delta (2,4)\Delta (3,5) \, G^I _{1,3,5,4,2} +(4 \leftrightarrow 5)
 \Big ) + \hbox{cycl}(1,2,3) 
\no \\
J_2 & = & - 2 \pi^4 \, \Psi _{10} \,  \omega_I(1)  \Delta(2,5) \Delta(3,4) \, G^I_{5,1,2}  + {\rm cycl}(1,2,3,4,5) 
\eea
These expressions are manifestly  single-valued. A useful alternative expression for $J_2$ is given in the canonical cyclic basis,
\bea
J_2 = 2 \pi^4 \Psi_{10} \om_I(1) \Delta (2,3) \Delta (4,5) \Big ( G^I_{1,2,3} + G^I_{1,4,5} \Big)
+ {\rm cycl}(1,2,3,4,5) 
\label{altJtwo}
\eea
Note that poles of $G^I_{1,2,3}$ in $(z_1 - z_3)$ and its cyclic permutations  are spurious since they do not occur in the summand in (\ref{J12}). Indeed, the coefficient of $g^I_{3,1}$ in (\ref{altJtwo}) is given by 
$( \om_I(1) \Delta(2,3) +  \om_I(3)\Delta(1,2)  )\Delta(4,5) = \omega_I(2) \Delta(1,3) \Delta(4,5)$
after assembling the cyclic orbit which cancels the pole in $(z_1 - z_3)$. One can similarly check that the expression for $J_1$ in (\ref{J12.6}) has no simple poles other than those from $g^I_{1,2}, \, g^I_{2,3}$, and $g^I_{3,1}$.

\subsection{Sums involving $S_\delta (q_1,q_2)$  }

The spin structure sums that involve the factor $S_\delta(q_1,q_2)$ times closed loops of Szeg\"o kernels at vertex points are the correlators that contribute to $\cF_{1\psi \psi}$, $\cF_{2x}$ and $\cF_5$ and involve the Wick contractions in subsection \ref{sec:241}. Those with three or fewer vertex points $I_1,I_2, I_3$ of (\ref{D2}) vanish; those with four vertex points $I_{11}$ and $I_{12}$ of (\ref{I11}) are given in (\ref{I11S}); and those with five vertex points are the sums, 
\bea
\label{J34}
J_3 (z_1, z_2, z_3; z_4, z_5) & = & \sum _\delta \cZ [\delta] S_\delta(q_1,q_2)  S_\delta (1, 2)S_\delta (2, 3)S_\delta (3, 1) S_\delta (4, 5)^2
\\
J_4 (z_1, z_2, z_3, z_4, z_5) & = & \sum _\delta \cZ [\delta] S_\delta(q_1,q_2)  S_\delta (1, 2)S_\delta (2, 3)S_\delta (3, 4) S_\delta (4, 5) S_\delta (5,1)
\no
\eea
where  $\cZ [\delta] $ was defined in (\ref{zeedelta}).
To evaluate the sums in $J_3,J_4$, we proceed by using the hyper-elliptic representation to reduce the product of three Szeg\"o kernels in $J_3$ using (\ref{tri1}) and the product of five Szeg\"o kernels in $J_4$ using (\ref{penta2}). The remaining spin structure sums may be carried out using the formulas for $I_{11}$ and $I_{12}$ given in (\ref{I11}). The resulting expressions may be simplified to give,
\bea
\label{J34.1}
J_3 & = &  -  \cZ_0 \, \rho_1 \, { x_{23}\, dx_1 \over  x_{31} x_{12} }+  \hbox{cycl} (1,2,3)
\no \\
J_4 & = &  \cZ_0  \, \rho _1 \, {   x_{25} \, dx_1 \over x_{51}x_{12} } 
+ \hbox{cycl} (1,2,3,4,5)
\eea
where we shall use the following notations throughout,\footnote{Note that the symbols $\rho_i$ and $ \rho$ used here differs from their definitions in \cite{DP6}, where they represented the product over three and four differentials, respectively.} 
\bea
\label{rhoi}
 \rho_i = \prod _{{j=1 \atop j \not = i}}^5 \varpi (z_j) 
 \hskip 1in 
  \rho = \prod _{j=1 }^5 \varpi (z_j) 
\eea
As in the case of $J_1,J_2$, these hyper-elliptic expressions may be recast in terms of derivatives of the prime form and ultimately in terms of the single-valued functions analogous to $G^I$ of (\ref{GI}), provided we generalize the definition of $G^I$ to allow for one of the points to be $q_\a$. The calculations are performed in appendix \ref{sec:E} and the results are given as follows,
\bea
\label{J34.2}
J_3 & = & - \cZ_0 \, \rho_1 \, \om _I (1) \, G^I_{q_\a, 2,1,3} 
+ \hbox{cycl}(1,2,3)
\no \\
J_4 & = &  \cZ_0 \, \rho_1 \, \om _I(1) \, G^I_{q_\a, 2,1,5} 
+ \hbox{cycl} (1,2,3,4,5)
\eea
where
\bea
G^I_{q_\a,  2,1,3} = g^I_{q_\a, z_2} + g^I_{z_2,z_1} + g^I_{z_1,z_3} + g^I _{z_3, q_\a}
\label{defgfour}
\eea
The combinations occurring in (\ref{J34.2}) have trivial monodromy in all variables, and the
definition (\ref{defgfour}) readily implies antisymmetry $G^I_{q_\a,  2,1,3}=-G^I_{q_\a,  3,1,2}$ as well
as zeros in $z_1{-}q_\a$ and $z_2{-}z_3$. The poles in $z_j - q_{\alpha}$ of the individual $G^I_{q_\alpha,j,\ldots}$ in (\ref{J34.2}) are spurious thanks to the zeros of the accompanying $\varpi(z_j)$ factors.

\subsection{Sums involving $S_\delta (q_1,z_i) S_\delta (z_j,q_2)$  }

The spin structure sums that involve an open chain of Szeg\"o kernels, beginning at the point  $q_1$ and ending at $q_2$ contribute to $\cF_{1xx}, \cF_{1n}$, $\cF_3$ and $\cF_4$ and involve the Wick contractions in subsection \ref{sec:243}. Those with four or fewer vertex points $I_4, \cdots, I_{10}$ of (\ref{D2}) all vanish; those with five vertex points are the sums, 
\bea
\label{J5-9}
J_5 (z_1,z_2;z_3,z_4,z_5) & = & \sum _\delta \cZ[\delta] S_\delta (q_1, 1) S_\delta (1,2) S_\delta (2,q_2) 
S_\delta (3, 4)S_\delta (4, 5)S_\delta (5, 3)
\no \\
J_6 (z_1;z_2,z_3;z_4,z_5) & = & \sum _\delta \cZ[\delta] S_\delta (q_1, 1) S_\delta (1,q_2) 
S_\delta (2, 3)^2 S_\delta (4, 5)^2
\no \\
J_7 (z_1;z_2,z_3,z_4,z_5) & = & \sum _\delta \cZ[\delta] S_\delta (q_1, 1) S_\delta (1,q_2) 
S_\delta (2, 3) S_\delta (3, 4)S_\delta (4, 5)S_\delta (5, 2)
\no \\
J_8 (z_1,z_2,z_3;z_4,z_5) & = & \sum _\delta \cZ[\delta] S_\delta (q_1, 1) S_\delta (1,2) S_\delta(2,3) S_\delta (3,q_2)  S_\delta (4, 5)^2
\no \\
J_9 (z_1,z_2,z_3,z_4,z_5) & = & \sum _\delta \cZ[\delta] S_\delta (q_1, 1) S_\delta (1,2) 
S_\delta(2,3) S_\delta (3, 4) S_\delta (4, 5) S_\delta (5,q_2) 
\quad
\eea
To evaluate $J_5$, we use the reduction formula (\ref{tri1}) for the product of the last three Szeg\"o kernels in the summand of $J_5$ and the vanishing of $I_6$ and $I_9$ in (\ref{D2}), and we readily find,
\bea
\label{J50}
J_5=0
\eea
Furthermore, $J_6$ and $ J_7$ are holomorphic in $z_2, z_3, z_4, z_5$ as a result of $I_4=I_5=I_{10}=0$, while the residues of their poles in $z_1$ at $q_1$ and $ q_2$  are given by $I_{11}$ and $I_{12}$, respectively. The functions $J_8$ and $J_9$ are holomorphic in $z_1,\cdots,  z_5$ as a result of $I_7=I_8=I_{9}=0$.

\sm

These spin structure sums are evaluated using the Riemann identities  in appendix \ref{sec:E}, and the final simplified results are as follows, 
\bea
\label{J6789new}
J_6& = & J_7 - \cZ_0 c_1^2 \p \varpi(q_1) \varpi(1) \Delta (2,4) \Delta (3,5)
\no \\
J_7 & = &  \cZ_0 c_1^2  \p \varpi (q_1) \varpi(1)^{-1} \Big (
\varpi (3) \varpi(5) \Delta (1,2) \Delta( 1,4)   + \varpi (2) \varpi(4) \Delta (1,3) \Delta( 1,5)  \Big )
\no \\
J_8 & = & \cZ_0  c_1^2 \p \varpi(q_1) \Big ( \varpi(1) \Delta (2,4) \Delta (3,5)  + \varpi(5) \Delta (1,4) \Delta (2,3) \Big )
\no \\
J_9 & = & 
- \cZ_0 c_1 ^2 \p \varpi(q_1)   \varpi(3) \Delta (1,4)  \Delta (2,5) 
\eea
The forms $c_1, c_2$ are given in (\ref{varpi3}) and related to $\varpi, \Delta $ and $\cZ_0$ by the relations of  
(\ref{varpi2}) and (\ref{varpi4}). In particular, the first relation in (\ref{varpi4})  makes it manifest that each function above is odd under swapping $q_1$ and $q_2$, as is expected from the definition of their spin structure sums in (\ref{J5-9}). Further consistency checks of the simplified formulae in (\ref{J6789new}) can be found
in appendix \ref{sec:cross}.

\subsubsection{Further identities and symmetrizations}

We note the following relation between $J_6$ and $J_7$,
\bea
J_6(1;2,3;4,5) = - J_7(1;2,5,3,4) + J_7(1;2,3,4,5) + J_7(1;2,4,5,3)
\eea
For later use, it will be convenient to introduce the symmetrized version of $J_6$ and $J_7$,
\bea
J_6 ^S (1;2,3;4,5) & = & { 1 \over 3} \Big ( J_6 (1;2,3;4,5) + J_6(1;2,4;5,3)+J_6(1;2,5;3,4) \Big )
\no \\
J_7 ^S (1;2,3,4,5) & = & { 1 \over 3} \Big ( J_7 (1;2,3,4,5) + J_7 (1;2,4,5,3) + J_7 (1;2,5,3,4) \Big )
\eea
From the relation (\ref{J6789new}) between $J_6$ and $J_7$ it is readily deduced that we have,
\bea
J_6 ^S (1;2,3;4,5)  = J_7 ^S (1;2,3,4,5) 
\eea
Their expression is given as follows,
\bea
\label{3.J7sym}
J_7^S & = & {1 \over 3} \cZ_0 { c_1^2 \p \varpi(q_1) \over \varpi (1)} \Big [ 
\Delta(1,2) \Delta (1,4) \varpi(3) \varpi(5) + \Delta(1,3) \Delta (1,5) \varpi(2) \varpi(4) 
\no \\ && \hskip 0.9in 
+ \Delta(1,2) \Delta (1,5) \varpi(4) \varpi(3) + \Delta(1,4) \Delta (1,3) \varpi(2) \varpi(5) 
\no \\ && \hskip 0.9in 
+ \Delta(1,2) \Delta (1,3) \varpi(5) \varpi(4)  + \Delta(1,5) \Delta (1,4) \varpi(2) \varpi(3) \Big ]
\eea
The different terms correspond to the 6 partitions of a set of 4 points $\{ 2,3,4,5\}$ into two pairs of 2 distinguishable points. Thus, the function $J^7_S$ is invariant under all permutations of the points $\{2,3,4,5\}$.   Next, we introduce the differences of $J_7$ and the symmetrized version, 
\bea
\tilde J_7 (1;2,3,4,5) = J_7 (1;2,3,4,5) - J^S_7 (1;2,3,4,5)
\eea
These functions may be computed in terms of basic objects,
\bea
\tilde J_7(1;2,3,4,5) = { 1 \over 3} \cZ_0 c_1^2 \p \varpi (q_1) \varpi (1) \Big ( \Delta(2,3) \Delta(4,5) + \Delta(2,5)\Delta(4,3) \Big )
\label{basicJ7}
\eea
which makes it manifest that $\tilde J_7$ is holomorphic in all vertex points $z_i$.
In summary,  $J_6$ and $J_7$ can be expressed as follows,
\bea
\label{3.J6J7}
J_6(1;2,3;4,5) & = &  J^S_7(1;2,3,4,5)  -2 \tilde J_7(1;2,5,3,4)
\no \\
J_7 (1;2,3,4,5) & = & J^S_7 (1;2,3,4,5) + \tilde J_7(1;2,3,4,5)
\eea
where the singularities in $z_1$ at $q_\a$ are entirely contained in $J^S_7$. These singularities are simple poles, whose residue is given by, 
\bea
J_7^S(1;2,3,4,5) =   { 2\cZ_0\, \rho_1 \, dz_1  \over z_1-q_1} - { 2\cZ_0\, \rho_1 \, dz_1  \over z_1-q_2} + \cO(1)
\eea
For later use, it will be convenient to express $J_9$ in the cyclic basis of holomorphic $(1,0)$-forms in five points, exhibited in (\ref{cycbasis}), 
\bea
\label{J9sym}
J_9 (1,2,3,4,5) & = &  \cZ_0 c_1^2 \p \varpi (q_1) 
\Big [ \varpi(2) \Delta (3,4) \Delta (5,1) + \varpi(4)  \Delta (5,1) \Delta (2,3) 
\no \\ && \hskip 1in
- \varpi(3) \Delta (4,5) \Delta (1,2)  \Big ]
\eea
The spin structure sum exhibits invariance under reversal, namely $(z_1,z_2) \leftrightarrow (z_5,z_4)$ leaving $z_3$ invariant, as expected from inspection of the original spin structure sum.

\subsection{Sums involving the fermion stress tensor }

The spin structure sums that involve the fermionic stress tensor $T_\psi$, defined in (\ref{ST}), inserted in closed loops of Szeg\"o kernels at vertex points, are the correlators that contribute to $\cF_{2\psi}$ and involve the Wick contractions in subsection \ref{sec:242}. Those with four or fewer vertex points $I_{13}, I_{14}, I_{15}, I_{16}$ of (\ref{I13}) are given by (\ref{I13S}); those with five vertex points are the sums, 
\bea
\label{J10-12}
J_{10} (w;z_1,z_2,z_3;z_4,z_5) & = & \sum _\delta \cZ[\delta]  
S_\delta (q_1,q_2)  \f [\delta] (w;4, 5)  S_\delta (4,5)  S_\delta(1,2) S_\delta(2,3) S_\delta (3,1)
\no \\
J_{11} (w;z_1,z_2,z_3;z_4,z_5) & = & \sum _\delta \cZ[\delta]  S_\delta (q_1,q_2)  \f [\delta] (w;1, 2)  S_\delta (2,3) 
S_\delta(3,1) S_\delta(4,5)^2 
\\
J_{12} (w;z_1,z_2,z_3,z_4,z_5) & = & \sum _\delta \cZ[\delta]  S_\delta (q_1,q_2)  \f [\delta] (w;1, 2)  S_\delta (2,3) 
S_\delta(3,4) S_\delta(4,5) S_\delta (5,1)
\qquad
\no
\eea
Here, we have expressed the correlator of the fermionic stress tensor with two $\psi$-fields in terms of the combination $\f[\delta]$ familiar from \cite{DP6}, 
\bea
 \< T_\psi (w) \psi^\mu_+(x) \psi^\nu_+(y) \>  =  \thalf  \eta^{\mu \nu} \f [\delta] (w;x,y)  
\eea
where $\f[\delta]$ is given in terms of the Szeg\"o kernel by, 
\bea
\f [\delta ] (w;x,y) = S_\delta (x, w) \p_w S_\delta (w,y) - S_\delta (y, w) \p_w S_\delta (w,x)
\eea
Using the Fay identity (\ref{Fay1}), we recast $\f[\delta]$ in an equivalent but more useful form,
\bea
\label{varphi}
\f [\delta] (w;x,y) = - { \tet[\delta] (x+y-2w) E(x,y) \over \tet[\delta](0) E(x, w)^2 E(y,w)^2}
\eea
The expression we shall need is actually the integral of $T_\psi$ against the Beltrami differential $\hat \mu $ of (\ref{hatmu})  for the deformation from the period matrix $\Omega$ to the super period matrix $\hat \Omega$. This deformation may be summarized in terms of the deformed Szeg\"o kernel $S'_\delta (z,w)$ defined by (\ref{defSprime}).
To perform the sum over spin structures, we extract the spin structure dependence of  the Beltrami differential $\hat \mu (w)= S_\delta (q_1,q_2) \mu(w)$ using (\ref{hatmu}), and assume that $\mu(w)$ is independent of $\delta$.  The deformed Szeg\"o kernel may then be expressed as follows, 
\bea
S'_\delta (x,y) =  - { 1 \over  4 \pi}  \int_\Sigma d^2 w \,  \mu(w)\,  \, S_\delta (q_1, q_2) \, \f[\delta] (w;x,y)
\eea
The dependence on the spin structure $\delta$ through the factor $S_\delta (q_1, q_2)$ must, of course, be included in the summand while carrying out the sum over spin structures, whence the form of the sums in (\ref{J10-12}).  We note for later use that both  $\f [\delta] $ and $S'_\delta$ preserve the antisymmetry under exchange of the points $x,y$,
\bea
S'_\delta (x,y)  = - S'_\delta (y,x) \hskip 1in  \f[\delta] (w;x,y) = - \f[\delta[ (w;y,x)
\eea
We shall proceed next to evaluating the spin structure sums in $J_{10}, J_{11}$ and $J_{12}$.

\subsubsection{Evaluating $J_{10}$}

To simplify the evaluation of $J_{10}$,  we use formula (\ref{tri1}) to reduce the product of the last three Szeg\"o kernels to squares of Szeg\"o kernels, which may be evaluated with the help of $I_{13}$ and $I_{16}$. The result is as follows, 
\bea
\label{J10}
J_{10} =  { dx_1 \, dx_2 \, dx_3 \over 2 x_{12}\, x_{23} \, x_{31}} \,  I_{13}(w;z_4,z_5)
- \left ( { x_{12} dx_3 \over 2 x_{23}x_{31}} \, I_{16} (w; z_4,z_5;z_1,z_2) + \hbox{cycl}(1,2,3) \right )
\eea
We shall refrain from expressing the hyper-elliptic combinations in terms of the Green function $G$, as the first term will be found to cancel upon integration against the Beltrami differential, and will therefore not be needed, see
appendix \ref{appF.J10} for details.

\subsubsection{Evaluating $J_{11}$ and $J_{12}$}

For $J_{11}$ and $J_{12}$, two further reductions in the number of $\tet$-functions in the summand may be achieved by use of the Fay identity (\ref{Fay1}) which allows the calculation of the remaining spin structure sums with the help of the Riemann identities (\ref{Riem2}). The detailed calculations are relegated to appendix \ref{sec:E}, and the results are as follows,
\bea
\label{J11-12}
J_{11} & = & { 1 \over 4} \Big ( - L_1(w;1,2;3,4,5) - L_1(w;1,2;4,5,3) - L_1(w;1,2;5,3,4) 
\no \\ && \hskip 0.25in 
+ L_2(w;1,2;3,4,5)+L_2(w;1,2;3,5,4) - L_2(w;1,2;4,5,3)
\no \\ && \hskip 0.25in
+ L_2(w;1,2;4,3,5) +L_2(w;1,2;5,3,4) - L_2(w;1,2;5,4,3) \Big )
\no \\
J_{12} & = & { 1 \over 4} \Big ( 
L_1(w;1,2;3,4,5) - L_1(w;1,2;4,5,3) + L_1(w;1,2;5,3,4)
\no \\ && \hskip 0.25in 
- L_2(w;1,2;3,4,5)- L_2(w;1,2;3,5,4) + L_2(w;1,2;4,3,5) 
\no \\ && \hskip 0.25in
-L_2(w;1,2;5,3,4)  - L_2(w;1,2;4,5,3)+ L_2(w;1,2;5,4,3) \Big )
\eea
where the functions $L_1, L_2$ are defined by,
\bea
L_1(w;1,2;3,4,5) & = & \sum _\delta \cZ[\delta] S_\delta (q_1,q_2) \f [\delta] (w;1,2)
R_\delta (1,3;4,5) R_\delta (2,3;4,5)
\no \\
L_2(w;1,2;3,4,5) & = & \sum _\delta \cZ[\delta] S_\delta (q_1,q_2) \f [\delta] (w;1,2)
R_\delta (1,3;4,5) R_\delta (2,4;5,3)
\label{evidsymm}
\eea
in terms of the combinations, 
\bea
R_\delta (z_1, z_2;w_1,w_2) = S_\delta (z_1,w_1) S_\delta(z_2,w_2) - S_\delta (z_1,w_2) S_\delta(z_2,w_1)
\eea
which occur in the Fay identity (\ref{Fay1}).  These functions are evaluated in appendix \ref{sec:E}, and are given as follows, 
\bea
\label{K2fin}
L_1 (w;1,2;3,4,5) & = & 2 \cZ_0  c_1^2   \varpi(w)^2 \, \cG_2(4,5,q_1;3,w) \,  \cG_3(1,2,3,q_1;4,5,w) 
\no \\ && 
+ (q_1 \leftrightarrow q_2) 
\no \\
L_2 (w;1,2;3,4,5) & = & 2 \, \cZ_0 \varpi(w)^2 \varpi(1) \varpi(2) G(3;2,4;q_1,w) G(4;1,3;q_1,w)  
\no \\ &&  \times  G(5;1,2;q_2,w)  + ( q_1 \leftrightarrow q_2)
\eea
An alternative expression for $L_2$ is given as follows,
\bea
L_2(w;1,2;3,4,5) & = & - 2 \, \cZ_0 \varpi(w)^2 \varpi(2) \varpi(5) G(3;2,4;q_1,w) G(4;1,3;q_1,w)  
\no \\ &&  \times  G(1;5,w;q_1, 2)  + ( q_1 \leftrightarrow q_2)
\eea
The Green functions $\cG_2$ and $\cG_3$ emerge naturally in the anti-commuting $b,c$ system analyzed in subsection \ref{B21}, where explicit formulas for the function $\cG_2(z_1,z_2,z_3;w_1,w_2)$ and $\cG_3(z_1,z_2,z_3,z_4;w_1,w_2,w_3)$ may also be found. Since they are separately permutation antisymmetric in the $z_i$ and $w_j$ and since $G(z;z_1,z_2;p_1,p_2)$ in (\ref{Green}) or (\ref{Green1}) is symmetric under $p_1\leftrightarrow p_2$ as well as antisymmetric under $z_1\leftrightarrow z_2$, the representation (\ref{K2fin})
of the $L_1$ and $L_2$ implies the following symmetry properties,
\bea
&& L_1 (w;1,2;3,4,5) = - L_1 (w;2,1;3,4,5) = L_1 (w;1,2;3,5,4) 
\no \\
&&
L_2 (w;1,2;3,4,5)  = - L_2 (w;2,1;4,3,5) 
\eea
which are evident from (\ref{evidsymm}) and the antisymmetry of $R_\delta (z_1, z_2;w_1,w_2) $
in both $z_1 \leftrightarrow z_2$ and $w_1 \leftrightarrow w_2$.
The above formulas for $J_{11}$ and $J_{12}$ are proven in appendix~\ref{sec:E}.

\newpage

%%%%%%%%%%%%%%%%%%%%%%%%%%%%%%%%%%%%%%%%%%%
%%%%%%%%%%%%%%%%%%%%%%%%%%%%%%%%%%%%%%%%%%%
\section{Fundamental simplifications and cancellations}
\label{sec:4}
\setcounter{equation}{0}
%%%%%%%%%%%%%%%%%%%%%%%%%%%%%%%%%%%%%%%%%%%
%%%%%%%%%%%%%%%%%%%%%%%%%%%%%%%%%%%%%%%%%%%

The summation over spin structures enforces the GSO projection, required for space-time supersymmetry, and produces major cancellations and simplifications to the chiral amplitude. In particular, these cancellations are required to render the amplitude properly independent of any intermediate gauge and slice choices. They will further guarantee the absence of certain low-energy interactions and thus lead to non-renormalization theorems. The importance of such cancellations is familiar from the amplitude for four external states in \cite{DP6}, but will be even more dramatic for the amplitude with five external NS states. In this section, we prove the required cancellations and simplifications in order of increasing complexity.

\subsection{Cancellation of $\cF_4$}
\label{sec:4.1}

The contribution $\cF_4$, obtained as the sum over spin structures of $\cY_4[\delta]$ given in (\ref{Ys}), vanishes. To prove this result, we use the fact that in $\cY_4[\delta]$ one vertex point coincides with $q_1$, another with $q_2$. The fermionic part of the correlators therefore involves a sum over contributions, each of which is formed out of a chain of Szeg\"o kernels from $q_1$ to $q_2$ with three vertex points being available for $\cY_4[\delta]$. But such contributions cancel upon summing over spin structures as follows (see appendix \ref{sec:D} for further details). The points $q_1$ and $q_2$ may be connected via zero vertex points which vanish by $I_1=I_2=I_3=0$, by one vertex point which vanish by $I_4=I_5=0$, by two points which vanishes by $I_6=0$ or by three points which vanishes by $I_7=0$.

\subsection{Cancellation of $\cF_{1\psi \psi}$}
\label{sec:4.2}

The contribution $\cF_{1\psi \psi}$, obtained as the sum over spin structures of $\cY_{1\psi \psi}[\delta]$ given in (\ref{Y1xx}), vanishes. To prove this result, we use the fact that the Wick contractions of the fermions are given by open chains spelt out in section \ref{sec:243} with $\mu$ and $\nu$ contracted with one another. The spin structure sums of $\cW^{\mu \nu}_1[\delta]$, $\cW^{\mu \nu}_{1,2}[\delta]$, $\cW^{\mu \nu}_{1,2,3}[\delta]$ and $\cW^{\mu \nu}_{1,2,3,4}[\delta]$ defined in (\ref{wlessmunu}) and (\ref{w1234munu}) vanish thanks to the vanishing of $I_4, \cdots, I_{10}$ in (\ref{D2}).
The remaining contribution to $\cF_{1\psi \psi }$ involves the chain of Szeg\"o kernels with five vertex points given in (\ref{Wmn5}). Expressing the resulting sums in terms of the functions $J_5, J_6,J_7,J_8,J_9$ of (\ref{J5-9}), we find that $\cF_{1 \psi \psi }$ is proportional to the following sum (omitting a factor of bosonic contractions),
\bea
\label{J9xx}
  \sum_{(i,j,k,\ell,m)} \!\!  2(f_i f_j f_k f_\ell f_m )\,  J_9(i,j,k,\ell,m)  
- \sum _{(i,j,k|\ell,m)} \!\! 2 (f_i f_j f_k) (f_\ell f_m) \, J_8(i,j,k;\ell,m)
\quad
\eea
where we recall from (\ref{Wmn5}) that the first sum is over all permutations $(i,j,k,\ell,m)$ of $(1,2,3,4,5)$ while the second sum is over all permutations $(i,j,k|\ell,m)$ modulo swapping $\ell, m$. 
In obtaining this result, we have used the facts that the spin structure sum of the first term in the second line of (\ref{Wmn5}) is proportional to $J_5$ which vanishes in (\ref{J50}), while  the contraction over $\mu$ and $\nu$ of the second term in the second line of (\ref{Wmn5}) vanishes since $f_i ^{\mu \mu}=0$. The last cancellation explains why $J_6$ and $J_7$ are absent from $\cF_{1\psi \psi}$.

\sm

Now recast the first term as the sum over pairs of terms with reversed assignments, 
\bea
&&
 (f_i f_j f_k f_\ell f_m )\,  J_9(i,j,k,\ell,m)  + (f_m f_\ell f_k f_j f_i )\,  J_9(m,\ell,k,j,i)  
\no \\ && \hskip 0.5in 
=    (f_i f_j f_k f_\ell f_m ) \Big (  J_9(i,j,k,\ell,m)  -  J_9(m,\ell,k,j,i)  \Big )
\eea
where we have used $(f_m f_\ell f_k f_j f_i ) = - (f_i f_j f_k f_\ell f_m )$ to combine both terms. Since we have  $J_9(i,j,k,\ell,m)  =  J_9(m,\ell,k,j,i) $ by (\ref{J6789new}) each term vanishes. A similar argument applies to the second term in the sum of (\ref{J9xx}), but this time applied to reversing only the entries $i,j,k$, 
\bea
&&
 (f_i f_j f_k) (f_\ell f_m) \, J_8(i,j,k;\ell,m) + (f_k f_j f_i) (f_\ell f_m) \, J_8(k,j,i;\ell,m) 
\no \\ && \hskip 0.5in
=   (f_i f_j f_k) (f_\ell f_m)  \Big (  J_8(i,j,k;\ell,m) - J_8(k,j,i;\ell,m) \Big )
\eea
where we have used $(f_k f_j f_i ) = - (f_i f_j f_k)$ to combine both terms. Inspection of $J_8$ in (\ref{J6789new}) and repeated use of the identities (\ref{omdel}) shows that $ J_8(i,j,k;\ell,m) = J_8(k,j,i;\ell,m) $ so that each term vanishes. This concludes our proof of the vanishing of $\cF_{1\psi \psi}$.

\subsection{The function $\Lambda$}

In preparation for the evaluation of the contributions $\cF_2$ of the stress tensors   in the later subsections, we review here the expression of the Beltrami differential $\mu$ in terms of the function $\Lambda$ which was introduced in \cite{DP6}. The starting point is the following key observation,
\bea
\label{4.mu1}
\int _\Sigma d^2 w \, \mu(w) \varpi (w) \om _I(w)=0
\eea
To prove it, we express $\varpi(w)$ in terms of $\Delta(q_\a,w)$ using (\ref{varpi2}), then express $\Delta (q_\a, w)$ in terms of $\ep^{IJ} \om_I(q_\a) \om_J(w)$ and finally use the form of the integral of $\mu$ against $\om_I(w) \om_J(w)$ using (\ref{muom}).  As a result, there exists a single-valued function $\Lambda (z)$ such that,
\bea
\label{4.mu2}
\mu(w) \varpi(w) = \pbw \Lambda (w) 
\eea
The function $\Lambda$ is unique up to the addition of a holomorphic function of $w$ which, on a compact surface, must be constant, and we have,
\bea
\label{4.Lambda1}
\Lambda(z)- \Lambda(z_0) = - { 1 \over 2 \pi} \int _\Sigma d^2w \, \mu (w) \varpi(w) G(w;z,z_0;p_1,p_2)
\eea
where we refer to (\ref{Green}) for the definition of $G$.
The integral on the right side does not depend on the points $p_1, p_2$, which are the zeros of $G$ as a function of $w$, since  a change in the points $p_1,p_2$ adds a single-valued well-defined holomorphic form in $w$  to $G$ whose integral vanishes in view of (\ref{4.mu1}). 

\sm

If $\Lambda $ were  smooth and  vanishing at $q_1$ and $q_2$, then  $\Lambda (w) /\varpi(w)$ would be a smooth vector field that produces a diffeomorphism of $\Sigma$. The part of $\mu$ and $\Lambda$ that corresponds to a genuine deformation of moduli is governed by the difference of $\Lambda$ at the points $q_1$ and $q_2$,\footnote{With this convention we have adopted in footnote 5,  the Green function $G$ satisfies  $\pbw G(w;x,y;p_1,p_2) = 2 \pi \delta(w-x) - 2 \pi \delta (w-y)$, a formula that will be used here and below.}
 \bea
\label{4.Lambda2}
\Lambda(q_1)- \Lambda(q_2) = - { 1 \over 2 \pi} \int _\Sigma d^2w \, \mu (w) \varpi(w) G(w;q_1,q_2;p_1,p_2)
\eea
The fact that only a single modulus suffers a deformation follows from the fact that, in unitary gauge, the difference of the period matrix and the super period matrix $\Omega - \hat \Omega$ is of rank 1 due to the linear dependence of $\om_I(q_1)$ and $\om_I(q_2)$.

\sm

The difference $\Lambda (q_1)-\Lambda (q_2)$ is an intrinsic quantity which was evaluated in equation (9.10) of \cite{DP6} and is given by,
\bea
\label{Lamminus}
\Lambda (q_1) - \Lambda (q_2) = { \zeta ^1 \zeta ^2 \over 8 \pi^2} \, { c_2 \over c_1} \, \p \varpi(q_2)
= - { \zeta ^1 \zeta ^2 \over 8 \pi^2} \, { c_1 \over c_2} \, \p \varpi(q_1)
\eea
Without loss of generality, we choose the additive constant so as to set the sum to zero,
\bea
\label{Lamplus}
\Lambda (q_1) + \Lambda (q_2) =0
\eea
in which case we have the simplified expression, 
\bea
\Lambda (q_\a) = - { \zeta ^1 \zeta ^2 \over 16 \pi^2} { c_\a^2 \p \varpi(q_\a) \over c_1 c_2}
\eea
We shall also make use of formula (\ref{varpi4}) which was  established in equation (9.29) of \cite{DP6}.

\subsection{Cancellation of the dependence on $\Lambda (z_i)$ and $\p \Lambda (z_i)$}
\label{sec:44}

We shall now obtain simple expressions for the dependence of the correlators $\cW'$ in section \ref{sec:242} on the gauge function $\Lambda$ at the vertex points $z_i$ which capture the contributions $\cF_{2 \psi}$ of the fermionic stress tensor. There will be further contributions to $\cW'$ arising from $\Lambda (q_\a)$ as pointed out in the preceding subsection, but they will be dealt with later on. We shall verify that these contributions are cancelled by the analogous contributions $\Lambda (z_i)$  and $\p \Lambda (z_i)$  from the bosonic stress tensor and from $\cF_5$.

\sm

As exhibited in subsection \ref{sec:242}, the contributions to $\cW'$ are given by a sum over closed cycles of Szeg\"o kernels, with a single substitution of $S'_\delta$ for $S_\delta$.  To obtain the dependence of $\cW'$ on $\Lambda$ at the vertex points, we evaluate its contribution to $S'_\delta(x,y)$ via (\ref{defSprime}),
\bea
S'_\delta (x,y) \eta^{\mu \nu}  =  { 1 \over 2 \pi} \, S_\delta (q_1,q_2)  
\int_\Sigma d^2w \,  \Lambda (w)  \pbw \left (  { 1 \over \varpi(w) } \, \< T_\psi(w) \psi^\mu_+(x) \psi^\nu_+(y)  \> \right )
\eea
Since $\varpi(w)$ is holomorphic it has precisely two zeros, namely at $q_1$ and $q_2$. Therefore, we may separate the  contributions involving $\Lambda (q_\a)$, which will be denoted with a superscript ${}^{(q)}$ throughout,  and those which involve only the dependence on $\Lambda$ at the vertex points, 
\bea
S'_\delta (x,y) \eta^{\mu \nu} =  S'_\delta(x,y) ^{(q)} 
\eta^{\mu \nu} + { 1 \over 2 \pi} \, S_\delta (q_1,q_2)  
\int_\Sigma d^2w \,  { \Lambda (w) \over \varpi(w)} \, \pbw  \< T_\psi(w)
 \psi^\mu_+(x) \psi^\nu_+(y) \>  
\eea
The derivative may be calculated with the help of the OPE of $T_\psi$ with $\psi_+$ given by,
\bea
T_\psi (w) \psi^\mu_+(x) = { \half \psi^\mu_+(x) \over (w-x)^2} + { \p_x \psi^\mu_+(x) \over w-x} + \hbox{ regular}
\eea
and we find, 
\begin{align}
S'_\delta (x,y)  =  S'_\delta(x,y) ^{(q)} &- S_\delta (q_1,q_2) \left (  { \p \Lambda (x) \over 2 \varpi(x)} -  { \Lambda (x) \p \varpi(x) \over 2 \varpi(x)^2}  + { \Lambda (x) \over \varpi(x)} \p_x \right ) S_\delta (x,y)  
 \\
&- S_\delta (q_1,q_2) \left (  { \p \Lambda (y) \over 2 \varpi(y)} -  { \Lambda (y) \p \varpi(y) \over 2 \varpi(y)^2}  + { \Lambda (y) \over \varpi(y)} \p_y \right ) S_\delta (x,y) \notag 
\end{align}
In any of the $\cW'$, it is always combinations of the form $S_\delta '(u,x) S_\delta (x,y) + S_\delta (u,x) S_\delta ' (x,y)$ that enter, and their contributions involving $\Lambda(x)$ are given by, 
\bea
S_\delta '(u,x) S_\delta (x,y) + S_\delta (u,x) S_\delta ' (x,y)
& \to &  - S_\delta (q_1, q_2) \, \p_x \left ( { \Lambda (x) \over \varpi(x)} S_\delta (u,x) S_\delta (x,y) \right )
\eea
where contributions involving $\Lambda (q_\a)$, $\Lambda(u), \Lambda (y)$ have been suppressed.
As a result, assembling all contributions, and including now the dependence on $\Lambda (z_i)$ and $\p \Lambda (z_i)$ for all the vertex points $z_i$, we find,
\bea
\cW_{1,\cdots, n} ' [\delta] = \cW_{1,\cdots, n} ' [\delta] ^{(q)} 
- S_\delta (q_1, q_2) \, \sum _{i=1}^n \p_{z_i} \left ( { \Lambda (z_i) \over \varpi(z_i)} \, \cW_{1,\cdots, n}[\delta]  \right )
\label{cwprime}
\eea
where $\cW_{1,\cdots, n} ' [\delta] ^{(q)}$ will be discussed in sections \ref{sec4.5} to \ref{sec4.7}.
There are two further contributions that yield dependences on $\Lambda (z_i)$ and $\p \Lambda (z_i)$, namely through the vertex operators $\cV^{(2)}_i$ which give rise to the contribution $\cY_5[\delta]$ and through the action of the bosonic part of the stress tensor in $\cY_{2x}[\delta]$. We discuss these in turn.

\subsubsection{Contributions in $\Lambda (z_i)$ from $\cY_5[\delta]$}

The contribution $\cY_5[\delta]$ is given in (\ref{Ys}), and may be expressed in terms of the vertex operator $\cV^{(0)}_i$ only, using (\ref{vertex}). Furthermore, we can extend the range of the partial derivative and complete the partial differential into a total differential $d_{z_i} = dz_i \p_{z_i} + d \bar z_i \p_{\bar z_i}$, as follows,\footnote{Henceforth, the range of the summation variable $i$  in sums and products over the labels of vertex points will be $1 \leq i \leq 5$ unless indicated otherwise. The summation over the index $\alpha$ on $q_\a$ will be $\alpha =1,2$.}
\bea
\label{Y5new}
\cY_5 [\delta] & = & 
- S_\delta (q_1, q_2) \sum _i d _{z_i} \left ( {  \Lambda (z_i) \over \varpi (z_i)} 
\, \left \< Q(p) \, \prod _{j} \cV_j ^{(0)} \right \> \right )
\no \\ &&
+ S_\delta (q_1, q_2) \sum _i d z_i \, \p_{z_i}  \left ( { \Lambda (z_i) \over \varpi (z_i)} 
\, \left \< Q(p) \, \prod _{j} \cV_j ^{(0)} \right \> \right )
\no \\ &&
+ 2 \pi S_\delta (q_1,q_2) \sum _i  \sum _\alpha \delta (z_i,q_\a) { \Lambda (q_\a) \, \over \p \varpi (q_\a)} 
\, \left \< Q(p) \, \prod _{j} \cV_j ^{(0)} \right \> 
\eea
The first term is an exact form and cancels upon pairing against anti-holomorphic or exact forms on the right-moving sector  and so may be omitted. The second term will cancel the derivatives in (\ref{cwprime}) from the
action of $\partial_{z_i}$ on $ \Lambda(z_i) / \varpi(z_i)$ or the fermionic variables in $\cV_j ^{(0)}$.
However, the second term of (\ref{Y5new}) additionally will give contributions from $\partial_{z_i}$ acting on
the bosonic variables whose cancellation against $\cY_{2x}[\delta]$ will be explained next. The last term
of (\ref{Y5new}) involves $\Lambda (q_\a)$ and thus will not contribute to the dependences on $\Lambda (z_i)$.

\subsubsection{Contributions in $\Lambda(z_i)$ from $\cY_{2x}[\delta]$}

The contribution from the bosonic  stress tensor $\cY_{2x}[\delta]$ is given by,
\bea
\cY _{2x} [\delta] 
& = &
 {1 \over 2 \pi } S_\delta (q_1, q_2) \left \< Q(p)\, \int_\Sigma d^2w \,  \mu (w)  T_x (w) ~ \prod _j \cV_j ^{(0)} \right \> _{(c)}
 \eea
The effect of the stress tensor insertion at the vertex points $z_i$ is to transform all bosonic operators according to the diffeomorphism Ward identities.  There will also be contributions from the zeros of $\varpi(w)$ which involve $\Lambda (q_\a)$ and will be discussed later.  The sum of the contributions involving $\Lambda (z_i)$  and $\p \Lambda (z_i)$ from $\cY_{2\psi}[\delta]$ and $\cY_{2x} [\delta]$ are precisely cancelled by the second term in (\ref{Y5new}). Hence, the entire dependence on $\Lambda (z_i)$ and $\p \Lambda (z_i)$ cancels, even before summing over spin structures, as is expected from diffeomorphism invariance on~$\Sigma$.

\subsection{$\Lambda(q_\a)$-dependence from $\cW'$ for $\leq 3$ vertex points}
\label{sec4.5}

Having now dispensed with the contributions involving $\Lambda$ and its derivative at the vertex points $z_i$, we proceed to collect the remaining contributions, namely those involving $\Lambda (q_\a)$, which arise from the poles at the points $q_\a$ and have been denoted by $\cW_{1, \cdots, n} ' [\delta]^{(q)}$ in (\ref{cwprime}).
Even for the case of four vertex points, the simplification of being allowed to discard the contributions involving $\Lambda (z_i)$ and $\p \Lambda (z_i)$ produces major simplifications whose advantage was not exploited in \cite{DP6}. For this reason, we shall start here by reviewing the contributions with four and fewer vertex points from this vantage point, and then proceed to the case of five vertex points. 

\sm

To obtain the dependence on $\Lambda(q_\a)$ for the cases of four or fewer vertex points, we need to extract the dependence on $\Lambda(q_\a)$ of the integrals against the Beltrami differential $\mu$  of the spin structure sums $I_{13}$, $I_{14}$, $I_{15}$, $I_{16}$ given in (\ref{I13}) and evaluated in (\ref{I13S}). It was proven in \cite{DP6} that the spin structure sums of the first three correlators vanish,
\bea
\sum _\delta \cZ[\delta] \cW'_1[\delta] = \sum _\delta \cZ[\delta] \cW'_{1,2}[\delta] = \sum _\delta \cZ[\delta] \cW'_{1,2,3}[\delta] = 0
\eea
The first vanishes in view of $\f[\delta](w;z_1,z_1)=0$, the second because the pairing of $I_{13}$ with $\mu$ vanishes, and the third because the cyclic sum of the pairing of $I_{14}$ with $\mu$ vanishes.

\subsection{$\Lambda(q_\a)$-dependence from $\cW'$ for 4 vertex points and cancellation}

To extract the dependence on $\Lambda(q_\a)$ of the spin structure sum of $\cW'_{1,2,3,4}[\delta]$ paired against $\mu$, we evaluate first the dependence on $\Lambda(q_\a)$ of the pairing against $\mu$ of $I_{15}$ and $I_{16}$, given by the following integrals for $a=15,16$, 
\bea
\cI_{a} (1,2,3,4) & = &  { 1 \over 2 \pi} \int _\Sigma  d^2w \, \mu (w) \,  I_{a} (w;1,2,3,4)
\label{enccalI}
\eea
Inspection of $I_{15}$ and $I_{16}$ in (\ref{I13S}) reveals that, as a function of $w$, they have poles at the vertex points which produce terms in $\Lambda (z_i)$ and its derivative, but those terms were already shown to cancel in the preceding subsection. Its values at $q_\a$ are finite, and may be determined by taking the following limit, 
\bea
\lim_{w \to q_\b} \varpi(w) G(z;p_1,p_2;q_\a,w) = 
c_\beta \p \varpi (q_\beta) \varpi(z) { \Delta (p_1,p_2)  \over \varpi(p_2) \varpi(p_1) }
\eea
when $\beta \not=\alpha$ and vanish when $\beta =\alpha$. 
Using this limit in the evaluation of $\cI_{15}$, along with (\ref{Lamminus}) and (\ref{varpi4}), we find, 
\bea
\cI_{15}^{(q)} (1,2,3,4)= {\zeta ^1 \zeta ^2 \over 8 \pi^2} \Delta (1,2) \Delta (3,4) 
\eea
Using the same procedure to evaluate $\cI_{16}$, we find, 
\bea
\cI_{16}^{(q)} (1,2;3,4) = { \zeta ^1 \zeta ^2 \over 8 \pi^2} 
\Big ( \Delta (1,3) \Delta (2,4) + \Delta (1,4) \Delta (2,3) \Big ) 
\eea
in agreement with the formulas (8.8), (8.12) and (8.14) given in \cite{DP6} for $\cI_{15}$ and $\cI_{16}$.
As a result, the spin structure sum is given as follows,
\bea
\sum _\delta \cZ[\delta] \cW_{1,2,3,4}' [\delta] 
&=& 
 \half \sum _{(i,j,k,\ell)_c} \!\! (f_if_jf_kf_\ell) \cI_{15}(i,j,k,\ell)
+ \half \sum_{(i,j|k,\ell)} \! \! (f_if_j)(f_kf_\ell) \cI_{16}(i,j;k,\ell)
\no \\ &=& 
{\zeta ^1 \zeta ^2 \over 16 \pi^2}\Big ( \Delta (1,3)\Delta(2,4) + \Delta (1,4) \Delta(2,3) \Big ) 
\no \\ && \qquad \times 
\Big ( 2  (f_1f_3f_2f_4)  + (f_1f_2) (f_3f_4) \Big ) 
+ {\rm cycl}(2,3,4)
\label{cancelw4A}
\eea
The number of permutations $(i,j,k,\ell)_c$ equals the number of partitions $(i,j|k,\ell)$ which allows us to group the terms as we did in the second line.

\subsubsection{Cancellation against the disconnected part $\cF^{(d4)}$}

The contribution $\cF^{(d4)}$ from the disconnected part for four vertex operators with fermion bilinears  was introduced in (\ref{d4andd5}). The spin structure sum due to fermion bilinears in external legs $1,2,3,4$ is given by,
\bea
\cF^{(d4)}_{1234}  & = &
\cN_5 \sum _\delta { \Xi_6[\delta] \tet[\delta](0)^4 \over 16 \pi^6 \Psi _{10}} \cW_{1,2,3,4}[\delta] 
\no \\ 
&=&  - {\cN_5 \over 16 \pi^2}\Big ( \Delta (1,3)\Delta(2,4) + \Delta (1,4) \Delta(2,3) \Big ) \Big ( 2  (f_1f_3f_2f_4)  + (f_1f_2) (f_3f_4) \Big ) 
\no \\ && 
+ {\rm cycl}(2,3,4)
\label{cancelw4B}
\eea
where $\cN_5$ is the chiral Koba-Nielsen factor arising from the contractions of the exponentials $\prod _i e^{i k_i \cdot x_+(z_i)}$ and the loop momentum operator $Q(p)$;  it will be given explicitly in (\ref{5.N5a}) and (\ref{5.N5b}). 
The two contributions (\ref{cancelw4A}) and (\ref{cancelw4B}) are seen to cancel one another exactly (after multiplying the second by $\zeta ^1 \zeta ^2$, or integrating the first over $\int d^2 \zeta$).  This cancellation persists inside any correlator in which only the bosonic field $x_+$ is contracted, i.e.\ after dressing (\ref{cancelw4A}) and (\ref{cancelw4B}) with the extra contribution from $\varepsilon_5^\mu \partial_{z_5} x_+^{\mu}(z_5)$ in the bosonic correlator.

\subsection{$\Lambda(q_\a)$-dependence from $\cW'$ for 5 vertex points}
\label{sec4.7}

The explicit expression for the insertion of the fermionic stress tensor is given by,
\bea
\cW'_{1,2,3,4,5}[\delta] & = & 
\sum_{(i,j,k|\ell,m)} i(f_if_jf_k) (f_\ell f_m) S_\delta(i,j) S_\delta(j,k) S_\delta(k,i) S'_\delta(\ell,m) S_\delta(m,\ell) 
\no \\ &&
+ \sum_{(i,j,k | \ell,m)} {i \over 2} (f_if_jf_k) (f_\ell f_m) S_\delta ' (i,j) S_\delta(j,k) S_\delta(k,i) S_\delta(\ell,m) S_\delta(m,\ell) 
\no \\ &&
- \sum _{(i,j,k,\ell,m)_r} i(f_if_jf_kf_\ell f_m) S_\delta '(i,j) S_\delta(j,k) S_\delta(k,\ell) S_\delta(\ell,m) S_\delta(m,i) 
\eea
The sum on the first   line is over all partitions $(i,j,k|\ell, m)$ of $(1,2,3,4,5)$ into $3+2$; the sum on the second line over all permutations $(i,j,k|\ell, m)$ modulo swapping $\ell,m$; while the last sum  is over all permutations $(i,j,k,\ell, m)$ of $(1,2,3,4,5)$ modulo reversal $(i,j,k,\ell,m) \to (m,\ell,k,j,i)$.  Performing the sum over spin structures, we obtain, 
\bea
\label{ZWp}
\cF_{2 \psi} ~ \to ~ \sum _\delta \cZ[\delta] \cW'_{1,2,3,4,5}[\delta] & = & 
 \sum_{(i,j,k|\ell,m)} {i \over 2} (f_if_jf_k) (f_\ell f_m) \, \cJ_{10} (i,j,k;\ell, m) 
\no \\ &&
+ \sum_{(i,j,k|\ell,m)} {i \over 4} (f_if_jf_k) (f_\ell f_m) \, \cJ_{11}^S(i,j,k;\ell,m)
\no \\ &&
+ \sum _{(i,j,k,\ell,m)_c} { i \over 2} (f_if_jf_kf_\ell f_m) \, \cJ_{12}^S (i,j,k,\ell,m)
\eea
The corresponding symmetrized functions are defined as follows, 
\bea
\cJ_{11} ^S(1,2,3;4,5) & = & \cJ_{11} ^{(q)} (1,2,3;4,5) + \cJ_{11} ^{(q)} (2,3,1;4,5) + \cJ_{11} ^{(q)} (3,1,2;4,5)
\no \\
\cJ_{12} ^S (1,2,3,4,5) & = & \cJ_{12} ^{(q)}  (1,2,3,4,5) + \cJ_{12} ^{(q)}  (2,3,4,5,1) + \cJ_{12} ^{(q)}  (3,4,5,1,2)
\no \\ && 
+ \cJ_{12} ^{(q)}  (4,5,1,2,3) + \cJ_{12} ^{(q)}  (5,1,2,3,4)
\label{symmJ11}
\eea
The sums in (\ref{ZWp}) on the first and second lines are over all partitions $(i,j,k|\ell,m)$ of $(1,2,3,4,5)$ into $3+2$, and the sum in the third line  is over all permutations $(i,j,k,\ell,m)$ of $(1,2,3,4,5)$ modulo cyclic permutations and reversal.  The functions $\cJ_{10}, \cJ_{11}, \cJ_{12}$ are the integrals against $\mu$ of the functions $J_{10}, J_{11}, J_{12}$, respectively, and are defined by, 
\bea
\label{calJdef}
\cJ_a (1,2,3,4,5) = { 1 \over 2 \pi} \int _\Sigma  d^2w \, \mu(w) J_a(w;1,2,3,4,5)
\eea
These functions are evaluated in appendix \ref{sec:F}, the last two  in terms of the pairings of the functions $L_1$ and $L_2$.  Having already shown the cancellations of any contributions involving $\Lambda (z_i)$ and $\p \Lambda (z_i)$, we retain here only the contributions that involve $\Lambda (q_\a)$, and use the freedom to shift $\Lambda$ by a constant to set $\Lambda (q_1) +\Lambda (q_2)=0$. The results from appendix \ref{sec:F} are as follows. For $\cJ_{10}$ we find agreement with the expression for $J_1$ in (\ref{J12.6}), 
\bea
\label{calJ10}
\cJ_{10}^{(q)} (1,2,3;4,5) & = &  { \zeta ^1 \zeta ^2 \over 32 \pi^6 \Psi _{10}}
J_1(1,2,3;4,5) 
\eea
For $\cJ_{11}^S$ we have, 
\bea
\label{calJ11main}
\cJ_{11} ^S (1,2,3;4,5) & = &
 { \zeta ^1 \zeta ^2 \over 8 \pi^2} \om_I(1) \Delta (2,4) \Delta (3,5) \Big ( g_{2,4}^I + g^I_{4,5} + g^I_{5,3} + B_2^I - B_3^I \Big ) + \hbox{ cycl}(1,2,3)
\no \\ &&
+ { \zeta ^1 \zeta ^2 \over 8 \pi^2} \om_I(1) \Delta (2,5) \Delta (3,4) \Big ( g_{2,5}^I + g^I_{5,4} + g^I_{4,3} + B_2^I - B_3^I \Big ) + \hbox{ cycl}(1,2,3)
\no \\ &&
+ { \zeta ^1 \zeta ^2 \over 16 \pi^6 \Psi_{10}} J_1(1,2,3;4,5) 
\eea
where the cyclic permutations are applied to the first two lines only and we have 
exposed a contribution of $J_1$ for later convenience. For $\cJ_{12}^S$ we find, 
\bea
\label{calJ12}
\cJ_{12}^S(1,2,3,4,5)
& = & 
- { \zeta ^1 \zeta ^2 \over 8 \pi^2} \, \om_I(1) \Delta (2,3) \Delta (4,5)
\Big (  g_{2,3}^I +  g^I_{4,5}  +B_2^I-B_3^I+B_4^I -B_5^I \Big ) 
\no \\ &&
- { \zeta ^1 \zeta ^2 \over 8 \pi^2} \, \om_I(1) \Delta (2,5) \Delta (3,4)
\Big (    g^I_{2,5} +B_2^I-B_5^I \Big ) 
 + \hbox{ cycl}(1,2,3,4,5)
 \no \\ && 
 + { \zeta^1 \zeta ^2 \over 8 \pi^6 \Psi_{10}} J_2(1,2,3,4,5)
\eea
where the cyclic permutations are applied to the first two lines only and the
identification of $J_2$ as given in (\ref{J12.6}) will be exploited in the next subsection. 
The functions $B^I_i$ were obtained in (\ref{defBI}), and are given by,
\bea
\label{defBIa}
B^I_i  & = & 
\sum _\a { c_\a \ep ^{IJ}  \over 2 \, \p \varpi (q_\a)} 
\Big (   \p_{q_\a} \om_J(q_\a) \tau_{z_i,z_0} (q_\a) - \om_J(q_\a)  \p_{q_\a} \tau_{z_i,z_0} (q_\a)  \Big )
\eea
where
\bea
\tau_{x,y}(z) = \partial_z \ln E(z,x) - \partial_z \ln E(z,y) = \omega_I(z) (g^I_{z,x} - g^I_{z,y})
\label{deftau}
\eea
Here $z_0$ is an arbitrary reference point which cancels out of the differences $B_i^I-B_j^I$. The functions $B_i^I$ are single-valued $(0,0)$-forms in the points $q_\a$ with double poles in $z_i$ at $q_1$ and $q_2$. 
An alternative presentation is in the cyclic basis,
\bea
\cJ_{12} ^S (1,2,3,4,5) & = &
{ \zeta ^1 \zeta ^2 \over 8 \pi^2} \om_I(1) \Delta (2,3) \Delta (4,5) \Big ( 
 2 g^I_{1,2} + 2 g^I_{5,1} + g^I_{4,5} + g_{2,3}^I + 3 g^I_{3,1} 
 \no \\ &&
+ 3 g^I_{1,4}  -B_2^I+2B_3^I-2B_4^I +B_5^I \Big ) + \hbox{ cycl}(1,2,3,4,5)
\eea

\subsubsection{Contribution from the disconnected part $\cF^{(d)}$}

The contribution $\cF^{(d5)}$ from the disconnected part for five vertex operators with a fermion bilinear
was defined in (\ref{d4andd5}) and is given by,
\bea
\cF^{(d5)} = \cN_5 \sum _{\delta} { \Xi_6[\delta] \tet[\delta](0)^4 \over 16 \pi^6 \Psi_{10}} \cW_{1,2,3,4,5}[\delta]
\eea
where $\cN_5$ is the chiral Koba-Nielsen factor given in (\ref{5.N5a}) and (\ref{5.N5b}).
Expressing the spin structure sums in terms of the functions $J_1, J_2$ we find, 
\bea
\cF^{(d5)} & = &
- { i \cN_5 \over 32 \pi^6 \Psi_{10}}  \sum_{(i,j,k|\ell,m)} (f_if_jf_k) (f_\ell f_m) \, J_1 (i,j,k;\ell, m) 
\no \\ &&
- { i \cN_5 \over 16 \pi^6 \Psi_{10}} \sum _{(i,j,k,\ell,m)_c}  (f_if_jf_kf_\ell f_m) \, J_2 (i,j,k,\ell,m)
\eea
where the sum on the first line is over all 10 partitions $(i,j,k|\ell,m)$ into $3+2$, while the sum on the second line is over all 12 permutations modulo cyclic and reversal.

\subsection{Combining $\cF^{(d)}$ and $\cF_{2 \psi}$}
\label{sec:4.8}

In combining the contributions from $\cF_{2 \psi}$ and $\cF^{(d)}$ for the five-point amplitude, we use the facts that (1) the contributions with three or fewer fermion bilinears from the vertex operators cancel separately for $\cF_{2 \psi}$ and $\cF^{(d)}$; (2) the contributions with four fermion bilinears from $\cF_{2 \psi}$ and $\cF^{(d)}$ precisely cancel one another as seen in (\ref{cancelw4A}) and (\ref{cancelw4B}), including when these four fermion bilinears occur in the five point function; (3) the contributions with five fermion bilinears are given in the above part. Assembling all contributions, the result is as follows, 
\bea
\label{4.FdF2}
\cF^{(d)} + \cF_{2 \psi} & = & 
{ i \cN_5 \over 32 \pi^2} \sum_{(i,j,k|\ell,m)} (f_if_jf_k) (f_\ell f_m) \, \cC_1 (i,j,k;\ell, m) 
\no \\ &&
- { i \cN_5 \over 16 \pi^2}  \sum _{(i,j,k,\ell,m)_c}  (f_if_jf_kf_\ell f_m) \, \cC_2 (i,j,k,\ell,m)
\eea
where the functions $\cC_1, \cC_2$ (not to be confused with the correlators in (\ref{c0andc2})) are given by,
\bea
\cC_1 (i,j,k;\ell,m) & = & -  { J_1(i,j,k;\ell,m) \over \pi^4 \Psi_{10}} 
+ 8 \pi^2 \int d^2 \zeta \, \Big ( 2 \cJ_{10} (i,j,k;\ell,m) + \cJ_{11}^S (i,j,k;\ell,m) \Big )
\no \\
\cC_2 (i,j,k,\ell,m) & = &{ J_2(i,j,k,\ell,m)  \over \pi^4 \Psi_{10}} - 8 \pi^2 \int d^2 \zeta \,  \cJ_{12}^S (i,j,k,\ell,m) 
\eea
The first term on each line arises from $\cF^{(d)}$ while the remaining terms arise from $\cF_{2\psi}$.
Inspection of (\ref{calJ10}) and (\ref{calJ11main}) reveals that the contributions in $J_1$ cancel, so that we are left with the following results for $\cC_1$,  
\bea
\label{4.C1}
\cC_1 (1,2,3;4,5) & = &  \om_I(1) \Delta (2,4) \Delta (3,5) 
\Big ( g_{2,4}^I + g^I_{4,5} + g^I_{5,3} + B_2^I - B_3^I \Big ) 
\no \\ &&
+ \hbox{ cycl}(1,2,3|4,5)
\eea 
where cycl$(1,2,3|4,5)$ stands for all six cyclic permutations of 1,2,3 combined with swaps of~4,5. Similarly, inspection of (\ref{calJ12}) reveals that the contributions in $J_2$ cancel, so that we are left with the following results for $\cC_2$,  
\bea
\cC_2 (1,2,3,4,5) & = & \om_I(1) \Delta (2,3) \Delta (4,5)
\Big (  g_{2,3}^I +  g^I_{4,5}  +B_2^I-B_3^I+B_4^I -B_5^I \Big ) 
\no \\ &&
+ \om_I(1) \Delta (2,5) \Delta (3,4)
\Big (    g^I_{2,5} +B_2^I-B_5^I \Big ) 
 + \hbox{ cycl}(1,2,3,4,5)
\eea
where cycl$(1,2,3,4,5)$ applies to both lines of $\cC_2$. 
A useful alternative presentation of $\cC_2$ in terms of the canonical cyclic basis is given by,
\bea
\label{4.C2}
\cC_2 (1,2,3,4,5) & = & \om_I(1) \Delta (2,3) \Delta (4,5)
\Big (  g_{2,3}^I +  g^I_{4,5} +g_{1,3}^I +g^I_{4,1} 
\no \\ && 
+ B_2^I-2B_3^I+2B_4^I -B_5^I \Big ) 
 + \hbox{ cycl}(1,2,3,4,5)
\eea
Note that, in view of the identity,
\bea
\om_I(1) \Delta (2,4) \Delta (3,5) g^I_{4,5} + \hbox{ cycl}(1,2,3|4,5) =0
\eea
the dependence on $g_{4,5}^I$ actually cancels out from (\ref{4.C1}). Similarly,
the overall coefficient of $g_{1,3}^I$ in the cyclic orbit in (\ref{4.C2}) is proportional to $\om_I(1) \Delta (2,3) 
+ \om_I(3) \Delta (1,2) =\om_I(2) \Delta (1,3)$. As a result, none of the $g_{a,b}^I$ terms in either $\cC_1$ or $\cC_2$ produces a singularity at coincident vertex points, and the expression is always accompanied by a factor $\Delta(a,b)$ which vanishes. Thus, the only singularities of $\cC_1, \cC_2$ arise from the double pole of $B_i^I$ in $z_i$ as $z _i \to q_\a$. The fact that all dependence on $B_i^I$ arises in the form $g_{a,b}^I +B_a^I - B^I_b$ guarantees the cancellation of monodromy in $z_a, z_b$.

\sm

The cancellation of these double poles requires contributions from $\cF_{1xx}$ and $\cF_{2x}$ that will be worked out in the next section. Therefore the cancellation of all double poles, along with the conversion of all $(0,1)$-form contributions into exact total differentials, will have to wait until section \ref{sec:6}.

 \newpage

%%%%%%%%%%%%%%%%%%%%%%%%%%%%%%%%%%%%%%%%%%%
%%%%%%%%%%%%%%%%%%%%%%%%%%%%%%%%%%%%%%%%%%%
\section{Assembling the chiral amplitude}
\label{sec:5}
\setcounter{equation}{0}
%%%%%%%%%%%%%%%%%%%%%%%%%%%%%%%%%%%%%%%%%%%
%%%%%%%%%%%%%%%%%%%%%%%%%%%%%%%%%%%%%%%%%%%

In this section, we shall assemble the various contributions to the chiral amplitude, and simplify their dependence on any remaining  choices of slice. In subsequent sections we shall show that all remaining slice-dependence cancels or results in the addition of exact differentials, whose contribution to the full integrated amplitudes automatically cancel.

\subsection{Combining the contributions of $\cF_{1xx}$ and $\cF_{2x}$}

We recall the definition of the contributions $\cF_{1xx}$ and $\cF_{2x}$ obtained by assembling (\ref{Y1xx}) and (\ref{Y2x}) and carrying out the integrations over the odd moduli $\zeta$, 
\bea
\cF_{1xx} & = & { 1 \over 16 \pi^2} \sum _\delta \cZ[\delta] S_\delta (q_1, q_2) 
\Big \< Q(p)\,  \p x _+^\mu (q_1) \p x_+ ^\mu (q_2)  \prod _i \cV_i ^{(0)} \Big \> _{(c)}
\no \\
\cF_{2x} & = & - {1 \over 4 \pi} \sum _\delta \cZ[\delta] S_\delta (q_1, q_2) \int d^2 \zeta \int \mu(w) 
\Big \< Q(p)\,  \p x _+^\mu (w) \p x_+ ^\mu (w)  \prod _i \cV_i ^{(0)} \Big \> _{(c)}
\eea
The contributions to $\cF_{2x}$ involving $\Lambda (z_i)$ and $\p \Lambda (z_i)$ have already been shown to cancel in subsection \ref{sec:44}. To obtain the contributions involving $\Lambda (q_\a)$, we express $\mu$ in terms of $\Lambda$ and $\varpi$, pick up the poles in $w$ at $q_\a$, and use (\ref{Lamminus}) and (\ref{Lamplus}) to obtain,
\bea
\cF_{2x} & = & - \sum _\a {c _\a^2 \over 32 \pi^2 c_1 c_2}  \sum _\delta \cZ[\delta] S_\delta (q_1, q_2) 
\Big \< Q(p)\,  \p x _+^\mu (q_\a) \p x_+ ^\mu (q_\a)  \prod _i \cV_i ^{(0)} \Big \> _{(c)}
\eea
Combining the two contributions gives,
\bea
\cF_{1xx}  + \cF_{2x} 
& = &
 - \sum _\delta {\cZ[\delta] S_\delta (q_1, q_2) \over 32 \pi^2c_1 c_2}
\Big \< Q(p)\,  \Big (c_1 \p x _+ (q_1) -c_2 \p x_+(q_2) \Big )^2
 \prod _i \cV_i ^{(0)} \Big \> _{(c)}
 \label{j123parts}
\eea
where the square of the large parentheses is understood in the sense of the square of the Lorentz vector, and the connectedness prescription $\langle \ldots \rangle_{(c)}$ excludes its self-contractions. The other contractions may be performed in terms of the Lorentz-vector-valued meromorphic $(1,0)$-form  $\cP(z)$  given by,\footnote{We shall frequently encounter the Lorentz-contractions $k_i \cdot \cP(z_i)$ and $\ep _i \cdot \cP(z_i)$ in which the kinematic relations $k_i^2= k_i \cdot \ep _i=0$ guarantee the absence of $\p_z \ln E(z,z_i)$ evaluated at $z =z_i$.}
\bea
\cP (z) = 2 \pi i p^I \om_I(z) + \sum _j k _j \, \p_z\ln E(z,z_j)
\label{pfield}
\eea
The contraction of $\p x_+(z)$ with $Q(p)$ and the exponential produce precisely $- i  \cP(z)$.
The contributions to the above sum may be organized into three parts, 
\bea
\cF_{1xx}+\cF_{2x} = \sum _{j=1}^3 \left ( \cF_{1xx}^{(j)}+\cF_{2x}^{(j)} \right )
\eea
 In the $j=1$ part, all five vertex operators have fermion bilinears, so that the two pre-factors $c_1 \p x _+ (q_1) -c_2 \p x_+(q_2)$ are contracted with $Q(p)$ and the exponential. In the $j=2$ part, one vertex has a bosonic $\p x_+$ contracted with $Q(p)$ and the exponential while the remaining four contribute fermion bilinears and the two pre-factors $c_1 \p x _+ (q_1) -c_2 \p x_+(q_2)$ are contracted with $Q(p)$ and the exponential. In the $j=3$ part, one vertex has a bosonic $\p x_+$ contracted with one of the pre-factors $c_1 \p x _+ (q_1) -c_2 \p x_+(q_2)$, the other factor being contracted with $Q(p)$ and the exponential. While the $j=1$ part is gauge-invariant by itself, it is only the combination of the $j=2,3$ parts that is gauge invariant.

\subsubsection{The $j=1$ part:  five vertices with fermion bilinears}

In terms of $\cP(z)$, the $j=1$ part of (\ref{j123parts}) is evaluated to be,
\bea
\cF_{1xx}^{(1)}   + \cF_{2x} ^{(1)}
=
 { \cN_5 \over 32 \pi^2 c_1c_2}  \Big ( c_1 \cP(q_1) - c_2 \cP(q_2) \Big ) ^2  \sum _\delta \cZ[\delta] S_\delta (q_1, q_2)   \cW_{1,2,3,4,5} [\delta]  
\eea
where we have used the following notation for the chiral Koba-Nielsen factor,\footnote{The chiral Koba-Nielsen factor $\cN_5$ was denoted by $\cI_5$ in \cite{DHoker:2020prr}. The notation was changed here to avoid confusion with the functions $\cI$ encountered in (\ref{enccalI}) and appendix \ref{sec:F}.}
\bea
\label{5.N5a}
\cN_5 = \Big \< Q(p) \prod _i e^{i k_i \cdot x_+(i)}  \Big \> 
\eea
Its evaluation gives,
\bea
\label{5.N5b}
\ln \cN_5 = i \pi \, \Omega _{IJ} p^I \cdot p^J + 2 \pi i  \sum _i k_i \cdot p^I \int ^{z_i}_{z_0}\om_I + \sum _{ i <j}
k_i \cdot k_j \ln E(z_i,z_j)
\eea
where the dependence on the endpoint $z_0$ drops out by momentum conservation. The function $\cP(z_i)$ is related to $\cN_5$ by,
\bea
\label{pN5}
\p_{z_i} \ln \cN_5 = k_i \cdot \cP(z_i)
\eea
Thanks to the first relation in (\ref{fund1}), the dependence on the loop momentum drops out of the combination $c_1  \cP(q_1) - c_2  \cP(q_2)$, and thanks to momentum conservation, we can use (\ref{fund3}) to simplify this combination as follows,
\bea
\label{5.cP}
c_1  \cP(q_1) - c_2  \cP (q_2) 
=  - c_1^2 \p \varpi(q_1) \sum _j k _j \, D(z_j,z_0)
\eea
where we have defined the following combination,
\bea
D(z,w) = { \Delta (z,w) \over \varpi (z) \varpi(w)} = - D(w,z)
\eea
The combination $D$ is a single-valued scalar in $z,w$ and a multiple-valued $(-2,0)$-form in $q_\a$ which satisfies $D(w,z)=-D(z,w)$. In view of momentum conservation, the expression in (\ref{5.cP}) is independent of the point $z_0$, as may be seen by considering the difference,
\bea
\label{sums}
D(z_j,z_0) = D(z_j,z_0') + D(z_0',z_0)
\eea
Putting all together using (\ref{Lamminus}) as well as the second and third formulas in (\ref{varpi4}) we find, 
\begin{align}
\label{twolines}
\cF_{1xx}^{(1)}   + \cF_{2x} ^{(1)}
 =  -  { \cN_5 \Dt \over 32 \pi^2 \cZ_0 \, \rho }  \sum _\delta \cZ[\delta] S_\delta (q_1, q_2)  
 \cW_{1,2,3,4,5} [\delta]  
\end{align}
where $\Dt$ is defined by, 
\bea
\label{5.Dt}
\Dt = \rho \sum _{i,j} k_i \cdot k_j D(z_i,z_0) D(z_j,z_0')
\eea
for arbitrary points $z_0$ and $z_0'$, and $\rho$ was defined in (\ref{rhoi}). Each factor of $D(z_i,z_0)$ has a simple pole as $z_i \to q_\a$, but the prefactor $\rho$ guarantees that all such poles cancel in $\Dt$. Thus,   $\Dt$ is a holomorphic $(1,0)$-form in each vertex point $z_i$, which is manifestly invariant under all simultaneous permutations of the pairs $(k_i, z_i)$. The arbitrary points $z_0$ and $z_0'$ may be eliminated at the cost of manifest permutation symmetry. Below we provide two alternative expressions for $\Dt$ that no longer involve $z_0$ and $z_0'$ and are manifestly holomorphic.
\begin{itemize}
\itemsep=0in
\item A first expression for $\Dt$, which has only manifest cyclic symmetry, is given by,
\bea
\label{7.Lem.1}
\Dt= - 2 k_3 \cdot k_4 \, \varpi (1) \Delta (2,3) \Delta (4,5) + \hbox{\rm cycl}(1,2,3,4,5)
\eea
\item A second  expression in which the point 1 is singled out is as follows,
\bea
\label{7.Lem.2}
\Dt & = & 2k_1 \cdot \Big [ k_2 \varpi(4) \Delta (1,3) \Delta (2,5)
+ k_3 \varpi(4) \Delta (1,2) \Delta (3,5)
+ k_4 \varpi(3) \Delta (1,2) \Delta (4,5) \Big ]
\no \\ && 
- 2\varpi(1) \Big [ k_3 \cdot k_4 \Delta (2,3) \Delta (4,5) + k_4 \cdot k_5 \Delta (3,4) \Delta (5,2) \Big ]
\eea
and cyclic permutations thereof. This expression has no manifest permutation symmetry left, but will turn out to be useful in the sequel in formulas where one vertex point is being singled out for other reasons.
\end{itemize}
To prove these formulas one may choose $z_0=z_0'=z_5$ in the definition (\ref{5.Dt}), then use the formulas of (\ref{omdel}) to convert all terms to expose either $\varpi(1)$, $\varpi(2)$ or $\varpi(5)$ upon which the remaining denominator $\varpi(5)$ is found to cancel due to momentum conservation. Finally, decomposing all holomorphic five-forms into the standard cyclic basis gives (\ref{7.Lem.1}). The structure of the right side has an obvious symmetry interpretation inside the pentagon. To obtain (\ref{7.Lem.2}) one again uses (\ref{omdel}) and momentum conservation to rearrange the sum in (\ref{7.Lem.1}).

\sm

The spin structure sum in the second line of (\ref{twolines}) is manifestly (space-time) gauge invariant. Carrying out the sum over spin structures in terms of the functions $J_3$ and $J_4$ defined in (\ref{J34}), 
\bea
\label{ZSW5}
\sum_\delta \cZ[\delta] S_\delta (q_1, q_2)   \cW_{1,2,3,4,5} [\delta]
& = & 
-{i \over 2}  \sum _{(i,j|k,\ell,m)} (f_i f_j) (f_k f_\ell f_m) J_3(k,\ell,m; i,j)
\no \\ &&
- i \sum _{(i,j,k,\ell,m)_c} (f_i f_j f_k f_\ell f_m) \, J_4 (i,j,k,\ell,m)
\eea
we obtain the following explicit form, 
\bea
\label{5.F1xxF2x}
\cF_{1xx}^{(1)}   + \cF_{2x} ^{(1)}
& = &
  {i \,  \cN_5 \, \Dt \over 64 \pi^2 \cZ_0 \rho } 
 \sum _{(i,j|k,\ell,m)} (f_i f_j) (f_k f_\ell f_m) J_3(k,\ell,m; i,j)
\no \\ &&
+  { i \, \cN_5 \,  \Dt \over 32 \pi^2 \cZ_0 \rho }  
 \sum _{(i,j,k,\ell,m)_c} (f_i f_j f_k f_\ell f_m) \, J_4 (i,j,k,\ell,m) 
 \qquad \quad
\eea
Note that neither $J_3$ nor $J_4$ has poles in $z_i$ as $z_i \to q_\a$, and
a simplified form of these spin structure sums can be found in (\ref{J34.2}).

\subsubsection{The $j=2$ part: one bosonic vertex contracted with the exponential}

The $j=2$ part of (\ref{j123parts}) arises from the contribution where one vertex operator is bosonic $\p x_+$ and contracted with $Q(p)$ and the exponential,  while the other four have fermion bilinears, and the two pre-factors $c_1 \p x _+ (q_1) -c_2 \p x_+(q_2)$ are contracted with $Q(p)$ and the exponential as in the case of part one. Putting all together, we get, 
\bea
\label{F2F1}
\cF_{1xx}^{(2)}   + \cF_{2x} ^{(2)}
& = & 
- { i \, \cN_5 \over 32 \pi^2 c_1 c_2} \Big ( c_1 \cP(q_1) - c_2 \cP(q_2) \Big )^2
 \sum _\ell \ep_\ell \cdot  \cP (z_\ell) 
 \no \\ && \qquad \times 
 \sum _\delta \cZ[\delta] S_\delta (q_1, q_2)  \cW_{1 \cdots \hat \ell \cdots 5} [\delta]
\eea
The sum over spin structures may be carried out using $I_{11}$ and $I_{12}$ defined in  (\ref{I11}) and given in (\ref{I11S}), and we find, 
\bea
\label{5.W4K}
\sum _\delta \cZ[\delta] S_\delta (q_1, q_2)  \cW_{1 \cdots \hat \ell \cdots 5} [\delta]
=  2 \cZ_0 \, \rho _\ell \, \mt_\ell
\eea
where $\rho_\ell$ was defined in (\ref{rhoi}) and $\mt_\ell$ was defined in (\ref{newsumm.5}) and (\ref{newsumm.5z}). Substituting this result into the expression in (\ref{F2F1}), we find,
\bea
\cF_{1xx}^{(2)}   + \cF_{2x} ^{(2)} = 
 -  { i\, \cN_5 \, \cZ_0 \over 16 \pi^2 c_1 c_2}\Big ( c_1 \cP(q_1) - c_2 \cP(q_2)  \Big )^2
 \sum _\ell \rho_\ell \, \mt _\ell  \, \ep_\ell \cdot \cP (z_\ell)  
\eea
Next, we use the relation (\ref{5.cP})  to convert the pre-factors, use the definition of $\Dt$ in (\ref{5.Dt}), as well as the relation $\rho= \rho_\ell \, \varpi(z_\ell)$ to obtain, 
\bea
\cF_{1xx}^{(2)}   + \cF_{2x} ^{(2)} = 
 { i \, \cN_5 \, \Dt \over 16 \pi^2 }  \sum _\ell  \mt_\ell  \, \ep_\ell \cdot  { \cP (z_\ell) \over \varpi(z_\ell)} 
\eea

\subsubsection{Part three: one bosonic vertex contracted with a prefactor}

In the third part of (\ref{j123parts}), the bosonic vertex $\p x_+(z_i)$ is contracted with one of the pre-factors $c_1 \p x _+(q_1) -c_2 \p x_+(q_2) $, the other pre-factor being contracted with  $Q(p)$ and the exponential. The contraction of the vertex $\p x_+$ with the prefactor is given by,
\bea
\left \< \p x_+^\mu (z_\ell)  (c_1 \p x _+^\nu  (q_1) -c_2 \p x_+^\nu (q_2) ) \right \> 
& = &
- \eta ^{\mu \nu} \p_{z_\ell} \Big ( c_1  \tau_{z_\ell, z_0''} (q_1) - c_2  \tau_{z_\ell,z_0''} (q_2) \Big ) 
\no \\
& = & c_1^2 \p \varpi (q_1)  \eta ^{\mu \nu} \p_{z_\ell} D(z_\ell,z_0'') 
\eea
where $\tau_{z_\ell, z_0''} (q_\alpha)$ is defined in (\ref{deftau}) and $z_0''$ is an arbitrary point, which is independent of $\ell$.  Also carrying out the sum over spin structures via (\ref{5.W4K}) we find, 
\bea
\cF_{1xx}^{(3)}  + \cF_{2x} ^{(3)}
=
{ i \cN_5  \over 8 \pi^2} 
 \sum_{j,\ell}   \rho_\ell  \, \mt_\ell \, (\ep_\ell \cdot k_j) \,  D(z_j,z_0') \p_{z_\ell} D(z_\ell, z_0'') 
\eea
This contribution may be decomposed into a total differential, plus the result of the integrations by parts,
\bea
\label{F3xx}
\cF_{1xx}^{(3)}  + \cF_{2x} ^{(3)}
& = &
  { i   \over 8 \pi^2} \sum_\ell d_\ell \left ( \cN_5 \, \rho_\ell \, \mt_\ell  
 \sum_j  (\ep_\ell \cdot k_j) D(z_j,z_0') D(z_\ell, z_0'') \right )
 \no \\ &&
+ \tilde \cF_3 - i { \cN_5 \over 8 \pi^2} \sum_{j,\ell} \rho_\ell \,  \mt_\ell \, k_\ell \cdot  \cP (z_\ell) \, (\ep_\ell \cdot k_j) \, D(z_j,z_0') D(z_\ell, z_0'') 
\eea
where the last term arises from differentiating $\cN_5$ with respect to $z_\ell$ and using (\ref{pN5}), and we have defined the following combination, which is a $(0,1)$-form in $z_\ell$,
\bea
\label{5.tilF3}
\tilde \cF_3 =   - { i \, \cN_5  \over 8 \pi^2} \sum_{j,\ell} \rho_\ell  \, \mt_\ell \, 
(\ep_\ell \cdot k_j) \,  D(z_j,z_0') \p_{\bar z_\ell} D(z_\ell, z_0'') 
 \eea
The total differential in $z_\ell$ on the first line of (\ref{F3xx}) may be omitted. Using, 
\bea
\label{pbzD}
\pbz D(z,w) = 2 \pi \sum_\a { c_\a \delta (z,q_\a) \over c_\a^2 \, \p \varpi(q_\a)} 
\eea
we shall relate the contribution $\tilde \cF_3$ to $\cY_3$ and $\cF_3$ in section \ref{sec:6}. Finally, in the last term we use the rearrangement $\ep ^\mu _\ell k^\nu _\ell  = f_\ell ^{\mu \nu} + \ep _\ell ^\nu k_\ell ^\mu$ to produce the following result, 
\bea
\cF_{1xx}^{(3)}  + \cF_{2x} ^{(3)} - \tilde \cF_3
& = &
-  {i \,  \cN_5 \over 8 \pi^2} \sum_{j,\ell} \rho_\ell \,  \mt_\ell \, k_j^\mu \, f_\ell ^{\mu \nu} \, \cP^\nu (z_\ell) \, 
 D(z_j,z_0') D(z_\ell, z_0'') 
 \no \\ &&
 -  { i \, \cN_5 \over 8 \pi^2} \sum_{j,\ell} \rho_\ell \,  \mt_\ell \, \ep _\ell \cdot \cP (z_\ell) \,  (k_j \cdot k_\ell) \, 
 D(z_j,z_0') D(z_\ell, z_0'')  
\eea
The first term is manifestly gauge invariant. Provided we set $z_0''=z_0'=z_0$, the second term will cancel those terms in $\cF_{1xx}^{(2)}  + \cF_{2x} ^{(2)}$ which have either $i=\ell$ or $j=\ell$. Adding these contributions, we have, 
\bea
\label{vanish?}
\cF_{1xx}^{(2)}  + \cF_{2x} ^{(2)}+\cF_{1xx}^{(3)}  + \cF_{2x} ^{(3)} - \tilde \cF_3
& = &   { i \, \cN_5 \over 16 \pi^2 }  \sum _{\ell} \rho_\ell \, \mt_\ell  \,  \ep_\ell \cdot  \cP (z_\ell) 
\sum _{i,j \not= \ell} k_i \cdot k_j D(z_i,z_0) D(z_j, z_0) 
\no \\ && 
 -  { i \, \cN_5 \over 8 \pi^2} \sum_{j,\ell} \rho_\ell \, \mt_\ell  \, k_j^\mu \, f_\ell ^{\mu \nu}  \, \cP^\nu (z_\ell) \, D(z_\ell, z_0)  \,  D(z_j,z_0) 
 \eea
 Although the first term on the right side involves a naked $\ep_\ell$, its gauge invariance is manifest since $\cP(z_\ell)$ and $\cN_5$ are the only factors that involve $z_\ell$ and a linearized gauge transformation on $\ep_\ell$ yields a total differential in view of (\ref{pN5}). The above formula is independent of the point $z_0$ at the expense of total differential terms. Indeed, replacing $z_0$ by $z_0'$ using (\ref{sums}) produces the following additional terms multiplying $D(z_0,z_0')$,
\bea
 { i \, \cN_5 \over 8 \pi^2 }  \sum _{\ell} \rho_\ell \, \mt_\ell  \cP^\nu (z_\ell)  \bigg (  \ep_\ell^\nu 
\sum _{i,j \not= \ell}  k_i \cdot k_j  D(z_j, z_0) 
 -    f_\ell ^{\mu \nu}  \sum_j k_j^\mu    \,  D(z_j,z_0) \bigg )
 \eea  
In the first term under the parentheses, the sum of $k_i$ over $i$  gives $-k_\ell$ which combined with the second term becomes $- \ep^\mu_\ell k^\mu _j k_\ell ^\nu$ and this term contracted with $\cP^\nu(z_\ell)$ and $\cN_5$ combines into a total differential in $z_\ell$. Hence we have independence of the point $z_0$ at the cost of a total differential.

\subsection{Calculation of $\cF_{1n}$}

The contribution $\cF_{1n}$ is obtained from $\cY_{1n}[\delta]$ in (\ref{Y1xx}) 
by summing over spin structures and integrating over odd moduli, 
\bea
\cF_{1n}&=& -  { 1 \over 16 \pi^2} 
\sum _\delta \cZ[\delta] \bigg \< Q(p) \psi _+^\mu (q_1) \p x_+^\mu (q_1) \psi _+^\nu (q_2) \p x_+^\nu (q_2)
\prod _i \cV^{(0)} _i \bigg \>_{(c)}
\eea
The connectedness prescription excludes any contractions between the fields $x_+$ and $\psi_+$ at the points $q_1, q_2$. All spin structure sums vanish unless all five vertex operators give a fermion bilinear, so that we have,
\bea
\label{6.F1n}
\cF_{1n}&=&
 { \cN_5 \over 16 \pi^2} \cP^\mu (q_1) \cP^\nu (q_2) \sum _\delta \cZ[\delta] \cW^{\mu \nu} _{1,2,3,4,5}[\delta]
\eea
To evaluate the spin structure sum, we consult the form of the expression for $\cW^{\mu \nu}[\delta]$ in (\ref{Wmn5}) and the spin structure sums in (\ref{J5-9}). Since the spin structure sum of the combination $\cW^{\mu \nu}_{i,j} [\delta] $ inside a correlator with five vertex points is proportional to $J_5$, which vanishes, it will not contribute to $\cF_{1n}$.  The remaining contributions give, 
\bea
\sum _\delta \cZ[\delta] \cW ^{\mu \nu} _{1,2,3,4,5}[\delta]
& = & 
- { i \over 4} \sum _{(i|j,k|\ell,m)} f_i ^{\mu \nu} (f_jf_k) (f_\ell f_m)  J_6 (i;j,k;\ell,m) 
\no \\ &&
+ i \sum _{(i|j,k,\ell,m)} f_i ^{\mu \nu} (f_j f_k f_\ell f_m)  J_7 (i;j,k,\ell,m)
\no \\ &&
- { i \over 2} \sum _{(i,j,k|\ell,m)} (f_i f_j f_k)^{\mu \nu} (f_\ell f_m) J_8 (i,j,k; \ell,m) 
\no \\ &&
-i \sum _{(i,j,k,\ell,m)} (f_i f_j f_k f_\ell f_m)^{\mu \nu} J_9 (i,j,k,\ell,m) 
\eea
The functions $J_6$ and $J_7$ both have simple poles in $z_i$ at $q_\a$, which may be isolated in terms of a symmetrized function $J_7^S$ plus holomorphic parts $\tilde J_7$, derived in (\ref{3.J6J7}). Recasting the above sum in terms of these objects, we recognize the coefficient of $J_7^S$ to be proportional to the kinematic combination $\mt_i$ in (\ref{newsumm.5}),
\bea
  i  \sum_i \mt _i f_i ^{\mu \nu} J_7^S(i;j,k,\ell,m) 
\eea
The labels $j,k,\ell,m$ are any permutation of the points in the set $\{1,2,3,4,5\} \setminus \{i\}$. Since $J^S_7(i;j,k,\ell,m)$ is invariant under all permutations of $\{j,k,\ell,m\}$, the choice  is immaterial.

\sm

Next, to work out the terms involving $\tilde J_7$,  we set $i=5$ and then include all cyclic permutations of the result. Using the expression (\ref{basicJ7}) for $\tilde J_7$ in terms of basics objects, we have the following contributions for $i=5$, to be multiplied by $i f_5 ^{\mu \nu}$,
\bea
&&
\sum _{(5|j,k,\ell,m)} \Big ( (f_j f_k f_\ell f_m) + \half  (f_jf_\ell) (f_m f_k)  \Big ) \tilde J_7(5;j,k,\ell,m) 
\no \\ && \hskip 0.5in 
= {1 \over 6} \cZ_0 c_1^2 \p \varpi(q_1) \varpi(5) \Delta (1,2) \Delta (3,4) C_T(1,2|3,4)
+ {\rm cycl}(2,3,4)
\eea
where the kinematic combination $C_T$ was introduced in \cite{DP1, DP6}, is given by,
\bea
\label{5.CT}
C_T (i,j|k,\ell) =   2 (f_i f_j f_k f_\ell) - 2 (f_i f_j f_\ell f_k) + (f_if_k) (f_jf_\ell) - (f_i f_\ell) (f_j f_k)
\eea
and has the following symmetry properties,
\bea
C_T (i,j|k,\ell)  = C_T (k,\ell | i,j) & = & - C_T (j,i|k,\ell) 
\no \\ 
C_T (i,j|k,\ell) + {\rm cycl}(j,k,\ell) & = & 0 
\eea
Putting all this together, we obtain the following result for the spin structure sum,
\bea
\label{sametensor}
\sum _\delta \cZ[\delta] \cW ^{\mu \nu} _{1,2,3,4,5}[\delta]
& = & 
  i  \sum_i \mt_i f_i ^{\mu \nu} J_7^S(i;j,k,\ell,m) 
\no \\ &&
+ {i \over 6} \cZ_0 c_1^2 \p \varpi (q_1) \sum _{(i|j,k|\ell,m)} f_i ^{\mu \nu} C_T(j,k|\ell,m) 
\varpi(i)  \Delta (j,k) \Delta(\ell,m)
\no \\ &&
- { i \over 2} \sum _{(i,j,k|\ell,m)} (f_i f_j f_k)^{\mu \nu} (f_\ell f_m) J_8 (i,j,k; \ell,m) 
\no \\ &&
-i \sum _{(i,j,k,\ell,m)} (f_i f_j f_k f_\ell f_m)^{\mu \nu} J_9 (i,j,k,\ell,m) 
\eea
The sum over $(i|j,k|\ell,m) $ on the second line is over all 15 permutations of $\{1,2,3,4,5\}$ modulo 
swapping $j,k$, swapping $\ell,m$ and swapping the pairs $(j,k)$ and $ (\ell,m)$. The expressions 
for $J_7^S$ and $J_8$, $J_9$ can be found in (\ref{3.J7sym}) and (\ref{J6789new}), respectively, and the additional simplifications upon insertion into (\ref{sametensor}) will be discussed in section \ref{sec:7}.

 \newpage

%%%%%%%%%%%%%%%%%%%%%%%%%%%%%%%%%%%%%%%%%%%
%%%%%%%%%%%%%%%%%%%%%%%%%%%%%%%%%%%%%%%%%%%
\section{Cancellation of $(0,1)$-forms and double poles}
\label{sec:6}
\setcounter{equation}{0}
%%%%%%%%%%%%%%%%%%%%%%%%%%%%%%%%%%%%%%%%%%%
%%%%%%%%%%%%%%%%%%%%%%%%%%%%%%%%%%%%%%%%%%%

There are three sources of $(0,1)$-forms that do not involve $\Lambda (z_i)$ (which were already dealt with in section \ref{sec:4}), given as follows,
\bea
\cY_3[\delta] & ~ = ~ & { 1 \over 2 \pi} \sum _i \Big \< Q(p) \int \chi S \, \cV^{(1)} _i 
\prod _{j \not= i}  \cV^{(0)}_j \Big \>
\no \\
\cY_5[\delta] & ~ \longrightarrow ~ & 2 \pi S_\delta(q_1, q_2) \sum_i \sum_\a \delta (z_i,q_\a) { \Lambda (q_\a) \over \p \varpi(q_\a)} 
\Big \< Q(p) \prod _j \cV^{(0)}_j \Big \>
\no \\
\tilde \cF_3 &~ = ~  &
- { i \cN_5 \over 4 \pi} \sum_{j,\ell} \sum_\a { \delta (z_\ell, q_\a) \over c_\a \p \varpi(q_\a)} \, 
\rho_\ell \, \mt_\ell \, \ep_\ell \cdot k_j \, D(z_j, z_0) 
\label{simpf3}
\eea
Here, the first  line is given by the full expression for $\cY_3[\delta]$  in (\ref{Ys}).
The second line is given by the $\Lambda(q_\a)$-dependent part of the contribution $\cY_5[\delta]$, obtained in (\ref{Y5new}) after all $\Lambda(z_i)$ and $\p \Lambda(z_i)$ parts have been cancelled. Finally, the third line is
given by the $(0,1)$-form part of the combination $\cF_{1xx}^{(3)} + \cF_{2x}^{(3)} $ obtained by  combining (\ref{5.tilF3}) and (\ref{pbzD}). We shall now make the expressions for these contributions more explicit, and prove that their sum cancels in the full chiral amplitude.

\subsection{The contribution $\cF_3$}
\label{sec:6.1}

The contribution $\cF_3$ is given by,
\bea
\cF_3 = \int d^2 \zeta \sum _\delta \cZ[\delta] \cY_3[\delta]
\eea
The integral over $\zeta$ may be readily evaluated using the expressions for $\cV_i^{(1)}$ in (\ref{vertex}), for the supercurrent $S$ in (\ref{ST}) and $\chi$ in (\ref{chiqq}), and we find, 
\bea
\label{6.V1i}
\int d^2 \zeta \int \chi S \, \cV_i ^{(1)} = 
 { 1 \over 4} \delta (z_i, q_1) \, \psi ^\mu _+(q_2) \p x_+^\mu (q_2)  \ep ^\nu _i \psi ^\nu _+(q_1) \, 
e^{i k_i \cdot x_+(z_i)} 
 - (q_1 \leftrightarrow q_2)
\eea
Therefore, $\cF_3$ is given by,
\bea
\cF_3 & = &  - { 1 \over 8 \pi} \sum _i \delta (z_i, q_2) \ep^\nu _i \sum _\delta \cZ[\delta] 
\Big \< Q(p) \p x_+^\mu (q_1) \, \psi ^\mu _+(q_1)  \psi ^\nu _+(q_2) \, e^{i k_i \cdot x_+ (z_i)}
\prod _{j \not= i}  \cV^{(0)}_j \Big \> 
\no \\ && + (q_1 \leftrightarrow q_2)
\quad
\eea
Note that $\cZ[\delta]$ reverses sign under $(q_1 \leftrightarrow q_2)$ as does (\ref{6.V1i}), whence the overall symmetrization instruction $(q_1 \leftrightarrow q_2)$. The contributions to the correlators may be grouped as follows.
\begin{enumerate}
\itemsep=-0.03in
\item Three or fewer vertex operators $\cV_j^{(0)}$ have a fermion bilinear. These contributions all vanish by the identities $I_a=0$ for $a=1,\cdots, 7$ in (\ref{D2}).
\item All four vertex operators $\cV_j^{(0)}$ for $j \not = i$ have fermion bilinears and the operators $\psi ^\mu _+(q_1)  \psi ^\nu _+(q_2)$ are not Wick contracted with one another. These contributions also vanish by the identities $I_8=I_9=I_{10}=0$ in (\ref{D2}).
\item All four vertex operators $\cV_j^{(0)}$ for $j \not = i$ have fermion bilinears and the operators $\psi ^\mu _+(q_1)  \psi ^\nu _+(q_2)$ are Wick contracted with one another. These contributions are non-vanishing in view of (\ref{I11}), (\ref{I11S}), and are proportional to $\rho_i \mt_i$.
\end{enumerate}
Collecting this information, we get, 
\bea
\label{6b3}
\cF_3 & = &  { 1 \over 8 \pi} \sum _i \delta (z_i, q_2) \, \ep^\mu _i \sum _\delta \cZ[\delta] S_\delta(q_1,q_2) 
\Big \< Q(p) \p x_+^\mu (q_1)  \, e^{i k_i \cdot x_+(z_i)}  \prod _{j \not= i}  \cV^{(0)}_j \Big \> + (q_1 \leftrightarrow q_2)
\no \\ &=&
 - { i \cN_5 \cZ_0 \over 4 \pi} \sum _i \rho_i \, \mt_i \, \ep_i^\mu 
\Big ( \delta (z_i, q_1) \cP^\mu(q_2) +\delta (z_i, q_2) \cP^\mu(q_1) \Big )
\eea
where we have used (\ref{5.W4K}) to go from the first to the second line, as well as the expression for the chiral Koba-Nielsen factor of (\ref{5.N5a}). Note that, because $\ep_i \cdot  k_i=0$, the terms $\delta (z_i,q_1) \ln E(z_i,q_2)$ produced by $\cP(q_2)$ are absent.

\subsection{The contribution $\cF_5$} 
\label{sec:6.2}

The contribution $\cF_5$ is given by,
\bea
\cF_5 = \int d^2 \zeta \sum _\delta \cZ[\delta] \cY_5[\delta]
\eea
The full combination $\cF_5$  receives contributions from terms involving $\Lambda (z_i)$ and $\p \Lambda (z_i)$ which were cancelled already in section \ref{sec:4}. The remaining contributions, which depend on  $\Lambda (q_\a)$ and were given in (\ref{simpf3}), will be collected here. All contributions to (\ref{simpf3}) involving three or fewer fermion bilinears from the vertex operators vanish, leaving contributions only from four or five fermion bilinears.
Thus, we shall separate those as follows, 
\bea
\cF_5 & ~ \longrightarrow ~ & \cF_5 ^{(a)} + \cF_5 ^{(b)}
\eea
where $\cF_5 ^{(a)}$ and $\cF_5^{(b)}$ are produced by the contributions with four and five fermion 
bilinears from the vertex operators, respectively, and are given by,
\bea
\label{6c3}
\cF_5 ^{(a)}  & = & { i \, \cN_5 \, \cZ_0 \over 4 \pi} \sum _\ell \rho_\ell \, \mt_\ell \, \ep_\ell ^\mu \left (
{ c_1 \over c_2} \delta (z_\ell, q_1) \cP^\mu (q_1) + { c_2 \over c_1} \delta (z_\ell, q_2) \cP^\mu (q_2) \right )
\no \\
\cF_5 ^{(b)} & = & 
- { \cN_5 \over 8 \pi} \sum _a \sum _\a { c_\a^2 \delta (z_a,q_\a) \over c_1 c_2} \sum _\delta \cZ[\delta] S_\delta(q_1,q_2) 
\cW_{1,2,3,4,5} [\delta]
\eea

\subsection{The contribution from $\tilde \cF_3$}
\label{sec:6.3}

The contribution $\tilde \cF_3$ arising from the reorganization of  $\cF_{1xx}^{(3)} + \cF_{2x}^{(3)} $ in (\ref{F3xx}) was simplified to the expression in (\ref{simpf3}). It will be convenient to rewrite the sum over $k D$ in 
terms of $\cP$ via (\ref{5.cP}) and to introduce $\cZ_0$ via (\ref{varpi4}), to obtain,
\bea
\label{6a1}
\tilde \cF_3 & = & 
 {i \, \cN_5 \, \cZ_0 \over 4 \pi} \sum_\ell \rho_\ell \, \mt_\ell \, \ep _\ell ^\mu \Big [ 
\delta (z_\ell, q_1) \cP^\mu (q_2) + \delta (z_\ell, q_2) \cP^\mu(q_1) 
\no \\ && \hskip 1.4in 
- { c_1 \over c_2} \delta (z_\ell, q_1) \cP^\mu (q_1) - { c_2 \over c_1} \delta (z_\ell, q_2) \cP^\mu(q_2) \Big ]
\eea

\subsection{Assembling $\cF_3, \cF_5$ and the $(0,1)$ part of $\cF_{1xx}^{(3)} + \cF_{2x}^{(3)} $}
\label{sec:6.4}

Assembling all contributions above we see, by inspection of (\ref{6a1}), (\ref{6b3}) and (\ref{6c3}), that the contributions of $\tilde \cF_3$ are cancelled by those of $\cF_3$ and $\cF_5^{(a)}$, so that the sum of the contributions of $ \cF_3$, $ \cF_5$, and $\tilde \cF_3$ simplifies as follows,
\bea
\cF_3 + \cF_5 + \tilde \cF_3 & ~ \longrightarrow ~ &  \cF_5 ^{(b)}
\eea
Thus, it remains to evaluate $\cF_5 ^{(b)}$ which we shall do in the remainder of this subsection.

\sm

To evaluate $\cF_5 ^{(b)}$ we use the spin structure sum of (\ref{ZSW5}) in terms of $J_3$ and $J_4$,
\bea
\sum_\delta \cZ[\delta] S_\delta (q_1, q_2)   \cW_{1,2,3,4,5} [\delta]
& = & 
-{i \over 2}  \sum _{(i,j,k|\ell,m)} (f_i f_j f_k) ( f_\ell f_m) J_3(i,j,k;\ell,m)
\no \\ &&
- i \sum _{(i,j,k,\ell,m)_c} ( f_i f_j f_k f_\ell f_m) \, J_4 (i,j,k,\ell,m)
\eea
Evaluating the quantity $\cF_5^{(b)}$ in (\ref{6c3}), and carrying out the sum over $a$,  requires evaluating each one of the arguments of $J_3$ or $J_4$ at the points $q_\a$.  To do so, we use the representation of these functions given in (\ref{J34.1}) but with the hyper-elliptic combination expressed in terms of the bi-holomorphic form $\Delta$ using (\ref{Deltaprime}) by, 
\bea
J_3(i,j,k;\ell,m) & = & - \cZ_0 \rho_i { \Delta (j,k) \Delta '(i) \over \Delta(i,j) \Delta (k,i)} + \hbox{cycl}(i,j,k)
\no \\
J_4(i,j,k,\ell,m) & = &  \cZ_0 \rho_i { \Delta (j,m) \Delta '(i) \over \Delta(i,j) \Delta (m,i)} + \hbox{cycl}(i,j,k,\ell,m)
\eea
The only non-zero contribution arises from $z_i = q_\a$ as otherwise the factor $\rho_i$ vanishes when $z_i \not = q_\a$. In particular, we have $J_3(i,j,k;q_\a,m)=0$ identically. The resulting evaluations simplify considerably and we have,
\bea
J_3(q_\a ,j,k;\ell,m) & = & - \cZ_0 \rho_i { \Delta (j,k) \Delta '(q_\a) \over \Delta(q_\a,j) \Delta (k,q_\a)} 
=  - \cZ_0 \, c_\a \p \varpi(q_\a) \, \rho_i D(j,k) 
\no \\
J_4(q_\a,j,k,\ell,m) & = &  \cZ_0 \rho_i { \Delta (j,m) \Delta '(q_\a) \over \Delta(q_\a,j) \Delta (m,q_\a)} 
= \cZ_0 \, c_\a \p \varpi(q_\a) \, \rho_i D(j,m)
\eea
Using the identity $ c_\a \Delta '(q_\a) = - \p \varpi (q_\a)$, derived from (\ref{varpi2}), the function $\cF_5^{(b)}$ takes the form,
\bea
\cF_5^{(b)} & = & 
-  { i\, \cN_5 \, \cZ_0 \over 16 \pi} \sum _\a { c_\a^3 \p \varpi (q_\a) \over c_1 c_2}   \sum _{(i,j,k|\ell,m)} 
(f_i f_j f_k) ( f_\ell f_m) \Big ( \rho_i \delta (z_i,q_\a) D(j,k) + {\rm cycl}(i,j,k) \Big )
 \\ &&
+  { i \,  \cN_5 \, \cZ_0 \over 8 \pi} \sum _\a { c_\a^3 \p \varpi (q_\a)  \over c_1 c_2} \sum _{(i,j,k,\ell,m)_c}
 (f_i f_j f_k f_\ell f_m) \Big ( \rho_i \delta (z_i,q_\a) D(j,m) + {\rm cycl}(i,j,k,\ell,m) \Big ) \quad
 \no 
\eea
Since the summand in the first line is invariant under swapping $\ell,m$ and $j,k$, we may undo the cyclic sum ${\rm cycl} (i,j,k)$ and extend the sum over all 10 partitions $(i,j,k |\ell,m)$ into a sum over all permutations $(i,j,k,\ell,m)$ upon including a factor of 1/2 for swapping $\ell,m$ and another factor of $1/2$ for swapping $j,k$. Similarly, since the summand in the second line is invariant under reversal of $(i,j,k,\ell,m)$, we may undo the cyclic sum ${\rm cycl} (i,j,k,\ell,m)$ and extend the sum over all 12 permutations in $(i,j,k,\ell,m)_c$ to a sum over all permutations $(i,j,k,\ell,m)$ upon including a factor of 1/2 for reversal. Finally, eliminating $\cZ_0$ using $\cZ_0 c_1 c_2 \p \varpi (q_1) \p \varpi (q_2) =1$ and $c_1^2 \p \varpi (q_1) = - c_2^2 \p \varpi (q_2)$ of (\ref{varpi4}), we find, 
\bea
\cF_5^{(b)} & = & 
 { i \, \cN_5  \over 64 \pi} \sum _\a  \sum _{(i,j,k,\ell,m)} { \delta (z_i, q_\a) \over c_\a \p \varpi (q_\a)}  D(j,k)
\rho_i (f_i f_j f_k) ( f_\ell f_m)
\no \\ &&
-  { i \, \cN_5  \over 16 \pi} \sum _\a    \sum _{(i,j,k,\ell,m)}  { \delta (z_i, q_\a) \over c_\a \p \varpi (q_\a)} D(j,m)
\rho_i (f_i f_j f_k f_\ell f_m) 
\eea
Next, we write this sum over $i$ of the $(0,1)$-form in $z_i$ (which is a $(1,0)$-form in all $z_j$ with $j \not= i$) in terms of a sum over $i$ of exact total differentials in $z_i$ plus the difference which is a $(1,0)$-form in all $z_i$. Care must be taken to include the $z_i$ dependence of $\cN_5$ in the process,
\bea
\cF^{(b)}_5 = \cF_5^{(1)} + \cF_5 ^{(2)} + \hbox{ exact total differential}
\eea
where the first two terms are given by,
\bea
\label{6.F51}
\cF_5^{(1)} & = &  { i \, \cN_5  \over 128 \pi^2}   \sum _{(i,j,k,\ell,m)}  D(j,k)
\rho_i (f_i f_j f_k) ( f_\ell f_m) k_i \cdot \cP(z_i) \sum _\a { \tau_{z_i,z_0} (q_\a) \over c_\a \p \varpi (q_\a)}
\no \\ &&
-  { i \,  \cN_5  \over 32 \pi^2}     \sum _{(i,j,k,\ell,m)}  
D(j,m) \rho_i (f_i f_j f_k f_\ell f_m)  k_i \cdot \cP(z_i) \sum_\a { \tau_{z_i,z_0} (q_\a)  \over c_\a \p \varpi (q_\a)} 
\no \\
\cF_5^{(2)} & = & 
 { i \, \cN_5  \over 128 \pi^2}   \sum _{(i,j,k,\ell,m)}   D(j,k)
\rho_i (f_i f_j f_k) ( f_\ell f_m) \sum _\a { \p_i \p_{q_\a} \ln E(z_i,q_\a) \over c_\a \p \varpi (q_\a)}
\no \\ &&
-  { i \, \cN_5  \over 32 \pi^2}     \sum _{(i,j,k,\ell,m)}  
D(j,m) \rho_i (f_i f_j f_k f_\ell f_m) \sum _\a {  \p_i \p_{q_\a} \ln E(z_i,q_\a)  \over c_\a \p \varpi (q_\a)} 
\eea
The exact differentials may be omitted as their integrals will vanish. The  contribution $\cF_5^{(1)}$ may be simplified using the relation (\ref{fund3}), and we find, 
\bea
\cF_5^{(1)} & = & -  { i \, \cN_5  \over 128 \pi^2}    \sum _{(i,j,k,\ell,m)} 
\rho_i (f_i f_j f_k) ( f_\ell f_m) \, k_i \cdot \cP(z_i) \, D(z_j,z_k) D(z_i,z_0)
\no \\ &&
+ { i \, \cN_5  \over 32 \pi^2}  \sum _{(i,j,k,\ell,m)}  
\rho_i (f_i f_j f_k f_\ell f_m) \, k_i \cdot \cP(z_i)  \, D(z_j,z_m) D(z_i,z_0) 
\eea
The function $\cF_5^{(1)}$ depends on  $z_0$ through an exact differential of the chiral Koba-Nielsen factor, which is immaterial and may be omitted. The function $\cF_5^{(2)}$ will cancel $\cF^{(d)}+ \cF_{2 \psi}$ as will be shown in the subsequent subsection.

\subsection{Cancellation of $\cF^{(d)}+ \cF_{2 \psi}+\cF_5^{(2)}$}
\label{sec:5.6}

The motivations  for combining these contributions is that they are,
\begin{itemize}
\itemsep=-0.02in
\item  the only ones remaining that have double poles in $z_i$ at $q_\a$, 
\item  the only ones remaining that involve kinematic invariants built out of concatenated products of field strengths $f$ without involving additional momentum factors,
\item expected to cancel in the final form of the amplitude on the basis of  non-renormalization theorems and predictions from S-duality and space-time supersymmetry.
\end{itemize}
To prove their cancellation, we begin by summing their expressions from (\ref{4.FdF2}), 
\bea
\cF^{(d)}+ \cF_{2 \psi}+\cF_5^{(2)} & = &
{ i \cN_5 \over 32 \pi^2} \sum_{(i,j,k|\ell,m)} (f_if_jf_k) (f_\ell f_m) \, \tilde \cC_1 (i,j,k;\ell, m) 
\no \\ &&
-{ i \cN_5 \over 16 \pi^2}  \sum _{(i,j,k,\ell,m)_c}  (f_if_jf_kf_\ell f_m) \, \tilde \cC_2 (i,j,k,\ell,m)
\label{majorcan}
\eea
where the functions $\tilde \cC_1$ and $\tilde \cC_2$ are given by, 
\bea
\label{deftildeC}
\tilde \cC_1(i,j,k;\ell,m) & = & \cC_1(i,j,k;\ell,m) 
 -\bigg[ \Delta(i,k) \varpi(\ell) \varpi(m) \sum_\a { \p_{z_j} \p_{q_\a} \ln E(z_j,q_\a) \over c_\a \p \varpi (q_\a)} 
 + \hbox{cycl}(i,j,k) \bigg]
\no \\ 
\tilde \cC_2(i,j,k,\ell,m) & = & \cC_2(i,j,k,\ell,m) 
- \bigg[ \Delta(i,k) \varpi(\ell) \varpi(m) \sum_\a { \p_{z_j} \p_{q_\a} \ln E(z_j,q_\a) \over c_\a \p \varpi (q_\a)}
\no \\ && \hskip 1.5in 
 + \hbox{cycl}(i,j,k,\ell,m) \bigg ]
\eea
The functions  $\cC_1$ and $\cC_2$ were given in (\ref{4.C1}) and (\ref{4.C2}), respectively.  By exposing the coefficients of the functions $B_j^I$ defined in (\ref{defBIa}), we arrive at the alternative presentation,
\bea
\cC_1 (i,j,k;\ell,m) & = &  \om_I(i) \Delta (j,\ell) \Delta (k,m) 
\Big ( g_{j,\ell}^I + g^I_{\ell,m} + g^I_{m,k}  \Big ) 
 \\ &&
- \half \Big ( \om_I(\ell) \Delta (j,m) + \om_I(m) \Delta (j,\ell) \Big ) \Delta (i,k) B_j^I
%\no \\ &&
+ \hbox{ cycl}(i,j,k|\ell,m)
\no \\
\cC_2 (i,j,k,\ell,m) & = &\om_I(i) \Delta (j,k) \Delta (\ell,m)
\Big (  g_{j,k}^I +  g^I_{\ell,m} +g_{i,k}^I +g^I_{\ell,i} \Big ) 
\no \\ &&  
-  \Big ( \om_I(\ell) \Delta (j,m) + \om_I(m) \Delta (j,\ell) \Big ) \Delta (i,k) B_j^I
%\no \\ &&
 + \hbox{ cycl}(i,j,k,\ell,m)
\no
\eea
For both formulas, the instruction to add cyclic permutations applies to all terms in the expression.  
The vanishing of the combinations $\tilde \cC_1$ and $\tilde \cC_2$ is demonstrated in appendix \ref{sec:H}. Thus, we have established the absence of double poles in the vertex points $z_i$ at the points $q_\a$ in the chiral amplitude, as well as the  relation, 
\bea
\cF^{(d)}+ \cF_{2 \psi}+\cF_5^{(2)} =0
\eea
both of which constitute a major conceptual check of our general methods and a significant simplification of the chiral amplitude.

\newpage

%%%%%%%%%%%%%%%%%%%%%%%%%%%%%%%%%%%%%%%%%%%
%%%%%%%%%%%%%%%%%%%%%%%%%%%%%%%%%%%%%%%%%%%
\section{Assembling and simplifying the chiral amplitude}
\label{sec:7}
\setcounter{equation}{0}
%%%%%%%%%%%%%%%%%%%%%%%%%%%%%%%%%%%%%%%%%%%
%%%%%%%%%%%%%%%%%%%%%%%%%%%%%%%%%%%%%%%%%%%

Having assembled the contributions to the chiral amplitude in section \ref{sec:5}, and proven the cancellation of all $(0,1)$-form parts and double poles in $z_i$ at the points $q_\a$ in section \ref{sec:6}, the remaining contributions to the chiral amplitude are as follows, (see figure \ref{fig:1} for an overview),
\begin{itemize}
\itemsep=-0.03in 
\item from (\ref{vanish?}) we obtain $\cF_{1xx}^{(2)}  + \cF_{2x} ^{(2)}+\cF_{1xx}^{(3)}  + \cF_{2x} ^{(3)}$;
\item from (\ref{5.F1xxF2x}) we obtain $\cF_{1xx}^{(1)}   + \cF_{2x} ^{(1)} $; 
\item from (\ref{6.F51}) we obtain $\cF_5 ^{(1)}$;
\item and from (\ref{sametensor}) we obtain $\cF_{1n}$. 
\end{itemize}
It will be convenient in subsequent calculations to factor out the normalization
$i/(64\pi^2)$ and the chiral Koba-Nielsen factor defined by (\ref{5.N5b}),
\bea
\label{defmFa}
\cF_{1xx}^{(2)}  + \cF_{2x} ^{(2)}+\cF_{1xx}^{(3)}  + \cF_{2x} ^{(3)}
& = &  { i \, \cN_5 \over 64 \pi^2} \left ( \mF_1+\mF_2 \right ) 
\no \\
\cF_{1xx}^{(1)}   + \cF_{2x} ^{(1)} 
& =  &
{ i \, \cN_5 \over 64 \pi^2} \left ( \mF_3+\mF_4 \right )
\no \\
\cF_5 ^{(1)} & = & { i \, \cN_5 \over 64 \pi^2} \left ( \mF_5+\mF_6 \right )
\no \\
\cF_{1n}  & =  & { i \, \cN_5 \over 64 \pi^2} \left ( \mF_7+\mF_8 + \mF_9 + \mF_{10} \right )
\eea 
The full chiral amplitude $\cF$ of (\ref{splittingF}) is then given in terms of $\mF$ as follows,
\bea
\label{defmFb}
\cF = { i \, \cN_5 \over 64 \pi^2} \, \mF + \hbox{exact differentials} 
\hskip 1in
\mF = \sum_{a=1}^{10} \mF_a
\eea
Here, we have indicated the presence of exact differentials in the vertex points $z_i$ that have been discarded in the process of cancelling all $(0,1)$-form contributions, and that do not contribute to the physical amplitudes (\ref{newsumm.1}) and (\ref{newsumm.1a}).

\subsection{The functions $\mF_a$}

It will be convenient to organize the individual functions $\mF_a$ according to the structure of their kinematic and worldsheet dependences. The functions $\mF_1$ and $\mF_2$ are the only ones that involve a naked polarization vector $\ep_\ell$ as opposed to its field strength $f_\ell$, and are given as follows,
\bea
\label{mF1mF2}
\mF_1 &=& 
4 \Dt \sum _{\ell} \mt_\ell   \, {  \ep_\ell \cdot \cP(z_\ell)  \over \varpi(z_\ell)} 
\no \\ 
 \mF_2 &=& 
 -8 \sum_{j,\ell} \rho_\ell \, \mt_\ell \, (k_j    \cdot \ep_\ell )  \, k_\ell \cdot \cP(z_\ell) \, D(z_\ell, z_0)  \,  D(z_j,z_0') 
\eea
where the permutation-invariant combination $\tilde {\cal D}$ was defined in (\ref{5.Dt}) and simplified in (\ref{7.Lem.1}) and (\ref{7.Lem.2}). The kinematic dependence of all remaining contributions is entirely through the linearized  field strengths~$f_\ell$. The individual functions $\mF_3, \cdots, \mF_6$ are given by,
\bea
\mF_3 & = & { \Dt \over \cZ_0 \, \rho}  \sum _{(i,j,k|\ell,m)} (f_i f_j f_k) (f_\ell f_m)   J_3(z_i,z_j,z_k;z_\ell,z_m) 
\no \\ 
\mF_4 & = &{ 2 \, \Dt \over \cZ_0 \, \rho} 
 \sum _{(i,j,k,\ell,m)_c} ( f_i f_j f_k f_\ell f_m) \, J_4 (z_i,z_j,z_k,z_\ell,z_m) 
\no \\
\mF_5 & = & -  \half  \sum _{(i,j,k,\ell,m)}  
\rho_i (f_i f_j f_k) ( f_\ell f_m) \, k_i \cdot \cP(z_i) \, D(z_j,z_k) D(z_i,z_0)
\no \\ 
\mF_6 &=&
2  \sum _{(i,j,k,\ell,m)}  
\rho_i (f_i f_j f_k f_\ell f_m) \, k_i \cdot \cP(z_i)  \, D(z_j,z_m) D(z_i,z_0) 
\label{mF1tomF6}
\eea
where  the functions $J_3$ and $J_4$ are given by (\ref{J34.2}),
\bea
J_3 (z_i,z_j,z_k;z_\ell, z_m) & = & - \cZ_0 \, \rho_i \, \om _I (z_i) \, G^I_{q_\a, j,i,k} 
+ \hbox{cycl}(i,j,k)
\no \\
J_4 (z_i,z_j,z_k,z_\ell, z_m)  & = &  \cZ_0 \, \rho_i \, \om _I(z_i) \, G^I_{q_\a, j,i,m} 
+ \hbox{cycl} (i,j,k,\ell,m)
\eea
The individual functions $\mF_7, \cdots, \mF_{10}$ are given by,
\bea
\mF_7 &=&
 4 \, \cP^\mu (q_1) \cP^\nu (q_2)   \sum_i \mt_i f_i ^{\mu \nu} J_7^S(z_i;z_j,z_k,z_\ell,z_m) 
\\ 
\mF_8 & = &
- 2 \, \cP^\mu (q_1) \cP^\nu (q_2)   \sum _{(i,j,k|\ell,m)} (f_i f_j f_k)^{\mu \nu} (f_\ell f_m) J_8 (z_i,z_j,z_k; z_\ell,z_m) 
\no \\ 
\mF_9 & = &
- 4 \, \cP^\mu (q_1) \cP^\nu (q_2)  \sum _{(i,j,k,\ell,m)} (f_i f_j f_k f_\ell f_m)^{\mu \nu} J_9 (z_i,z_j,z_k,z_\ell,z_m) 
   \no \\ 
\mF_{10} &= &
 {2 \over 3} \, \cP^\mu (q_1) \cP^\nu (q_2)  \sum _{(i|j,k|\ell,m)} f_i ^{\mu \nu} C_T(j,k|\ell,m)  \cZ_0 c_1^2 \p \varpi (q_1)  \varpi(z_i)  \Delta (z_j,z_k) \Delta(z_\ell,z_m)
\no
\eea
The kinematic factor $C_T$ was defined in (\ref{5.CT}), the function $J_7^S$ was obtained in (\ref{3.J7sym}), and the functions $J_8, J_9$ were given in   (\ref{J6789new}),
\bea
J^S_7 (z_i;z_j,z_k,z_\ell, z_m) & = & 
{ 1 \over 3} \cZ_0 c_1^2 \p \varpi(q_1)  \varpi(z_i)^{-1}
\sum_{(j,k|\ell,m)} \Delta (z_i,z_j) \Delta (z_i, z_k) \varpi(z_\ell) \varpi(z_m) 
 \\
J_8 (z_i,z_j,z_k;z_\ell, z_m)  & = & \cZ_0  c_1^2 \p \varpi(q_1) 
\big ( \varpi(z_i) \Delta (z_j,z_\ell) \Delta (z_k,z_m)  
+ \varpi(z_m) \Delta (z_i,z_\ell) \Delta (z_j,z_k) \big )
\no \\
J_9 (z_i,z_j,z_k,z_\ell, z_m)  & = & - \cZ_0 c_1 ^2 \p \varpi(q_1)   \varpi(z_k) \Delta (z_i,z_\ell)  \Delta (z_j,z_m)
\no
\eea
The sum $(j,k|\ell,m)$ in $J^S_7$ is over all 6 partitions into  inequivalent pairs of $\{ 1,2,3,4,5\} \setminus \{i \}$. Finally, note that despite its asymmetrical presentation above, $J_8$ satisfies,
\bea
J_8 (z_i,z_j,z_k;z_\ell, z_m)  = J_8 (z_i,z_j,z_k;z_m, z_\ell)  = J_8 (z_k,z_j,z_i;z_\ell, z_m) 
\eea
In each one of the functions $\mF_7, \cdots, \mF_{10}$, the  product $\cP^\mu (q_1) \cP^\nu (q_2)$ of one-forms (\ref{pfield}) enters anti-symmetrically in $\mu$ and $\nu$. This leads us to introduce the following convenient combination, 
\bea
\cB^{\mu \nu} = \cZ_0 c_1^2 \p \varpi (q_1) \Big ( \cP^\mu (q_1) \cP^\nu (q_2) - \cP^\nu (q_1) \cP^\mu (q_2) \Big )
\eea
In terms of $\cB^{\mu \nu}$ and the expressions for $J_7^S, J_8, J_9$, we obtain the following explicit formulas, 
\bea
\label{7.mF78910}
\mF_7 &=& {2 \over 3} \cB^{\mu \nu} \sum_{(i|j,k|\ell,m)}  \mt_i f_i ^{\mu \nu}  {1 \over \varpi(z_i)}
\Delta (z_i,z_j) \Delta (z_i, z_k) \varpi(z_\ell) \varpi(z_m)
\no \\ 
\mF_8 & = &
-  \cB^{\mu \nu}   \sum _{(i,j,k|\ell,m)} (f_i f_j f_k)^{\mu \nu} (f_\ell f_m) \Big (  \varpi(z_i) \Delta (z_j,z_\ell) \Delta (z_k,z_m)  + \varpi(z_m) \Delta (z_i,z_\ell) \Delta (z_j,z_k)  \Big ) 
\no \\ 
\mF_9 & = &
 2 \cB^{\mu \nu}  \sum _{(i,j,k,\ell,m)} (f_i f_j f_k f_\ell f_m)^{\mu \nu}  \varpi(z_k) \Delta (z_i,z_\ell)  \Delta (z_j,z_m)
   \no \\ 
\mF_{10} &= &
{1 \over 3} \cB^{\mu \nu}   \sum _{(i|j,k|\ell,m)} f_i ^{\mu \nu} C_T(j,k|\ell,m)  \varpi(z_i)  \Delta (z_j,z_k) \Delta(z_\ell,z_m)
\eea
where we remind the reader that $(i|j,k|\ell,m)$ refers to 15 permutations of $1,2,3,4,5$ modulo swapping $j,k$,
swapping $\ell,m$ and swapping the pairs $(j,k)$ and $(\ell,m)$.

\subsection{Simplifying $\cB^{\mu \nu}$}

To simplify $\cB^{\mu \nu}$ we make use of formula (\ref{5.cP}) to eliminate $\cP(q_2)$ in favor of $\cP(q_1)$, use the third formula in (\ref{varpi4}) to eliminate $\cZ_0$,  and  obtain the following formula independent on the choice of $\alpha=1,2$ which contains only a single factor of $\cP$,
\bea
\label{calB1}
\cB^{\mu \nu} = c_\a \sum _a \Big ( k^\mu _a \cP^\nu(q_\a) - k^\nu _a \cP^\mu(q_\a) \Big ) D(z_a,z_0)
\eea
To simplify this formula further, we decompose the meromorphic form $\cP^\mu (z)$ onto the basis of holomorphic $(1,0)$-forms $\om_I(z)$, and thereby define the meromorphic homology shift invariant functions $\mP^I(z)$ as follows,
\bea
\label{7.mPa}
\cP (z) = \om_I(z) \mP^I(z)
\hskip 1in 
\mP^I (z)= 2 \pi i p^I + \sum _a k_a\,  g^I_{z, z_a}
\eea
for an arbitrary point $z$. Substituting the left side expression into (\ref{calB1}), using the identity,
\bea
\label{7.cD}
c_1 \om_I (q_1) D(z_a, z_0) 
= { \om_I(z_a) \over \varpi(z_a)} - { \om_I(z_0) \over \varpi(z_0)}
\eea
and omitting the contribution from the second term on the right side of (\ref{7.cD}) in view of overall momentum conservation in the sum over $a$, we recast (\ref{calB1}) in the following form, 
\bea
\label{7.mPb}
\cB^{\mu \nu} = \mP^{I  \nu } (q_\a)  \sum _a { \om_I(z_a) \over \varpi(z_a) } \, k^\mu _a - (\mu \leftrightarrow \nu)
\eea
The advantage of this formula  is that $\mP^{I \nu} (q_\a) $ factors out of the sum over $a$, leaving us to deal only with the simpler summation over $a$ with manifestly $z_0$ independent summand. In later steps of the computation, we will frequently use the following corollary of (\ref{7.mPa}),
\bea
\label{7.PG}
\mP^I (z) - \mP^I (q_\a) =  \sum_a k_a \,  G^I_{z, z_a , q_\a}
\eea
which makes it manifest that $\mP^I (z_i) - \mP^I (q_\a)$ is independent of the loop momenta $p^I$.

\subsection{Bases of holomorphic and meromorphic  forms in 5 points}
\label{sec:7.2}

Inspection of the contributions $\mF_1, \cdots, \mF_{10}$ in (\ref{mF1mF2}), (\ref{mF1tomF6}) and (\ref{7.mF78910}) to the chiral amplitude reveals that these $(1,0)$-forms in the five vertex points $z_i$ are built out of a few simple holomorphic and meromorphic $(1,0)$-forms in the five points $z_i$. For the holomorphic forms it is readily proven by enumeration and the techniques of appendix \ref{sec:B1} that one may choose the following cyclic basis,\footnote{In the sequel of this section we shall freely move indices $I,J,\ldots$ between subscripts and superscript and not distinguish between $\omega_I$ and $\omega^I$ or $ \mP^\nu _I$ and $ \mP^{I \nu}$, in order to avoid cumbersome index 
configurations.}
\bea
W^I _1 & = &  \om^I (z_1) \Delta(z_2,z_3) \Delta (z_4,z_5)
\no \\
\Wt _1 & = &  \varpi(z_1) \Delta(z_2,z_3) \Delta (z_4,z_5)
\label{defwws}
\eea
and cyclic permutations thereof to obtain $W^I_a$ and $\Wt_a$ for $a=1,2,3,4,5$.  It will  be useful to introduce also the following  meromorphic forms, 
\bea
W_{a;b} ^I = { \om^I(z_a) \over \varpi(z_a)} \, \Wt _b
\hskip 1in 
W_{a;a} ^I = W_a^I
\label{basforms}
\eea
The vector space generated by the 25 meromorphic forms $W_{a;b}$ has dimension 20. A basis may be chosen to consist of the forms $W_{a;b}$ with $ b \not= a$, leaving the holomorphic forms $W_a$ expressible as linear combinations of $W_{a;b} $ with $b \not=a$. These linear combinations are generated by the identity,
\bea
W_{1;2}^I+W_{1;5}^I = W_2^I+W_5^I
\eea 
and cyclic permutations thereof. These relations may be solved for $W_a^I$ as follows, 
\bea
\label{7.Wdiag}
2W_1^I & = & W_{1;2}^I +W_{1;5}^I +W_{2;1}^I + W_{2;3}^I +W_{5;1}^I  + W_{5;4}^I
\no \\ &&
 - W_{3;4}^I  - W_{3;2}^I - W_{4;5}^I - W_{4;3}^I
\eea
and cyclic permutations thereof. Finally, in expressing $\Dt$ and other combinations in terms of the  bases of holomorphic forms, we shall encounter the following ubiquitous combinations,
\bea
&&
k_2 \, \varpi(4) \Delta (1,3) \Delta (2,5)
+ k_3 \, \varpi(4) \Delta (1,2) \Delta (3,5)
+ k_4 \, \varpi(3) \Delta (1,2) \Delta (4,5)
\no \\  && \qquad =
(k_2+k_3+k_4) \Wt_3 - k_2 \Wt_4 + (k_2+k_3) \Wt_5
\label{7.omcon} \\ &&
k_2 \, \om_I(4) \Delta (1,3) \Delta (2,5)
+ k_3 \, \om_I(4) \Delta (1,2) \Delta (3,5)
+ k_4 \, \om_I(3) \Delta (1,2) \Delta (4,5)
\no \\ && \qquad =
(k_2+k_3+k_4) W_3^I - k_2 W_4^I + (k_2+k_3) W_5^I
\no \\ &&
\om_I(1) \varpi(1)^{-1} \Big [ k_2 \, \varpi(4) \Delta (1,3) \Delta (2,5)
+ k_3 \, \varpi(4) \Delta (1,2) \Delta (3,5)
+ k_4 \, \varpi(3) \Delta (1,2) \Delta (4,5) \Big ]
\no \\ && \qquad =
(k_2+k_3+k_4) W_{1;3}^I - k_2 W_{1;4}^I + (k_2+k_3) W_{1;5}^I
\no
\eea

\subsection{Simplifying $\mF_1+ \mF_2 $}

We shall now derive a simplified representation $\mF_1+\mF_2=\mF_A+\mF_A'+\mF_A''$
for $\mF_1$ and $ \mF_2 $ in (\ref{mF1mF2}) where the three terms on the right-hand side 
are given as follows,
\bea
\label{7.mFA}
\mF_A & = & 8 \mt_1    \ep_1^\mu \,   \cP^\mu (1) 
\Big [  k_4 \cdot k_5 \Delta(2,5)  \Delta(3,4)  - k_3 \cdot k_4  \Delta(2,3) \Delta (4,5)     \Big ]
  + {\rm cycl}(1,2,3,4,5)
\no \\
\mF_A' & = & 8 \mt_1  f_1^{\mu \nu}  \mP^\nu _I (q_\a)   
\Big [  k_5^\mu  W_{1;3}^I + k_2^\mu W_{1;4}^I + (k_4^\mu+k_5^\mu)  W_{1;5}^I  \Big ]
 + {\rm cycl}(1,2,3,4,5)
 \\
 \mF_A'' & = & 8 \mt_1   f_1^{\mu \nu}  \Big ( \mP^\nu _I (z_1) - \mP^\nu _I (q_\a) \Big )  
 \Big [  k_5^\mu  W_{1;3}^I + k_2^\mu W_{1;4}^I + (k_4^\mu+k_5^\mu)  W_{1;5}^I  \Big ]
 + {\rm cycl}(1,2,3,4,5)
\no
 \eea
To do so, we use (\ref{7.Lem.2}) to recast $\mF_1$ as follows, 
\bea
\mF_1-\mF_A & = & 8 \mt_1 \,  { \ep_1 \cdot \cP(1) \over \varpi(1)} 
\Big [ k_1 \cdot k_2 \, \varpi(4) \Delta(1,3) \Delta (2,5) + k_1 \cdot k_3 \, \varpi(4) \Delta(1,2) \Delta (3,5) 
\no \\ && \hskip 1in 
+ k_1 \cdot k_4 \, \varpi(3) \Delta(1,2) \Delta (4,5)  \Big ] + {\rm cycl}(1,2,3,4,5)
\eea
To rearrange $\mF_2$, we choose $z_0'=z_5$, select an adapted point $z_0$ for each term so as to match the terms in $\mF_1-\mF_A$, and multiply through by $\rho_1$, as follows, 
\bea
 \mF_2 & = & 
-8 \mt_1    \, {  k_1\cdot \cP(1) \over \varpi(1)}  
\Big [ \ep_1 \cdot k_2\,  \varpi(4) \Delta (1, 3) \Delta (2,5) + \ep_1 \cdot k_3 \, \varpi(4) \Delta (1, 2) \Delta (3,5) 
\no \\ && \hskip 1.3in 
+ \ep_1 \cdot k_4 \, \varpi(3) \Delta (1, 2) \Delta (4,5) \Big ]
+ {\rm cycl}(1,2,3,4,5)
\eea
We observe that $\mF_1+\mF_2- \mF_A$ is gauge-invariant and use $\ep_1^\nu k_1^\mu - \ep_1^\mu k_1^\nu = - f_1^{\mu \nu}$ to combine the terms.  Expressing the result in terms of the basis forms $W_{a;b}$ using (\ref{7.omcon}) and the decomposition of $\cP(1)$ given in (\ref{7.mPa}),  we obtain, 
 \bea
\mF_1+\mF_2-\mF_A & = & 
8 \mt_1    f_1^{\mu \nu}    \mP^\nu_I (1)    \Big [ 
k_5^\mu  W_{1;3}^I +k_2^\mu W_{1;4}^I + (k_4^\mu+k_5^\mu)  W_{1;5}^I  \Big ]
 + {\rm cycl}(1,2,3,4 ,5)
 \quad
\eea
Here we have made use of momentum conservation and orthogonality to $k_1$ to re-express $(k_2+k_3+k_4) \to - k_5$ and $(k_2+k_3) \to -(k_4+k_5)$. For later convenience we have further decomposed $\mF_1+\mF_2-\mF_A = \mF_A'+\mF_A''$ to give formula (\ref{7.mFA}). Note that $\mF_A$ is gauge-invariant, and independent of $q_\a$.

\sm

A final rearrangement may be carried out for $\mF_A''$ by recognizing that the difference $\mP^\nu _I (z_i) - \mP^\nu _I (q_\a)$ is independent of the loop momenta $p^\nu _I$ and may be re-expressed in terms of the functions $G^I$  by means of (\ref{7.PG}). Using this expression in $\mF_A''$ and the rearrangement,
\bea
G^I_{q_\a, j,i,k}  = g^I _{q_\a, j} + g^I _{j,i} + g^I_{i,k} + g^I_{k,q_\a}
=  G^I_{q_\a, i,k} - G^I_{q_\a, i,j} 
\label{rearrange}
\eea
we obtain, 
\bea
 \mF_A'' & = & 
8 \mt_1  \, f_1^{\mu \nu} 
\Big [ k_2 ^\nu G^I_{1,2,q_\a,5} +  k_3 ^\nu G^I_{1,3,q_\a,5}  +  k_4 ^\nu G^I_{1,4,q_\a,5}   \Big ] 
\\ && \qquad  \times
\Big [ k_5^\mu  W_{1;3}^I +k_2^\mu W_{1;4}^I + (k_4^\mu+k_5^\mu)  W_{1;5}^I  \Big ]
+ {\rm cycl}(1,2,3,4,5)
\no 
\eea
where the term proportional to $k_1^\nu$ contracts to zero into $f_1^{\mu \nu}$ and may therefore be omitted.

\subsection{Simplifying $\mF_3+ \mF_4$}

To simplify $\mF_3+\mF_4$, we begin by recasting $\mF_3,\mF_4$ in (\ref{mF1tomF6}) as follows, 
\bea
\mF_3 & = & 
- \Dt \sum _{(i,j,k|\ell,m)} (f_i f_j f_k) (f_\ell f_m) \left (  { \om _I (i) \over \varpi(i)} \, G^I_{q_\a, j,i,k} 
+ \hbox{cycl}(i,j,k)
 \right )
\no \\ 
\mF_4 & = &
2 \Dt \sum _{(i,j,k,\ell,m)_c} ( f_i f_j f_k f_\ell f_m) \left ( { \om _I(i) \over \varpi(i) } \, G^I_{q_\a, j,i,m} 
+ \hbox{cycl} (i,j,k,\ell,m)
 \right ) 
\eea
We note the following antisymmetry property,
\bea
G^I_{q_\a, j, i, k} = - G^I_{q_\a,k,i,j}
\eea
which lines up with the symmetry properties of the corresponding kinematic factors above.
In~$\mF_3$, we may extend the sum $(i,j,k|\ell,m)$ to the sum over all permutations upon including a factor of $\half$ to account for the permutations on $\ell, m$, as well as another factor of $\half$ to complete the summation over all cyclic permutations of $(i,j,k)$  to all permutations $(i,j,k)$, and then omitting the permutation instruction on $(i,j,k)$ inside the parentheses in the summand. In $\mF_4$, we may similarly extend the sum $(i,j,k,\ell,m)_c$ to the sum over all permutations by omitting the cyclic permutations instruction on $(i,j,k,\ell,m)$ inside the parentheses of the summand, upon including an extra factor of 2 to undo the reversal quotient in $(i,j,k,\ell,m)_c$. The result is as follows,
\bea
\mF_3 & = & 
- {1 \over 4} \Dt \sum _{(i,j,k,\ell,m)}  (f_i f_j f_k) (f_\ell f_m)  { \om _I (i) \over \varpi(i)} \, G^I_{q_\a, j,i,k} 
\no \\ 
\mF_4 & = &
 \Dt \sum _{(i,j,k,\ell,m)}  ( f_i f_j f_m f_\ell f_k) { \om _I(i) \over \varpi(i) } \, G^I_{q_\a, j,i,k} 
\eea
In $\mF_3$, the factor $(f_if_jf_k)= - (f_k f_j f_i)$ is manifestly anti-symmetric under swapping any pair of indices.  In $\mF_4$, we use the fact that, for a given pair $i,j$, the sum over all permutations of the complementary indices $k,\ell,m$ produces the same matrix in both brackets in the summand of $\mF_4$. Putting all together and using the rearrangement formula (\ref{rearrange}), 
\bea
\mF_3 & = & 
{ 1 \over 8}  \Dt \sum _{(i,j,k,\ell,m)}  ([f_i, f_j]  f_k) (f_\ell f_m)  \left ( { \om _I (i) \over \varpi(i)} + { \om _I (j) \over \varpi(j)} \right ) \, G^I_{q_\a, i,j} 
\no \\ 
\mF_4 & = &
- \half \Dt \sum _{(i,j,k,\ell,m)}   ( [f_i, f_j]  f_m f_\ell f_k)  \left ( { \om _I(i) \over \varpi(i) } + { \om _I(j) \over \varpi(j) } \right )
\, G^I_{q_\a, i,j} 
\eea
Using the definition of $t_8$ of (\ref{newsumm.5}), and the fact that the cyclic permutations required to define $t_8$ are being automatically added in the sum for $\mF_4$, we may recast the entire expression
for $\mF_3{+}\mF_4$ in terms of $t_8$ involving the commutator $[f_i, f_j]$, 
\bea
\label{7.def.T}
\mt_{12} & = &  t_8([f_1, f_2], f_3, f_4, f_5) = 
([f_1,f_2]f_3f_4f_5) -{1 \over 4} ([f_1,f_2]f_3)(f_4f_5) + {\rm cycl}(3,4,5)
\eea
and permutations thereof. This kinematic building block is manifestly symmetric in $3,4,5$ and antisymmetric in $1,2$. In the expressions for $\mF_3$ and $\mF_4$ we sum over all permutations $(k,\ell,m)$, so there is a factor of 2 coming out from this observation, 
\bea
\mF_3 + \mF_4 = 
- \Dt \, \sum _{ i \not= j}   \mt_{ij}  \left ( { \om _I(i) \over \varpi(i) } + { \om _I(j) \over \varpi(j) } \right )
\, G^I_{q_\a, i,j} 
= 
- 2 \Dt \sum _{ i \not= j}   \mt_{ij}  { \om _I(i) \over \varpi(i) } \, G^I_{q_\a, i,j} 
\eea
Finally, we recast the result as a sum over cyclic permutations, 
\bea
\mF_3 + \mF_4 
=  - 2 \Dt  \, { \om _I(1) \over \varpi(1) } \, \sum _{ j \not= 1}   \mt_{1j} G^I_{q_\a, 1,j} + {\rm cycl}(1,2,3,4,5)
\eea
It will be useful in later steps to eliminate $\mt_{15}$ by means of the four-term identity
\bea
\mt_{12} + \mt_{13} + \mt_{14} + \mt_{15} = 0
\label{4termid}
\eea
which readily follows from (\ref{7.def.T}). Based on (\ref{4termid}), the identity $G_{q_\a, 1,2}^I - G^I_{q_\a, 1,5}= G^I_{1,2,q_\a,5}$ and  cyclic permutations of $(2,3,4)$ thereof, we obtain,
\bea
\mF_3 + \mF_4 
& = & 
- 2 \Dt  \, { \om _I(1) \over \varpi(1) } \Big [  
\mt_{12}  G^I_{1,2, q_\a, 5} + \mt_{13}  G^I_{1,3, q_\a, 5} + \mt_{14}  G^I_{1,4, q_\a, 5}  \Big ]
+ {\rm cycl}(1,2,3,4,5)
\quad
\eea
Using the decomposition of (\ref{7.Lem.2}) for $\Dt$ and the basis forms $W_{a;b}$ in (\ref{basforms}), we have, 
\bea
\label{7.F3F4}
\mF_3+\mF_4 = \mF_B'+\mF_B''
\eea
where
\bea
\label{7.FBB}
\mF_B '
& = & 
4    \Big [ k_3 \cdot k_4  W_1^I  + k_4 \cdot k_5  (W_2^I + W_5^I) \Big ] 
\no \\ && \quad \times 
 \Big [  
\mt_{12} \, G^I_{1,2, q_\a, 5} + \mt_{13} \, G^I_{1,3, q_\a, 5} + \mt_{14} \, G^I_{1,4, q_\a, 5}  \Big ]
+ {\rm cycl}(1,2,3,4,5)
\no \\ 
\mF_B'' & = &  4  \Big [   k_1 \cdot k_5  W_{1;3}^I +  k_1 \cdot k_2 W_{1;4}^I +  k_1 \cdot (k_4+k_5)  W_{1;5}^I  \Big ]
\no \\ && \quad \times  
 \Big [   \mt_{12} \, G^I_{1,2, q_\a, 5} + \mt_{13} \, G^I_{1,3, q_\a, 5} +  \mt_{14} \, G^I_{1,4, q_\a, 5}  \Big ]
+ {\rm cycl}(1,2,3,4,5)
\eea
These contributions will be collected with other terms linear in $\mt_{ij}$ in subsection \ref{sec:tijterms}.

\subsection{Simplifying $\mF_5+ \mF_6$}

We shall now simplify the expressions for both of $\mF_5, \mF_6$ in (\ref{mF1tomF6}) and then combine both using $t_8$. Decomposing the summand of $\mF_5$ using $D(z_j,z_k) = D(z_j,z_0') - D(z_k,z_0')$ for an arbitrary point $z_0'$, and using the anti-symmetry $(f_i f_j f_k)=-(f_j f_i f_k)$, we recast the formula as follows,
\bea
\mF_5 =
-  \half \sum _i \rho_i \, k_i \cdot \cP(z_i) \, D(z_i,z_0)  \sum_{j \not=i} D(z_j,z_0') 
\sum_{ (k,\ell,m) \not =i,j} ([f_i, f_j]   f_k) ( f_\ell f_m)  
\eea
where the  sum  is over  all mutually distinct $(k,\ell,m)$ not equal to $i$ or $j$. Similarly decomposing the summand in $\mF_6$ using $D(z_j,z_m) = D(z_j,z_0') - D(z_m,z_0')$, we obtain, 
\bea
\mF_6 =
2  \sum_i \rho_i \, k_i \cdot \cP(z_i) \, D(z_i,z_0)  \sum _{j\not=i}  D(z_j,z_0') 
\sum_{(k,\ell,m) \not= i,j} \Big [ (f_i f_j f_k f_\ell f_m) - (f_j f_i f_m f_k f_\ell ) \Big ]  
\quad
\eea
The sum is over all mutually distinct $(k,\ell,m)$ not equal to $i$ or $j$. As a result, the corresponding sums over the matrices $f_k f_\ell f_m$ and $f_m f_k f_\ell$ are equal to one another, and we have, 
\bea
\mF_6 =
2  \sum_i \rho_i \, k_i \cdot \cP(z_i) \, D(z_i,z_0)  \sum _{j\not=i}  D(z_j,z_0') 
\sum_{(k,\ell,m) \not= i,j}  ([f_i, f_j]  f_k f_\ell f_m) 
\eea
Adding $\mF_5$ and $\mF_6$, we recover the
$t_8$ contraction of commutators in (\ref{7.def.T}),
\bea
\mF_5 + \mF_6 =
4 \sum_{i \not= j}  \rho_i \, k_i \cdot \cP(z_i) \, D(z_i,z_0)   D(z_j,z_0') \,  \mt_{ij}
\eea
Decomposing also this sum into cyclic permutations, and choosing $z_0'=z_5$, we have, 
\bea
\mF_5 + \mF_6 =
4  \rho_1 \, k_1 \cdot \cP(z_1) \, D(z_1,z_0)  \sum_{j =2}^4  D(z_j,z_5) \,  \mt_{1j}
+ {\rm cycl}(1,2,3,4,5)
\eea
Moreover, we choose the point $z_0$ adapted to each one of the three terms, 
\bea
\mF_5 + \mF_6 & = &
4   { k_1 \cdot \cP (1) \over \varpi(1)} 
\Big [ \varpi(4) \Delta (1,3) \Delta (2,5) \mt_{12} + \varpi(4) \Delta (1,2) \Delta (3,5) \mt_{13} 
\no \\ && \hskip 0.8in 
+ \varpi(3) \Delta (1,2) \Delta (4,5) \mt_{14} \Big ] + {\rm cycl}(1,2,3,4,5)
\eea
Finally, decomposing these forms onto the basis $W_{a;b}^I$, we obtain, 
\bea
\label{7.F5F6}
\mF_5 + \mF_6 & = &
4   k_1 \cdot \mP _I(1)  
\Big [ (W^I_{1;3}-W^I_{1;4}+W^I_{1;5}) \mt_{12} + (W^I_{1;3}+W^I_{1;5}) \mt_{13} + W^I_{1;3} \mt_{14} \Big ] 
\no \\ && 
+ {\rm cycl}(1,2,3,4,5)
\eea
These contributions will be collected with other terms linear in $\mt_{ij}$ in subsection \ref{sec:tijterms}.

\subsection{Simplifying $\mF_7$}

We may recast $\mF_7$ as given in (\ref{7.mF78910}) in the following way,
\bea
\mF_7 =
 {2 \over 3} \cB^{\mu \nu} \sum_i \mt_i f_i ^{\mu \nu}  {X_i \over \varpi(z_i)}
\eea
where $X _i$ is defined by 
\bea
X_1 & = & 
   \Delta (1,2)\Delta(1,3) \varpi(4) \varpi(5) + \Delta (1,3)\Delta(1,4) \varpi(2) \varpi(5) 
   \no \\ &&
+ \Delta (1,2)\Delta(1,4) \varpi(3) \varpi(5) + \Delta (1,3)\Delta(1,5) \varpi(2) \varpi(4) 
   \no \\ &&
+ \Delta (1,2)\Delta(1,5) \varpi(3) \varpi(4) + \Delta (1,4)\Delta(1,5) \varpi(2) \varpi(3) 
\eea
and cyclic permutations thereof. Expressed in this manner, the contributions to $\mF_7$ involve  simultaneous poles arising from $(\varpi(1) \varpi(2))^{-1}$ and  $(\varpi(1) \varpi(5))^{-1}$ in the product of $\cB^{\mu \nu}$ given in (\ref{7.mPb}) with $X_i/\varpi(z_i)$ but these simultaneous poles will be proven to be spurious below.
To do so, we substitute  the expression (\ref{7.mPb}) for $\cB^{\mu \nu}$ to obtain, 
\bea
\mF_7 =  {4 \over 3} \, \mt_1 f_1 ^{\mu \nu}  \mP^\nu _I(q_\a)  {X_1 \over \varpi(1)} 
\sum_{a=2}^5 k_a^\mu   {\om_I(a) \over \varpi(a)}  
 + {\rm cycl}(1,2,3,4,5)
\eea
Expressing $k_5$ in terms of the other momenta by momentum conservation and using the fact that $k_1^\mu f_1^{\mu \nu}=0$, we obtain, 
\bea
\mF_7 & = &
  {4 \over 3} \, \mP^\nu _I(q_\a)  \mt_1 f_1 ^{\mu \nu}    \left (
  k_2^\mu \left \{  {\om_I(2) \over \varpi(2)}   - {\om_I(5) \over \varpi(5)} \right \}   {X_1 \over \varpi(1)}
 +{\rm cycl}(2,3,4) \right )
 \no \\ &&
 + {\rm cycl}(1,2,3,4,5)
 \qquad
 \label{spupoles}
\eea
By decomposing the differential form inside the parentheses  onto $\om_I(z_i)/\varpi(z_i)$, we obtain,
\bea
\label{7.omX}
&&
\left \{  {\om_I(2) \over \varpi(2)}   - {\om_I(5) \over \varpi(5)} \right \}   {X_1 \over \varpi(1)}
\no \\ && \hskip 0.3in =
3  \left [ { \om_I(1) \over \varpi(1)} - \half { \om_I(2) \over \varpi(2)} - \half { \om_I(5) \over \varpi(5)}\right ] \Delta(2,5) \Big ( \Delta (1,3) \varpi(4) + \Delta(1,4) \varpi(3) \Big )
\no \\ && \hskip 0.5in
- \half \left [ { \om_I(2) \over \varpi(2)} - { \om_I(5) \over \varpi(5)} \right ]
\Big ( \Delta(2,4) \Delta(3,5) + \Delta(2,3) \Delta(4,5) \Big ) \varpi (1) 
\eea
where the spurious  simultaneous poles $(\varpi(1) \varpi(2))^{-1}$ and  $(\varpi(1) \varpi(5))^{-1}$ present in (\ref{spupoles}) have been now removed.

\sm

The expressions involving $z_2 \leftrightarrow z_3, z_4$ may be obtained from (\ref{7.omX}) by cyclic permutations of $2,3,4$ on both sides. Decomposing the result onto the basis $W_{a;b}$  and then converting the diagonal parts 
$W_{a;a}$ onto the basis of off-diagonal parts $W_{a;b}$ with $b \not= a$ via (\ref{7.Wdiag}), we obtain, 
\bea
\mF_7 & = &
{4 \over 3} \, \mP^\nu _I(q_\a)  \mt_1 f_1 ^{\mu \nu}    \Big [
k_2^\mu \Big \{ 
{-}3 W^I_{1; 2} - 6 W^I_{1; 4} - 2 W^I_{2; 3} + 3 W^I_{2; 4} - W^I_{2; 5} 
\no \\ && \hskip 1.3in 
+ 
 2 W^I_{3; 2} + 4 W^I_{3; 4} + 2 W^I_{4; 3} - W^I_{5; 1} - W^I_{5; 4}
 \Big \}
\no \\ && \hskip 1in 
+ k_3^\mu \Big \{ 2 W^I_{1; 5} -2 W^I_{4; 5}-W^I_{1; 2}  - W^I_{3; 1} - W^I_{3; 5} \Big \}
\no \\ && \hskip 1in
+ k_4^\mu \Big \{ 2 W^I_{1; 2}- 2 W^I_{3; 2}  - W^I_{1; 5}  - W^I_{4; 1} - W^I_{4; 2} \Big \} 
 \no \\ && \hskip 1in 
 + k_5^\mu \Big \{ 
{-} 3 W^I_{1; 5} -6 W^I_{1; 3}- 2 W^I_{5; 4}  + 3 W^I_{5; 3}  - W^I_{5; 2}   \no \\ && \hskip 1.3in 
+ 2 W^I_{4; 5} + 4 W^I_{4; 3} + 2 W^I_{3; 4}  - W^I_{2; 1} - W^I_{2; 3} 
 \Big \}
  \Big ]
  \no \\ &&\quad
  + {\rm cycl} (1,2,3,4,5)
  \label{finalmf7}
\eea
where we have used $f_1^{\mu \nu}\sum_{a=2}^5 k_a^\mu = 0$ to simplify the expression. Note that the coefficients of $k_2, k_3, k_4,k_5$ are not related by cyclic permutations of the labels of $W_{a;b}$ since the underlying forms (\ref{7.omX}) and $z_2\leftrightarrow z_3,z_4$ are not images of each other under cyclic permutations of $1,2,3,4,5$. Instead, (\ref{finalmf7}) follows from separately reducing the right-hand side of (\ref{7.omX}) and its relabelling $z_2\leftrightarrow z_3,z_4$ into the basis of $W_{a;b}$. Fortunately, many of the complicated contributions to $\mF_7$ will cancel against counterparts in $\mF_8+\mF_9+\mF_{10}$ to which we now turn our attention.

\subsection{Simplifying  $\mF_8+ \mF_9 + \mF_{10}$}
\label{sec:simp8910}

The starting point is the expression for $\mF_8, \, \mF_9, \, \mF_{10}$ given in (\ref{7.mF78910}), which we recast in terms of the functions $D$ and the differential $\rho$ as follows,
\bea
\label{7.F890}
\mF_8 & = &
- \half \rho \cB^{\mu \nu}   \sum _{(i,j,k,\ell,m)} (f_i f_j f_k)^{\mu \nu} (f_\ell f_m) \Big ( D (z_j,z_\ell) D(z_k,z_m)  + D (z_i,z_\ell) D (z_j,z_k)  \Big ) 
\no \\
\mF_9 & = &
 2 \rho \cB^{\mu \nu}  \sum _{(i,j,k,\ell,m)} (f_i f_j f_k f_\ell f_m)^{\mu \nu} D (z_i,z_\ell) D (z_j,z_m)
  \no \\
 \mF_{10} &= &
 {1 \over 3} \rho \cB^{\mu \nu}   \sum _{(i|j,k|\ell,m)} f_i ^{\mu \nu} C_T(j,k|\ell,m) D(z_j,z_k) D(z_\ell,z_m)
\eea
In $\mF_8$ we have included a factor of 1/2 to account for the fact that the formula is written as a sum over all permutations $(i,j,k,\ell,m)$.  Introducing an arbitrary intermediate point $z_0$ to split the $D$-functions using the relation $D(z_i,z_j) = D(z_i, z_0) - D(z_j,z_0)$, and recasting the sums in terms of  the functions $D(z_i,z_0) D(z_j,z_0)$,  we may express the sum $\mF_8+\mF_9+\mF_{10}$ as follows after properly  symmetrizing in $i$ and $j$,
\bea
\mF_8+\mF_9+\mF_{10} =  \cB^{\mu \nu} \cL^{\mu \nu}
\hskip 1in 
\cL^{\mu \nu} = \rho \sum_{i \not= j} D(z_i, z_0) D(z_j, z_0) S_{ij} ^{\mu \nu}
\label{matchwith}
\eea
The anti-symmetric tensors $S_{ij}^{\mu \nu}$ are built out of the field strengths $f$ without further dependence on the momenta $k$, whose explicit expressions are given in appendix \ref{sec:I}, but will not be needed here. The only information needed here is the following properties, 
\bea
\label{7.Sa}
S^{\mu \nu} _{ij} = S^{\mu \nu} _{ji} = - S^{\nu \mu} _{ij} 
\hskip 1in 
\sum _{j \not= i} S_{ij}^{\mu \nu} =0
\eea
which in particular guarantee that the combination above is independent of the arbitrary point $z_0$. The following non-trivial kinematic relation, which is proven in appendix \ref{sec:I}, 
\bea
\label{7.kS}
k_i ^\mu S_{ij}^{\mu \nu} = - k_i ^\nu \mt_{ij} + k_i^\mu f_j^{\mu \nu} \mt_j
- {1 \over 3} k_i^\mu \sum _{\ell \not=i,j}  f_\ell^{\mu \nu} \mt_\ell
\eea
will be  important in simplifying $\mF_8+ \mF_9 + \mF_{10}$ and obtaining a useful explicit formula.

\subsubsection{Re-expressing $\cL^{\mu \nu}$}

We shall now re-express $\cL^{\mu \nu} $ in terms of the five holomorphic basis forms $\Wt_i$
in (\ref{defwws}). To do so, we choose $z_0=z_5$ in $\cL^{\mu \nu}$ and derive the following identities from  (\ref{7.Sa}), 
\bea
\label{7.Ssums}
S_{12} + S_{13} + S_{14} + S_{23} + S_{24} + S_{34}& = & 0
\no \\
S_{12}+S_{13}+S_{23} - S_{45} & = & 0
\no \\
S_{23}+S_{24}+S_{34} -  S_{15} & = & 0
\eea
We subtract from $\cL^{\mu \nu}$ the product of the first identity times $2 \rho D(1,5) D(4,5)$,  the second identity times $ 2 \rho D(1,5) D(3,4)$ and the third identity times $2 \rho D(2,1) D(4,5)$ to obtain, 
\bea
\cL^{\mu \nu} & = & 
- 2 \rho S_{23}^{\mu \nu} D(1,2) D(3,4) 
- 2 \rho S_{34}^{\mu \nu} D(2,3) D(4,5) 
- 2 \rho S_{45}^{\mu \nu} D(3,4) D(5,1) 
\no \\ &&
 - 2 \rho S_{51}^{\mu \nu} D(4,5) D(1,2)  
 - 2 \rho S_{12}^{\mu \nu} D(5,1)  D(2,3)  
\eea
This identity is parallel to (\ref{7.Lem.1}) in the sense that this expression involves  only those $S_{ij}$ for which $i$ and $j$ are ``nearest neighbors", namely related by $j=i\pm 1 \ {\rm mod} \ 5$.
Finally, along with its cyclic permutations, we have,
\bea
\cL^{\mu \nu} =
 -2 \Wt_1 \, S_{34}^{\mu \nu} 
 - 2 \Wt_2 \, S_{45}^{\mu \nu} 
 - 2 \Wt_3 \, S_{51}^{\mu \nu} 
 - 2\Wt_4 \, S_{12}^{\mu \nu} 
 - 2\Wt _5 \, S_{23}^{\mu \nu}
\eea
 Using the following shorthand for nearest neighbors $S_{ij}$,
\bea
\cS_1^{\mu \nu}  = S_{34}^{\mu \nu} , \qquad
\cS_2^{\mu \nu}  = S_{45}^{\mu \nu} , \qquad
\cS_3^{\mu \nu}  = S_{51}^{\mu \nu} , \qquad
\cS_4^{\mu \nu}  = S_{12}^{\mu \nu} , \qquad
\cS_5^{\mu \nu}  = S_{23}^{\mu \nu} 
\eea
we have the more compact formula,
\bea
\label{7.LS}
\cL^{\mu \nu}  = - 2 \sum_a \Wt_a \, \cS_a^{\mu \nu} 
\eea

\subsubsection{Contraction with $\cB^{\mu \nu}$}

We now consider the contraction of $\cL^{\mu \nu}$ in (\ref{7.LS}) with the expression for $\cal B^{\mu \nu}$ in (\ref{7.mPb}), and use the anti-symmetry of $\cL^{\mu \nu}$ in $\mu, \nu$ to simplify the expression, 
\bea
\label{7.Wab}
\mF_8+\mF_9+ \mF_{10} =  -4 \, \mP^\nu _I(q_\a)  \sum _{a,b} W_{a;b}^I \,  k^\mu _a   \cS_b ^{\mu \nu}
\eea
As discussed in subsection \ref{sec:7.2}, the 25 meromorphic $(1,0)$-forms generate a vector space of dimension 20 for which a basis may be chosen of off-diagonal $W^I_{a;b}$ with $b \not= a$. Using the explicit relations of (\ref{7.Wdiag}) we may eliminate all diagonal $W_{a;a}^I$ in terms of the off-diagonal entries.  Thus, $\mF_8+\mF_9+\mF_{10}$ may be expressed as follows,
\bea
\mF_8+\mF_9+\mF_{10} =  -2 \, \mP^\nu _I(q_\a) \sum _{a \not= b} W_{a;b}^I \,  C_{a;b}^\nu
\label{tot8910}
\eea
where the kinematic factors $C_{a;b}^\nu$ are defined by the decomposition of (\ref{7.Wab})  into the basis of $W^I_{a;b}$ with $a\neq b$, and a factor of 2 has been absorbed into  their normalization for later convenience.

\sm

Our next task is to calculate $C_{a;b}^\nu$ explicitly. Inspection of the expression (\ref{7.Wdiag}) for the diagonal entries $W^I_{a;a}$ in terms of the off-diagonal entries, we see that $W^I_{a;a}$ involves $W^I_{a;b}$ for $b \not=a$ where $a$ and $b$ are nearest neighbors. Thus, the calculation of $C_{a;b}^\nu$ may be separated into two cases: either $a\not =b$ are ``nearest neighbors" (mod 5), or they are not. When $a,b$ are not nearest neighbors we have,
\bea
C_{1;3}^\nu = 2 k_1^\mu \cS_3^{\mu \nu} = 2 k_1^\mu S_{15}^{\mu \nu} \, , \ \ \ \
C_{1;4}^\nu = 2 k_1^\mu \cS_4^{\mu \nu} = 2 k_1^\mu S_{12}^{\mu \nu} 
\label{someCab}
\eea
and their respective cyclic permutations. Both cases are of the form $ k_i^\mu S_{ij}^{\mu \nu}$
and thereby amenable to the rewriting (\ref{7.kS}) in terms of $t_8$ tensors.

\sm

The evaluation of $C_{a;b}^\nu$ for the case where $a,b$ are nearest neighbors is more complicated as one now also needs to pick up the contributions from the diagonal parts $k_a ^\mu S_a^{\mu \nu}$. We begin by observing that the diagonal contributions from $k_a^\mu \cS_a^{\mu \nu}$ to $C_{1;2}$ and to $C_{1;5}$ are equal to one another. Therefore, their difference evaluates as follows,
\bea
C_{1;2}^\nu  - C_{1;5}^\nu = 2 k_1 ^\mu ( S_{45}-S_{23})^{\mu \nu} 
= k_1 ^\mu (S_{12}+S_{13}-S_{14}-S_{15}) ^{\mu \nu} 
\qquad
\eea
along with its cyclic permutations. Their sum may also be readily obtained and we have,
\bea
C_{1;2}^\nu  + C_{1;5}^\nu  
=
2 k_1 ^\mu (S_{34}+ S_{45} +S_{23})^{\mu \nu}  + 2 k_2 ^\mu S_{45}^{\mu \nu}   
- 2 k_3 ^\mu S_{15}^{\mu \nu}  - 2 k_4 ^\mu S_{12}^{\mu \nu}  + 2 k_5 ^\mu S_{23}^{\mu \nu} 
\eea
Combining the terms proportional to $S_{23}$ and $S_{45}$, using momentum conservation on both coefficients and  the summation relations in (\ref{7.Sa}) for $i=2$ and $i=5$ to re-express the combinations inside the parentheses, and combining this result with the formula for the difference, we obtain, 
\bea
C_{1;2}^\nu  & = &   
-  (k_4+k_5)^\mu S_{45}^{\mu \nu}  -  (k_2+k_3) ^\mu S_{23}^{\mu \nu}   -  (k_2+k_5) ^\mu S_{25}^{\mu \nu}  
\no \\ && 
+  k_3 ^\mu  S_{35}^{\mu \nu}    +  k_4^\mu  S_{24}^{\mu \nu}   +k_1 ^\mu (S_{12}  -S_{14})^{\mu \nu} 
\no \\
C_{1;5}^\nu  & = &   
-  (k_4+k_5)^\mu S_{45}^{\mu \nu}   -  (k_2+k_3) ^\mu S_{23}^{\mu \nu}   -  (k_2+k_5) ^\mu S_{25}^{\mu \nu}  
\no \\ && 
+  k_3 ^\mu S_{35}^{\mu \nu}    +  k_4 ^\mu  S_{24}^{\mu \nu}    + k_1 ^\mu (S_{15} -S_{13})^{\mu \nu} 
\label{moreCab}
\eea
along with their respective cyclic permutations. Note that every single term in $C_{1;2}$ and $C_{1;5}$ is of
the form of $k_i^\mu S^{\mu \nu}_{ij}$ and may be expressed in terms of $t_8$ tensors by virtue of (\ref{7.kS}).
In summary, by inserting (\ref{someCab}), (\ref{moreCab}) and their cyclic permutations into (\ref{tot8910}),
the kinematic factors in $\mF_8{+}\mF_9{+}\mF_{10} $ boil down to gauge invariant combinations of
$\mt_i f_i^{\mu \nu}$ and $\mt_{ij}$.

\subsection{Combining all terms proportional to $\mt_i$}

By the results of the previous subsections, the kinematic dependence of the entire chiral amplitude, proportional to  $\mF$ in (\ref{defmFb}), is expressed in terms of $\mt_i$ and $\mt_{ij}$ defined in (\ref{newsumm.5}) and (\ref{7.def.T}), respectively. In this subsection, we combine all the contributions proportional to $\mt_1 $ which, by addition of cyclic 
permutations, will give us all contributions proportional to $\mt_i $ (as opposed to $\mt_{ij}$). 
In subsequent equations, the restriction $|_{\mt_i}$ stands for retaining all terms that involve 
either $\mt_i \, \ep_i$ or $\mt_i \, f_i$, relegating the contributions of $\mt_{ij}$ to section \ref{sec:tijterms}.

\sm

No contributions of the form $\mt_i $ arise from $\mF_3+\mF_4$ and $\mF_5 + \mF_6$ since all the kinematic factors in (\ref{7.FBB}) and (\ref{7.F5F6}) are of the form $\mt_{ij}$. Thus, here we collect all $\mt_i$ contributions from $\mF_1+ \mF_2+\mF_7+\mF_8+\mF_9 + \mF_{10}$. Note that we have reorganized $\mF_1+\mF_2=\mF_A+ \mF_A'+\mF_A''$ and that $\mF_A$ is manifestly independent of $q_\a$, see (\ref{7.mFA}). Our first goal will be to show independence of $q_\a$, so we shall concentrate on the contribution from $\mF_A'+\mF_A''$.  Using {\tt MAPLE}, one readily combines the following contributions,
\bea
\label{7.com1}
\Big ( \mF_A'+\mF_7+\mF_8+\mF_9 + \mF_{10} \Big ) \Big |_{\mt_i }
& = &  8 \mt_1 f_1 ^{\mu \nu} \mP_I^\nu (q_\a)
\Big ( k_5^\mu W_3^I + k_2^\mu W_4^I + (k_4^\mu +k_5^\mu ) W_5^I \Big )
\no \\ &&
+ {\rm cycl}(1,2,3,4,5)
\eea
where the poles from $\varpi(a)^{-1}$ due to $W^I_{a;b}$ with $a\neq b$ cancel. In order to
combine with the remaining terms of $\mF_1{+}\mF_2$, we decompose $\mP^\nu _I(q_\a) = \mP^\nu _I(1) - \big ( \mP^\nu _I(1) {-} \mP^\nu _I(q_\a)  \big )$ in (\ref{7.com1}). Like this, the last term has the same
prefactor as $\mF_A''$ in (\ref{7.mFA}), and the differences of $\mP^\nu _I$ can be rewritten using (\ref{7.PG})
in both cases. We thus arrive at,
\bea
\mF \, \Big |_{t_i} =  \mF_A+ \mF_C + \mF_C' 
\label{tidec}
\eea
where $\mF_A$ is kept in the form of (\ref{7.mFA}) and $\mF_C$ is obtained from 
(\ref{7.com1}) by replacing $\mP_I^\nu (q_\a) \rightarrow \mP_I^\nu (1)$
and re-expressing $W_a^I$ in terms of $\om_I$ and $\Delta$
\bea
\mF_C &=& - 8 \mt_1 f_1 ^{\mu \nu} \mP_I^\nu (1)
\Big [ k_2^\mu \om_I(4) \Delta (1, 3) \Delta (2,5) 
+ k_3^\mu \om_I (4) \Delta (1, 2) \Delta (3,5) 
\no \\ && \hskip 1.1in 
+ k_4^\mu \om_I (3) \Delta (1, 2) \Delta (4,5) \Big ] + {\rm cycl}(1,2,3,4,5)
\eea
While $\mF_A$ and $\mF_C$ are manifestly independent of $q_\alpha$, the third
contribution $\mF_C'$ to (\ref{tidec}) carries the leftover $q_\alpha$-dependence from
$ \mP^\nu _I(1)  - \mP^\nu _I(q_\a) $ of both $\mF_A''$ and (\ref{7.com1}),
 \bea
\mF_C' & = & 8 \mt_1 f_1^{\mu \nu} \Big [ k_2 ^\nu G^I_{1,2,q_\a,5} +  k_3 ^\nu G^I_{1,3,q_\a,5}  +  k_4 ^\nu G^I_{1,4,q_\a,5}  \Big ] 
 \no \\ && \qquad \times
\Big [  k_2^\mu \left ( W^I_{1;3} - W^I_{1;4} +W^I_{1;5} - W_3^I +W^I_4 - W_5^I  \right ) 
\no \\ && \qquad \quad
+ k_3^\mu \left ( W^I_{1;3} + W^I_{1;5} - W_3^I - W_5^I  \right ) + k_4^\mu \left ( W^I_{1;3} - W_3^I \right )  \Big ]
\no \\ && 
+ {\rm cycl}(1,2,3,4,5)
\label{fcprime}
 \eea
As will be shown in section \ref{sec:7.summ} and appendix \ref{sec:J}, the $q_\alpha$-dependent contribution $\mF_C' $ will eventually cancel terms with kinematic factors $\mt_{ij}$ that we shall discuss next.

 \subsection{Combining all terms proportional to $\mt_{ij}$}
 \label{sec:tijterms}

In this subsection, we combine all the contributions proportional to $\mt_{ij}$ (as opposed to
the kinematic factors $\mt_i$ in the simplified form (\ref{tidec})). Based on the results of 
section \ref{sec:simp8910}, a straightforward collection of terms involving 
$\mt_{ij}$  in $\mF_7+\mF_8+\mF_9+\mF_{10}$ shows that we have,
\bea
\Big ( \mF_7+\mF_8+\mF_9+\mF_{10} \Big ) \Big | _{\mt_{ij}} 
& = & 
- 4 k_1 \cdot \mP_I(q_\a) \Big [  \mt_{12} \left ( W^I_{1;3} - W^I_{1;4} +W^I_{1;5} - W_3^I +W^I_4 - W_5^I  \right ) 
\no \\ && \qquad
+ \mt_{13} \left ( W^I_{1;3} + W^I_{1;5} - W_3^I - W_5^I  \right )
+ \mt_{14} \left ( W^I_{1;3} - W_3^I \right ) \Big ]
\qquad
\no \\ &&
+ {\rm cycl}(1,2,3,4,5)
\eea
where the restriction $|_{\mt_{ij}}$ stands for retaining all terms that involve $\mt_{ij}$ rather than $\mt_i$.
Combining the above contribution  with $\mF_5+\mF_6$ from (\ref{7.F5F6}) and splitting up the 
combination $\mP^\nu _I(q_\a) = \mP^\nu _I(1) - \big ( \mP^\nu _I(1)  - \mP^\nu _I(q_\a)  \big )$ as we did earlier,  mechanical simplifications lead to,
\bea
\Big ( \mF_5+\mF_6+ \mF_7+\mF_8+\mF_9+\mF_{10} \Big ) \Big | _{\mt_{ij}} 
=  \mF_D + \mF_D'  
\label{terms6to10}
\eea
In the first piece $\mF_D$ all the $W_{a;b}^I$ with $a{\neq}b$ and thus all the poles
$\varpi(a)^{-1}$ cancel, so we find
 \bea
\mF_D & = & 
 4 k_1 \cdot \mP^I (1)  \Big [ 
 \mt_{12} \, \om_I(4) \Delta (1,3) \Delta (2,5) 
 + \mt_{13} \, \om_I(4) \Delta (1,2) \Delta (3,5)  
 \no \\ && \hskip 0.8in 
 + \mt_{14} \, \om_I(3) \Delta (1,2) \Delta (4,5)  \Big ] 
 + {\rm cycl}(1,2,3,4,5)
 \label{mfdd}
 \eea
after reconverting its $W_a^I$ to the basis of $\om_I$ and $\Delta$. The $q_\alpha$-dependent
terms $\mF_D'$ in (\ref{terms6to10}) are given by,
\bea
\mF_D' &=& 
4 k_1^\mu \Big [k_2^\mu G^I_{1,2,q_\a,5} + k_3^\mu G^I_{1,3,q_\a,5} + k_4^\mu G^I_{1,4,q_\a,5}  \Big ]
\no \\ && \qquad \times 
\Big [ 
\mt_{12} \left ( W^I_{1;3} - W^I_{1;4} +W^I_{1;5} - W_3^I +W^I_4 - W_5^I  \right )
\no \\ && \hskip 0.5in 
+ \mt_{13} \left ( W^I_{1;3} + W^I_{1;5} - W_3^I - W_5^I  \right )
+ \mt_{14} \left ( W^I_{1;3} - W_3^I \right ) \Big ]
\no \\ &&
+ {\rm cycl}(1,2,3,4,5)
 \eea
In order to complete the assembly of the terms $\mt_{ij}$ in the chiral amplitude, we combine (\ref{terms6to10})
with $\mF_3+\mF_4=\mF_B'+\mF_B''$ from (\ref{7.F3F4}) and obtain
\bea
\mF\, \Big | _{\mt_{ij}} 
= \mF_B' + \mF_B'' + \mF_D + \mF_D'  
\eea
with $\mF_B'$ and $\mF_B''$ given by (\ref{7.FBB}).

\subsubsection{Rearranging the $q_\alpha$-dependent terms}

The combination of  $\mF_B''$ and $\mF_D'$ may be rearranged as follows, 
\bea
\mF_B''+\mF_D' & = & \mF_E' + \mF_E''
\eea
where a first piece is again free of $\varpi(a)^{-1}$,
\bea
\mF_E' & = & 
- 4 k_1^\mu \Big [  \mt_{12} G^I_{1,2,q_\a,5} + \mt_{13} G^I_{1,3,q_\a,5} + \mt_{14} G^I_{1,4,q_\a,5} \Big ]
\no \\ && \qquad \times 
\Big [ k_2^\mu (W_3^I-W_4^I+W_5^I) 
+ k_3^\mu (W_3^I+W_5^I) + k_4^\mu  W_3^I \Big ] 
\no \\ &&
+ {\rm cycl}(1,2,3,4,5)
\eea
and the leftover terms are arranged as follows for later convenience,
\bea
\mF_E'' & = & 
4 \left ( W^I_{1;3} - W^I_{1;4} +W^I_{1;5} - W_3^I +W^I_4 - W_5^I  \right )
\no \\ && \qquad \times 
\Big [ (k_1 \cdot k_3 \mt_{12} - k_1 \cdot k_2 \mt_{13} ) G^I_{1,3,q_\a,5}
+ (k_1 \cdot k_4 \mt_{12} - k_1 \cdot k_2 \mt_{14} ) G^I_{1,4,q_\a,5} \Big ]
\no \\ &&
+ 4 \left ( W^I_{1;3} + W^I_{1;5} - W_3^I - W_5^I  \right )
\label{mfepp} \\ && \qquad \times 
\Big [ (k_1 \cdot k_2 \mt_{13} - k_1 \cdot k_3 \mt_{12} ) G^I_{1,2,q_\a,5}
+ (k_1 \cdot k_4 \mt_{13} - k_1 \cdot k_3 \mt_{14} ) G^I_{1,4,q_\a,5} \Big ]
\no \\ &&
+ 4 \left ( W^I_{1;3} - W_3^I \right )
\Big [ (k_1 \cdot k_2 \mt_{14} - k_1 \cdot k_4 \mt_{12} ) G^I_{1,2,q_\a,5}
%\no \\ && \hskip 1.3in
+ (k_1 \cdot k_3 \mt_{14} - k_1 \cdot k_4 \mt_{13} ) G^I_{1,3,q_\a,5} \Big ]
\no \\ &&
+ {\rm cycl}(1,2,3,4,5)
\no
\eea
The separation into $\mF_E'$ and $\mF_E''$ is motivated by the subsequent simplifications
of the combination,
\bea
\mF_B= \mF_B'+\mF_E'
\eea
We will use the identity, 
\bea
\label{8.cycl}
&&
k_3 \cdot k_4 \, \om_I(1) \Delta (2,3) \Delta (4,5) + k_4 \cdot k_5 \, \om_I(1) \Delta (3,4) \Delta (5,2)
\no \\ && \qquad =
\Big ( k_1 \cdot k_2 \, \om_I(4) \Delta (5,1) \Delta (2,3) + {\rm cycl}(1,2,3,4,5) \Big )
\no \\ && \qquad \quad 
+ k_1 \cdot k_2 \, \om_I(4) \Delta (1,3) \Delta (2,5) 
+ k_1 \cdot k_3 \, \om_I(4) \Delta (1,2) \Delta (3,5) 
\no \\ && \qquad \quad 
+ k_1 \cdot k_4 \, \om_I(3) \Delta (1,2) \Delta (4,5) 
\eea
analogous to the equality of $\Dt$ in equations (\ref{7.Lem.1}) and (\ref{7.Lem.2}).
In fact, it will be convenient to introduce the shorthand
\bea
{\cal D}_I= - 2 k_3 \cdot k_4 \, \omega_I(1) \Delta (2,3) \Delta (4,5) + \hbox{\rm cycl}(1,2,3,4,5)
\label{defdt}
\eea
for the permutation invariant which follows from the replacement $\varpi(a) \rightarrow
\omega_I(a)$ in the expression (\ref{7.Lem.1}) for $\Dt$. With this definition and
the identity (\ref{8.cycl}), the sum $\mF_B$ is readily combined as follows,
\bea
\mF_B & = & 
-2 {\cal D}_I \Big [  \mt_{12} \, G^I_{1,2,q_\a,5} + \mt_{13} \, G^I_{1,3,q_\a,5} + \mt_{14} \, G^I_{1,4,q_\a,5} + {\rm cycl}(1,2,3,4,5) \Big ]
\eea
Using also the relation $G^I_{1,2,q_\a,5} = G^I_{1,2, q_\a} - G_{1,5,q_\a}^I$ and
the four-term identity (\ref{4termid}) among $\mt_{ij}$ we obtain equivalently, 
\bea
\mF_B & = & 
-2  {\cal D}_I  \Big [  \mt_{12} \, G^I_{1,2,q_\a} + \mt_{13} \, G^I_{1,3,q_\a} + \mt_{14} \, G^I_{1,4,q_\a} 
+ \mt_{15} \, G^I_{1,5,q_\a} + {\rm cycl}(1,2,3,4,5) \Big ]
\eea
By further use of (\ref{4termid}), it is readily shown that all $q_\alpha$-dependence cancels out 
between different terms of the cyclic orbit, so that we have,
\bea
\mF_B & = & 
-2 {\cal D}_I
\sum_{i \not = j} \mt_{ij} \, g^I _{i,j}
= - 4 {\cal D}_I \sum_{1\leq i<j}^5 \mt_{ij} \, g^I _{i,j}
= - 4 {\cal D}_I \sum_{2\leq i<j}^5 \mt_{ij} \, G^I _{1,i,j}
\label{mfbb}
\eea
The last step is based on (\ref{4termid}) to recast this permutation invariant combination of
$\mt_{ij} \, g^I _{i,j}$ in terms of the single-valued $G^I_{a,b,c}$ combinations.

\sm

In summary, the contributions to $\mF$ involving $\mt_{ij}$ are given by,
\bea
\mF\, \Big |_{\mt_{ij}} =   \mF_B + \mF_D  + \mF_E''
\label{parttij}
\eea
with $q_\alpha$-independent pieces $ \mF_D$ and $ \mF_B$ given by (\ref{mfdd}) and
(\ref{mfbb}), respectively, as well as a leftover $q_\alpha$-dependence in $\mF_E''$ given by (\ref{mfepp}).

\subsection{Summary}
\label{sec:7.summ}

In this subsection, we assemble all parts of $\mF$ based on
the organization of the $\mt_i$ and $\mt_{ij}$ dependent terms
in (\ref{tidec}) and (\ref{parttij}),
\bea
\mF = \mF\, \Big |_{\mt_{i}} + \mF\, \Big |_{\mt_{ij}} =\mF_A+ \mF_C + \mF_C'  + \mF_B + \mF_D  + \mF_E''
\eea
The only leftover $q_\alpha$-dependence resides in the parts $ \mF_C' $
and $\mF_E''$ given by (\ref{fcprime}) and (\ref{mfepp}), respectively.
In fact, it is proven in appendix \ref{sec:J} that the $q_\alpha$-dependent terms
entirely cancel,
\bea
\mF_C'+\mF_E'' =0
\eea
using highly non-trivial identities for kinematic factors and worldsheet dependences that will both be given in the same appendix.  Taking this result into account, we now arrive at a manifestly $q_\alpha$-independent representation of $\mF$,
\bea
\mF = \mF_A + \mF_B + \mF_C+\mF_D 
\label{summallf}
\eea
The remaining functions are recalled as follows
\bea
\mF_A & = & - 8 \mt_1    \ep_1\cdot   \cP(1) 
\Big [  k_4 \cdot k_5 \Delta(5,2)  \Delta(3,4)  + k_3 \cdot k_4  \Delta(2,3) \Delta (4,5)     \Big ]
  + {\rm cycl}(1,2,3,4,5)
\no \\
\mF_B & = & 
4 \Big [ k_1 \cdot k_2 \, \om_I(4) \Delta (5,1) \Delta (2,3) +  {\rm cycl}(1,2,3,4,5) \Big ]
\sum_{i \not = j} \mt_{ij} g^I _{i,j}
\no \\
\mF_C &=& - 8 \mt_1 f_1 ^{\mu \nu} \mP^I_\nu (1)
\Big [ k_2^\mu \om_I(4) \Delta (1, 3) \Delta (2,5) 
+ k_3^\mu \om_I (4) \Delta (1, 2) \Delta (3,5) 
 \label{otherform} \\ && \hskip 1.1in 
+ k_4^\mu \om_I (3) \Delta (1, 2) \Delta (4,5) \Big ] + {\rm cycl}(1,2,3,4,5)
\no \\
\mF_D & = & 
 4 k_1 \cdot \mP^I (1) \Big [ 
 \mt_{12} \om_I(4) \Delta (1,3) \Delta (2,5) 
 + \mt_{13} \om_I(4) \Delta (1,2) \Delta (3,5)  
 \no \\ && \hskip 0.8in 
 + \mt_{14} \om_I(3) \Delta (1,2) \Delta (4,5)  \Big ] 
 + {\rm cycl}(1,2,3,4,5)
 \no
\eea
where $ \cP(1) $ and $\mP^I (1)$ are defined in (\ref{pfield}) and (\ref{7.mPa}), respectively.

\subsubsection{Alternative presentation}

A further rearrangement of $\mF_A+\mF_C$ will lead to an alternative presentation of 
the manifestly $q_\alpha$-independent chiral amplitude $\mF$ in (\ref{summallf}).
The identity (\ref{8.cycl}) may be used to re-express $\mF_A$ as follows,
\bea
\mF_A& = & 
- 8 \mt_1    \ep_1 \cdot  \mP^I (1) 
\Big ( k_1 \cdot k_2 \, \om_I(4) \Delta (5,1) \Delta (2,3) + {\rm cycl}(1,2,3,4,5) \Big )
+ {\rm cycl}(1,2,3,4,5)
\no \\ &&
- 8 \mt_1 \ep_1 \cdot \mP^I(1) \Big [ 
k_1 \cdot k_2 \, \om_I(4) \Delta (1,3) \Delta (2,5) 
+ k_1 \cdot k_3 \, \om_I(4) \Delta (1,2) \Delta (3,5) 
\no \\ && \hskip 1in 
+k_1 \cdot k_4 \, \om_I(3) \Delta (1,2) \Delta (4,5)  \Big ] + {\rm cycl}(1,2,3,4,5)
\eea
Using the relations $f_1^{\mu \nu} k_2^\mu + \ep_1 ^\nu k_1 \cdot k_2 = \ep_1 \cdot k_2 \, k_1^\nu$, we observe that the terms on the second and third lines above combine with $\mF_C$ to produce the combinations $k_1 \cdot \mP^I(1)$, and we have,
\bea
\mF_A +\mF_C & = & 
- 8 \mt_1    \ep_1 \cdot   \mP^I (1) 
\Big ( k_1 \cdot k_2 \, \om_I(4) \Delta (5,1) \Delta (2,3) + {\rm cycl}(1,2,3,4,5) \Big )
\no \\ &&
- 8 \mt_1 k_1 \cdot \mP^I(1)  \ep_1^\nu \Big [ 
k_2^\nu \, \om_I(4) \Delta (1,3) \Delta (2,5) 
+ k_3^\nu  \, \om_I(4) \Delta (1,2) \Delta (3,5) 
\no \\ && \hskip 1.1in 
+ k_4^\nu \, \om_I(3) \Delta (1,2) \Delta (4,5)  \Big ] + {\rm cycl}(1,2,3,4,5)
\eea 
where the instruction to add cyclic permutations applies to all terms.
Further combining this result with the other terms gives the following alternative formula, 
\bea
\mF & = & 
- 8 \mt_1    \ep_1 \cdot   \mP^I (1) 
\Big ( k_1 \cdot k_2 \, \om_I(4) \Delta (5,1) \Delta (2,3) + {\rm cycl}(1,2,3,4,5) \Big )
\no \\ &&
+ 4 k_1 \cdot \mP^I(1) \Big [ 
(\mt_{12} -2 \mt_1 \ep_1 \cdot k_2 ) \om_I(4) \Delta (1,3) \Delta (2,5) 
\no\\ && \hskip 0.9in 
+ (\mt_{13} -2 \mt_1 \ep_1 \cdot  k_3 ) \om_I(4) \Delta (1,2) \Delta (3,5) 
 \label{finalform} \\ && \hskip 0.9in 
+ (\mt_{14} -2 \mt_1 \ep_1 \cdot k_4 ) \om_I(3) \Delta (1,2) \Delta (4,5)  \Big ] 
 \no \\ &&
+ 8 k_1 \cdot k_2 \, \om_I(4) \Delta (5,1) \Delta (2,3)  \sum_{1\leq i < j\leq 5} \mt_{ij} g^I _{i,j}
  + {\rm cycl}(1,2,3,4,5) \no
\eea
where the instruction to add cyclic permutations applies to all terms in $\mF$.

\sm

The final result of the RNS computation in this work, given by $\mF$ in (\ref{finalform}) combined with the chiral Koba-Nielsen factor in (\ref{defmFb}), leads to the expression for the chiral amplitude $\cF$ advertised
in section \ref{sec:int.1}. The new form (\ref{finalform}) manifests different properties of the chiral amplitudes as compared to the representations obtained from the pure-spinor computation \cite{DHoker:2020prr, DHoker:2020tcq}
to be reviewed in the next section. For instance, (\ref{finalform}) exhibits an interesting echo of the chiral five-point amplitude (\ref{newsum.13}) at genus one since the shorthand ${\cal D}_I$ in (\ref{defdt}) condenses the first and last line to,
\bea
4 {\cal D}_I \bigg\{ \sum_{j=1}^5  \mt_j    \ep_j \cdot \mP^I (j)
-  \sum_{1\leq i < j\leq 5} \mt_{ij} g^I _{i,j} \bigg\}
\eea
see section \ref{sec:int.2} for further details.

\newpage

%%%%%%%%%%%%%%%%%%%%%%%%%%%%%%%%%%%%%%%%%%%
%%%%%%%%%%%%%%%%%%%%%%%%%%%%%%%%%%%%%%%%%%%
\section{Matching with results of {\tt 2006.05270} and {\tt 2008.08687}}
\label{sec:int.3}
\setcounter{equation}{0}
%%%%%%%%%%%%%%%%%%%%%%%%%%%%%%%%%%%%%%%%%%%
%%%%%%%%%%%%%%%%%%%%%%%%%%%%%%%%%%%%%%%%%%%

We shall now compare the simplified forms (\ref{newsumm.2}) and (\ref{Falt}) of the genus-two chiral amplitude for five massless external NS states with the parity-even part of the bosonic components obtained in \cite{DHoker:2020prr, DHoker:2020tcq} using a combination of chiral splitting and pure-spinor methods. Among the multiple representations of the chiral amplitude given in \cite{DHoker:2020prr, DHoker:2020tcq}, we will work with the manifestly homology shift invariant correlator in section 5.3 of \cite{DHoker:2020prr}  together with the effective NS components of \cite{DHoker:2020tcq}. This choice will furnish  the most convenient starting point to demonstrate agreement with the RNS result in this work, since our presentation of the RNS result is manifestly homology shift invariant.

%%%%%%%%%%%%%%%%%%%%%%%%%%%%%%%%%%%%%%%%%%%
%%%%%%%%%%%%%%%%%%%%%%%%%%%%%%%%%%%%%%%%%%%
\subsection{The bosonic components of the pure-spinor results}
\label{sec:int.3a}

The parity-even components of the genus-two five-point chiral amplitude obtained in \cite{DHoker:2020prr, DHoker:2020tcq}, which is to be compared with the result  (\ref{newsumm.2}), or equivalently (\ref{finalform}), of  the RNS computation carried out in this work, may conveniently be organized as follows,
\begin{align}
\mF = \mF_{p.\varepsilon} + \mF_{p.k} + \mF_{\rm scalar}
\label{summ.2}
\end{align}
where we have again stripped off the ubiquitous chiral Koba-Nielsen factor ${\cal N}_5$ in (\ref{5.N5b}).
The three contributions $\mF_{p.\varepsilon}$, $\mF_{p.k}$ and $\mF_{\rm scalar}$ refer to different ways of contracting the loop-momentum dependent factors of $ {\cal P}_\mu(i)$ defined in (\ref{pfield}). Using the prescription for effective NS components in \cite{DHoker:2020tcq},  the three polarization-dependent ingredients in (\ref{summ.2}) are given by,
\begin{align}
\mF_{p.\varepsilon} &= - 8 
\big[ {\cal P}_\mu(1) \Delta(2,3) \Delta(4,5) k_3 \cdot k_4  + {\rm cycl}(1,2,3,4,5) \big] \notag \\
& \ \ \ \ \times
\Big\{ \varepsilon_1^\mu t_8(f_2,f_3,f_4,f_5) + {\rm cycl}(1,2,3,4,5) \Big\}
\notag \\
\mF_{p.k} &= -4i {\cal P}_\mu(1) \Delta(2,3) \Delta(4,5) 
 \Big\{
 k_5^\mu (R_{5;1|2,3,4} + R_{5;2|1,3,4}) + k_1^\mu (R_{1;5|2,3,4} + R_{1;2|3,4,5}) \notag \\
&\ \ \ \ + k_2^\mu (R_{2;5|1,3,4} + R_{2;1|3,4,5}) +2 k_3^\mu R_{3;4|1,2,5} +2 k_4^\mu R_{4;3|1,2,5}
 \Big\}
 + {\rm cycl}(1,2,3,4,5)
 \label{summ.4} \\
\mF_{\rm scalar} &= i \omega_I(1) \Delta(2,3) \Delta(4,5) 
   \Big\{
 G^I_{1,2,5} T^{\rm eff}_{25,1|3,4}
 + G_{1,2,3}^I S^{\rm eff}_{2;3|4|5,1}
 + G_{1,2,4}^I S^{\rm eff}_{2;4|3|5,1} \notag \\
 &\ \ \ \ 
 + G_{1,5,3}^I S^{\rm eff}_{5;3|4|1,2}
 + G_{1,5,4}^I S^{\rm eff}_{5;4|3|1,2}
 \Big\}
 + {\rm cycl}(1,2,3,4,5)
  \notag
\end{align}
where the single-valued function $G^I_{a,b,c}$ is defined in (\ref{GI}) and the cyclic permutations act on the external-particle labels of both the differential forms and of $k_i, \varepsilon_i, f_i$. The kinematic dependence of both $\mF_{p.k}$ and $\mF_{\rm scalar}$ is exclusively built from the following combinations, 
\bea
R_{a;b|c,d,e} = i(\varepsilon_a \cdot k_b) t_8(f_b,f_c,f_d,f_e) - \frac{i}{2} t_8([f_a,f_b],f_c,f_d,f_e)
\label{summ.6}
\eea
symmetric in $c,d,e$ which also featured in the simplifications of appendix \ref{sec:J} of this paper.
The expression for $\mF_{\rm scalar}$ in (\ref{summ.4}) depends on the polarizations through the building blocks,
\begin{align}
T^{\rm eff}_{ab,c|d,e} &= (8k_d\cdot k_e - 4 k_a\cdot k_b)(R_{a;b|c,d,e}- R_{b;a|c,d,e})
+ 4 k_a\cdot k_b (R_{b;c|a,d,e}- R_{a;c|b,d,e}) \notag \\
S^{\rm eff}_{a;b|c|d,e} &=(8k_d\cdot k_e - 4 k_a\cdot k_b) R_{a;b|c,d,e} 
- 4 k_a\cdot k_b   R_{a;c|b,d,e} 
\label{summ.7} \\
&\ \ \ \ + 8(k_c\cdot k_d R_{d;e|a,b,c} - k_d\cdot k_e R_{d;c|a,b,e})
+ 8(k_c\cdot k_e R_{e;d|a,b,c} - k_d\cdot k_e R_{e;c|a,b,d}) 
\notag
\end{align}
symmetric in $d,e$ with further permutation properties spelt out in \cite{DHoker:2020tcq}.
Note that the five-forms of $\mF_{\varepsilon \cdot p}$ in (\ref{summ.4}) naturally
generalize the combinations of $\Delta(a,b)$ and momenta in the chiral four-point amplitude 
which is proportional to $ \Delta(1,2) \Delta(3,4) k_2 \cdot k_3 + {\rm cycl}(1,2,3,4)$.

\sm

Finally, the superspace result of \cite{DHoker:2020prr} naturally unifies the parity-even
bosonic components in (\ref{summ.2}) with the parity-odd ones in (\ref{summ.10}) whose
derivation in the RNS formalism is left for future work.

%%%%%%%%%%%%%%%%%%%%%%%%%%%%%%%%%%%%%%%%%%%
%%%%%%%%%%%%%%%%%%%%%%%%%%%%%%%%%%%%%%%%%%%
\subsection{Agreement with the RNS computation}
\label{sec:int.3b}

The agreement of the chiral amplitude in (\ref{summ.2}), obtained in \cite{DHoker:2020prr, DHoker:2020tcq} from chiral splitting and pure-spinor methods, with the result (\ref{finalform}) of the RNS computation will be demonstrated in the following five steps,
\begin{itemize}
\itemsep=-0.03in
\item[(i)] match the loop-momentum dependent terms with contractions $\varepsilon_a \cdot p^I$
\item[(ii)] organize the $\varepsilon_a \cdot p^I$-independent terms from (i) and (\ref{finalform}) into
$R_{a;b|c,d,e}$ and $\mt_{ab}$
\item[(iii)] show agreement of the terms $\mt_{ab}$ including those from (ii)
\item[(iv)] match the $k_a \cdot p^I$ parts of remaining terms $R_{a;b|c,d,e}$
\item[(v)] match the $k_a \cdot k_b$ parts of remaining terms $R_{a;b|c,d,e}$
\end{itemize}

\subsubsection{Step (i)}

We begin by comparing the first line of (\ref{finalform}) with the
$\varepsilon_a \cdot p^I$ terms $\mF_{p.\varepsilon}$ in (\ref{summ.4}).
Based on ${\cal P}_\mu(a) = \mP_\mu^I(a) \omega_I(a)$ and the definition 
(\ref{defdt}) of ${\cal D}_I$, their difference is given by
\begin{align}
\mF_{\rm diff} &= \mF_{p.\varepsilon} 
- 4 {\cal D}_I \Big( \mt_1    \ep_1^\mu \,   \mP_\mu^I (1) + {\rm cycl}(1,2,3,4,5) \Big )
\notag \\
&= 8 \mt_1 \varepsilon_1^\mu \Big\{
\big[ \mP_\mu^I(1) - \mP_\mu^I(2) \big] \omega_I(2) \Delta(3,4) \Delta(5,1) k_4\cdot k_5 \notag\\
&\hspace{1.31cm} +\big[ \mP_\mu^I(1) - \mP_\mu^I(3) \big] \omega_I(3) \Delta(4,5) \Delta(1,2) k_5\cdot k_1 \phantom{\Big\}}\label{step1.1}\\
&\hspace{1.31cm} +\big[ \mP_\mu^I(1) - \mP_\mu^I(4) \big] \omega_I(4) \Delta(5,1) \Delta(2,3) k_1\cdot k_2 \phantom{\Big\}} \notag\\
&\hspace{1.31cm} +\big[ \mP_\mu^I(1) - \mP_\mu^I(5) \big] \omega_I(5) \Delta(1,2) \Delta(3,4) k_2\cdot k_3
\Big\} + {\rm cycl}(1,2,3,4,5)
\notag
\end{align}
and can be simplified to the following expression that no
longer depends on the loop momentum via permutations of 
(\ref{7.PG})
\begin{align}
\mF_{\rm diff} &=- 8 \mt_1 \varepsilon_1^\mu \Big\{
\big[ G^I_{1,2,3}  k_3^\mu
+G^I_{1,2,4}  k_4^\mu
+G^I_{1,2,5}  k_5^\mu \big] \omega_I(2) \Delta(3,4) \Delta(5,1) k_4\cdot k_5 \notag\\
&\hspace{1.56cm} +\big[ G^I_{1,3,2}  k_2^\mu
+G^I_{1,3,4}  k_4^\mu
+G^I_{1,3,5}  k_5^\mu \big] \omega_I(3) \Delta(4,5) \Delta(1,2) k_5\cdot k_1 \phantom{\Big\}} \notag \\
&\hspace{1.56cm} +\big[ G^I_{1,4,2}  k_2^\mu
+G^I_{1,4,3}  k_3^\mu
+G^I_{1,4,5}  k_5^\mu \big] \omega_I(4) \Delta(5,1) \Delta(2,3) k_1\cdot k_2 \phantom{\Big\}} \label{step1.2} \\
&\hspace{1.56cm} +\big[ G^I_{1,5,2}  k_2^\mu
+G^I_{1,5,3}  k_3^\mu
+G^I_{1,5,4}  k_4^\mu \big] \omega_I(5) \Delta(1,2) \Delta(3,4) k_2\cdot k_3
\Big\}  \notag \\
& \ \ + {\rm cycl}(1,2,3,4,5)
\notag
\end{align}

\subsubsection{Step (ii)}

All the remaining terms $ \mF_{p.k} ,\mF_{\rm scalar}$ in the target expression
(\ref{summ.4}) are written in terms of the tensor structure $R_{a;b|c,d,e} $ defined
in (\ref{summ.6}). We shall therefore organize the kinematic factors of (\ref{step1.2})
and the last four lines of (\ref{finalform}) into $R_{a;b|c,d,e} $ and $\mt_{ab}$ using
\bea
\mt_1\, \varepsilon_1 \cdot k_2 = - i R_{1;2|3,4,5} + \tfrac{1}{2} \mt_{12}
 \label{step2.1}
\eea
In the difference (\ref{step1.2}), this leads to the unique decomposition
\begin{align}
\mF_{\rm diff} &= \mF_{\rm diff} \, \big|_{R}+ \mF_{\rm diff} \, \big|_{\mt_{ab}} \notag \\
\mF_{\rm diff} \, \big|_{R}& = \mF_{\rm diff} \, \big|_{\mt_a \varepsilon_a \cdot k_b \rightarrow -i R_{a;b|c,d,e}}
 \label{step2.2} \\
 \mF_{\rm diff} \, \big|_{\mt_{ab}}& =  \mF_{\rm diff} \, \big|_{\mt_a \varepsilon_a \cdot k_b \rightarrow \frac{1}{2} \mt_{ab}}
 \notag
\end{align}
In the final result (\ref{finalform}) of the RNS computation, the decomposition 
(\ref{step2.1}) casts the sum of the second, third and fourth line into the compact form
\begin{align}
\mF_R &= 8i k_1 \cdot \mP^I(1) \bigg\{ R_{1;2|3,4,5} \omega_I(4) \Delta(1,3) \Delta(2,5)
+ R_{1;3|2,4,5} \omega_I(4) \Delta(1,2) \Delta(3,5) \notag \\
&\hspace{2.43cm}
+ R_{1;4|2,3,5} \omega_I(3) \Delta(1,2) \Delta(4,5) \bigg\} + {\rm cycl}(1,2,3,4,5)
 \label{step2.3}
\end{align}

\subsubsection{Step (iii)}

As a key benefit of the reorganization (\ref{step2.2}) of (\ref{step1.2}),
all the terms $\mt_{ab}$ cancel between $ \mF_{\rm diff}$ and the last line of 
(\ref{finalform}) since
\begin{align}
 \mF_{\rm diff} \, \big|_{\mt_{ab}} &=   8 \omega_I(1) \Delta(2,3) \Delta(4,5) k_3 \cdot k_4
 \sum_{1\leq i<j}^5 \mt_{ij} g^I_{i,j}  + {\rm cycl}(1,2,3,4,5)
  \label{step3.1}
 \end{align}
where we have used the following corollary of the identity (\ref{4termid}) among $\mt_{ij} $,
\bea
\sum_{2\leq i<j}^5 \mt_{ij} G^I_{1,i,j} =  \sum_{1\leq i<j}^5 \mt_{ij} g^I_{i,j}
\eea
On these grounds, the difference between $\mF_{p.\varepsilon}$
in (\ref{summ.4}) and the complete RNS result (\ref{finalform}) entirely boils
down to combinations of $R_{a;b|c,d,e} $ in (\ref{summ.6}),
\begin{align}
\mF_{p.\varepsilon} - \mF
= \mF_{\rm diff} \, \big|_{R} - \mF_R
  \label{step3.2}
\end{align}
with $\mF_{\rm diff} \, \big|_{R}$ and $\mF_R$ given by (\ref{step2.2}) and (\ref{step2.3}).
In order to finish the proof of equivalence between the pure-spinor and RNS
computations, it therefore remains to show that,
\bea
 \mF_R -  \mF_{\rm diff} \, \big|_{R} =  \mF_{p.k} + \mF_{\rm scalar}
   \label{step3.3}
\eea
where also the right-hand side is expressed in terms of $R_{a;b|c,d,e} $ by virtue of (\ref{summ.4}).

\subsubsection{Step (iv)}

It is convenient to first consider the loop-momentum dependent terms in proving (\ref{step3.3}). For this purpose, we rearrange the cyclic orbit in (\ref{summ.4}) and eliminate $R_{1;5|2,3,4}$ using the extension
$R_{1;2|3,4,5} + (2\leftrightarrow 3,4,5) = 0$ of the four-term identity (\ref{4termid}) to write,
\begin{align}
 \mF_R - \mF_{p.k} &= 8i R_{1;2|3,4,5} k_1^\mu
\Big\{ \mP_\mu^I(1) \Delta(1,3) \Delta(2,5) \omega_I(4) 
+ {\cal P}_\mu(4) \Delta(5,1) \Delta(2,3) \notag \\
&\ \ \ \ \ \ \ \ 
- {\cal P}_{\mu}(3) \Delta(4,5) \Delta(1,2)
+ \tfrac{1}{2} {\cal P}_{\mu}(2) \Delta(3,4) \Delta(5,1) 
-  \tfrac{1}{2} {\cal P}_{\mu}(5) \Delta(1,2) \Delta(3,4)  \Big\} \notag \\
&+8i R_{1;3|2,4,5} k_1^\mu
\Big\{  \mP_\mu^I(1) \Delta(1,2) \Delta(3,5) \omega_I(4) 
- {\cal P}_{\mu}(3) \Delta(4,5) \Delta(1,2) \label{step4.1}  \\
&\ \ \ \ \ \ \ \ 
+ \tfrac{1}{2} {\cal P}_{\mu}(2) \Delta(3,4) \Delta(5,1) 
-  \tfrac{1}{2} {\cal P}_{\mu}(5) \Delta(1,2) \Delta(3,4)   \Big\} \notag \\
&+ 8i R_{1;4|2,3,5} k_1^\mu
\Big\{  \mP_\mu^I(1) \Delta(1,2) \Delta(4,5) \omega_I(3) 
- {\cal P}_{\mu}(3) \Delta(4,5) \Delta(1,2)  \Big\}   \notag \\
&+ {\rm cycl}(1,2,3,4,5) \notag
\end{align}
After decomposing $\cP_\mu(a) = \mP_\mu^I(a) \omega_I(a)$, reducing the five-forms
to their cyclic basis and exploiting the vanishing of $k_1 \cdot \mP^I(1)\omega_I(1)
= \partial_{z_1} \ln {\cal N}_5$ upon integration, all the loop momenta drop out. Based
on (\ref{7.PG}) and $R_{1;2|3,4,5} + (2\leftrightarrow 3,4,5) = 0$, we find
\begin{align}
 \mF_R - \mF_{p.k} &= 4i ( R_{1;2|3,4,5} +R_{1;3|2,4,5} ) \notag \\
 &\ \ \ \ \ \ \times \Big\{
  \Delta(3,4) \Delta(5,1) \omega_I(2) (k_1\cdot k_3 G^I_{1,2,3} + k_1\cdot k_4 G^I_{1,2,4} + k_1\cdot k_5G^I_{1,2,5} )\notag \\
 & \ \ \ \ \ \ \ \ - \Delta(1,2) \Delta(3,4) \omega_I(5) (k_1\cdot k_2 G^I_{1,5,2} + k_1\cdot k_3 G^I_{1,5,3} + k_1\cdot k_4G^I_{1,5,4} )
  \Big\} \label{step4.2} \\
&\quad +  8i R_{1;2|3,4,5}  \Delta(5,1) \Delta(2,3) \omega_I(4) (k_1\cdot k_2 G^I_{1,4,2} + k_1\cdot k_3 G^I_{1,4,3} + k_1\cdot k_5G^I_{1,4,5} )\notag \\
%%%
&\quad+ 8i R_{1;5|2,3,4} \Delta(4,5) \Delta(1,2) \omega_I(3) (k_1\cdot k_2 G^I_{1,3,2} + k_1\cdot k_4 G^I_{1,3,4} + k_1\cdot k_5G^I_{1,3,5} ) \notag \\
 &\quad + {\rm cycl}(1,2,3,4,5)
 \notag
\end{align}

\subsubsection{Step (v)}

The final step in proving (\ref{step3.3}) is to show that the simplified expression (\ref{step4.2})
is equal to $ \mF_{\rm diff} \, \big|_{R} + \mF_{\rm scalar}$. By (\ref{summ.4}), 
(\ref{summ.7}) and
\begin{align}
\mF_{\rm diff} \, \big|_{R}  &=8i  \Big\{
\big[ G^I_{1,2,3} R_{1;3|2,4,5}
+G^I_{1,2,4} R_{1;4|2,3,5}
+G^I_{1,2,5} R_{1;5|2,3,4} \big] \omega_I(2) \Delta(3,4) \Delta(5,1) k_4\cdot k_5 \notag\\
&\hspace{0.7cm} +\big[ G^I_{1,3,2} R_{1;2|3,4,5}
+G^I_{1,3,4} R_{1;4|2,3,5}
+G^I_{1,3,5} R_{1;5|2,3,4} \big] \omega_I(3) \Delta(4,5) \Delta(1,2) k_5\cdot k_1 \phantom{\Big\}} \notag \\
&\hspace{0.7cm} +\big[ G^I_{1,4,2} R_{1;2|3,4,5}
+G^I_{1,4,3} R_{1;3|2,4,5}
+G^I_{1,4,5} R_{1;5|2,3,4} \big] \omega_I(4) \Delta(5,1) \Delta(2,3) k_1\cdot k_2 \phantom{\Big\}} \notag \\
&\hspace{0.7cm} +\big[ G^I_{1,5,2} R_{1;2|3,4,5}
+G^I_{1,5,3} R_{1;3|2,4,5}
+G^I_{1,5,4} R_{1;4|2,3,5} \big] \omega_I(5) \Delta(1,2) \Delta(3,4) k_2\cdot k_3
\Big\}  \notag \\
& \quad + {\rm cycl}(1,2,3,4,5)
 \label{step5.1}
\end{align}
each term under consideration is of the form $R_{a;b|c,d,e} G^I_{i,j,k}$. In order to
verify (\ref{step3.3}), the single-valued functions have to be brought into a six-element basis 
of $G^I_{1,a,b}$ with $2{\leq} a{<}b{\leq} 5$ via permutations of $G^I_{2,3,4} = G^I_{1,2,3}
{+}G^I_{1,3,4}{+}G^I_{1,4,2}$. Moreover, we need the relation (\ref{8.oli}) among the
gauge invariant combinations $R^{\rm inv}_{1;2,3} = i k_1\cdot k_2 R_{1;3|2,4,5}
-i k_1\cdot k_3 R_{1;2|3,4,5}$. With these manipulations of both the worldsheet
functions and the kinematic factors, it is a long but mechanical task to confirm
(\ref{step3.3}) which we did with the help of {\tt Mathematica}. This concludes 
our proof of equivalence of the chiral amplitude
(\ref{finalform}) computed from the RNS formalism and the bosonic components
(\ref{summ.2}) of the pure-spinor computation in \cite{DHoker:2020prr, DHoker:2020tcq}.

\subsection{Bootstrapping the odd parity five-point amplitude}
\label{sec:boots}

The odd parity part of the chiral amplitude was constructed in \cite{DHoker:2020prr, DHoker:2020tcq} using a combination of chiral splitting and pure-spinor methods. No corresponding calculation has been carried out from first principles in the RNS formulation to date. This is due, in large part, to the fact that the zero modes of the worldsheet fermion fields complicate the structure of the super moduli space $\mM_{2,-}$ of compact genus two super Riemann surfaces  with odd spin structures. In particular, it has been argued that $\mM_{2,-}$ is not projected \cite{Donagi:2013dua}, a property which is presumably related to the fact that no simple super period matrix exists  in the odd spin structure case \cite{DHoker:1989cxq}. Now that a well-motivated proposal for the odd parity part is available from \cite{DHoker:2020prr, DHoker:2020tcq}, it is urgent to obtain a first principles calculation in the RNS formalism for the odd spin structure sector and to understand the supermoduli space $\mM_{2,-}$ in greater detail than had been available thus far. Such a detailed derivation is relegated to future work.

\sm

In this subsection we shall argue that the genus-two odd parity chiral amplitude may be partially conjectured from symmetry arguments and the structure of the correlators of vertex operators for massless NS states in the RNS formulation. The construction hinges on simple assumptions regarding the role of the odd spin structure supermoduli space $\mM_{2,-}$, the power counting of loop momenta and the types of tensors contracting the one-forms $\omega_I(z_i)$ in the chiral amplitude.  To begin with, the integrated vertex operators are identical to those given for even spin structure in (\ref{V}) and (\ref{vertex}). Furthermore, there are two odd moduli $\zeta ^\a$ and therefore the gravitino slice $\chi$ is linear in $\zeta ^\a$ while the Beltrami differential $\hat \mu$ may be chosen to be bilinear in $\zeta ^\a$.

\subsubsection{Zero-mode counting in RNS}

For odd spin structures, each chiral worldsheet fermion field $\psi _+^\mu$ has one zero mode, providing a total of 10 zero modes. In order to saturate the zero modes in the  functional integral over $\psi_+^\mu$, we need at least 10 insertions of the field $\psi_+^\mu$. At most 12 are available from the 5 vertex operators, the insertion of two supercurrents $S$, and the insertion of one stress tensor~$T_\psi$. The contribution from the fermion bilinears in all 5 vertex operators being saturated by zero modes cancels since $\eps_{10}(f_1, f_2, f_3, f_4, f_5)=0$ by momentum conservation. A Lorentz-invariant and gauge-invariant contraction of all five field strengths is obtained by forming the  combination $\eps_{10}(p^I, \ep_1, f_2, f_3, f_4, f_5)$ and its cyclic permutations, which is non-zero. 

\sm

The first assumption we shall make regarding the net effect of the integration over supermoduli space is that all contributions with two or more powers of loop momenta in the chiral amplitude cancel. This assumption is consistent with the structure of the corresponding supergravity amplitude \cite{Carrasco:2011mn, Mafra:2015mja} and with the results of the construction of \cite{DHoker:2020prr, DHoker:2020tcq}.

\subsubsection{Uniqueness of the kinematic factor}

Invariance of $\eps_{10}(p^I, \ep_1, f_2, f_3, f_4, f_5)$ under all permutations of $2,3,4,5$ is manifest. To show invariance under all permutations of $1,2,3,4,5$, it suffices to prove that,
\bea
\label{7.eps}
\eps_{10}  (p^I, \ep_1, f_2, f_3, f_4, f_5) =\eps_{10}  (p^I, \ep_2, f_1, f_3, f_4, f_5) 
\eea
since invariance under all other permutations then follows from combining this transposition with the invariance under all permutations of $2,3,4,5$. To prove (\ref{7.eps}),  we express $f_2$ in terms of $\ep_2$ and $k_2$,
\bea
\eps_{10}  (p^I, \ep_1, f_2, f_3, f_4, f_5) = 2 \eps_{10} ^{ \mu_1 \nu_1 \mu_2 \nu_2 \cdots \mu_5 \nu_5} \, 
(p^I)^ {\mu_1}  \ep_1^{\nu_1} \ep_2^{\mu _2} k_2^{\nu_2}   f_3^{\mu _3 \nu_3} f_4^{\mu _4 \nu_4} f_5 ^{\mu_5 \nu_5}
\eea
Using momentum conservation on $k_2=-k_1-k_3-k_4-k_5$ we see that the contributions from $k_3,k_4,k_5$ cancel by the Bianchi identities for $f_3,f_4,f_5$, leaving only the contribution from $-k_1$, and we obtain, 
\bea
\eps_{10}  (p^I, \ep_1, f_2, f_3, f_4, f_5) = - 2 \eps_{10} ^{ \mu_1 \nu_1 \mu_2 \nu_2 \cdots \mu_5 \nu_5} \, 
(p^I)^{\mu_1}  \ep_2^{\mu _2}  \ep_1^{\nu_1} k_1^{\nu_2}   f_3^{\mu _3 \nu_3} f_4^{\mu _4 \nu_4} f_5 ^{\mu_5 \nu_5}
\eea
The identity (\ref{7.eps}) follows upon properly rearranging the indices and expressing the combination in terms of $f_1$ and $\ep_2$.

\subsubsection{Uniqueness of the odd parity chiral amplitude}

The second assumption we shall make is that the chiral amplitude $\cF_{{\rm odd}}$ is the product of the universal chiral Koba-Nielsen factor $\cN_5$ times a linear combination of products of $(1,0)$-forms $\om_{I_i}(z_i)$ for $i=1,\cdots, 5$ with constant coefficients, i.e.\ independent of $z_i$ and $\Omega_{IJ}$. Indeed, contractions with $g^I_{a,b}$ or analogous $z_a$-dependent objects obtained from higher derivatives of prime forms are ruled out by our assumption of holomorphicity and homology shift invariance. So this second assumption really boils down to requiring the tensors contracting the $\om_{I_i}(z_i)$ to be independence of $\Omega_{IJ}$.

\sm

Using the $GL(2,\ZZ)$ subgroup of $Sp(4,\ZZ)$ under which $\om_I$ transforms linearly  (see appendix \ref{sec:A}), it may be established that the only invariant tensors of the modular weight $-2$ of five factors of $\omega_{I_i} (z_i)$ lead to linear combinations of the five-fold holomorphic forms $\om_I(i) \Delta(j,k) \Delta(\ell,m)$. Specifically, this form for $\cF_{{\rm odd}}$ is required by the branching rules for the tensor product ${\bf 2}^{\otimes 5}$ into a single ${\bf 2}$ together with the modular weight~$-2$. As a result of the assumptions, the only holomorphic form that is invariant under all permutations, linear in loop momenta, and that has no extra factors of external momenta is given as follows,
\bea
 \eps_{10} (p^I, \ep_1, f_2, f_3, f_4, f_5) \, \om_I(1) \Delta(2,3) \Delta(4,5)
 \label{unitens}
\eea
plus all possible permutations. But the permutations include the subgroup of cyclic permutations of (3,4,5) upon which the sum over such permutations vanishes.  Hence no permutation invariant exists without including extra factors of external momentum. Investigating the possible contractions of the bosonic fields $x_+^\mu$ in the correlators of vertex operators, supercurrents, and stress tensor $T$, it is manifest that at most two additional 
factors of external momenta can be produced beyond $ \eps_{10} (p^I, \ep_1, f_2, f_3, f_4, f_5)$.
(Additional factors of $k_i \cdot p^J$ or $p^J \cdot p^K$ would violate our earlier assumption
on the power counting of loop momenta.) For example, the bosonic stress tensor can be contracted with two exponential factors in two different vertex operators, and analogously, the fields $\partial x_+^\mu$ of the two supercurrents can each be contracted onto a single exponential.

\sm

Thus, we consider combinations of five-forms that involve one power of $k_i \cdot k_j$, and begin with the term that includes $\om_I(1)$. Since the total expression must be invariant under all permutations, it must in particular be invariant under all permutations of 2,3,4,5 that leave 1 invariant. The corresponding combination is unique and generalizes the chiral four-point amplitude written in a manner that does not make use of overall momentum conservation (since this is different for the cases of 4 or 5 external states). Thus, the expression is unique and given by, 
\bea
&&
 \eps_{10} (p^I, \ep_1, f_2, f_3, f_4, f_5) \om_I(1)  \Big [ 
(k_2-k_3) \cdot (k_4-k_5) \Delta(2,3) \Delta(4,5)
 \\ && \hskip 1.9in 
+ (k_2-k_4) \cdot (k_5-k_3) \Delta (2,4) \Delta (5,3) 
\no \\ && \hskip 1.9in 
+ (k_2 - k_5) \cdot (k_3-k_4) \Delta(2,5) \Delta (3,4) \Big ]
+ {\rm cycl}(1,2,3,4,5)
\no
\eea
To simplify it, we may expand each term in the canonical basis of five-forms given by the cyclic orbit of $W^I_1 = \om^I(1)\Delta(2,3) \Delta(4,5)$, 
\bea
&&
\eps_{10}  (p_I, \ep_1, f_2, f_3, f_4, f_5)  \Big [ 
(k_2-k_3) \cdot (k_4-k_5) W_1^I
 \\ && \hskip 1.6in 
+ (k_2-k_4) \cdot (k_5-k_3) (-W_1^I +W_2^I+W_5^I)
\no \\ && \hskip 1.6in 
+ (k_2 - k_5) \cdot (k_3-k_4) (-W_2^I-W_5^I)  \Big ]
+ {\rm cycl}(1,2,3,4,5)
\no
\eea
Rearranging the cyclic permutations so as to expose the contribution proportional to $W_1^I$ and applying momentum conservation, the preceding expression reduces as follows,
\bea
-12   \eps_{10} (p^I, \ep_1, f_2, f_3, f_4, f_5)   (k_3 \cdot k_4)\om_I(1) \Delta(2,3) \Delta(4,5) + {\rm cycl}(1,2,3,4,5)
\eea
which coincides, up to an overall multiplicative factor, with the result (\ref{summ.10}) we have obtained from the pure-spinor calculation.

\newpage

%%%%%%%%%%%%%%%%%%%%%%%%%%%%%%%%%%%%%%%%%%%
%%%%%%%%%%%%%%%%%%%%%%%%%%%%%%%%%%%%%%%%%%%
\section{Conclusions and future directions}
\label{sec:concl}
\setcounter{equation}{0}
%%%%%%%%%%%%%%%%%%%%%%%%%%%%%%%%%%%%%%%%%%%
%%%%%%%%%%%%%%%%%%%%%%%%%%%%%%%%%%%%%%%%%%%

In this work, we have computed the contributions from even spin structures to the genus-two amplitudes for five massless NS-NS states of Type II superstrings and five massless NS states of Heterotic superstrings from first principles in the RNS formulation. Our simplified result in (\ref{newsumm.1}) and (\ref{Falt}) reproduces the bosonic components of the massless genus-two five-point amplitudes in pure-spinor superspace \cite{DHoker:2020prr, DHoker:2020tcq}. On the one hand, our results provide the highest order in the number of  loops and legs where string amplitude computations in the RNS and pure-spinor formalism are explicitly shown to agree. On the other hand, the superspace result of \cite{DHoker:2020prr} combined the zero-mode structure of the non-minimal pure-spinor formalism \cite{Berkovits:2005bt} with a bootstrap strategy involving  BRST cohomology,  homology shift invariance and locality.   Hence, our present work confirms the result of \cite{DHoker:2020prr} by providing  a first-principles computation, complementing a variety of earlier checks through the  supergravity limit \cite{DHoker:2020prr} and S-duality of Type IIB superstrings \cite{DHoker:2020tcq}.

\sm

Our final expressions for the chiral amplitude (\ref{newsumm.2}) and (\ref{Falt}) provide a more compact and more symmetrical alternative to its numerous superspace representations in \cite{DHoker:2020prr, DHoker:2020tcq} and expose an interesting similarity in structure to its  genus-one counterpart. Straightforward loop integration over the pairing of left and right chiral amplitudes yields Type II and Heterotic superstring amplitude in terms of convergent integrals over the moduli space of compact  genus-two Riemann surfaces. The moduli-space integrand is itself an integral over vertex points of combinations of Abelian differentials and the scalar Green function.

\sm

Given that the even parity genus-two chiral amplitude for five external NS bosons has now been calculated in the RNS formulation, it becomes urgent to provide a complete calculation of the odd parity part as well. While its general kinematic structure was developed here in subsection \ref{sec:boots}, it still remains to understand how the measure on the supermoduli space $\mM_{2,-}$ reduces to the odd parity amplitude on $\cM_2$, and this will be studied in future work. Equally urgent is a derivation of the R-sector of the chiral amplitude in the RNS formulation, presumably using the results on the super period matrix in  \cite{Witten:2015hwa, DHoker:2015gwa} and the spin-field correlators in \cite{Atick:1986es, Haertl:2010qlb}. 

\sm

For genus two amplitudes with more than five external states, perhaps the greatest challenge to overcome  is the summation over spin structures of the various concatenated products of Szeg\"o kernels. Producing a method that is more systematic and more efficient than the diverse fauna of methods used here will be of great importance for this goal. Once the even spin structure sums have been performed, the general construction used here should generalize to chiral amplitudes with an arbitrary number of external massless NS states.

\newpage

\appendix

%%%%%%%%%%%%%%%%%%%%%%%%%%%%%%%%%%%%%%%%%%%
%%%%%%%%%%%%%%%%%%%%%%%%%%%%%%%%%%%%%%%%%%%
\section{Function theory for arbitrary genus }
\label{sec:A}
\setcounter{equation}{0}
%%%%%%%%%%%%%%%%%%%%%%%%%%%%%%%%%%%%%%%%%%%
%%%%%%%%%%%%%%%%%%%%%%%%%%%%%%%%%%%%%%%%%%%

In this appendix we present a summary of relevant definitions and formulas for holomorphic and meromorphic differentials on compact Riemann surfaces of genus $h \geq 1$, in terms of  Riemann $\tet$-functions and related objects (for standard references see for example \cite{fay,RMP}).

\subsection{Compact Riemann surfaces}
\label{sec:A.RS}

A Riemann surface is a connected orientable complex manifold of dimension two over $\RR$. The topology of a compact Riemann surface $\Sigma$ without boundary, the only case needed in this paper, is completely specified by its genus $ h$ which equals the number of handles of~$\Sigma$. 
On a compact Riemann surface $\Sigma$ of genus $h\geq 1$ we choose a basis of homology 1-cycles 
$\mA_I, \mB_J\in H_1(\Sigma, \ZZ)$ for $I,J=1, \cdots, h$ with canonical anti-symmetric intersection pairing $\mJ$,
\bea
\label{1b1}
\mJ (\mA_I, \mA_J) & = & \phantom{+}\hspace{-0.063cm} \mJ (\mB_I, \mB_J) =0  
\no \\ 
\mJ (\mA_I, \mB_J) & = & - \mJ (\mB_J, \mA_I) = \delta _{IJ}
\eea
Since $\Sigma$ is a complex manifold, its cotangent bundle is the direct sum of the holomorphic canonical bundle and its complex conjugate, whose sections are $(1,0)$ and $(0,1)$-forms, respectively.  As a result, the first cohomology group $H^1(\Sigma,\ZZ)$ is the direct sum of the Dolbeault cohomology groups $H^1(\Sigma,\ZZ)=H^{(1,0)} (\Sigma, \ZZ) \oplus H^{(0,1)} (\Sigma, \ZZ)$ of holomorphic $(1,0)$-forms and their complex conjugates, respectively. We normalize a canonical basis  $\om_I \in H^{(1,0)} (\Sigma,\ZZ)$ of holomorphic $(1,0)$-forms dual to the canonical basis of 1-cycles $\mA_I, \mB_I$ by,
\bea
\label{1b5}
\oint _{\mA_I} \om_J = \delta _{IJ} \hskip 1in \oint _{\mB_I} \om_J = \Omega _{IJ}
\eea
The period matrix $\Omega $ is symmetric and its imaginary part $\Im \Omega$ is positive definite  by the Riemann bilinear relations, and therefore takes values in the Siegel upper half space of rank $h$, to be defined in the next subsection. A modular transformation acts by an invertible linear map $M$ with integer entries on the canonical homology cycles, represented below as column matrices $\mA$ and $\mB$ with entries $\mA_I$ and $\mB_I$ for $I=1,\cdots, h$ respectively,  
\bea
\label{1b2}
\left ( \bma \mB \cr \mA \ema \right )  \to \left ( \bma \tilde \mB \cr \tilde \mA \ema \right ) =
M \left ( \bma \mB \cr \mA \ema \right )
\eea
As  a map between canonical homology bases, the transformation $M$  preserves the canonical pairing $\mJ$, 
so that the associated transformation matrix $M$ belongs to $Sp(2h,\ZZ)$ and satisfies, 
\bea
\label{1b4}
M^t \mJ M = \mJ
\hskip 0.8in 
\mJ = \left ( \begin{matrix}0 & -I_h \cr I_h & 0 \cr \end{matrix} \right )
\hskip 0.8in 
M = \left ( \begin{matrix}A & B \cr C & D \cr \end{matrix}  \right ) 
\eea
where $\mJ$ is the symplectic pairing matrix, $I_h$ denotes the $h\times h$ unit matrix and $A,B,C,D$ are $h \times h$ blocks. Under a modular transformation $M$ the column vector of holomorphic $(1,0)$-forms $\om$ and period matrix $\Omega$ transform as follows, 
\bea
\label{1b6}
\om & \to & \tilde \om =  (\Omega C^t + D^t)^{-1} \om  
\no \\
\Omega  & \to & \ti \Omega = (A \Omega +B) (C \Omega +D)^{-1}
\eea 
The periods of an arbitrary homology cycle in $H_1 (\Sigma, \ZZ)$ span a lattice $\ZZ^h + \Omega \ZZ^h$ whose associated torus is the Jacobian variety $J(\Sigma) = \CC^h / (\ZZ^h + \Omega \ZZ^h)$. The Abel map is a holomorphic map of an arbitrary number $d$ of copies of $\Sigma$ into $J(\Sigma)$. The Abel map of  a divisor of degree $d$ of points  $z_1, \cdots , z_d \in \Sigma$, symbolically denoted by $z_1+ \cdots + z_d$, is defined by, 
\bea
z_1+ \cdots + z_d ~~ \to ~~ \int ^{z_1}_{z_0} \om _I + \cdots + \int ^{z_d}_{z_0} \om _I
\eea
where $z_0$ is an arbitrarily chosen base point in $\Sigma$. The map is multiple-valued in $\CC^h$  but becomes single-valued in $J(\Sigma)$.

\subsection{The Siegel upper half space}
\label{sec:A.siegel}

The rank $h$ Siegel upper half space $\cH_h$  may be defined as the space of all $h \times h$ symmetric matrices with complex-valued entries whose imaginary part is a positive definite matrix,
\bea
\cH_h = \left \{ \Omega  \in \CC^{h \times h} \,  | \, \Omega^t=\Omega, \,   Y=\Im \Omega >0  \right \}
\eea 
More geometrically, $\cH_h$ is the coset of $Sp(2h,\RR)$ by its maximal compact subgroup $U(h)$,
\bea
\cH_h = Sp(2h,\RR)/U(h)
\eea
The presence of a $U(1)$ factor in the stability group implies that $\cH_h$ is a K\"ahler manifold.  The $Sp(2h,\RR)$-invariant K\"ahler metric on $\cH_h$ is given by,
\bea
ds^2 = \sum _{I,J,K,L=1}^h (Y^{-1} )^{IK} (Y^{-1})^{JL} \, d \Omega _{IJ} \, d\bar \Omega _{KL}
\eea
The Siegel upper half space $\cH_1$ is the upper half complex plane with the Poincar\'e metric. The quotient of $\cH_1$ by the modular group $SL(2,\ZZ)$ is given by $SL(2,\ZZ) \backslash \cH_1$ and represents the moduli space of compact Riemann surfaces of genus one.  

\sm

The period matrix $\Omega$, defined in (\ref{1b5}) for a compact Riemann surface of arbitrary genus $h\geq 1$, takes values in the Siegel upper half space $\cH_h$ for an arbitrary Riemann surface. But the converse is false for $h \geq 4$ as may already  be seen from the dimension formulas, $\half h(h+1)$ for $\cH_h$ and $3h-3$ for $\cM_h$ and $h \geq 2$ which agree for $h=1,2,3$ but differ for $h \geq 4$. 

\sm

For  $h=2$, the case used in this paper, the quotient of $\cH_2$ by the modular group $Sp(4,\ZZ)$ is $Sp(4,\ZZ) \backslash \cH_2$ and represents  the moduli space $\cM_2$ of compact Riemann surfaces of genus two, provided we remove the divisor of diagonal matrices $\Omega$ corresponding to the union of two disconnected tori.  Since the metric $ds^2$ is invariant under $Sp(4,\ZZ)$, it pulls back to a well-defined metric on the quotient $Sp(4,\ZZ) \backslash \cH_2$ and thus on moduli space $\cM_2$. From this metric, one obtains the $Sp(4,\ZZ)$-invariant volume form on $\cM_2$, 
\bea
{|d^3 \Omega |^2 \over (\det \Im \Omega )^3 }
\eea
which arises in the physical amplitudes (\ref{newsumm.1}) and (\ref{newsumm.1a}) upon integration
over loop momenta $p^I$ and forming modular invariant measures for integration over the vertex points.

\sm

For $h=3$, the dimensions of $\cH_3$ and $\cM_3$ coincide, but obtaining $\cM_3$ from $\cH_3$  requires taking the quotient by the involution of hyper-elliptic Riemann surfaces. For genus $h \geq 4$, the dimensions of $\cH_h$ and $\cM_h$ differ, and specifying $\cM_h$ inside $\cH_h$ requires Schottky relations.

\subsection{The Riemann $\tet$-function}
\label{sec:A.theta}

The Riemann $\tet$-function of rank $h$, with given characteristics $\kap$, is a holomorphic function $\tet : \CC^h \times \cH_h \mapsto \CC$ defined by,\footnote{Throughout, when no confusion is expected to arise, we shall use a notation in which the contraction of two column vectors, such as $n$ and $\zeta$,  is denoted by $n \zeta = n^t \zeta$ and similarly $n \Omega n = n^t \Omega n$.} 
\bea
\tet [\kap] (\zeta | \Omega) 
= 
\sum _{n  \in \mathbb Z^h  } 
\exp \Big (i \pi (n+\kap')  \Omega (n+\kap') + 2\pi i (n+\kap')   (\zeta + \kap '') \Big ) 
\eea
where $\zeta = (\zeta _1, \cdots , \zeta _h)^t \in \CC^h$, $\Omega \in \cH_h$, and $\kap$ is an array $\kap = \left ( \kap', \kap '' \right )$ where $\kap', \kap '' \in \CC^h$ are thought of as column matrices. For our purposes, $\kap$ will be  ``half-characteristics" specifying spin structures of line bundles on a Riemann surface so that the entries of $\kap'$ and $ \kap '' $ take the values 0 or $\half$ (mod~1).  The parity of the $\tet$-functions depends on $\kap$ and is even  or odd depending on whether the integer $4 \kap ' \cdot \kap ''$ is even or odd. The corresponding spin structure is then also referred to as even or odd.   

\sm

The standard $\tet$-function is defined by $\tet (\zeta| \Omega)= \tet[0] (\zeta| \Omega)$, and is related to $\tet [\kap]$ by
\bea
\label{charrel}
\tet [\kap ] (\zeta | \Omega)
=
\tet (\zeta + \kap '' + \Omega \kap ' | \Omega)
\ \exp \{\pi i \kap ' \Omega \kap ' +  2 \pi i \kap '(\zeta + \kap '') \}
\eea
We have the following periodicity relation for $\tet[\kap ](\zeta |\Omega)$ where $m,n \in \ZZ^h$ are column matrices,
\bea
\label{tetper}
\tet [\kap ] (\zeta + m + \Omega n| \Omega )
& = &
\tet [\kap ](\zeta | \Omega) \ \exp \{ -i \pi n \Omega n - 2 \pi i n (\zeta +\kap '') + 2 \pi i \kap ' m \}
\no \\
\tet [ \kap' +n, \kap ''+m](\zeta|\Omega) & = & \tet [\kap] (\zeta |\Omega) \, \exp \{ 2 \pi i m \kap ' \}
\eea
The following formulas for special cases of the above will be useful,
\bea
\label{periodshift}
\tet [\kap] (\zeta -\kap | \Omega) & = & \tet(\zeta | \Omega) \, e^{- i \pi \kap ' \Omega \kap ' + 2 \pi i \kap ' \zeta}
\no \\
\tet [\kap] (\zeta +\kap | \Omega) & = & \tet(\zeta | \Omega) \, e^{- i \pi \kap ' \Omega \kap ' - 2 \pi i \kap ' \zeta + 4 \pi i \kap ' \kap ''}
\no \\
\tet [\kap] (\zeta + 2\lambda | \Omega) & = & \tet[\kap](\zeta | \Omega) \, \< \lambda |\kap\> \, e^{- 4 \pi i \lambda ' (\Omega \lambda' + \zeta)}
\eea
where $\kap$ and $\lambda$ are even or odd characteristics. The signature symbol is defined by,
\bea
\label{signature}
\< \lambda | \mu \>
= \exp \{ 4 \pi i ( \kap ' \lambda '' - \lambda ' \kap ''  ) \}
\eea
for $\kap, \lambda$ both half-integer characteristics, and we have $\< \kap |\lambda \> = \pm 1$.

\subsubsection{Modular transformations on $\tet$-functions and characteristics}

A modular transformation $M \in Sp(2h,\ZZ)$, parametrized in (\ref{1b4}), transforms $\Omega$ and $\zeta$ by, 
\bea
\Omega & \to & \tilde \Omega = (A\Omega +B) (C \Omega +D)^{-1}
\no \\
\zeta & \to & \tilde \zeta =  (\Omega C^t +D^t) ^{-1} \zeta
\eea
The transformation rule for $\Omega$ coincides, of course, with the expression given for the period matrix in  (\ref{1b6}). The spin structure $\kap = (\kap ' | \, \kap '')$ transforms inhomogeneously as follows,
\bea
\left ( \begin{matrix} \tilde \kap ' \cr \tilde \kap '' \end{matrix} \right )
=
\left ( \begin{matrix} D & -C \cr -B & A \end{matrix} \right )
\left ( \begin{matrix} \kap ' \cr \kap '' \end{matrix} \right )
+ \half  \left ( \begin{matrix} {\rm diag}(CD^t) \cr {\rm diag}(AB^t) \end{matrix} \right ) 
\hskip .7 in
M= \left ( \begin{matrix} A & B \cr C & D \end{matrix} \right )
\eea
where ${\rm diag}(A)$ is the column matrix whose entries are the diagonal elements $A_{II}$ of the square matrix $A$ with entries $A_{IJ}$.  Finally, the $\tet$-function transforms as follows (see \cite{Igusa}, page 85), 
\bea
\label{thetamod}
\tet [\tilde \kap ] ( \tilde  \zeta | \tilde \Omega) & = &
\epsilon (\kap, M) \, \det (C\Omega + D)^{\half} \tet [\kap ](\zeta | \Omega) 
\no \\ && \times 
\exp \left \{ \pi i \sum _{I,J} \zeta _I \zeta _J \big [ (C \Omega +D)^{-1} C \big ]^{IJ}  \right \} 
\eea
The factor  $\eps(\kap, M)$ is independent of $\zeta$ and $\Omega$ and satisfies $\eps(\kap, M)^8=1$. Its explicit expression is complicated and may be found in \cite{Igusa}, but will not be needed here.

\subsubsection{The Riemann identities}

The  Riemann identity for an arbitrary spin structure $\mu$ is given as follows,\footnote{When the dependence on $\Omega$  is clear from the context, we shall simply write $\tet[\lambda](\zeta) = \tet [\lambda] (\zeta |\Omega)$.}
\bea
\label{Riem1}
\sum _{\lambda} \< \kap | \lambda \> \prod _{a=1}^4 \tet [\lambda] (\zeta _a) 
= 4 \prod _{a=1}^4 \tet [\kap ] ( \zeta _a^+) 
\eea
where $\zeta^+_a$ is given in (\ref{RiemM}). The sum in $\lambda$ is over all even and odd spin structures $\lambda$.  The Riemann identities for an arbitrary spin structure $\kap$, but this time for the sum over even spin structures $\delta$ only, or the sum over odd spin structures $\nu$ only,  are given by, 
\bea
\label{Riem2}
\sum _{\delta \, {\rm  even}} \< \kap | \delta \> \prod _{a=1}^4 \tet [\delta] (\zeta _a) 
& = & 2 \prod _{a=1}^4 \tet [\kap ] (\zeta _a^+) + 2 \prod _{a=1}^4 \tet [\kap ] (\zeta _a^-)
\no \\
\sum _{\nu \, {\rm  odd}} \< \kap | \nu \> \prod _{a=1}^4 \tet [\nu] (\zeta _a) 
& = & 2 \prod _{a=1}^4 \tet [\kap ] (\zeta _a^+) - 2 \prod _{a=1}^4 \tet [\kap ] (\zeta _a^-)
\eea
where the relations between $\zeta _a$ and $ \zeta _a^\pm$ are given by,
\bea
\label{RiemM}
\left ( \begin{matrix} \zeta _1 ^\pm \cr \zeta _2 ^\pm \cr \zeta _3^\pm \cr \zeta _4^\pm \end{matrix} \right )
= \Pi \left ( \begin{matrix} \pm \zeta _1  \cr \zeta _2  \cr \zeta _3 \cr \zeta _4 \end{matrix} \right )
\hskip 0.7in 
\Pi = \half \left ( \begin{matrix} 1 & 1 & 1 & 1 \cr 1 & 1 & -1 & -1 \cr 1 & -1 & 1 & -1 \cr 1 & -1 & -1 & 1 \cr  \end{matrix} \right )
\eea
Note that we have $\Pi^t=\Pi$ and $\Pi^2=I$.

\subsection{The Riemann vanishing Theorem}
\label{sec:A.vanishing}

The definitions and results of subsections \ref{sec:A.siegel} and \ref{sec:A.theta} were given for arbitrary points $\Omega \in \cH_h$, not necessarily corresponding to the period matrix of a Riemann surface. In this subsection, as well as the subsequent subsection of this appendix, we shall specialize to definitions and results that hold only when $\Omega$ is the period matrix of a compact Riemann surface $\Sigma$, so that $\Omega$ is given by (\ref{1b5})  as defined in subsection \ref{sec:A.RS}.

\sm

The Riemann vanishing Theorem states that for $\Omega$ the period matrix of a compact Riemann surface $\Sigma$, and $\zeta \in J(\Sigma)$, the relation $\tet (\zeta | \Omega )=0$ holds if and only if there exist $h-1$ points $p_1, \cdots , p_{h-1} \in \Sigma$, such that,
\bea
\label{zetaDel}
\zeta_I  = \sum_{i=1}^{h-1} \int ^{p_i}_{z_0} \! \om _I  - \Delta_I(z_0)  
\hskip 1in I=1,\cdots, h
\eea
where $\Delta_I(z_0) \in \mathbb C^h$ is the Riemann vector with base-point $z_0$ given by,\footnote{It should go without saying that the Riemann vector $\Delta_I(z_0)$ is not to be confused with the bi-holomorphic form $\Delta$ introduced in (\ref{Delta}).}
 
\bea
\label{Riemvec}
\Delta _I(z_0)  = -\half - \half \Omega _{II} + \sum _{J\not= I} \oint _{\mA_J} \om_J(z) \int ^z _{z_0} \om _I
\eea 
All dependence on $z_0$ cancels out of the combination $\zeta_I$ in (\ref{zetaDel}). The points $p_i$ need not be distinct. An important application of the Riemann vanishing theorem is to the existence of the following holomorphic $(1,0)$-form, defined for any odd spin structure $\nu$ by, 
\bea
\label{omnu}
\om_\nu(z) = \sum _I \om _I(z) \p^I \tet [\nu] (0)
\eea
Its $2(h-1)$ zeros are all double zeros at points $p_i$ such that $\zeta_I=\nu_I$ in (\ref{zetaDel}).  Therefore, $\om_\nu(z)$ admits a holomorphic and single-valued square root $h_\nu(z)$ such that $\om_\nu(z) = h_\nu(z)^2$. For each odd spin structure $\nu$, the holomorphic $(\half, 0)$-form $h_\nu$ is unique up to a sign.

\subsection{The  prime form, the Fay form, and the Szeg\"o kernel}
\label{sec:A.Fay}

The prime form is defined in \cite{fay} by,
\bea
\label{prime}
E(z,w) = { \tet [\nu] (z-w) \over h_\nu(z) \, h_\nu (w) }
\eea
where $z-w$ stands for the Abel map of the divisor $z-w$. The prime form $E(z,w)$ has weight $(-\half, 0)$ and is holomorphic  in $z$ and~$w$,  independent of the odd spin structure $\nu$,  and multiple-valued on $\Sigma$. Closely related to the prime form is Fay's form $\sigma$ defined in \cite{fay} by,
\bea
\label{sigmaz}
\sigma (z) = \exp \Big \{ - \sum _{I=1}^h \oint _{\mA_I} \om_I(y) \ln E(z,y) \Big \}
\eea
The form $\sigma(z)$ has weight $({h \over 2} ,0)$, is nowhere vanishing, and multiple-valued on $\Sigma$.   It satisfies the following relation for $p_1, \cdots ,p_h$ arbitrary points in $\Sigma$,
\bea
\label{sigmaE}
{\sigma (z) \over \sigma(w) } = { \tet (p_1 + \cdots + p_h -z - \Delta) E(w, p_1) \cdots E(w,p_h)
\over \tet (p_1 + \cdots + p_h -w - \Delta) E(z, p_1) \cdots E(z,p_h)}
\eea
The Szeg\" o kernel $S_\delta (z,w)$ for even spin structure $\delta$ is a meromorphic  $(\half,0)$-form on $\Sigma$ in both $z$ and $w$, with a single simple pole at $z=w$ and is given by the following expression, 
\bea
S_\delta (z,w) = 
{\vartheta [\delta ]\big (z-w  \big ) \over  \vartheta [\delta ]\big ( 0 \big ) \ E(z,w)} 
\eea
The definition of the Szeg\" o kernel for an odd spin structure $\nu$ is complicated by the existence of the holomorphic $(\half, 0)$-form, and will not be needed in this paper.

\sm

Additional properties of the prime form and Fay form are as follows.
On a simply-connected domain $\Sigma '$ obtained by cutting $\Sigma$ along $ \mA_I$ and $\mB_I$ cycles with a common base-point, we may define a single-valued $E(x,y)$ and a single-valued $\sigma(z)$. On $\Sigma'$ the prime form vanishes only on the diagonal, 
\bea
E(z,w) = z-w + \cO \big ( (z-w)^3 \big )
\eea 
The monodromy of $E(z,w)$ and $\sigma(z)$ around $\mA_I$ cycles is trivial, while around $\mB_I$ cycles  it is given as follows, 
\bea
\label{sigmamonod}
E(z+\mB_I, w) & = &  E(z,w) \, \exp \Big \{i \pi  -i \pi \Omega _{II} - 2 \pi i \int ^z _w \om_I \Big \}
\no \\
\sigma (z+\mB_I) & = & \sigma (z) \, \exp \left \{ i \pi (h-1) \Omega _{II} + 2 \pi i  V_I(z) \right \}
\eea
where
\bea 
\label{VIz}
V_I(z) = (h-1) \int ^z _{z_0} \om _I - \Delta _I(z_0) 
\eea
The monodromies of the meromorphic $(1,0)$-forms derived from the prime form are,
\bea
\label{Emon}
\p_z \ln E(z+\mA_I, w) & = & \p_z \ln  E(z,w)
\no \\
\p_z \ln E(z+\mB_I, w) & = &  \p_z \ln  E(z,w)  - 2 \pi i  \om_I (z)
\no \\
\p_z \ln E(z, w+\mB_I) & = &  \p_z \ln  E(z,w)  + 2 \pi i  \om_I (z)
\eea
The Abelian differential with a double pole $\p_z \p_w \ln E(z,w)$ is invariant. Under modular transformations $E(z,w)$ and $\sigma(z) $ transform as follows,
\bea
E(z,w) & \to & E(z,w) \, \exp \left \{ i \pi \sum_{I,J} \int ^z _w \om_I \left [ (C \Omega +D)^{-1} C \right ]^{IJ}  \int ^z _w \om_J \right \}
\no \\
\sigma (z) & \to & \sigma (z) \, \exp \left \{ - { i \pi \over h-1} \sum_{I,J} \, V_I(z) \left [ (C \Omega +D)^{-1} C \right ]^{IJ} V_J(z) \right \}
\eea
where $V_I(z)$ was defined in (\ref{VIz}). The transformation rule for $E(z,w)$ was given in \cite{fay}, while the transformation rule for $\sigma(z)$ is presented above up to a multiplicative factor which is independent of $z$,
and clearly cancels out in the ratio $\sigma (z)/\sigma (w)$ given in (\ref{sigmaE}).

\subsection{The Fay trisecant  identity}
\label{revszegoe}

The Fay trisecant identity \cite{fay} holds for the period matrix $\Omega$ of a Riemann surface $\Sigma$, four points $z_1, z_2, w_1, w_2$ on $\Sigma$, and arbitrary real characteristics $\kap = [\kap ', \kap '']$ with $\kap', \kap '' \in \RR^h / \ZZ^h$,
\bea
\label{Fay-id}
&&
\tet [\kap] (z_1+z_2-w_1-w_2) \tet [\kap] (0) E(z_1,z_2) E(w_1,w_2) 
\no \\ && \hskip 0.3in
=
\tet [\kap] (z_1-w_1) \tet [\kap] (z_2-w_2) E(z_2,w_1) E(z_1,w_2)
\no \\ && \hskip 0.45in
- \tet [\kap] (z_2-w_1) \tet [\kap] (z_1-w_2) E(z_1,w_1) E(z_2,w_2)
\eea
The identity is automatically satisfied for every odd half characteristic $\kap$ (or spin structure) for which the left side vanishes identically, and the terms on the right side are manifestly equal upon using the expression for the prime form (\ref{prime}) by choosing $\nu=\kap$ in (\ref{prime}).

\sm

For an arbitrary even spin structure $\kap = \delta$ the Fay identity (\ref{Fay-id}) may be expressed in terms of Szeg\"o kernels, prime forms, and a single ratio of $\tet$-functions,
\bea
\label{Fay1}
S _\delta (z_1, w_2) S_\delta(z_2,w_1) - (w_1 \leftrightarrow w_2)
= { \tet[\delta] (z_1+z_2-w_1-w_2) E(z_1,z_2)E(w_1,w_2) \over \tet[\delta](0) E(z_1, w_1) E(z_1,w_2)E(z_2,w_1) E(z_2,w_2)}
\quad
\eea
Taking the limit of coincident points $w_2 \to z_2$, we find an important  special case,
\bea
\label{Fay2}
S _\delta (z_1, z_2) S_\delta(z_2,z_3)
= S_\delta(z_1,z_3) \p_{z_2} \ln { E(z_2,z_3) \over E(z_2,z_1)} - \sum_I \om _I (z_2) 
{ \p^I \tet [\delta] (z_1-z_3) \over \tet[\delta](0) E(z_1,z_3)}
\eea
Taking yet another limit gives the further special case, 
\bea
\label{Fay3}
S_\delta (z,w)^2 = \p_z \p_w \ln E(z,w) + \sum_{I,J}\, \om_I(z) \om _J(w) {\p^I \p^J \tet [\delta] (0) \over \tet [\delta ](0)}
\eea
The Fay identities will play a crucial role in this paper by simplifying the spin structure dependence of products of Szeg\"o kernels.

\subsection{The scalar Green functions}

The Riemann bilinear relations show that $\Omega _{IJ} = \Omega _{JI}$, as well as the positivity of the integral,
\bea
\int _\Sigma d^2 z \, \om_I(z) \overline{ \om _J(z)} = 2 Y_{IJ} 
\hskip 1in 
Y_{IJ} = \Im \Omega _{IJ}
\eea
where we recall that the measure and Dirac delta-function are normalized as in footnote 5,
\bea
\int d^2 z \, \delta (z,w) f(z) = f(w) \hskip 1in d^2 z = i \, dz \wedge d \bar z
\eea
This allows us to define the canonical volume form $\kappa(z)$ on $\Sigma$ as the pull-back under the Abel map of the canonical K\"ahler form on the Jacobian $J(\Sigma)$, and we have,  
\bea
\kappa (z) = { 1 \over 2h}  \sum_{I,J} \, \om_I(z) (Y^{-1}) ^{IJ} \,  \overline{ \om_J (z)} \, 
\hskip 0.7in 
\int _{\Sigma} d^2 z \, \kappa (z)=1
\eea
The string scalar Green function $G_s(z,w)$ is symmetric in $z,w$ and obeys the following differential equation, 
\bea
\label{1b10}
\p_{\bar z} \p_z G_s(z,w) =  - 2 \pi \delta (z,w) + 2 \pi \, h \, \kappa (z) 
\eea
The solution of the differential equation (\ref{1b10}) is given as follows,
\bea
G_s(z,w) = - \ln |E(z,w)|^2 + 2 \pi  \sum_{I,J} \, 
(Y^{-1})^{IJ} \left ( \Im \int ^z _w \om_I \right ) \left ( \Im \int ^z _w \om_J \right )
\eea 
Note that the integral of the right side of (\ref{1b10}) equals $2(h-1) \pi$, and therefore defines a single-valued function on $\Sigma \times \Sigma$ only for $h=1$. For $h \geq 2$, the function $G_s$ is well-defined only on a cut surface $\Sigma' $ and  transforms inhomogeneously under conformal  transformations in $z,w$. However,  the mixed derivative, 
\bea
\pbz \p_w G_s(z,w) =   2 \pi \delta (z,w) -  \pi  \sum_{I,J} \overline{ \om_I (z)} \, (Y^{-1}) ^{IJ} \, \om_J(w)
\eea
obeys an equation whose right side integrates to zero against an arbitrary holomorphic $(1,0)$-form in $z$ and an arbitrary anti-holomorphic $(0,1)$-form in $w$. 

\sm

The Arakelov Green function is defined to satisfy the differential equation 
\bea
\label{1b15}
\p_{\bar z} \p_z G(z,w) =  - 2 \pi \delta (z,w) + 2 \pi \, \kappa(z)
\eea
along with the condition $\int _\Sigma d^2 z \, \kappa(z)  G(z,w)=0$. The right side of (\ref{1b15}) properly integrates to zero against constant functions for all $h$. An explicit formula for the Arakelov Green function may be constructed as follows from the string Green function $G_s$,
\bea
G(z,w) = G_s(z,w) - \gamma_s (z) - \gamma_s(w) + \gamma _0
\eea
where,
\bea
\gamma _s(z) = \int _{\Sigma '} d^2 w \, \kappa (w) G_s(z,w) 
\hskip 0.7in 
\gamma  _0 = \int _{\Sigma '} d^2 z \, \kappa (z) \gamma_s(z)
\eea
The Arakelov Green function $G$ is single-valued on $\Sigma \times \Sigma$, independent of the how the surface is being cut into $\Sigma'$,  and conformal invariant.

\newpage

%%%%%%%%%%%%%%%%%%%%%%%%%%%%%%%%%%%%%%%%%%%
%%%%%%%%%%%%%%%%%%%%%%%%%%%%%%%%%%%%%%%%%%%
\section{Function theory for genus two}
\label{sec:B}
\setcounter{equation}{0}
%%%%%%%%%%%%%%%%%%%%%%%%%%%%%%%%%%%%%%%%%%%
%%%%%%%%%%%%%%%%%%%%%%%%%%%%%%%%%%%%%%%%%%%

The function theory of a genus-two Riemann surface exhibits special relations many of which are used in the calculation of the genus-two string amplitudes. Some of these relations are more easily established in terms of modular geometry as will be shown in subsection \ref{sec:B1} while others more naturally make use of the hyper-elliptic nature of genus-two Riemann surfaces, as will be shown in subsection \ref{sec:B2} of this appendix.

\subsection{Multi-holomorphic 1-forms}
\label{sec:B1}

A number of multi-holomorphic $(1,0)$-forms are special to genus two and will be discussed here. 
The ubiquitous bi-holomorphic $(1,0)$-form was introduced in \cite{DP6} and is given by,
\bea
\Delta (z,w) = \om_1(z) \om_2(w) - \om_2(z) \om _1(w)
\eea
It is a holomorphic $(1,0)$-form in both $z$ and $w$, anti-symmetric in $z,w$ (viewed as the coefficient of the coordinate $(1,0)\otimes (1,0)$-form $dz \, dw$), and satisfying the following relations on tri- and quadri-holomorphic forms, 
\bea
\label{omdel}
 \om _I(z_1) \Delta (z_2,z_3) + \om _I(z_2) \Delta (z_3,z_1) +\om _I(z_3) \Delta (z_1,z_2) & = & 0
\no \\ 
\Delta (z_1,z_2) \Delta (z_3,z_4) + \Delta (z_1,z_3) \Delta (z_4,z_2) + \Delta (z_1,z_4) \Delta (z_2,z_3) & = & 0
\eea
Further ubiquitous combinations of significance for the five-point function are the vector-valued $(1,0)$-forms in five points $\om_I(z_i) \Delta (z_j,z_k) \Delta (z_\ell,z_m)$, for which a {\sl cyclic basis} may be chosen of the form,
\bea
\label{cycbasis}
\om_I(z_1) \Delta(z_2,z_3) \Delta (z_4,z_5) \hskip 1in \hbox{and 4 cyclic permutations in $1,2,3,4,5$}
\eea
To show that all forms $\om_I(z_i) \Delta (z_j,z_k) \Delta (z_\ell,z_m)$ may be decomposed in this cyclic basis, we use a cyclic permutation to set $i=1$, producing  the three forms $\om_I(z_1) \Delta (z_2,z_3) \Delta (z_4,z_5)$,  $\om_I(z_1) \Delta (z_2,z_5) \Delta (z_3,z_4)$, and $\om_I(z_1) \Delta (z_2,z_4) \Delta (z_5,z_3)$. The first form is already of the desired form; the third form is related to the first two by the second identity in (\ref{omdel}); and the second form may be reduced to a linear combination in the cyclic basis by implementing the first identity of (\ref{omdel}) on the product $\om_I(z_1) \Delta (z_2,z_5) $.

\subsection{The hyper-elliptic formulation}
\label{sec:B2}

Every compact genus-two Riemann surface $\Sigma$ is hyper-elliptic: it possesses a holomorphic involution $\cI$ and may be represented as a double cover of the sphere $\hat \CC$ ramified over 6 branch points  $u_1 , \cdots, u_6 \in \CC$. The surface $\Sigma$ may be parametrized by $z=(x,s)$ where,
\bea
\label{shyper}
s^2 = \prod _{i=1}^6 (x-u_i)
\eea
The involution $\cI$ acts by $\cI (x, s) = (x, -s)$ where $s=s(x)$ is given by the above relation. The action of the modular group reduces to the group $\mS_6$ of permutations of the branch points. The $SL(2,\CC)$ conformal automorphism of $\hat \CC$ allows one to fix three of the six branch points at arbitrary  points leaving the remaining three branch points to parametrize the complex moduli of a genus-two surface. 

\sm

A basis for holomorphic $(1,0)$-forms in the hyper-elliptic formulation is given by,
\bea
{ dx \over s(x)} \hskip 1in { x \, dx \over s(x)}
\eea
The normalized holomorphic $(1,0)$-forms $\om_I$  may be expressed in the hyper-elliptic basis with the help of a matrix $\sigma_{IJ}$ as follows, 
\bea
\om_I (z) =  \sum_{J} \sigma _{IJ} { x^{J-1} dx \over s(x)}
\hskip 1in 
\delta _{IK} =  \sum_{J} \sigma _{IJ} \oint _{\mA_K} { x^{J-1} dx \over s(x)} 
\eea
where the coefficients $\sigma_{IJ}$ depend on moduli but not on $z$. The bi-holomorphic form $\Delta$  may be expressed in terms of the hyper-elliptic basis for $z_i=(x_i, s(x_i))$,
\bea
\label{DelHyper}
\Delta (z_1,z_2) = (\det \sigma) \, { (x_1-x_2)  \, dx_1 \, dx_2 \over s(x_1) \, s(x_2)}
\eea
Therefore, as a function of $z_1=(x_1, s(x_1))$, the two zeros of the holomorphic $(1,0)$-form $\Delta (z_1,z_2)$ are at $z_2=(x_2,s(x_2)) $  and its image under involution $\cI z_2=(x_2,-s(x_2))$.

\subsection{The chiral bosonic partition function}
\label{apponchiralZ}

There exists a one-to-one correspondence between the six odd spin structures $\nu_i $ and the branch points $u_i$  for $i=1,\cdots, 6$  given by a relation on the Abel map,
\bea
(\nu _i)_I = \int _{z_0} ^{u_i} \om _I - \Delta _I(z_0)
\eea
where $\Delta _I(z_0)$ is the Riemann vector of (\ref{Riemvec}), considered here for $h=2$. For each branch point $u_i$ with $i=1,\cdots, 6$ there exists a holomorphic $(1,0)$-form $\om _{\nu_i}$ with a unique double zero at that branch point $u_i$ given by (\ref{omnu}) and a holomorphic $(\half, 0)$-form $h_{\nu_i}(z)$ with a unique simple  zero at $u_i$ which enters the construction (\ref{prime}) of the prime form.

\sm

For genus two, the nowhere vanishing  form $\sigma(z)$ of (\ref{sigmaz}) is a $(1,0)$-form with non-trivial monodromy. The relation (\ref{sigmaE}) may be expressed as the fact that the combination, 
\bea
\label{sigma3}
 { \tet(u+v-z -\Delta) \over E(z,u) E(z,v) \sigma (z)}
 \eea
 is independent of $z$. Taking the limit $z\to v$, we obtain an interesting equivalent formula,
 \bea
 \label{sigma2}
  { \tet(u+v-z -\Delta) \over E(z,u) E(z,v) \sigma (z)} 
  =   { \om_I(v) \p^I \tet(u -\Delta) \over E(u,v)  \sigma (v)}
  \eea
Recalling the expression for the chiral bosonic partition function $Z$ from \cite{DP6},
\bea
\label{Z3}
Z^3 = { \tet(u+v-z -\Delta) E(u,v)  \sigma (u) \sigma (v)  \over E(z,u) E(z,v) \sigma (z) \Delta (u,v)} 
\eea 
and combining the two formulas, we obtain the following simple relation,
\bea
 \Delta (u,v) & = &  Z^{-3} \, \sigma (u)  \om_I (v) \p^I \tet(u -\Delta) 
 \label{andbelow}
\eea
This formula proves that the right side is single-valued and odd under swapping $u,v$. Under a modular transformation $M\in Sp(4,\ZZ)$, the bi-holomorphic form $\Delta$ transforms by,
\bea
\tilde \Delta (u,v) =  \det (C \Omega +D)^{-1} \Delta (u,v)
\eea
see (\ref{1b4}) for the decomposition of $M$  into $2\times 2$ matrices $A,B,C,D$ and (\ref{1b6}) for its action on $\om_I$. The modular transformation of $Z$ is given as follows,
\bea
\label{Zmod}
Z(\tilde \Omega)  = \epsilon_0 \, \det(C \Omega +D)^\half  \, Z(\Omega)
\eea
where $\epsilon_0^{24}=1$, and $\epsilon_0^3=\eps(0,M)$ occurs in the transformation law of the $\tet$-function of (\ref{thetamod}).

\subsection{Meromorphic $(1,0)$-forms and the  $b,c$ system}
\label{app.bc}

An Abelian differential of the third kind is a meromorphic $(1,0)$-form with two simple poles of residues $\pm 1$ at the points $z_1$ and $z_2$, respectively. This condition does not define the differential uniquely, as one may add a linear combination of the two  holomorphic Abelian differentials. To specify the differential uniquely, we fix two of its zeros to be at points $p_i$ with $i=1,2$. The resulting differential $G(z;z_1,z_2;p_1,p_2)$ satisfies, 
\bea
\label{Gthird}
\pbz G(z;z_1,z_2;p_1,p_2) & = & 2 \pi \delta (z,z_1) - 2 \pi \delta (z,z_2)
\no \\ 
G(p_i; z_1, z_2;p_1,p_2) & = & 0  \hskip 1in i=1,2
\eea
Here, $G$ is a $(1,0)$-form in $z$ and a $(0,0)$-form in $z_1, z_2, p_1, p_2$. It may be interpreted as a Green function for the Cauchy-Riemann operator acting on $(1,0)$-forms and may be represented explicitly in terms of  $\tet$-functions and the prime form by,
\bea
\label{Green}
G(z;z_1,z_2;p_1,p_2) =
{ \tet(z-z_1-z_2+p_1+p_2 - \Delta) \tet(-z+p_1+p_2-\Delta) E(z_1, z_2) 
\over \tet(-z_1+p_1+p_2-\Delta) \tet(-z_2+p_1+p_2-\Delta) E(z,z_1)  E(z,z_2) }
\quad
\eea
An alternative solution to the differential equation in (\ref{Gthird}) is given by,
\bea
\label{tau}
\tau_{1,2} (z) = \p _z \ln  E(z,z_1)- \p_z \ln E(z,z_2)
\eea
The form $\tau_{1,2}$ is a single-valued $(1,0)$-form in $z$ and a multiple-valued $(0,0)$-form in $z_1, z_2$. 
Under a modular transformation $\tau_{1,2}(z)$ behaves as follows,
\bea
\tilde \tau _{1,2}(z) = \tau_{1,2}(z) + 2 \pi i \sum_{I,J} \om_I(z) \left [ (C \Omega +D)^{-1} C \right ]^{IJ} \int ^{z_2} _{z_1} \om _J
\eea
In terms of $\tau$, the Green function $G$ is obtained by requiring zeros at $p_1, p_2$, and is given by,
\bea
\label{Green1}
G(z;z_1,z_2;p_1,p_2) = \tau_{1,2} (z) 
+ { \Delta (p_2,z) \over \Delta (p_1,p_2) } \, \tau_{1,2} (p_1)
+ { \Delta (z,p_1) \over \Delta (p_1,p_2) } \, \tau_{1,2} (p_2)
\eea
Both (\ref{Green}) and (\ref{Green1}) show that $G(z;z_1,z_2;p_1,p_2) $ is invariant under $p_1\leftrightarrow p_2$ and changes sign  under $z_1\leftrightarrow z_2$. The monodromy in $z_1, z_2$  of the three individual  terms given by (\ref{Emon}) cancels in the sum in view of the first relation of (\ref{omdel}). Similarly, the shifts in $\tau_{1,2}$ produced by modular transformations in each individual term in (\ref{Green1}) cancel in the sum, and we find that $G$ is modular invariant.

\subsubsection{The $b,c$ system for weights $(1,0)$ and $(0,0)$}
\label{B21}

A useful unification of various formulas for holomorphic and meromorphic forms is provided by the conformal field theory correlators of anti-commuting fields $b$ and $c$ of weights $(1,0)$ and $(0,0)$, respectively.\footnote{We note that the  $b,c$ system discussed in this appendix differs from the $b,c$ ghost system associated  with diffeomorphism invariance, whose  weights are $(2,0)$ and $(-1,0)$, respectively.} We shall normalize the OPE of the fields by,
\bea
b(z) \, c(w) \sim { 1 \over z-w}
\eea
For genus two all non-vanishing correlators have one more $b$ than $c$ operators and we define, 
\bea
 \cG_n (z_1, \cdots, z_{n+1}; w_1, \cdots, w_n) =  \< b(z_1) \cdots b(z_{n+1}) c(w_1) \cdots c(w_n) \>
\eea
The form $\cG_n$ is odd under swapping $z_i$ and $ z_j$ with $i\not=j$ and under swapping $w_a$ and $ w_b$ with $a \not=b$. Therefore, viewed as a function of $z_i$, it has simple zeros at all points $z_j$ with $j \not=i$, and viewed as a function of $w_a$ it has simple zeros  at all points $w_b$ with $b \not= a$. Finally, viewed as a function of $z_i$ the form $\cG_n$ has simple poles at all  points~$w_a$, governed by the OPE and, for $n \geq 2$, given by,
\bea
\cG_n (z_1,  \cdots, w_n)  \sim 
{ (-)^{n+a-i} \over z_i-w_a} \, \cG_{n-1} ( z_1, \cdots,  \widehat{z_i}, \cdots, z_{n+1}; w_1, \cdots , \widehat{w_a}, \cdots, w_n)
\label{polebc}
\eea
where the wide hat on $z_i$ instructs the omission of the point $z_i$, and similarly for $w_a$.
Quantum field theory allows us to evaluate $\cG_n$ by the rules of Wick contraction. However, an equivalent but more useful evaluation is closely related to chiral bosonization, 
\bea
\cG_n (z_1,  \cdots, w_n) & = & { \tet \left ( \sum _i z_i - \sum _a w_a -\Delta \right ) \prod _i \sigma (z_i) 
\over Z^3 \prod _{i,a} E(z_i,w_a) \prod _a \sigma (w_a) }  
\prod _{i<j} E(z_i,z_j) \prod _{a<b} E(w_a,w_b) 
\label{defcgn}
\quad
\eea
where $i,j =1, \cdots, n+1$ and $a,b=1,\cdots, n$.  The standard overall normalization factor produced by chiral bosonization is $Z^{-1}$ instead of $Z^{-3}$. The latter has been chosen here instead to simplify  subsequent formulas as well as the behavior of $\cG_n$ under modular transformations. With the factor of $Z^{-3}$ the correlator transforms under modular transformations by,
\bea
\tilde \cG_n(z_1, \cdots, w_n) = \det (C \Omega +D)^{-1} \cG_n(z_1, \cdots, w_n)
\label{trflawG}
\eea
Note the absence in this transformation law of the 24-th root of unity factor $\eps_0$ in (\ref{Zmod}), which is present in the transformation law for $\tet$ and also appears for the transformation law of $\cG_n$ if we had used the standard normalization obtained from replacing $Z^{-3} $ by $  Z^{-1}$.
For  $n=1$, one may use (\ref{Z3}) to obtain the explicit form
\bea
\label{cG1}
\cG_1(z_1,z_2;w)  
= { \tet (z_1+z_2-w-\Delta) E(z_1,z_2) \sigma (z_1) \sigma (z_2) \over Z^3 E(w,z_1) E(w,z_2) \sigma (w)} 
=  \Delta (z_1,z_2)
\eea
which manifests the transformation law (\ref{trflawG}). For  $n=2$ the correlator $\cG_2$ is proportional to the Green function $G$ of (\ref{Green}), 
\bea
\label{cG2a}
\cG_2(z,p_1,p_2;w_1, w_2) = \Delta (p_1,p_2) \, G(z;w_1,w_2;p_1,p_2) 
\eea 
and the relation (\ref{Green1}) now takes on a more symmetrical form in terms of $\cG_2$,
\bea
\label{cG2b}
\cG_2(z,p_1,p_2;z_1, z_2) 
=
\Delta (p_1,p_2) \tau_{1,2}(z) + \Delta (z,p_1) \tau_{1,2}(p_2) + \Delta (p_2,z) \tau_{1,2}(p_1)
\quad
\eea
where $\tau_{1,2}$ was defined in (\ref{tau}). The three terms on the right side are obtained from one another by cyclic permutations of $z,p_1,p_2$ thereby exhibiting the full permutation antisymmetry  of $\cG_2$ between these points. Moreover, the monodromies of the individual terms as $z_{1,2} \rightarrow z_{1,2} + \mB_I$ again cancel in view of the first identity in (\ref{omdel}). The explicit formula (\ref{cG2b}) may alternatively be obtained from Wick contractions of the $b,c$ correlator.

\subsection{Compendium of formulas for genus two and unitary gauge}
\label{app:comp}

In this subsection we collect formulas for holomorphic and meromorphic forms on an arbitrary genus-two Riemann surface in which the pair of points $q_1, q_2$ enters. Recall that the points $q_1, q_2$ are the zeros of a
holomorphic $(1,0)$-form $\varpi(z)$, and are therefore related to one another by the divisor relation,
\bea
 q_1+q_2- 2 \Delta = 2 \kappa
\eea
where $\kappa=(\kappa', \kappa'')$ is an arbitrary half-integer characteristic given  by $\kappa = \Omega \kappa ' + \kappa''$ and $\Delta$ denotes the Riemann vector (\ref{Riemvec}). We have the following relations for $\a=1,2$,
\bea
\label{varpi}
\varpi (z) = c_\a \Delta (q_\a,z)  
\eea
or equivalently, 
\bea
\varpi(z) =   (-)^{\a-1} \, e^{2 \pi i \kappa ' (q_\a-\Delta)}  \sum_I \om_I (z) \p^I \tet (q_\a-\Delta) 
\eea
and 
\bea
c_\a \sigma (q_\a) & = & (-)^{\a-1} Z^3 \, e^{2 \pi i \kappa '(q_\a-\Delta)}
\eea
The objects $c_1, c_2$ are holomorphic $(-1,0)$-forms with non-trivial monodromy in $q_1$ and $q_2$, respectively, and satisfy the following relations (see (\ref{zeedelta1}) for $\cZ_0$),
\bea
\label{fund1}
c_1 \, \om_I(q_1) - c_2 \, \om _I (q_2) & = & 0
\no \\
c_1^2 \, \p \varpi(q_1) + c_2^2 \, \p \varpi (q_2) & = & 0
\no \\
\cZ_0 \, c_1 c_2 \, \p \varpi (q_1) \, \p \varpi(q_2) & = & 1
\eea
as well as the identity,
\bea
\label{fund3}
\varpi (x) \varpi (y) \Big ( c_1 \tau_{x,y}(q_1) - c_2 \tau_{x,y}(q_2) \Big ) = 
- c_1^2 \p \varpi(q_1) \Delta (x,y)  
\eea
The value of $\Delta '(z)$ at the zeros $q_1, q_2$ of the holomorphic $(1,0)$-form $\varpi(z)$ is given by, 
\bea
\label{B.delom}
c_\a \Delta '(q_\a) = - \p \varpi(q_\a)
\eea
In particular, the meromorphic form evaluated at this point gives, 
\bea
 { \Delta(z_1,z_2) \, \Delta '(q_\a) \over \Delta (z_2,q_\a) \Delta (q_\a,z_1)} = 
{ c_\a  \p \varpi(q_\a) \Delta(z_1,z_2) \over  \varpi(z_1)  \varpi(z_2) }  
\eea
Further useful formulas are as follows, 
 \bea
\tet(q_1-\Delta +z-w) \tet(-q_1+\Delta +z-w) = - \varpi(z) \varpi (w) E(z,w)^2 e^{- 4 \pi i \kappa ' (q_1-\Delta)}
\label{remove2tet}
\eea
and
\bea
{ \varpi(x) \over \varpi(y)} 
= { E(x,q_1) E(x,q_2) \sigma (x)^2 \over E(y,q_1) E(y,q_2) \sigma (y)^2} \, e^{- 4 \pi i \kappa ' (x-y)}
\eea
and 
\bea
\label{C.Ginter}
\varpi(z_1) G(z_5;z_1,z_2;w,q_2) = - \varpi(z_5) G(z_1;z_5,w;z_2,q_1)
\eea
The latter formula is proven by expressing both sides in terms of (\ref{Green}) and then using the following relation between $\tet$-functions, 
\bea
{ \tet(z_5+w+q_2-z_1-z_2-\Delta)  \over \tet(z_1+z_2+q_1-z_5-w-\Delta)} = 
 \exp \big \{  4 \pi i \kappa '(\kappa + z_1+z_2-z_5-w-q_2-\Delta) \big \}
 \qquad
\eea

\newpage

%%%%%%%%%%%%%%%%%%%%%%%%%%%%%%%%%%%%%%%%%%%
%%%%%%%%%%%%%%%%%%%%%%%%%%%%%%%%%%%%%%%%%%%
\section{Reducing products of Szeg\"o kernels}
\label{sec:C}
\setcounter{equation}{0}
%%%%%%%%%%%%%%%%%%%%%%%%%%%%%%%%%%%%%%%%%%%
%%%%%%%%%%%%%%%%%%%%%%%%%%%%%%%%%%%%%%%%%%%

In this appendix, we shall reduce certain products of concatenated genus-two Szeg\"o kernels for even spin structures to a small set of standard forms. The goal is to obtain a simplified dependence on the spin structure, so that the sum over all even spin structures is facilitated. 

\sm

To do so we use the hyper-elliptic representation of the Szeg\"o kernel, which is obtained as follows. Each one of the 10 even spin structures $\delta$ uniquely corresponds to a partition $(A|B)=(B|A)$ of the 6 branch points $u_i$ into two disjoint sets $A,B$ of 3 branch points each,
\bea
A \cup B = \{ 1, \cdots, 6\}  \hskip 0.5in A \cap B = \emptyset  \hskip 0.5in \# A = \# B=3
\eea
To a partition $(A|B)$, we associate the degree 3 polynomials $s_A(x)^2$ and $ s_B(x)^2$ defined by,
\bea
s_A(x)^2 = \prod _{i\in A} (x-u_i) \hskip 1in s_B(x)^2 = \prod _{i\in B} (x-u_i) 
\eea
such that the product is $s_A(x) s_B(x) = s(x)$ given in (\ref{shyper}).  The Szeg\"o kernel for an even spin structure $\delta$ specified by the partition $(A|B)$ of the points $z_i=(x_i,s(x_i))$ is given by,
\bea
S_\delta(z_i,z_j) = \half \, {  s_A(x_i) s_B(x_j)   + s_A(x_j) s_B(x_i)  \over x_{ij}} \, 
\left ( { dx_i \, dx_j \over s(x_i) \, s(x_j)} \right )^\half 
\eea
where we use the standard notation $ x_{ij} = x_i - x_j$. We note that no holomorphic $(\half, 0)$-forms exist, for an arbitrary even spin structure, and throughout the genus-two moduli space.

\subsection{Product of two Szeg\"o kernels}

The square of the Szeg\"o kernel is an Abelian differential of second kind, namely with one double pole. It may be expressed as follows,
\bea
\label{Sze-2}
S_\delta (z_i, z_j)^2 =   S_{ij}^2 { dx_i dx_j \over 4\, x_{ij}^2}
\hskip 0.7in
S_{ij}^2 = 2 + { s_A(x_i)^2 s_B(x_j)^2 +s_A(x_j)^2 s_B(x_i)^2 \over s(x_i) \, s(x_j)}
\eea
The numerator of the ratio in $S_{ij}^2$ is a polynomial of degree three in $x_i$ and in $x_j$, which depends on the spin structure through the partition $(A|B)$, while all other ingredients  in the formula are independent of the partition $(A|B)$.

\subsection{Product of three Szeg\"o kernels}

The product of three concatenated Szeg\"o kernels forming a closed loop may be similarly decomposed in powers of $S_{ij}^2$ by using only polynomial algebra,
\bea
\label{Sze-3}
S_\delta (z_1, z_2) S_\delta(z_2,z_3) S_\delta (z_3,z_1)  
= { dx_1 dx_2 dx_3 \over 8 \, x_{12}\, x_{23} \, x_{31}} \Big ( S_{12}^2+ S_{23}^2+ S_{31}^2  - 4 \Big )
\quad
\eea
Recasting $S_{ij}^2$ in terms of $S_\delta(z_i,z_j)$, we find the following reduction formula,
\bea
\label{tri1}
S_\delta (z_1, z_2) S_\delta(z_2,z_3) S_\delta (z_3,z_1) 
= - { dx_1 dx_2 dx_3 \over 2\, x_{12}\, x_{23} \, x_{31}}
+  \left ({ x_{12} \,  dx_3 \over 2\, x_{23}\, x_{31}} S_\delta (z_1,z_2)^2 + \hbox{cycl} (1,2,3) \right )
\quad
\eea
The reduced product of three Szeg\"o kernels to a sum of squares will be fundamental in our ability to perform various sums over spin structures.

\subsection{Product of four Szeg\"o kernels}

There are two distinct products of four Szeg\"o kernels forming closed loops. The first is the product of two one-loop contributions, of the form $S_\delta(z_1,z_2)^2 S_\delta (z_3,z_4)^2$ and permutations thereof. Its simplified expression is obtained by taking the product of two copies of (\ref{Sze-2}).

\sm

The product of four concatenated Szeg\"o kernels forming a single closed loop may be similarly decomposed in powers of $S_{ij}^2$ by using polynomial algebra,
\bea
&&
S_\delta (z_1, z_2) S_\delta(z_2,z_3) S_\delta (z_3,z_4) S_\delta(z_4,z_1)
\no \\ && \hskip 0.3in
= { dx_1 \, dx_2 \, dx_3 \, dx_4 \over 32\, x_{12} \, x_{23} \, x_{34} \, x_{41}}
\Big ( - 16 + 4 S_{13}^2 + 4 S_{24}^2 
+ S_{12}^2 S_{34}^2 +  S_{23}^2 S_{41}^2 - S_{13}^2 S_{24}^2  \Big )
\eea
In terms of the original Szeg\"o kernel, this expression becomes, 
\bea
&&
S_\delta (z_1, z_2) S_\delta(z_2,z_3) S_\delta (z_3,z_4) S_\delta(z_4,z_1)
\no \\ && \qquad 
=
- {  dx_1 \, dx_2 \, dx_3 \, dx_4 \over  2 \, x_{12} \, x_{23} \, x_{34} \, x_{41}}
+  {  x_{13}^2  \, dx_2 \,  dx_4 \over  2 \, x_{12} \, x_{23} \, x_{34} \, x_{41}}  S_\delta (z_1,z_3)^2 
 + {   x_{24}^2 \, dx_1  \, dx_3  \over  2 \, x_{12} \, x_{23} \, x_{34} \, x_{41}}  S_\delta (z_2,z_4)^2 
\no \\ && \hskip 0.5in 
+ {  x_{12} x_{34} \over 2\,  x_{23} \,x_{41}} 
   S_\delta(z_1,z_2)^2  S_\delta(z_3,z_4)^2 
 +  {  x_{23} x_{41}   \over 2\,  x_{12} \, x_{34}} 
 S_\delta(z_2,z_3)^2 S_\delta(z_4,z_1)^2 
 \no \\ && \hskip 0.5in 
 -  {  x_{13}^2 x_{24}^2 \over 2\,  x_{12} \, x_{23} \, x_{34} \, x_{41}}   S_\delta(z_1,z_3)^2  S_\delta(z_2,z_4)^2  
\eea
The relative sign factors of the last three terms depend on whether the pairs of points that occur as the argument of the Szeg\"o kernels are adjacent or not.

\subsection{Product of five Szeg\"o kernels}

There are two distinct products of  five Szeg\"o kernels forming closed loops. The first one involves two loops and reduces with the help of (\ref{tri1}) as follows,
\bea
\label{penta1}
&&
S_\delta (z_1, z_2)S_\delta (z_2, z_3)S_\delta (z_3, z_1) S_\delta (z_4, z_5)^2
\\ && \qquad 
= -  { dx_1 dx_2 dx_3 \over 2\, x_{12} \, x_{23} \, x_{31}} S_\delta (z_4, z_5)^2 
+  \left ( { x_{12} \,  dx_3 \over 2\, x_{23} \, x_{31}} S_\delta (z_1,z_2)^2 S_\delta (z_4,z_5)^2 
+  \hbox{cycl} (1,2,3) \right )
\no
\eea
The second involves a single loop and reduces as follows,
\bea
&&
S_\delta (z_1, z_2)S_\delta (z_2, z_3)S_\delta (z_3, z_4) S_\delta (z_4, z_5) S_\delta (z_5,z_1)
 \\ &&  \hskip 0.4in 
 =
{ dx_1 \, dx_2 \, dx_3 \, dx_4 \, dx_5 \over 64 \, x_{12} \, x_{23} \, x_{34} \, x_{45} \, x_{51}} \Big [
- 16 +   \sum _{1\leq i < j \leq 5} \pm 4 \, S_{ij}^2
\no \\ && \hskip 2.1in 
+ S_{12}^2 S_{34}^2 
+ S_{23}^2 S_{45}^2
+ S_{34}^2 S_{51}^2 
+ S_{45}^2 S_{12}^2
+ S_{51}^2 S_{23}^2
\no \\ && \hskip 2.1in  
+ S_{12}^2 S_{35}^2 
+ S_{23}^2 S_{41}^2
+ S_{34}^2 S_{52}^2 
+ S_{45}^2 S_{13}^2
+ S_{51}^2 S_{24}^2  
\no \\ && \hskip 2.1in 
- S_{13}^2 S_{24}^2
- S_{24}^2 S_{35}^2
- S_{35}^2 S_{41}^2 
- S_{41}^2 S_{52}^2 
- S_{52}^2 S_{13}^2    
\Big ]
\no
\eea
The signs in the sum over 10 pairs $(i,j)$ on the first line on the right side correspond to whether $i,j$ are nearest neighbors (minus sign) or not (plus sign). The last 15 terms correspond to all possible inequivalent partitions of the five points into two pairs of points each along with a single remaining point. Their sign corresponds to whether the points in the two pairs are interlaced (minus sign) or not (plus sign). 

\sm

Expressing the result in terms of the original Szeg\"o kernels we obtain,
\bea
\label{penta2}
&&
S_\delta (z_1, z_2)S_\delta (z_2, z_3)S_\delta (z_3, z_4) S_\delta (z_4, z_5) S_\delta (z_5,z_1)
\no \\ && \qquad =
- { dx_1 \, dx_2 \, dx_3 \, dx_4 \, dx_5 \over 20 \, x_{12} \, x_{23} \, x_{34} \, x_{45} \, x_{51}}
- { x_{12}^2 dx_3 dx_4 \, dx_5 \over 4 x_{12} \cdots x_{51}} S_\delta (z_1, z_2)^2 
+ { x_{13}^2 dx_2 dx_4 dx_5 \over 4 \, x_{12} \cdots x_{51}} S_\delta (z_1,z_3)^2 
\no \\ && \qquad \quad
+ { x_{23}^2 x_{45}^2 dx_1 \over 4 \, x_{12} \cdots x_{51}} S_\delta (z_2,z_3)^2 S_\delta (z_4, z_5)^2 
+ { x_{25}^2 x_{34}^2 dx_1 \over 4 \, x_{12} \cdots x_{51}} S_\delta (z_2,z_5)^2 S_\delta (z_3, z_4)^2 
\no \\ && \qquad \quad
- { x_{24}^2 x_{35}^2 dx_1 \over 4 \, x_{12} \cdots x_{51}} S_\delta (z_2,z_4)^2 S_\delta (z_3, z_5)^2 + \hbox{ cycl}(1,2,3,4,5)
\eea
where the instruction to add cyclic permutations applies to the entire expression on the right side of the equality. Note that the relative signs of the last three terms prior to the instruction of cyclic symmetrization again relate to whether pairs of points are interlaced (minus sign) or not interlaced (plus sign).

 \subsection{Alternative expressions for the ubiquitous meromorphic form}
 \label{sec:C5}

In each one of the above reductions, the coefficients of the squares of the Szeg\"o kernels and their products are expressed in terms of differential $(1,0)$-forms in the hyper-elliptic formulation. For example, the coefficient in the second term of (\ref{tri1}), 
\bea
{ x_{12} \, dx_3 \over x_{23} x_{31} } = - { dx_3 \over x_{23} } - { dx_3 \over x_{31}}
\eea
is a meromorphic $(1,0)$-form in $z_3$ (where $z_i=(x_i, s(x_i))$) and a scalar in $z_1$ and $z_2$, with four poles in $z_3$, namely simple poles at $z_1$, $z_2$ and at their images under involution $\cI z_1, \cI z_2$, with residues $\pm 1$. Its six zeros are simple zeros at the six branch points $u_i$. It may be expressed in terms of standard differentials with the help of the form $\Delta(x,y)$, and the following holomorphic $(3,0)$-form derived from it, 
\bea
\Delta'(z) = \p_w \Delta (w,z) \Big |_{w=z}
\eea
Its expression in the hyper-elliptic formulation with $z=(x,s(x))$ is derived from (\ref{DelHyper}),
\bea
\Delta'(z) = (\det \sigma) { dx^3 \over s(x)^2}
\eea
which makes it clear that this holomorphic $(3,0)$-form has simple zeros at the six branch points. Again using (\ref{DelHyper}) it is now straightforward to see that we have,
\bea
\label{C.third}
{ x_{12} \, dx_3 \over x_{23} \, x_{31} } = { \Delta(z_1,z_2) \, \Delta '(z_3) \over \Delta (z_2,z_3) \Delta (z_3,z_1)}
\eea
This result allows us to express all the reduced forms of the concatenated products of Szeg\"o kernels in two different ways: Either in terms of the hyper-elliptic formulation as was done in the first subsection of this appendix, or in terms of the canonical differentials $\om_I$ through the bi-holomorphic form $\Delta$, as we have done in this subsection.

\newpage

%%%%%%%%%%%%%%%%%%%%%%%%%%%%%%%%%%%%%%%%%%%
%%%%%%%%%%%%%%%%%%%%%%%%%%%%%%%%%%%%%%%%%%%
\section{Summary of spin structure sums up to four points}
\setcounter{equation}{0}
\label{sec:D}
%%%%%%%%%%%%%%%%%%%%%%%%%%%%%%%%%%%%%%%%%%%
%%%%%%%%%%%%%%%%%%%%%%%%%%%%%%%%%%%%%%%%%%%

 In this appendix, we summarize the spin structure sums for up to four vertex points derived in \cite{DP6}. In the next appendix we shall derive the additional spin structure sums that are needed for the five-point string amplitude with external NS states and even spin structures. Throughout, the points $z_i$ will refer to vertex points, while $q_1,q_2 $ will refer to the zeros of a holomorphic differential $\varpi(w)$. The chiral partition function $\cZ[\delta]$ was calculated in \cite{DP6}, 
\bea
\cZ[\delta] & = & \cZ_0 \, E(q_1,q_2) \, e^{ 4 \pi i \kappa ' \Omega \kappa '} \, \< \kappa |\delta \> \, \tet [\delta ](0)^4
\label{defcZ} \\
\cZ_0 & = & { Z^{12} \over \pi^{12} \, \Psi _{10} \, E(q_1, q_2)^2 \sigma (q_1)^2 \sigma (q_2)^2} \notag
\eea 
where the chiral scalar partition function $Z$ can be found in (\ref{Z3}) and \cite{DP4}, 
$q_1+q_2-2\Delta = 2 \kappa$ and $\Psi_{10}$ is the Igusa cusp form (\ref{defigusa}).

\subsection{Vanishing spin structure sums}

The following spin structure sums over even spin structures $\delta$ vanish.
\bea
\label{D2}
I_1 & = & \sum _\delta \cZ [\delta] ~ S_\delta (q_1, q_2)
\no \\
I_2 & = & \sum _\delta \cZ [\delta] ~ S_\delta (q_1, q_2) S_\delta (z_1,z_2)^2
\no \\
I_3 & = & \sum _\delta \cZ [\delta] ~ S_\delta (q_1, q_2) S_\delta (z_1,z_2)
S_\delta (z_2,z_3) S_\delta (z_3,z_1)
\no \\
I_4 & = & \sum _\delta \cZ [\delta] ~ S_\delta (q_1, z_1) S_\delta (z_1,q_2)
\no \\
I_5 & = & \sum _\delta \cZ [\delta] ~ S_\delta (q_1, z_1) S_\delta (z_1,q_2)
S_\delta (z_2,z_3)^2 
\no \\
I_6 & = & \sum _\delta \cZ [\delta] ~ S_\delta (q_1, z_1) S_\delta (z_1,z_2)
S_\delta (z_2,q_2)
\no \\
I_7 & = & \sum _\delta \cZ [\delta] ~ S_\delta (q_1, z_1) S_\delta (z_1,z_2)
S_\delta (z_2,z_3) S_\delta (z_3,q_2)
\no \\
I_8 & = & \sum _\delta \cZ [\delta] ~ S_\delta (q_1, z_1) S_\delta (z_1, z_2)
S_\delta (z_2, z_3) S_\delta (z_3, z_4) S_\delta (z_4, q_2)
\no \\
I_9 & = & \sum _\delta \cZ [\delta] ~ S_\delta (q_1, z_1) S_\delta (z_1, z_2)
S_\delta (z_2, q_2) S_\delta (z_3, z_4)^2 
\no \\
I_{10} & = & \sum _\delta \cZ [\delta] ~ S_\delta (q_1, z_1) S_\delta (z_1, q_2)
S_\delta (z_2, z_3) S_\delta (z_3, z_4) S_\delta (z_4, z_2)
\eea

\subsection{More spin structure sums involving $S_\delta(q_1, q_2)$}

The following spin structure sums involve a product of $S_\delta (q_1,q_2)$ times loops of Szeg\"o kernels. 
They may be evaluated with the use of the Riemann relations~(\ref{Riem2}),
\bea
\label{I11}
I_{11} (z_1,z_2;z_3,z_4)
& = &
\sum _\delta \cZ [\delta] S_\delta (q_1, q_2)
S_\delta (z_1, z_2)^2 S_\delta (z_3, z_4)^2
\no \\
I_{12} (z_1,z_2,z_3,z_4)
& = &
\sum _\delta \cZ [\delta] S_\delta (q_1, q_2)
S_\delta (z_1, z_2) S_\delta (z_2, z_3) S_\delta (z_3, z_4) S_\delta (z_4, z_1)
\eea
and are given by,
\bea
\label{I11S}
I_{11}  (z_1,z_2;z_3,z_4) & = &  -2\cZ_0 \varpi(z_1) \varpi(z_2) \varpi(z_3) \varpi(z_4) 
\no \\
I_{12}  (z_1,z_2,z_3,z_4) & = &  -2\cZ_0 \varpi(z_1) \varpi(z_2) \varpi(z_3) \varpi(z_4) 
\eea
The holomorphic $(1,0)$-form $\varpi (z)$ is proportional to both $\Delta (q_1, z)$ and $\Delta (q_2,z)$. It was introduced in \cite{DP6} and is given explicitly in (\ref{varpi2}), (\ref{varpi3}) as well as in the compendium of formulas in (\ref{varpi}).

\sm

The following spin structure sums involve the insertion $\varphi [\delta] (w;z_i,z_j)$ 
of a fermion stress tensor defined by (\ref{varphi}), 
\bea
\label{I13}
I_{13} (w; z_1,z_2)
& = &
\sum _\delta \cZ [\delta] S_\delta (q_1, q_2) \varphi [\delta] (w;z_1,z_2) S_\delta (z_2, z_1)
\no \\
I_{14} (w; z_1,z_2,z_3)
& = &
\sum _\delta \cZ [\delta] S_\delta (q_1, q_2) \varphi [\delta] (w;z_1,z_2)
S_\delta (z_2, z_3) S_\delta (z_3,z_1)
\no \\
I_{15} (w; z_1,z_2,z_3,z_4)
& = &
\sum _\delta \cZ [\delta] S_\delta (q_1, q_2) \varphi [\delta] (w;z_1,z_2)
S_\delta (z_2, z_3) S_\delta (z_3, z_4) S_\delta (z_4, z_1)
\no \\
I_{16} (w; z_1,z_2; z_3, z_4)
& = &
\sum _\delta \cZ [\delta] S_\delta (q_1, q_2) \varphi [\delta] (w;z_1,z_2)
S_\delta (z_2, z_1) S_\delta (z_3, z_4)^2
\eea
and produce the following simple results,
\bea
\label{I13S}
I_{13}(w;z_1,z_2) & = & 4\cZ_0\varpi(z_1)\varpi(z_2)\varpi(w)^2
\no \\
I_{14} (w;z_1,z_2,z_3) & = & 2 \cZ_0 \varpi (z_1) \varpi (z_2) \varpi (w)^2 
G(z_3;z_1,z_2;q_1 ,w) + (q_1 \leftrightarrow q_2)
\eea
as well as the following more involved expressions,
\bea
I_{15} (w;z_1,z_2,z_3,z_4) & = & 
 \cZ_0 \varpi (z_1) \varpi (z_2) \varpi (w)^2 \big \{ 
 G(z_3;z_1,z_2;q_1,w)  G(z_4;z_1,z_2;q_2,w) 
\no \\ && \hskip 1.45in
+ G(z_3;z_4,z_1;q_1,w)  G(z_4;z_3,z_2;q_1,w)
\no \\ && \hskip 1.45in
- G(z_3;z_4,z_2;q_1,w)  G(z_4;z_3,z_1;q_1,w)
 \no \\ && \hskip 1.45in + (q_1 \leftrightarrow q_2)  \big \}
\no \\ 
I_{16} (w;z_1,z_2;z_3,z_4)
& = &
-  \cZ_0 \varpi (z_1) \varpi (z_2) \varpi (w)^2
\big \{ G(z_3;z_1,z_2;q_1,w)  G(z_4;z_1,z_2;q_2,w)
\no \\ && \hskip 1.55in
+ G(z_3;z_4,z_1;q_1,w)  G(z_4;z_3,z_2;q_1,w)
\no \\ && \hskip 1.55in
+ G(z_3;z_4,z_2;q_1,w)  G(z_4;z_3,z_1;q_1,w)
 \no \\ && \hskip 1.55in + (q_1 \leftrightarrow q_2)  \big \}
\eea
see (\ref{Green}) or (\ref{Green1}) for the Green function $G(z;z_1,z_2;p_1,p_2)$.

\subsection{Spin structure sums involving $\Xi_6[\delta]$}

The following sums involve $\Xi_6[\delta]$ given by (\ref{Xi6}) and vanish identically,
\bea
\label{I17}
I_{17} & = & \sum _\delta \Xi _6 [\delta] \tet [\delta](0)^4 
\no \\
I_{18} (z_1,z_2) & = & \sum _\delta \Xi _6 [\delta] \tet [\delta](0)^4 S_\delta (z_1, z_2)^2
\no \\
I_{19} (z_1,z_2,z_3) & = & \sum _\delta \Xi _6 [\delta] \tet [\delta](0)^4 S_\delta (z_1, z_2)S_\delta (z_2, z_3)S_\delta (z_3, z_1)
\eea
Finally, the following sums involve $\Xi_6[\delta]$,
\bea
\label{I20}
I_{20} (z_1,z_2;z_3,z_4) & = & \sum _\delta \Xi _6 [\delta] \tet [\delta](0)^4 S_\delta (z_1, z_2)^2 S_\delta (z_3, z_4)^2
\no \\
I_{21} (z_1, z_2, z_3, z_4) & = & \sum _\delta \Xi _6 [\delta] \tet [\delta](0)^4 S_\delta (z_1, z_2)S_\delta (z_2, z_3)S_\delta (z_3, z_4) S_\delta (z_4,z_1)
\eea
 and were also evaluated in \cite{DP6},
\bea
\label{I20S}
I_{20} (z_1,z_2;z_3,z_4) & = & 
- 4 \pi^4 \Psi _{10} \Big ( \Delta (z_1, z_3) \Delta (z_2,z_4) + \Delta (z_1, z_4) \Delta (z_2,z_3) \Big ) 
\no \\
I_{21} (z_1,z_2,z_3,z_4) & = & 
2 \pi^4 \Psi _{10} \Big ( \Delta (z_1, z_2) \Delta (z_3,z_4) - \Delta (z_1, z_4) \Delta (z_2,z_3) \Big )
\eea
We note the identity $I_{20}(z_1,z_3;z_2,z_4) = - 2 I_{21}(z_1, z_2, z_3, z_4)$.

\newpage

%%%%%%%%%%%%%%%%%%%%%%%%%%%%%%%%%%%%%%%%%%%
%%%%%%%%%%%%%%%%%%%%%%%%%%%%%%%%%%%%%%%%%%%
\section{Evaluation of spin structure sums with five points}
\setcounter{equation}{0}
\label{sec:E}
%%%%%%%%%%%%%%%%%%%%%%%%%%%%%%%%%%%%%%%%%%%
%%%%%%%%%%%%%%%%%%%%%%%%%%%%%%%%%%%%%%%%%%%

In this appendix we present the calculations and simplifications of the spin structure sums involving various combinations of Szeg\"o kernels anchored at five vertex points, complementing the discussion in section \ref{sec:3}.

\subsection{Simplification of $J_1$ and $J_2$}
\label{appE.1}

The starting point for the simplification of $J_1$ and $J_2$ is the formula obtained in (\ref{J2.1}) in the hyper-elliptic formulation. To convert the expression for $J_1$ into prime forms, we begin by matching its poles by $ -2 \pi^4 \Psi_{10} \, j_1^n$ where $j_1^n$ is given by,
\bea
j^n_1 = 
\big[ \partial_3 \ln E(3,2) - \partial_3 \ln E(3,1) \big]
\Big ( \Delta (1,4)\Delta (2,5) + \Delta (1,5) \Delta (2,4) \Big ) 
+ \hbox{ cycl}(1,2,3)
\quad
\label{J1.3}
\eea
While $-2 \pi^4 \Psi_{10} \, j_1^n$ matches the poles and residues of $J_1$, has vanishing monodromy under $\mA_I$ cycles in all points,  and is single-valued in $z_4,z_5$, it  has non-trivial monodromy in the points $z_1, z_2, z_3$, whereas $J_1$ is single-valued. The monodromy transformation of $j_1^n$ under $z_1 \to z_1 + \mB_I$ is given by, 
\bea
j_1 ^n  \to  j_1^n - 2 \pi i \, \Big  ( \om _I (4) \Delta (1,5)  +  \om_I(5) \Delta (1,4) \Big ) \Delta (2,3)
\eea
The monodromy in $z_1$, as well as in the points $z_2,z_3$, may be readily cancelled by the addition to $j_1^n$ of the following holomorphic combination, 
\bea
j_1^f & = & 
+ \p_1 \ln E(1,4) \Delta (2,3) \Delta (4,5) + \p_4 \ln E(4,1) \Delta (2,3) \Delta (1,5)
\no \\ &&
+ \p_1 \ln E(1,5) \Delta (2,3) \Delta (5,4) + \p_5 \ln E(5,1) \Delta (2,3) \Delta (1,4)
+ \hbox{cycl}(1,2,3)
\eea
The monodromy of $j_1^f$ in $z_4, z_5$ vanishes, and  its monodromy in $z_1, z_2, z_3$ cancels the monodromy of $j_1^n$. Thus, the combination $-2 \pi^4 \Psi _{10} (j_1^n + j_1^f)$ is single-valued in all its variables and  has exactly the same poles and residues  as $J_1$, so that its difference with $J_1$ must be single-valued and holomorphic in all variables. Its general form must be,
 \bea
 \sum _{I,J,K,L,M} T^{I,J,K,L,M} \, \om_I(1)  \om_J(2)  \om_K(3)  \om_L(4)  \om_M(5) 
 \eea
 for some modular invariant tensor $T$. The subgroup of $Sp(4, \ZZ)$ which leaves the splitting into $\mA$ and $\mB$-cycles invariant is $GL(2,\ZZ)$, and contain the element $-I$ under which $\om_I$ is odd. But there can be no odd-rank invariant tensor $T$, so this tensor must vanish and we have $J_1=-2 \pi^4 \Psi _{10} (j_1^n + j_1^f)$. The result may be simplified in terms of the functions $g^I_{i,j}$ introduced in (\ref{gI}), and regrouped in terms of the single-valued combinations $G^I$ of (\ref{GI}). An analogous calculation gives also the simplified form of $J_2$ and we recover the final results for $J_1$ and $J_2$ given in (\ref{J12.6}).

\subsection{Simplification of $J_3$ and $J_4$}

The procedure used to simplify the expressions for $J_3$ and $J_4$ in (\ref{J34.1}) is analogous to the one
used to simplify $J_1$ and $J_2$, so we shall be brief here. We match the poles and residues  of (\ref{J34.1}) with an expression $-\cZ_0 j_3^n$ in terms of derivatives of the prime form, 
\bea
j_3^n = \Big ( \p_1 \ln  E(1,3) -\p_1 \ln  E(1,2) \Big ) \prod _{i=2,3,4,5} \varpi (i) + \hbox{cycl}(1,2,3)
\eea
This object is single-valued in $z_4,z_5$ but has non-trivial monodromy in $z_1, z_2, z_3$. Specifically, under $z_1 \to z_1 +\mB_I$, we have, 
\bea
j_3^n \to j_3^n + 2 \pi i \Big ( \om_I(2) \varpi (3) - \om_I(3) \varpi(2) \Big ) \prod _{i=1,4,5} \varpi(i)
\eea
To match its monodromy by a holomorphic form, we use the relation between $\varpi$ and $\Delta$,
\bea
\label{c1om}
\varpi (z) & = & c_1 \Delta (q_1, z) = c_2 \Delta (q_2, z) 
\no \\
 c_\a & = & (-)^{\alpha-1} Z^3 \sigma (q_\a)^{-1} e^{2 \pi i \kappa ' (q_\a-\Delta)}
\eea
This allows us to recast the monodromy using (\ref{omdel}) as follows for either value of $\alpha$, 
\bea
\om_I(2) \varpi (3) - \om_I(3) \varpi(2) = c_\a \, \om _I (q_\a) \Delta (2,3)
\eea
The monodromy of $j_3^n$ may be compensated by adding to $j^n_3$ the following holomorphic combination
whose monodromy is opposite to that of $j_3^n$, i.e.\ $J_3= -\cZ_0( j_3^n+j_3^f)$ with,
\bea
j_3^f = - c_\a \, \p _{q_\a} \ln E(q_\a, 1) \varpi (1) \Delta (2,3) \varpi(4) \varpi (5) + \hbox{cycl}(1,2,3)
\eea
Using the relation (\ref{omdel}) for the points $z_2, z_3, q_\a$, and reconverting $\Delta (q_\a, z_i)$ to $\varpi(z_i)$, we find, 
\bea
j_3 ^f =  \sum_I \om _I (1)  (g_{q_\a, 2}^I + g_{3, q_ \a}^I ) \prod _{i=2}^5 \varpi(i) + \hbox{cycl}(1,2,3)
\eea
Proceeding analogously for $J_4$ and assembling all contributions  in terms of the functions $g^I_{a,b}$, we recover the expressions for $J_3$ and $J_4$ in (\ref{J34.2}).

\subsection{Calculation of $J_6,J_7,J_8$ and $J_9$}

The functions $J_6,J_7,J_8,J_9$ are defined in (\ref{J5-9}) in terms of spin structure sums over a product of six Szeg\"o kernels. The Riemann identities (\ref{Riem2}) evaluate spin structure sums with four $\tet$-functions. To  evaluate the spin structure sums in $J_6,J_7,J_8,J_9$ using the Riemann identities, we need to reduce the number of $\delta$-dependent $\tet$-functions in the summand with the help of the Fay identities (\ref{Fay1}), (\ref{Fay2}) and (\ref{Fay3}). Moreover, we will frequently make use of the results on spin structure sums $I_1,\ldots, I_{21}$ with four or fewer vertex points reviewed in appendix~\ref{sec:D}.

\subsubsection{Calculation of $J_6$}

To compute $J_6$ we shall use the Fay identity (\ref{Fay3}). Upon substituting this identity for the squares of both Szeg\"o kernels in the spin structure sum for $J_6$, we see that all the contributions from the first term on the right side of the identity (\ref{Fay3}) cancel in view of $I_4=I_5=0$. Making the  $\delta$-dependence of $\cZ[\delta]$ explicit, we have,
\bea
J_6 (z_1;z_2,z_3;z_4,z_5)  = {\cZ_0 E(q_1, q_2) \, e^{4 \pi i \kappa ' \Omega \kappa '}  \over E(q_1, z_1) E(z_1, q_2)} \, \sum_{I,J,K,L} \om _I(z_2) \om_J (z_3) \om_K (z_4) \om _L (z_5) \, \mJ^{IJ;KL}_6
\quad
\eea
where the modular tensor $\mJ_6$ is given by, 
\bea
\mJ^{IJ;KL} _6= 
\sum _\delta \< \kappa | \delta \> \, \tet [\delta] (q_1-z_1) 
\, \tet [\delta ] (z_1-q_2) \, \p^I \p^J \tet [\delta ](0) \, \p^K \p^L \tet [\delta ] (0)
\eea
To carry out the sum, we use the Riemann relations of (\ref{Riem2})  for a sum over even spin structures $\delta$.
We begin by evaluating,
\bea
\mJ_6 = \sum _\delta \< \kappa | \delta \> \, \tet [\delta] (q_1-z_1) 
\, \tet [\delta ] (z_1-q_2) \, \tet [\delta ](2\zeta_3) \,  \tet [\delta ] (2\zeta _4)
\eea
in terms of which $\mJ^{I,J;K,L}_6$ is given by,
\bea
\mJ ^{I,J;K,L}_6 = { 1 \over 16} \, { \p^4  \, \mJ_6 \over \p \zeta _{3I} \, \p  \zeta _{3J}\,  \p \zeta _{4K} \, \p \zeta _{4L}}  \Bigg | _{\zeta _3=\zeta _4=0} 
\eea
We evaluate $\mJ_6$ using the Riemann identities (\ref{Riem2}) with the following values of $\zeta ^\pm _a$,
\begin{align}
\zeta _1 ^+ & = \mq + \zeta _3 + \zeta _4 & \zeta _1^- & = \mz + \zeta _3 +\zeta _4
\no \\
\zeta _2 ^+ & = \mq - \zeta _3 - \zeta _4 & \zeta _2^- & = \mz - \zeta _3 - \zeta _4
\no \\
\zeta _3 ^+ & = -\mz  + \zeta _3 - \zeta _4& \zeta _3^- & = -\mq  + \zeta _3 -\zeta _4
\no \\
\zeta _4 ^+ & = -\mz  - \zeta _3 + \zeta _4 & \zeta _4^- & = -\mq  - \zeta _3  + \zeta _4
\end{align}
with
\bea
\mq = q_1 - \Delta -\kappa \hskip 1in \mz = z_1 - \Delta - \kappa
\eea
When $\zeta_3=\zeta_4=0$, each factor $\tet [\kappa] (\zeta _a^\pm)$ vanishes by the Riemann vanishing Theorem of appendix \ref{sec:A.vanishing}, so that in computing the 4-fold derivative at $\zeta_3=\zeta_4=0$, we need to apply precisely one derivative to each factor. Doing so we find,
\bea
\mJ^{I,J;K,L} _6& = & 
\p^I \tet [\kappa] (\mz)  \p^J \tet [\kappa] (\mz)  \p^K \tet [\kappa] (\mq)  \p^L \tet [\kappa] (\mq) 
\no \\ &&
+ \p^I \tet [\kappa] (\mq)  \p^J \tet [\kappa] (\mq) \, \p^K \tet [\kappa] (\mz)  \p^L \tet [\kappa] (\mz) 
\no \\ &&
- \big [ \p^I \tet [\kappa] (\mq)  \p^J \tet [\kappa] (\mz) + \p^I \tet [\kappa] (\mz)  \p^J \tet [\kappa] (\mq) \big ]
\no \\ && \qquad \qquad \times
\big [ \p^K \tet [\kappa] (\mq)  \p^L \tet [\kappa] (\mz) + \p^K \tet [\kappa] (\mz)  \p^L \tet [\kappa] (\mq) \big ]
\eea
Contracting with the holomorphic differentials, we use,
\bea
\varpi(z) = \sum_I \om_I(z) \p^I \tet [\kappa] (\mq) \, e^{ i \pi \kappa ' \Omega \kappa '} 
= \sum_I \om_I(z) \p^I \tet (q_1-\Delta) \, e^{2 \pi i \kappa ' (q_1-\Delta)}
\eea
The remaining combination,
\bea
 \sum_{I,J} \om_I(z_i) \p^I \tet [\kappa] (\mz) \, \om_J (z_j) \p^J \tet [\kappa] (\mz) 
{ E(q_1,q_2) \, e^{ 2 i \pi \kappa ' \Omega \kappa '} \over E(q_1,z_1) E(z_1,q_2)} 
\eea
is a $(1,0)$-form in $z_1, z_i, z_j$ which is holomorphic and symmetric in $z_i,z_j$, and is a scalar and odd under swapping $q_1$ and $q_2$. As a function of $z_1$, it vanishes at $z_i,z_j$ and  has simple poles  at $q_1, q_2$ with opposite residues $\pm \varpi(z_i) \varpi(z_j)$. These properties are uniquely matched by,
\bea
\label{ZG}
&&
{ E(q_1,q_2) \, e^{ 2 i \pi \kappa ' \Omega \kappa '} \over E(q_1,z_1) E(z_1,q_2)} \sum_{I,J} \om_I(z_i) \p^I \tet [\kappa] (\mz)  \om_J (z_j) \p^J \tet [\kappa] (\mz) 
\no \\ && \qquad 
= - G(z_1;q_1,q_2;z_i,z_j) \varpi(z_i) \varpi(z_j)
\eea
where $G$ is the Green function defined in (\ref{Green}). The simple poles of $G$ in $z_i$ and $z_j$ are cancelled by the factors  $\varpi(z_i) \varpi(z_j)$ so that the expression is holomorphic in $z_i, z_j$ as required. In terms of these functions, we have,
\bea
\label{oldJ6}
J_6 (z_1;z_2,z_3;z_4,z_5) & = &   \cZ_0 \, \rho_1 \,   \big ( 
{-} G(z_1;q_1,q_2;z_2,z_3)  - G(z_1;q_1,q_2;z_4,z_5)  
\\ && \hskip 0.5in
+ G(z_1;q_1,q_2;z_2,z_4) + G(z_1;q_1,q_2;z_3,z_4)  
\no \\ && \hskip 0.5in
+ G(z_1;q_1,q_2;z_2,z_5)  + G(z_1;q_1,q_2;z_3,z_5) 
\big ) 
\no \eea
where $\rho_i$ was defined in (\ref{rhoi}). The result of (\ref{oldJ6}) will be simplified in subsection \ref{sec:J6789}.

\subsubsection{Calculation of $J_7$}

To calculate $J_7$ we  evaluate the following auxiliary quantity in two different ways,
\bea
\hat J_7 = \sum _\delta \cZ[\delta] S_\delta (q_1, z_1) S_\delta (z_1,q_2) 
\Big ( S_\delta (z_2, z_3) S_\delta (z_4, z_5) - S_\delta(z_2,z_5)S_\delta(z_4,z_3) \Big )^2
\eea
On the one hand, by expanding the square and identifying with the functions $J_i$ we find,
\bea
\label{hatJ7}
\hat J_7 = J_6(z_1;z_2,z_3;z_4,z_5) + J_6(z_1;z_2,z_5;z_3,z_4) - 2 J_7(z_1;z_2,z_3,z_4,z_5)
\eea
On the other hand, evaluating the combination in parentheses using the Fay identity (\ref{Fay1}),
and factoring out the remaining spin structure independent factors, we have,
\bea
\hat J_7 = { \cZ_0 \, e^{ 4 \pi i \kappa ' \Omega \kappa'} E(q_1,q_2) E(z_2,z_4)^2 E(z_3,z_5)^2 \, \mJ_7
\over E(q_1,z_1) E(z_1,q_2) E(z_2,z_3)^2 E(z_2,z_5)^2 E(z_4,z_3)^2E(z_4,z_5)^2}
\eea
where
\bea
\mJ_7= \sum _{\delta} \< \kappa|\delta\> \tet[\delta](q_1-z_1) \tet[\delta] (z_1-q_2) \tet [\delta ](z_2+z_4-z_3-z_5)^2
\eea
This sum may be evaluated using Riemann identities (\ref{Riem2}) with,
\begin{align} 
\zeta _1^+ & = q_1-\Delta -\kappa + z_2+z_4-z_3-z_5 & \zeta _1^- & = z_1-\Delta -\kappa + z_2+z_4-z_3-z_5
\no \\
\zeta _2^+ & = q_1-\Delta -\kappa - z_2-z_4+z_3+z_5 & \zeta _2^- & = z_1-\Delta -\kappa - z_2-z_4+z_3+z_5 
\no \\
\zeta _3^+ & = \Delta +\kappa - z_1 & \zeta _3^- & = q_2-\Delta -\kappa
\no \\
\zeta _4^+ & = \Delta +\kappa - z_1 & \zeta _4^- & = q_2-\Delta -\kappa 
\end{align} 
Since we have $\tet [\kappa] (\zeta _3^\pm) =\tet [\kappa] (\zeta _4^\pm) = 0$, we readily have $\mJ_7=\hat J_7=0$.
Using this result in conjunction with (\ref{hatJ7}) we obtain a formula for $J_7$ in terms of $J_6$,
 \bea
J_7(z_1;z_2,z_3,z_4,z_5) = \half J_6(z_1;z_2,z_3;z_4,z_5) + \half J_6(z_1;z_2,z_5;z_3,z_4) 
\eea
Using the expression (\ref{oldJ6}) for $J_6$, we notice some simplifications, 
\bea
J_7(z_1;z_2,z_3,z_4,z_5) =
 \cZ_0 \, \rho_1 \, \big (   G(z_1;q_1,q_2;z_2,z_4)   + G(z_1;q_1,q_2;z_3,z_5)  \big ) 
 \label{oldJ7}
\eea
The result of (\ref{oldJ7}) will be simplified in subsection \ref{sec:J6789}.

\subsubsection{Calculation of $J_8$}

To compute $J_8$ we evaluate the following auxiliary combination in two different ways,
\bea
\hat J_8& = & \sum _\delta \cZ[\delta] \Big ( S_\delta (z_4, z_5)^2 - \p_{z_4} \p_{z_5} \ln E(z_4,z_5) \Big )
\Big ( S_\delta (q_1, z_1) S_\delta (z_1,z_2) - \tau_{q_1, z_2}(z_1) S_\delta(q_1,z_2) \Big )
\no \\ && \hskip 0.5in \times 
\Big ( S_\delta(z_2,z_3) S_\delta (z_3,q_2) - \tau_{z_2, q_2}(z_3) S_\delta(z_2,q_2) \Big )
\eea
On the one hand, by expanding the summand term by term and expressing the sums in terms of $I$ and $J$-functions, we see that the contribution proportional to $\p_{z_4} \p_{z_5} \ln E(z_4,z_5)$ cancels using $I_7=0$, and the contributions proportional to $\tau$ cancel in view of $I_9=0$. The remaining contribution is exactly $J_8$  so that $\hat J_8=J_8$. On the other hand, using  the Fay identity (\ref{Fay3}) for the first  factor and (\ref{Fay2}) for the second and third factors, we arrive at the following expression,  
\bea
J_8 (z_1,z_2,z_3;z_4,z_5) = 
{ \cZ_0 e^{4 \pi i \kappa ' \Omega \kappa '} E(q_1,q_2) \over E(q_1,z_2) E(z_2,q_2)}
\sum_{I,J,K,L} \om_K(z_1) \om_L(z_3) \om_I(z_4) \om _J(z_5)    \, \mJ^{I,J; K,L}_8
\quad
\eea
where
\bea
\mJ^{I,J;K,L} _8= \sum _\delta \< \kappa |\delta\> \, 
\tet[\delta ](0) \, \p^I \p^J \tet [\delta] (0) \, \p^K \tet[\delta](q_1-z_2) \, \p^L \tet [\delta] (z_2-q_2) 
\eea
To calculate the spin structure sum, we first compute the auxiliary quantity,
\bea
\mJ_8= \sum _\delta \< \kappa |\delta\> \, \tet[\delta ](0) \,  \tet [\delta] (2 \xi_2) \,  
\tet[\delta](q_1-z_2+ 2 \xi_3 ) \,  \tet [\delta] (z_2-q_2 + 2 \xi_4) \eea
so that,
\bea
\mJ^{I,J;K,L}_8 = { 1 \over 16} \, {\p^4 \mJ_8 \over \p \xi_{2I} \, \p \xi_{2J} \, \p \xi_{3K} \, \p \xi_{4L}} \, \Bigg |_{\xi_2=\xi_3=\xi_4=0}
\eea
Evaluating the spin structure sum using the Riemann relations (\ref{Riem2}) with,
\begin{align}
\zeta ^\pm _1 & = q_1 -\Delta -\kappa + \xi_2 +   \xi_3 + \xi_4  
\no \\
\zeta ^\pm _2 & = q_2 -\Delta -\kappa + \xi_2 -   \xi_3 - \xi_4 
\no \\
\zeta ^\pm _3 & =  \Delta + \kappa -z_2- \xi _2  +\xi_3 - \xi_4
\no \\
\zeta ^\pm _4 & = z_2  - \Delta - \kappa - \xi _2 -\xi_3 + \xi_4
\end{align}
we note that $\tet[\kappa](\zeta^\pm _a)=0$ for each $a=1,2,3,4$ when $\xi_2=\xi_3=\xi_4=0$ by the Riemann vanishing Theorem of appendix \ref{sec:A.vanishing}. Hence, when taking the four derivatives, it must be that exactly one derivative ends up on each factor. It will be convenient to convert $\tet[\kappa]$ to $\tet$-functions without characteristics. Carrying out the  derivatives in $\xi_2,\xi_3,\xi_4$ and setting these variables to zero we find, 
\bea
\mJ^{I,J;K,L}_8 & = & 
\Big ( \p^I \tet (q_1-\Delta) \p^J \tet (q_1-\Delta) \p^K \tet (z_2-\Delta) \p^L \tet (z_2-\Delta)
\\ && 
- \p^K \tet (q_1-\Delta) \p^L \tet (q_1-\Delta) \p^I \tet (z_2-\Delta) \p^J \tet (z_2-\Delta) \Big )
\, e^{- 4 \pi i \kappa ' (\Omega \kappa ' - q_1-z_2+2 \Delta)}
\no
\eea
We now use the formulas (\ref{varpi}) and  (\ref{ZG}), observe that all exponential dependence on  $\kappa$ cancels
as expected, and we find,
\bea
J_8 (z_1, z_2, z_3; z_4, z_5) = \cZ_0 \, \rho_2 \, \big ( G(z_2;q_1,q_2;z_4,z_5) - G(z_2;q_1,q_2;z_1,z_3)  \big )
\label{oldJ8}
\eea
The poles of the Green functions in $z_1,z_3,z_4,z_5$ at $q_1$ and $q_2$ are cancelled by the prefactor
$\rho_2$ which vanishes there, while the poles in $z_2$ at $q_1, q_2$ are cancelled in the subtraction of the two Green functions, so that $J_8$ is indeed holomorphic in all $z_i$ as expected.  The result of (\ref{oldJ8}) will be simplified in subsection \ref{sec:J6789}.

\subsubsection{Calculation of $J_9$}

We relate $J_9$ to $J_6$ and $J_8$ by evaluating the following auxiliary quantity in two different ways,
\bea
\hat J_9 & = & \sum _\delta \cZ[\delta] 
\Big ( S_\delta(q_1,z_1) S_\delta(z_2,z_3) - S_\delta(q_1,z_3) S_\delta(z_2,z_1) \Big ) S_\delta(z_1,z_2) 
\no \\ && \hskip 0.5in \times
\Big ( S_\delta(z_3,z_4) S_\delta(z_5,q_2) - S_\delta(z_3,q_2) S_\delta(z_5,z_4) \Big ) S_\delta(z_4,z_5) 
\eea
On the one hand, by expanding both parentheses and identifying the result with the $J$-functions, we have,
\bea
\label{hatJ9}
\hat J_9 & = &
 J_9(z_1,z_2,z_3,z_4,z_5)+ J_6(z_3;z_1,z_2;z_4,z_5)  
 \no \\ &&
+ J_8(z_1,z_2,z_3;z_4,z_5)  + J_8(z_3,z_4,z_5;z_1,z_2) 
\eea 
On the other hand, by using the Fay identity (\ref{Fay1})  for each parenthesis,
and factoring out the sum over spin structures, we obtain, 
\bea
\hat J_9 = { \cZ_0 \, e^{ 4 \pi i \kappa \Omega \kappa'} \, E(q_1,q_2) E(q_1,z_2) E(z_1,z_3) E(z_3,z_5) E(z_4,q_2) \, \mJ_9
\over E(q_1,z_1) E(q_1,z_3) E(z_2,z_3) E(z_3,z_4) E(z_3,q_2) E(z_5,q_2) E(z_1,z_2)^2 E(z_4,z_5)^2}
\eea
where,
\bea
\mJ_9 = \sum_\delta \< \kappa |\delta\> 
\tet[\delta](z_1-z_2) 
\tet[\delta](z_4-z_5)
\tet [\delta] (q_1+z_2-z_1-z_3) 
\tet [\delta] (z_3+z_5-z_4-q_2)
\qquad
\eea
We calculate this spin structure sum using  the Riemann identities (\ref{Riem2}) with, 
\begin{align}
\zeta _1^+ & =  q_1-\Delta -\kappa   & \zeta ^-_1 & = q_1-\Delta -\kappa -z_1+z_2 
\no \\
\zeta _2^+ & = z_1+z_4-z_2-z_5-q_1+\Delta +\kappa    & \zeta ^-_2 & = z_4-z_5-q_1+\Delta +\kappa 
\no \\
\zeta _3^+ & =  \Delta + \kappa -z_3  & \zeta ^-_3 & = z_2-z_1-z_3+\Delta +\kappa 
\no \\
\zeta _4^+ & =  z_1+z_3+z_5-z_2-z_4 -\Delta -\kappa   & \zeta ^-_4 & = z_3+z_5-z_4-\Delta -\kappa
\end{align}
The Riemann vanishing Theorem reviewed in appendix \ref{sec:A.vanishing} implies that
$\tet[\kappa] (\zeta ^+_a)=0$ for $a=1,3$ which causes the product involving $\zeta ^+$  to vanish. Converting $\tet[\kappa]$ to $\tet$-functions without characteristics, the product becomes,
\bea
\prod _{a=1}^4 \tet [\kappa ] (\zeta _a^-) & = & 
e^{ 4 \pi i \kappa '(z_3-\Delta)}
\tet(q_1+z_2-z_1-\Delta ) \tet(q_2+z_4-z_5 -\Delta )
\no \\ && \times  \tet(z_3+z_1-z_2-\Delta ) \tet(z_3+z_5-z_4-\Delta ) 
\eea
We use (\ref{sigma2}) to express the $\tet$-functions as follows,
\bea
\tet(q_\a+z_j-z_i-\Delta ) & = & 
  (-)^\a \varpi(z_j) { E(z_i,q_\a) E(z_i,z_j) \sigma (z_i) \over E(z_j,q_\a) \sigma (z_j)} \, 
  e^{- 2 \pi i \kappa ' (q_\a-\Delta)} 
\no \\
\tet(z_j+z_k-z_i-\Delta ) & = & 
\sum_I \om _I (z_k) \p^I \tet (z_j-\Delta) { E(z_i,z_j) E(z_i,z_k) \sigma (z_i) \over E(z_j,z_k) \sigma (z_k)}
 \label{removetet}
\eea
and taking the product of all four $\tet$-functions we  use the variant of (\ref{ZG})
\begin{align}
&\sum_{I,J} \omega_I(z_1) \partial^I \tet(z_3-\Delta) \omega_J(z_5) \partial^J \tet(z_3-\Delta) \frac{ E(q_1,q_2) }{E(q_1,z_3) E(z_3,q_2) }   \notag
\\ 
& \quad
= - G(z_3;q_1,q_2;z_1,z_5) \varpi(z_1) \varpi(z_5) e^{4\pi i  \kappa'(\Delta - z_3)}
\end{align}
 to obtain,
\bea
\hat J_9 =   2 \cZ_0 \,  \rho_3  \, G(z_3;q_1,q_2;z_1,z_5)
\eea
Combining this result with (\ref{hatJ9}) and using the expressions (\ref{oldJ6})
and (\ref{oldJ8}) for $J_6$ and $J_8$, we find, 
\bea
\label{oldJ9}
 J_9(z_1,z_2,z_3,z_4,z_5)& = & \cZ_0 \, \rho_3 \, \big (
 G(z_3;q_1,q_2;z_1,z_2)   +  G(z_3;q_1,q_2;z_4,z_5) 
   \no \\ && \hskip 0.35in 
 +  G(z_3;q_1,q_2;z_1,z_5) -G(z_3;q_1,q_2;z_1,z_4)  
 \no \\ && \hskip 0.35in
 -  G(z_3;q_1,q_2;z_2,z_5)   -  G(z_3;q_1,q_2;z_2,z_4)  \big )
 \no \\ && 
+ \cZ_0 \rho_2 \big ( G(z_2;q_1,q_2;z_1,z_3) - G(z_2;q_1,q_2;z_4,z_5)  \big )
  \no \\ && 
+ \cZ_0 \rho_4 \big ( G(z_4;q_1,q_2;z_3,z_5) - G(z_4;q_1,q_2;z_1,z_2)  \big )
 \eea
The poles in $z_1, z_2, z_4, z_5$ at $q_1,q_2$ are cancelled by the prefactor $\rho_3$, while the poles in $z_3 $ at $q_1, q_2$ cancel in the sum of the six Green functions, so that $J_9$ is holomorphic in $z_1, \cdots, z_5$ as expected. The result of (\ref{oldJ9}) will be simplified in subsection \ref{sec:J6789}.

\subsubsection{Simplifications of $J_6,J_7,J_8,J_9$} 
\label{sec:J6789}

The expressions obtained in (\ref{oldJ6}), (\ref{oldJ7}), (\ref{oldJ8}) and (\ref{oldJ9}) may be simplified further by using the following relation
\bea
\label{Gq1q2}
\varpi(z) \varpi(x) \varpi(y) G(z;q_1,q_2;x,y) = c_1^2 \p \varpi (q_1) \Delta (z,x) \Delta (z,y)
\eea
To prove this formula, we compare the properties of both sides which are single-valued in $x,y,z$. The  left side is a holomorphic $(2,0)$-form in $z$ since the poles in $z$ at $q_1, q_2$ are cancelled by $\varpi (z)$, and a holomorphic $(1,0)$-form in $x$ and $y$, since their poles at $q_1,q_2$ are cancelled by $\varpi (x) \varpi(y)$. Furthermore, the left-hand side is symmetric in $x,y$ and vanishes at $z=x$  and $z=y$. Therefore it must be proportional to $\Delta(z,x)\Delta (z,y)$. The factor of proportionality must be independent of $x,y,z$ and may be determined by taking the limit $z \to q_1$.

\sm

Applying (\ref{Gq1q2}) to $J_7$ readily gives its expression in (\ref{J6789new}). In the remaining cases, namely $J_6-J_7$ and $ J_8, J_9$, the factor of the inverse of $\varpi$ cancels out using the relations of (\ref{omdel}), in agreement with the fact that those spin structure sums are holomorphic in all the variables. Their resulting simplified expressions are given in (\ref{J6789new}).

\subsubsection{Consistency checks for $J_6$ to $J_9$}
\label{sec:cross}

We note some simple checks between the defining formulas for the spin structure sums $J_6$ to $J_9$ in (\ref{J5-9}) and the expressions obtained in (\ref{J6789new}). From their definitions in (\ref{J5-9}) one observes that $J_6$ and $J_7$ take the same value as we let either $z_2 \to z_4$, or $z_3 \to z_5$, in agreement with the first formula of (\ref{J6789new}). As we let $z_1 \to z_4$ (or similarly when $z_2 \to z_5$), it is seen from the definition in (\ref{J5-9}) that $J_9$ tends to $J_5$, which vanishes and this is borne out in the last formula of (\ref{J6789new}). Finally, as we set $z_3=q_1$ or $q_2$ in the definition of $J_9$ in (\ref{J5-9}) we obtain $J_5$, in agreement with the last formula of (\ref{J6789new}).

\sm

The residue of $J_7$ as $z_1 \to q_1$ is related to $I_{12}$ by inspection of the spin structure sums,
\bea
J_7(1;2,3,4,5) \approx - { 1 \over z_1 - q_1} \, I_{12}(2,3,4,5) = { 2 \cZ_0 \rho_1 \over z_1-q_1}
\eea
The rightmost expression is produced by evaluating the pole in $J_7$ starting from (\ref{J6789new}) as well as by using (\ref{I11S}) for $I_{12}$. Further checks relate $J_8$ and $J_6$ by letting $z_3\to z_1$,  and $J_9$ to $J_7$ by letting $z_5 \to z_1$ in the expressions for these functions  given in (\ref{J6789new}),
\bea
J_8(1,2,1;4,5) & = & - J_6(1;2,1;4,5) 
= \cZ_0 c_1^2 \p \varpi(q_1) \Big [ \varpi(1) \Delta (1,5) \Delta(2,4) 
- \varpi(5) \Delta (1,2) \Delta(1,4) \Big ]
\no \\
J_9(1,2,3,4,1) & = &\phantom{+} J_7(1,2,3,4,1) 
= \cZ_0 c_1^2 \p \varpi(q_1) \Delta (1,2) \Delta (1,4) \varpi(3)
\eea
Finally, one relates $J_3$ to $J_8$ and $J_4$ to $J_9$ by letting $z_1 \to q_2$, 
\bea
\label{J3J4q}
J_3(q_2,2,3;4,5) & = & J_8(q_2,2,3;4,5) = - \cZ_0 c_2 \p \varpi(q_2) \Delta (2,3) \varpi(4) \varpi(5)
\no \\
J_4(q_2,2,3,4,5) & = & J_9(q_2,2,3,4,5) = \cZ_0 c_2 \p \varpi(q_2) \Delta (2,5) \varpi (3) \varpi(4)
\eea
Both equalities on the left follow from the definitions of $J_3, J_4$ in (\ref{J34}) and the definitions of $J_8, J_9$ in (\ref{J5-9}) in terms of spin structure sums. Both equalities on the right are obtained from their general expressions in (\ref{J6789new}) by setting $z_1=q_2$. To evaluate $J_3(q_2,2,3;4,5) $ and $J_4(q_2,2,3,4,5) $   it is convenient to use the expressions of (\ref{J34.1}). The contributions from the cyclic permutations vanish automatically, since $\rho_i(q_2)=0$ for $i=2,3,4,5$, so that only the terms proportional to $\rho_1$ remains. To evaluate those, we use (\ref{C.third}) and (\ref{B.delom}) and obtain the right side of both relations in (\ref{J3J4q}). 

\subsection{Calculation of $J_{11}$ and $J_{12}$}

To evaluate the spin structure sums in $J_{11}$ and $J_{12}$ we perform two further reductions of the number of Szeg\"o kernels in the summand on which the Riemann identities (\ref{Riem2}) can be used. To do so, we introduce the following combinations, 
\bea
R_\delta (z_1, z_2;w_1,w_2) = S_\delta (z_1,w_1) S_\delta(z_2,w_2) - S_\delta (z_1,w_2) S_\delta(z_2,w_1)
\eea
which satisfy the following symmetry properties,
\bea
R_\delta (z_2, z_1;w_1,w_2) & = & - R_\delta (z_1, z_2;w_1,w_2)
\no \\
R_\delta (z_1, z_2;w_2,w_1) & = & - R_\delta (z_1, z_2;w_1,w_2)
\no \\
R_\delta (w_1,w_2;z_1, z_2) & = & R_\delta (z_1, z_2;w_1,w_2)
\eea
On the one hand, by the Fay identity (\ref{Fay1}), it is given in terms of the following ratio of $\tet$-functions and prime forms,
\bea
\label{Rdef}
R_\delta(z_1, z_2;w_1,w_2) = - { \tet [\delta] (z_1+z_2-w_1-w_2) E(z_1, z_2) E(w_1, w_2) 
\over \tet[\delta] (0) E(z_1, w_1) E(z_1,w_2) E(z_2, w_1) E(z_2,w_2) }
\eea
On the other hand,  listing  the three possible combinations involving four points, 
\bea
R_\delta (z_1,z_2;z_3,z_4) & = & S_\delta (z_1,z_3) S_\delta (z_2,z_4) - S_\delta (z_1,z_4) S_\delta (z_2,z_3)
\no \\
R_\delta (z_1,z_3;z_4,z_2) & = & S_\delta (z_1,z_4) S_\delta (z_3,z_2) - S_\delta (z_1,z_2) S_\delta (z_3,z_4)
\no \\
R_\delta (z_1,z_4;z_2,z_3) & = & S_\delta (z_1,z_2) S_\delta (z_4,z_3) - S_\delta (z_1,z_3) S_\delta (z_4,z_2)
\eea
shows that the product of two Szeg\"o kernels may be simply expressed in terms of $R_\delta$,
\bea
\label{SSdec}
S_\delta(z_1,z_2) \, S_\delta (z_3,z_4) = \half R_\delta (z_1,z_2;z_3,z_4) - \half R_\delta (z_1,z_3;z_4,z_2) - \half R_\delta (z_1,z_4;z_2,z_3)
\quad
\eea
We shall now introduce the following spin structure sums involving the combinations $R_\delta$,
\bea
\label{E.L1L2}
L_1(w;1,2;3,4,5) & = & \sum _\delta \cZ[\delta] S_\delta (q_1,q_2) \f[\delta] (w;1,2)
R_\delta (1,3;4,5) R_\delta (2,3;4,5)
\no \\
L_2(w;1,2;3,4,5) & = & \sum _\delta \cZ[\delta] S_\delta (q_1,q_2) \f[\delta] (w;1,2)
R_\delta (1,3;4,5) R_\delta (2,4;5,3)
\eea
Applying the decomposition (\ref{SSdec}) of products of two Szeg\"o kernels into the products $S_\delta(2,3)S_\delta(4,5)$ and $S_\delta(1,3)S_\delta(4,5)$ in the summand of $J_{11}$, and to the products $S_\delta(2,3)S_\delta(4,5)$ and $S_\delta(3,4)S_\delta(5,1)$ in the summand of $J_{12}$, one verifies that  the  functions $J_{11}$ and $J_{12}$ are given by the simple linear combinations of these building blocks in  (\ref{J11-12}). It remains to evaluate $L_1$ and $L_2$ which is done in the subsequent two subsections.

\subsubsection{Evaluating $L_1$}

Substituting the expressions (\ref{varphi}) and (\ref{Rdef}) for $\f[\delta]$ and $R_\delta$, respectively, into $L_1$, we obtain, 
\bea
L_1= -{  \cZ_0  \, E(1,2) E(1,3) E(2,3) E(4,5)^2 \, \mL_1 \over E(1,w)^2 E(2,w)^2 E(1,4)E(1,5)E(2,4)E(2,5)  E(3,4)^2 E(3,5)^2}
\eea
where $\mL_1$ is given by, 
\bea
\mL_1 & = & e^{4 \pi i \kappa ' \Omega \kappa '} \sum _\delta \< \kappa |\delta\> \tet [\delta] (q_1-q_2) \tet [\delta](z_1+z_2-2w) 
\no \\ && \hskip 0.8in \times
\tet[\delta] (z_1+z_3-z_4-z_5) \tet[\delta] (z_2+z_3-z_4-z_5) 
\eea
To apply the Riemann relations (\ref{Riem2}), we use,
\begin{align}
\zeta ^\pm _1 & = \pm (q_1 -\Delta - \kappa ) + z_1+z_2+z_3-z_4-z_5-w
\no \\
\zeta ^\pm _2 & = \pm (q_1 -\Delta - \kappa ) + z_4+z_5-z_3-w
\no \\ 
\zeta ^\pm _3 & = \pm (q_1 -\Delta - \kappa ) + w-z_2
\no \\
\zeta ^\pm _4 & = \pm (q_1 -\Delta - \kappa ) +w-z_1
\end{align}
Since $-(q_1-\Delta -\kappa) = q_2 -\Delta -\kappa$, the Riemann relations give,
\bea
\mL_1 = 2 \, e^{4 \pi i \kappa ' \Omega \kappa '} \prod _{a=1}^4 \tet [\kappa] (\zeta _a^+) + ( q_1 \leftrightarrow q_2)
\eea
Using the relation,
\bea
\label{tetkappa}
\tet [\kappa] (\zeta - \kappa) = e^{- i \pi \kappa ' \Omega \kappa ' + 2 \pi i \kappa ' \zeta}\tet (\zeta)
\eea
we find after various simplifications, 
\bea
\label{cK1}
\mL_1 & = & 2 \, e^{8 \pi i \kappa ' (q_1-\Delta) } \tet (q_1 -\Delta + z_1+z_2+z_3-z_4-z_5-w) 
\no \\ && \quad \times 
\tet(q_1 -\Delta + z_4+z_5-z_3-w)  \tet (q_1 -\Delta  +w-z_2)
\no \\ && \quad \times 
 \tet (q_1 -\Delta  +w-z_1) + ( q_1 \leftrightarrow q_2)
\eea
Expressing the $\tet$-functions  in terms first of the Green functions $\cG_3$ and $\cG_2$ discussed in appendix \ref{app.bc}, we recover the result for $L_1$ in (\ref{K2fin}). More specifically, intermediate steps are based
on (\ref{defcgn}), (\ref{varpi3}) and (\ref{removetet}).

\subsubsection{Evaluating $L_2$}

Substituting the expressions (\ref{varphi}) and (\ref{Rdef}) for $\f[\delta]$ and $R_\delta$, respectively, into $K_2$, we obtain, 
\bea
L_2 (w;1,2;3,4,5) = -  { \cZ_0 \, E(1,2) E(1,3) E(2,4)  \, \mL_2 \over E(1,w)^2 E(2,w)^2 E(1,4)E(1,5)E(2,3)E(2,5) E(3,4)^2 }
\eea
where $\mL_2$ is given by, 
\bea
\mL_2 & = & e^{4 \pi i \kappa ' \Omega \kappa '} \sum _\delta \< \kappa |\delta\> \tet [\delta] (q_1-q_2) \tet [\delta](z_1+z_2-2w) 
\no \\ && \hskip 0.8in \times
\tet[\delta] (z_1+z_3-z_4-z_5) \tet[\delta] (z_2+z_4-z_3-z_5) 
\eea
To apply the Riemann relations (\ref{Riem2}), we use,
\begin{align}
\zeta ^\pm _1 & = \pm (q_1 -\Delta - \kappa ) + z_1+z_2-z_5-w
\no \\
\zeta ^\pm _2 & = \pm (q_1 -\Delta - \kappa ) + z_5-w
\no \\ 
\zeta ^\pm _3 & = \pm (q_1 -\Delta - \kappa ) + w+z_3 -z_2- z_4
\no \\
\zeta ^\pm _4 & = \pm (q_1 -\Delta - \kappa ) +w+z_4 -z_1-z_3
\end{align}
Since $-(q_1-\Delta -\kappa) = q_2 -\Delta -\kappa$, the Riemann relations give,
\bea
\mL_2 = 2 \, e^{4 \pi i \kappa ' \Omega \kappa '} \prod _{a=1}^4 \tet [\kappa] (\zeta _a^+) + ( q_1 \leftrightarrow q_2)
\eea
Using the relation (\ref{tetkappa}) we find, 
\bea
\mL_2 & = & 2 \, e^{8 \pi i \kappa ' (q_1-\Delta) } \, \tet (q_1 -\Delta  +z_5-w) \, \tet (q_1 - \Delta +z_1 + z_2-z_5-w) 
\no \\ && \times 
\tet(q_1 -\Delta + z_3 + w -z_2-z_4) \, \tet (q_1 -\Delta + z_4 + w -z_1-z_3) 
+ ( q_1 \leftrightarrow q_2)
\qquad
\eea
Expressing this combination in terms of the scalar Green function $G$ in (\ref{Green}) we recover the expression for $L_2$ given in (\ref{K2fin}) using intermediate steps based on (\ref{remove2tet}) and (\ref{C.Ginter}).

\subsubsection{Consistency checks for $L_1$ and $L_2$}

By comparing the defining relations for $L_1$ and $L_2$ in (\ref{E.L1L2}) with the evaluations given in (\ref{K2fin}) in various limits we obtain consistency checks on our results, especially regarding the overall signs. We shall use the facts that $\f[\delta] (w;z_1,z_2)$ tends to zero as $z_2 \to z_1$; $R_\delta(z_1, z_2;w_1,w_2)$ tends to zero as $z_2 \to z_1$ and as $w_2 \to w_2$; and that $R_\delta(z_1, z_2;w_1,w_2)$ has a simple pole as $w_i \to z_j$ for $i,j=1,2$, whose residue is given by,
\bea
R_\delta (z_1, z_2;w_1,w_2) \approx { 1 \over w_1-z_2} S_\delta(z_1,w_2)
\eea
and permutations thereof. 

\sm

From these observations and its definition in (\ref{E.L1L2}), it is clear that $L_1\to 0$ as $z_1\to z_2$; as $z_1 \to z_3$; as $z_2 \to z_3$; and as $z_4 \to z_5$, and these results are indeed borne out by the corresponding limits of the explicit solutions given in (\ref{K2fin}). Furthermore, $L_1$ has a double pole as $z_4 \to z_3$, whose residue may be read off from inserting the limit of both factors $R_\delta$ into the definition of $L_1$ in (\ref{E.L1L2}), using the fact that the resulting sum is given by $- I_{14}(w;1,2;5)$, and then using the evaluation of $I_{14}$ given in (\ref{I13}),
\bea
L_1(w;1,2;3,4,5) \Big |_{{\small (\ref{E.L1L2})}}  \approx - { 2 \cZ_0 \varpi(1) \varpi(2) \over (z_3-z_4)^2}   \varpi(w)^2 G(5;1,2;q_1,w) + ( q_1 \leftrightarrow q_2)
\eea
On the other hand, the evaluation of the same limit from (\ref{K2fin}) gives, 
\bea
L_1(w;1,2;3,4,5) \Big |_{{\small (\ref{K2fin})}}  \approx  { 2 \cZ_0 \varpi(5) \varpi(2)  \over (z_3-z_4)^2}   \varpi(w)^2  G(1;5,w;2,q_1) + ( q_1 \leftrightarrow q_2)
\eea
The overall sign of $L_1$ may be checked by comparing the residues of these formulas at the pole in $z_1-z_5$, 
which are both given by $4 \cZ_0 \varpi(1) \varpi(2) \varpi(w)^2$ and thus agree. Actually, the entire residues agree in view of formula (\ref{C.Ginter}).

\sm

Similarly, it is clear from its definition in (\ref{E.L1L2}) that $L_2\to 0$ as $z_1\to z_2$; as $z_1 \to z_3$; and as $z_2 \to z_4$, all of which are borne out by the corresponding limits of (\ref{K2fin}). Furthermore, $L_2$ has a double pole in $z_3-z_4$, and we have, 
\bea
L_2(w;1,2;3,4,5) \Big |_{{\small (\ref{E.L1L2})}}  \approx 
- { 2 \cZ_0 \varpi(1) \varpi(2) \over (z_3-z_4)^2}   \varpi(w)^2 G(5;1,2;q_1,w) + ( q_1 \leftrightarrow q_2)
\eea
On the other hand, the evaluation of the same limit from (\ref{K2fin}) gives, 
\bea
L_2(w;1,2;3,4,5) \Big |_{{\small (\ref{K2fin})}}  \approx  - { 2 \cZ_0 \varpi(1) \varpi(2)  \over (z_3-z_4)^2}   \varpi(w)^2  G(5;1,2;q_2,w) + ( q_1 \leftrightarrow q_2)
\eea
These expressions manifestly agree with one another.

\newpage

%%%%%%%%%%%%%%%%%%%%%%%%%%%%%%%%%%%%%%%%%%%
%%%%%%%%%%%%%%%%%%%%%%%%%%%%%%%%%%%%%%%%%%%
\section{$\Lambda (q_\a)$-dependence of pairing $J_{10}, J_{11},J_{12}$ against $\mu$}
\setcounter{equation}{0}
\label{sec:F}
%%%%%%%%%%%%%%%%%%%%%%%%%%%%%%%%%%%%%%%%%%%
%%%%%%%%%%%%%%%%%%%%%%%%%%%%%%%%%%%%%%%%%%%

In this appendix, we shall derive the pairing integrals of the spin structure sums $J_{10}, J_{11}, J_{12}$ 
in (\ref{J10-12}) against the Beltrami differential $\mu$ and denote the resulting integrals by $\cJ_a$ for $a=10,11,12$, given in (\ref{calJdef}). The results will be expressed in terms of the function $\Lambda$ introduced in (\ref{4.mu2}) and its derivative. Two different types of contributions arise, namely those  which depend on $\Lambda (z_i)$ and $\p \Lambda (z_i)$ and those which involve only $\Lambda (q_\a)$. The contributions involving $\Lambda (z_i)$ and $\p \Lambda (z_i)$ were already shown to cancel against similar contributions from the stress tensor of the bosonic fields in section \ref{sec:44} and do not need to be retained here. 
Instead, we will only evaluate the contributions involving $\Lambda (q_\a)$ as will be reflected by
superscripts $^{(q)}$ in the notation. Throughout, we shall make heavy use of the  formulas given in the compendium (\ref{fund1}) and (\ref{fund3}), and originally established in \cite{DP6}.

\subsection{Integral of $J_{10}$ against the Beltrami differential}
\label{appF.J10}

Using (\ref{J10}), the fact that the integral of $I_{13}$ against $\mu$ vanishes, and retaining only the dependence on $\Lambda (q_\a)$, we readily have,
\bea
\cJ_{10}^{(q)} (1,2;3,4,5) =  - { x_{12} \, dx_3 \over 2 x_{23} x_{31}} \, \cI_{16}^{(q)} (4,5;1,2) +  {\rm cycl}(1,2,3)
\eea
The expression for $\cI_{16}^{(q)}$  is obtained from formulas (8.8), (8.12) and  (8.14) of \cite{DP6}, 
\bea
\cI_{16}^{(q)} (4,5;1,2) = { \zeta ^1 \zeta ^2 \over 8 \pi^2} \Big (\Delta (1,4) \Delta (2,5) + \Delta (1,5) \Delta (2,4) \Big )
\eea
Hence we have,
\bea
\cJ_{10}^{(q)} (1,2;3,4,5) =  - { \zeta ^1 \zeta ^2 \over 16 \pi^2}  { x_{12} \, dx_3 \over  x_{23} x_{31}} \,
\Big (\Delta (1,4) \Delta (2,5) + \Delta (1,5) \Delta (2,4) \Big ) + {\rm cycl}(1,2,3)
\quad
\eea
This is proportional to the expression (\ref{J2.1}) for $J_1$,
\bea
\cJ_{10}^{(q)} (1,2,3;4,5) =  { \zeta ^1 \zeta ^2 \over 32 \pi^6 \Psi _{10}} \, J_1(1,2,3;4,5)
\eea
Recasting $J_1$ in terms of $G^I$-functions as in (\ref{J12.6}),  we find, 
\bea
\cJ_{10}^{(q)} (1,2,3;4,5) & = &  -{ \zeta ^1 \zeta ^2 \over 16 \pi^4} \sum_I
\Big (  \om_I(1)  \Delta (2,4) \Delta (3,5)  G^I_{1,3,5,4,2} + (4 \leftrightarrow 5) \Big )
 +{\rm cycl}(1,2,3) 
 \quad
\eea

\subsection{Integrals of $L_1$ and $L_2$ against $\mu$}

The functions $J_{11}$ and $J_{12}$ are given in (\ref{J11-12}) as linear combinations with constant coefficients  of the building blocks $L_1$ and $L_2$ with simplified expressions in (\ref{K2fin}). Thus, to evaluate the integrals of $J_{11}$ and $J_{12}$ against the Beltrami differential $\mu$, it suffices to compute the integrals of $L_1,L_2$ against $\mu$, 
\bea
\cL_a (1,2;3,4,5) = { 1 \over 2 \pi} \int _\Sigma \mu(w) L_a(w;1,2;3,4,5) 
\eea
Since we have already dealt with the contributions involving $\Lambda$ at the vertex points, we shall retain here only the contributions involving $\Lambda(q_\a)$ and denote those by $\cL^{(q)}_a$. Expressing $\mu$ in terms of $\Lambda$, 
using the choice of additive constant by the vanishing of the sum $\Lambda (q_1) + \Lambda(q_2)=0$ as given in (\ref{Lamplus}), and the expression for the difference in (\ref{Lamminus}), we obtain the following simplified expression, 
\bea
\cL_a^{(q)}  (1,2;3,4,5) =  {\zeta ^1 \zeta ^2 \over 16 \pi^2c_1 c_2}  \sum _{\a=1,2} c_\a^2 L_a (q_\a; 1,2,3,4,5)
\eea
In carrying out the above simplifications, we have assumed that the limit of $L_a(w;1,2,3,4,5)$ as $w \to q_\a$ exists, but we shall show by explicit calculation below that this is indeed the case for both $L_1$ and $L_2$.

\subsubsection{Evaluating $\cL_1^{(q)}$} 

To evaluate $\cL_1^{(q)}$ we use (\ref{K2fin}) for $L_1$ and the behavior of $\cG_2$ and $\cG_3$ near these poles in $w$ at $q_\a$, which may be read off from their representation in terms of the $b,c$ system, see (\ref{polebc}),
\bea
\cG_2 (4,5,q_1;3,w)  \sim  { \Delta (4,5) \over w-q_1} 
\hskip 0.8in
\cG_3(1,2,3,q_1;4,5,w)  \sim   - { \cG_2(1,2,3;4,5) \over w-q_1} 
\eea
Since $\varpi(w) =(w-q_\a) \partial \varpi(q_\a)+{\cal O}((w-q_\a)^2)$ has a simple  zero in $w$ at $q_\a$,  $L_1$ has a smooth limit as $w\to q_\a$, and (\ref{fund1}) implies that, 
\bea
\cL_1 ^{(q)}(1,2;3,4,5) = { \zeta ^1 \zeta ^2 \over 4 \pi^2} \, \cG_2(1,2,3;4,5) \Delta (4,5)
\eea
This expression is nicely independent of $q_1, q_2$.  Finally, we may use (\ref{cG2b}) to express $\cG_2$ in terms of $\tau$, and then express that result in terms of the functions $G^I$,
\bea
\cL_1 ^{(q)}(1,2;3,4,5) = { \zeta ^1 \zeta ^2 \over 4 \pi^2} \sum_I \om _I(1) \Delta (2,3)  \Delta (4,5) G^I_{1,4,5} + \hbox{ cycl}(1,2,3)
\label{finalcL1}
\eea

\subsubsection{Evaluating $\cL_2^{(q)}$}

To evaluate $\cL_2^{(q)}$, we need the limit of $L_2(w;1,2,3,4,5)$ as $w \to q_\a$. To this end, it will be useful to recast the Green functions $G$ in $L_2$ of (\ref{K2fin}) in terms of the Green function $\cG_2$, and simplify the extra factors with the prefactor $\varpi(w)^2$ in $L_2$ to obtain,
\bea
L_2 (w;1,2;3,4,5) & = & 
2  c_1c_2 \cZ_0{ \varpi(1) \varpi(2) \over \varpi(w)} \Big [ c_1\cG_2(3,q_1,w;2,4) \cG_2(4,q_1,w;1,3)   \cG_2(5,q_2,w;1,2)  
\no \\ && \hskip 0.5in
\ \ \ + c_2  \cG_2(3,q_2,w;2,4) \cG_2(4,q_2,w;1,3)   \cG_2(5,q_1,w;1,2)  \Big ]
\eea
As $w \to q_1$, both terms have a simple pole in $w$ from the prefactor $\varpi(w)^{-1}$. The first term has a double zero from the first two Green functions while the third Green function is regular, so that the first term vanishes. The first two Green functions in the second term are regular as $w \to q_1$ while the third Green function, \bea
\cG_2(5,q_2,w;1,2) = \frac{w-q_2}{c_2} \partial \varpi(q_2) G(5;1,2;q_2,q_2) + {\cal O} \left ((w-q_2)^2 \right )
\eea 
has a simple zero which cancels the pole in the prefactor. As $w \to q_2$, the roles of the terms are reversed. Therefore, $L_2$ has a smooth limit as $w \to q_\a$ which, after reconverting the third Green function $\cG_2$ back to $G$, is given by,
\bea
L_2(q_\a;1,2,3,4,5) & = &
2 {c_1^2 c_2^2 \over c_\a^2} \cZ_0 \varpi (1) \varpi(2) \cG_2(3,q_1,q_2;2,4) 
\no \\ && \qquad 
\times \cG_2(4,q_1,q_2;1,3)  G(5;1,2;q_\a, q_\a)
\eea
This expression may be simplified considerably by using the following identity, 
\bea
\label{GGid}
\varpi(1) \varpi(2) \, \cG_2(3,q_1,q_2;2,4) \, \cG_2(4,q_1,q_2;1,3) 
=  -  \p \varpi(q_1)  \p \varpi(q_2)  \Delta (1,3)  \Delta (2,4) 
\eea
Both sides are single-valued holomorphic $(1,0)$-forms in $z_1, z_2, z_3, z_4$ since the poles in all these points at $q_1, q_2$ cancel on the left side, namely the pole in $z_3$ at $z_2$ cancels by (\ref{cG1}) and similarly for the pole in $z_3$ at $z_4$. To evaluate this combination in terms of more standard objects, we use the expression for $\cG_2$ in terms of $\tau$ of (\ref{cG2b}) in the following form,
\bea
\cG_2(z,q_1,q_2;x,y) ={ \varpi(z)  \over c_1 c_2} \big( c_1 \tau_{x,y}(q_1) -c_2 \tau_{x,y}(q_2)  \big)
\eea
Using formula (\ref{fund3}) to simplify this expression and taking the product of the two copies of $\cG_2$ on the left side of (\ref{GGid}) then establishes  the formula. Using (\ref{GGid}), the expression for $L_2$ simplifies, and we have,
\bea
L_2(q_\a;1,2,3,4,5) =
- 2 { c_1 c_2 \over c_\a^2} \Delta (1,3)  \Delta (2,4)  G(5;1,2;q_\a, q_\a)
\eea
so that,
\bea
\cL_2^{(q)}(1,2,3,4,5) = - { \zeta ^1 \zeta ^2 \over 8 \pi^2} \Delta (1,3)  \Delta (2,4)  \sum _\a G(5;1,2;q_\a, q_\a)
\eea

\subsubsection{Simplifying $\cL_2^{(q)}$} 

The poles in $z_5$ at $z_1$ and $z_2$ are independent of $q_\a$, while the poles in $z_1$ and $z_2$ at $q_\a$ are holomorphic in $z_5$, and may be isolated as follows,
\bea
\half \sum_\a G(5;1,2;q_\a,q_\a) & = & \tau_{1,2} (5) -  \sum_I \om_I(5) ( B_1 ^I - B_2^I) 
\eea
where the functions $B_i^I$ appeared in the discussion of
section \ref{sec4.7} and are defined as follows, 
\bea
\label{defBI}
B^I_i  & = & 
 \sum _{\a,J} { c_\a \ep ^{IJ}  \over 2 \, \p \varpi (q_\a)} 
\Big (   \p_{q_\a} \om_J(q_\a) \tau_{z_i,z_0} (q_\a) - \om_J(q_\a)  \p_{q_\a} \tau_{z_i,z_0} (q_\a)  \Big )
\eea
Here $z_0$ is an arbitrary reference point which has been introduced via $\tau_{1,2}(q_\alpha) = \tau_{1,z_0}(q_\alpha)-\tau_{2,z_0}(q_\alpha)$ and cancels out of the differences $B_i^I-B_j^I$. The functions $B_i^I$ are single-valued $(0,0)$-forms in the points $q_\a$ with double poles in $z_i$ at $q_1$ and $q_2$. Neither $\tau_{1,2}(5)$ nor $B^I_1-B^I_2$ is single-valued in $z_1,z_2$, but their combination above is single-valued. The monodromy of $B_j^I$ under $z_j \to z_j +\mB_K$  is given by 
\bea
B_j^I \to B_j^I +  2 \pi i \delta _K^I 
\eea
which guarantees that $\cL_2^{(q)} (1,2,3,4,5) $ is single-valued, and takes the following form,
\bea
\cL_2^{(q)}(1,2,3,4,5) =  { \zeta ^1 \zeta ^2 \over 4 \pi^2} \Delta (1,3)  \Delta (2,4) \sum_I \om_I(5) 
\Big ( g^I_{1,5}+g^I_{5,2}  - B_2^I  + B_1 ^I \Big )
\label{finalcL2}
\eea

\subsection{Evaluating $\cJ_{11}^{(q)}$ and $\cJ_{11}^S$}

The expression for $\cJ_{11}$ in terms of $\cL_1$ and $\cL_2$ restricted to only the contributions from $\cL_1^{(q)}$ and $\cL_2^{(q)}$ may be read off from (\ref{J11-12}) and  is given by,
\bea
\cJ_{11} ^{(q)}  (1,2,3;4,5) & = & { 1 \over 4} \Big [  \, 
- \cL_1^{(q)} (1,2;3,4,5)  - \cL_1^{(q)} (1,2;4,5,3)  - \cL_1^{(q)} (1,2;5,3,4) 
\no \\ && \hskip 0.15in 
+ \cL_2^{(q)} (1,2;3,4,5)+ \cL_2 ^{(q)} (1,2;3,5,4) - \cL_2 ^{(q)} (1,2;4,5,3)
\no \\ && \hskip 0.15in
+ \cL_2 ^{(q)} (1,2;4,3,5) + \cL_2 ^{(q)} (1,2;5,3,4) - \cL_2 ^{(q)} (1,2;5,4,3) 
\Big ]
\eea
Substituting the corresponding expressions (\ref{finalcL1}) and (\ref{finalcL2}) for $\cL_1^{(q)}$ and $\cL_2^{(q)}$,
\bea
\cL_1^{(q)} (1,2;3,4,5) & = &
{ \zeta ^1 \zeta ^2 \over 4 \pi^2} \, \cG_2(1,2,3;4,5) \Delta (4,5)
\no \\ &=&
{\zeta ^1 \zeta ^2 \over 4 \pi^2} \sum_I \om_I(1) \Delta(2,3) \Delta(4,5) (g^I_{1,4}-g^I_{1,5} ) + \hbox{ cycl}(1,2,3)
\no \\
\cL_2^{(q)} (1,2;3,4,5) & = &  { \zeta ^1 \zeta^2 \over 4 \pi^2}  \sum_I \om_I(5) \Delta (1,3) \Delta (2,4) ( g^I_{1,5} - g^I_{2,5} +B_1^I-B_2^I)
\eea
The definition of $B_a^I$ may be found in (\ref{defBI}) respectively. Working out the combinatorics for the cyclically symmetrized version ${\cal J}_{11}^S$ in (\ref{symmJ11}),  we find, 
\bea
\label{calJ11}
\cJ_{11} ^S (1,2,3;4,5) & = & 
 {\zeta ^1 \zeta ^2 \over 8 \pi^2} \bigg \{ 
\sum_I  \om_I(1)  \Delta (2,4) \Delta (3,5) 
\Big ( g_{1,2}^I + 2 g_{2,4}^I + 2 g^I_{4,5} +  2 g_{5,3}^I + g^I_{3,1} 
\no \\ && \hskip 0.8in 
+B_2^I - B_3^I \Big ) + (4 \leftrightarrow 5)  \bigg \}
+ \hbox{ cycl}(1,2,3)
\eea
The result may be related to the final expression (\ref{J12.6}) for $J_1$ as follows,
\bea
\cJ_{11} ^S (1,2,3;4,5) & = &
 { \zeta ^1 \zeta ^2 \over 8 \pi^2} \sum_I  \om_I(1) \Delta (2,4) \Delta (3,5) \Big ( g_{2,4}^I + g^I_{4,5} + g^I_{5,3} + B_2^I - B_3^I \Big ) +  {\rm cycl}(1,2,3)
\no \\ &&
+ { \zeta ^1 \zeta ^2 \over 8 \pi^2} \sum_I \om_I(1) \Delta (2,5) \Delta (3,4) \Big ( g_{2,5}^I + g^I_{5,4} + g^I_{4,3} + B_2^I - B_3^I \Big ) + {\rm cycl}(1,2,3)
\no \\ &&
+ { \zeta ^1 \zeta ^2 \over 16 \pi^6 \Psi_{10}} J_1(1,2,3;4,5) 
\eea

\subsection{Evaluating $\cJ_{12}^{(q)}$ and $\cJ_{12}^S$}

The expression for $\cJ_{12}$ in terms of $\cL_1$ and $\cL_2$ restricted to only the contributions from $\cL_1^{(q)}$ and $\cL_2^{(q)}$ may be read off from (\ref{J11-12}) and is given by,
\bea
\cJ_{12} ^{(q)}  (1,2,3,4,5) & = &  { 1 \over 4} \Big [  \, 
\cL_1^{(q)} (1,2;3,4,5)  - \cL_1^{(q)} (1,2;4,5,3)  + \cL_1^{(q)} (1,2;5,3,4) 
\no \\ && \hskip 0.2in 
- \cL_2^{(q)} (1,2;3,4,5)- \cL_2 ^{(q)} (1,2;3,5,4) + \cL_2 ^{(q)} (1,2;4,3,5)
\no \\ && \hskip 0.2in
- \cL_2 ^{(q)} (1,2;4,5,3) - \cL_2 ^{(q)} (1,2;5,3,4) + \cL_2 ^{(q)} (1,2;5,4,3) 
\Big ]
\eea
We shall now substitute the expressions (\ref{finalcL1}) and (\ref{finalcL2}) for $\cL_1^{(q)}$ and $\cL_2^{(q)}$ into the cyclically symmetrized combination $\cJ_{12}^S$ in (\ref{symmJ11}) that actually enters the calculation of the amplitude. Converting this result into the functions $g^I_{a,b}$, and decomposing onto the canonical cyclic basis of five-fold holomorphic $(1,0)$-forms, we find the following simplified result,
\bea
\cJ_{12} ^S (1,2,3,4,5) & = &
{ \zeta ^1 \zeta ^2 \over 8 \pi^2} \sum_I \om_I(1) \Delta (2,3) \Delta (4,5) \Big ( 
2 g^I_{5,1} +  2 g^I_{1,2} + g_{2,3}^I + 3 g^I_{3,1} + 3 g^I_{1,4} +  g^I_{4,5} 
 \no \\ && \hskip 0.5in
 -B_2^I+2B_3^I-2B_4^I +B_5^I \Big ) + \hbox{ cycl}(1,2,3,4,5)
\eea
Part of $\cJ_{12}^S$ may be expressed in terms of $J_2$ given by (\ref{altJtwo}),
\bea
\cJ_{12}^S(1,2,3,4,5)
& = & { \zeta^1 \zeta ^2 \over 8 \pi^6 \Psi_{10}} J_2(1,2,3,4,5)
\\ &&
- { \zeta ^1 \zeta ^2 \over 8 \pi^2} \sum_I  \om_I(1) \Delta (2,3) \Delta (4,5)
\Big (  g_{2,3}^I +  g^I_{4,5} +   g^I_{1,3} +  g^I_{4,1} 
\no \\ && \hskip 0.7in 
 +B_2^I-2B_3^I+2B_4^I -B_5^I \Big ) 
 + \hbox{ cycl}(1,2,3,4,5)
 \no
\eea
where the cyclic permutations are taken only of the last two lines. The poles of the terms in $g^I_{1,3}$ and $g^I_{4,1}$ are cancelled upon cyclic permutation, and one may write a manifestly pole-free expression, as follows,
\bea
\cJ_{12}^S(1,2,3,4,5)
& = & { \zeta^1 \zeta ^2 \over 8 \pi^6 \Psi_{10}} J_2(1,2,3,4,5)
\\ &&
- { \zeta ^1 \zeta ^2 \over 8 \pi^2}  \sum_I  \om_I(1) \Delta (2,3) \Delta (4,5)
\Big (  g_{2,3}^I +  g^I_{4,5}  +B_2^I-B_3^I+B_4^I -B_5^I \Big ) 
\no \\ &&
- { \zeta ^1 \zeta ^2 \over 8 \pi^2}  \sum_I  \om_I(1) \Delta (2,5) \Delta (3,4)
\Big (    g^I_{2,5} +B_2^I-B_5^I \Big ) 
 + \hbox{ cycl}(1,2,3,4,5)
 \no
\eea
where the cyclic permutations are applied only to the last two lines.

\newpage

%%%%%%%%%%%%%%%%%%%%%%%%%%%%%%%%%%%%%%%%%%%
%%%%%%%%%%%%%%%%%%%%%%%%%%%%%%%%%%%%%%%%%%%
\section{The vanishing of $\tilde \cC_1$ and $\tilde \cC_2$}
\setcounter{equation}{0}
\label{sec:H}
%%%%%%%%%%%%%%%%%%%%%%%%%%%%%%%%%%%%%%%%%%%
%%%%%%%%%%%%%%%%%%%%%%%%%%%%%%%%%%%%%%%%%%%

We shall now demonstrate the vanishing of the differential forms $\tilde \cC_1$ and $\tilde \cC_2$ in (\ref{deftildeC}).  This result will establish another major cancellation within the chiral five-point amplitude among the contributions in (\ref{majorcan}) indicated by the blue arrows in figure \ref{fig:1}. It will be useful to re-express these combinations as follows, 
\bea
\tilde \cC_1(i,j,k;\ell,m) & = & 
\sum_I \om_I(i) \Delta(j,\ell) \Delta(k,m) \Big ( g_{j,\ell}^I + g_{\ell,m}^I + g_{m,k}^I \Big ) 
\no \\ && \quad 
- \Delta (i,k) \mH(j,\ell,m) + {\rm cycl}(i,j,k)
\no \\
\tilde \cC_2(i,j,k,\ell,m) & = & 
\sum_I \om_I(i) \Delta(j,k) \Delta(\ell,m) \Big ( g_{j,k}^I + g_{\ell,m}^I + g_{i,k}^I + g^I_{\ell,i} \Big ) 
\no \\ && \quad 
- \Delta (i,k) \mH(j,\ell,m) + {\rm cycl}(i,j,k,\ell,m)
\eea
The instructions to add cyclic permutations apply to all terms in each expression. The form $\mH(j,\ell,m)$ contains all the double poles of the vertex points at $q_\a$, and is given by,
\bea
\label{6.H2}
\mH(j,\ell,m) =  \sum_I \Big ( \om_I(\ell) \Delta (j,m) + \om_I(m) \Delta (j,\ell) \Big )B_j^I 
+ \varpi(\ell) \varpi(m) \sum_\a { \p_{z_j} \p_{q_\a} \ln E(z_j,q_\a) \over c_\a \p \varpi (q_\a)}
\quad
\eea

\subsection{Properties of the differential forms $\tilde \cC_1$ and $\tilde \cC_2$}

The differential forms $\tilde \cC_1$ and $\tilde \cC_2$ obey the following properties,
\begin{enumerate}
\itemsep=-0.05in
\item $\tilde \cC_1$ and $\tilde \cC_2$ are  single-valued $(1,0)$-forms in each vertex point $z_i$;
\item $\tilde \cC_1$ and $\tilde \cC_2$ are single-valued $(0,0)$-forms in $q_\a$;
\item $\tilde \cC_1(i,j,k;\ell,m)$ is invariant under cyclic permutations of $i,j,k$ and $\ell,m$;
\item $\tilde \cC_1(i,j,k;\ell,m)$ is odd under the interchange of $i,j$;
\item $\tilde \cC_2(i,j,k,\ell,m)$ is invariant under cyclic permutations of $i,j,k,\ell,m$;
\item $\tilde \cC_2(i,j,k,\ell,m)$ is odd under reversal $(i,j,k,\ell,m) \to (m,\ell,k,j,i)$;
\item $\tilde \cC_1$ and $\tilde \cC_2$ are holomorphic in each $z_i$;
\item as a consequence of the previous items, $\tilde \cC_1 = \tilde \cC_2=0$.
\end{enumerate}

Single-valuedness in items 1 and 2 was achieved by construction. The symmetry properties in items 3 to 6 may be shown  from the explicit formulas for $\tilde \cC_1$ and $\tilde \cC_2$ in (\ref{deftildeC}) and the transformation properties of $g_{a,b}^I$ and $B_i^I$. Next, we shall show prove item 7 that $\tilde \cC_1$ and $\tilde \cC_2$ are holomorphic in the vertex points $z_j$, namely that there are no singularities as two vertex points collide, and there are no singularities as $z_j \to q_\a$.

\subsection{Holomorphicity of $\tilde \cC_1$ and $\tilde \cC_2$}

To show that $\tilde \cC_1$ has no singularities at coincident vertex points, we use the fact that $\mH(j,\ell,m)$ has no such singularities, that the contribution from $g^I_{\ell,m}$ to $\tilde \cC_1$  cancels out, and that the poles 
in $g_{j,\ell}^I + g_{m,k}^I$ are cancelled by the prefactor $\Delta(j,\ell) \Delta(k,m)$. Similarly, the poles in $\tilde \cC_2$ arising from  $g_{j,k}^I + g_{\ell,m}^I$ are cancelled by the prefactor $\Delta(j,k) \Delta(\ell,m)$, while those of 
$g_{i,k}^I + g^I_{\ell,i}$ are cancelled upon adding the cyclic permutations. This leaves only the poles of $\mH(j,\ell,m)$ and their cyclic permutations as $z_j \to q_\a$. 

\sm

To show the absence of the remaining singularities in $\tilde \cC_1$ and  $\tilde \cC_2$  we use the following formulas to extract the double and simple poles as $z_i \to q_\a$,
\bea
\p_{q_\a} \ln E(z_j,q_\a) & = & - { 1 \over z_j-q_\a} + \hbox{reg}
\no \\
\p_{z_j} \p_{q_\a} \ln E(z_j,q_\a) & = & { 1 \over (z_j-q_\a)^2} + \hbox{reg}
\no \\
\p_{q_\a}^2 \ln E(z_j,q_\a) & = & - { 1 \over (z_j-q_\a)^2} + \hbox{reg}
\eea
The terms in $\tilde \cC_1$ and $\tilde \cC_2$ with poles in $z_j$ at $q_\a$ are proportional to $\mH(j,\ell,m)$ defined in (\ref{6.H2}). Using the expression for $B_j^I$ we obtain the forms of the double and simple poles,
\bea
\sum_I \om_I(a) B_j^I & = & 
- \sum_\a { \varpi(a) \over 2  \p \varpi(q_\a) (z_j-q_\a)^2}
- \sum _{\a,I,J} { c_\a \om_I(a) \ep ^{IJ} \p_{q_\a} \om _J(q_\a) \over 2 \p \varpi(q_\a) (z_j-q_\a)} + \hbox{reg}
\eea
The double pole in $z_j $ at $q_\a$ cancels  in view of the identity $c_\a \Delta (q_\a,a) = \varpi (a)$, while  the simple pole in $z_j$ at $q_\a$ is given by, 
\bea
\mH(j,\ell,m) & = & 
{ -1 \over z_j-q_\a} 
\sum _\a \bigg (  { \varpi(\ell) \p_{q_\a} \Delta (q_\a,m) + \varpi(m) \p_{q_\a} \Delta (q_\a,\ell) \over 2 \p \varpi (q_\a)} 
\no \\ && \hskip 0.8in 
+ \sum_{I,J} { c_\a \big (\varpi(\ell) \om_I(m) + \varpi(m) \om_I(\ell) \big ) \ep ^{IJ} \p_{q_\a} \om _J(q_\a) \over 2 \p \varpi(q_\a) } \bigg ) + \hbox{reg}
\eea
also cancels in view of $c_\a \Delta (q_\a,\ell) = \varpi (\ell)$. In summary, we have the following Lemma.

\subsection{The vanishing of $\tilde \cC_1$ and $\tilde \cC_2$}

To prove item 8 we shall begin by combining the implications of the properties established in items 1 to 7.  Since $\tilde \cC_1(i,j,k;\ell,m)$ is a single-valued holomorphic $(1,0)$-form in each vertex point $z_i$, it may be expressed in the basis of holomorphic $(1,0)$-forms $\om_I(z_i)$ for each $z_i$, 
\bea
\tilde \cC_1 (i,j,k;\ell,m) = \sum_{I,J,K,L,M} \om_I(i) \om_J(j) \om_K(k) \om_L(\ell) \om_M(m) \, \cC_1 ^{I,J,K;L,M}
\eea
The modular tensor  $\cC_1^{I,J,K;L,M}$ is a single-valued scalar in $q_\a$ in view of item 2, and independent of all $z_i$ in view of item 7. It has cyclic symmetry in $I,J,K$ and $L,M$ in view of item 3, and is odd under swapping $I,J$ in view of item 4. Because of the last property, its components may be parametrized in terms of a rank three tensor $A^{K;L,M}$ as follows,
\bea
\cC_1 ^{I,J,K;L,M} = \ep^{IJ} A^{K;L,M} + \ep^{JK} A^{I;L,M} +\ep^{KI} A^{J;L,M} 
\eea
As a result, we have, 
\bea
\tilde \cC_1 (i,j,k;\ell,m) = \sum_{K,L,M}  \om_L(\ell)  \om_M(m) \, A^{K;L,M} \Big ( 
\Delta(i,j)  \om_K(k)  +{\rm cycl}(i,j,k)  \Big ) 
\eea
but this combination vanishes by the fundamental identity (\ref{omdel}) of the bi-holomorphic form $\Delta$. This completes the proof of item 8 for $\tilde \cC_1$.

\sm

Since $\tilde \cC_2(i,j,k,\ell,m)$ is a single-valued holomorphic $(1,0)$-form in each vertex point $z_i$ in view of item 1 it may be expressed in the basis $\om_I(z_i)$ for each $z_i$, so that we have,
\bea
\tilde \cC_2 (i,j,k,\ell,m) = \sum_{I,J,K,L,M} \om_I(i) \om_J(j) \om_K(k) \om_L(\ell) \om_M(m) \, \cC_2 ^{I,J,K,L,M} 
\eea
The tensor  $\cC_2^{I,J,K,L,M}$ is a single-valued scalar in $q_\a$ in view of item 2, which is independent of all $z_i$ in view of item 7. It has cyclic symmetry in $I,J,K,L,M$ in view of item 5 and is odd under reversal $(I,J,K,L,M) \to (M,L,K,J,I)$  in view of item 6. Since the indices $I,J,K,L,M$ can take the values $1,2$, the implications of its behavior under cyclic permutations and reversal are readily analyzed on its 32 components, for $J \not=I$,
\bea
\cC_2 ^{IIIII} & = & - \cC_2 ^{IIIII}
\no \\
\cC_2 ^{IIIIJ} & = & - \cC_2 ^{JIIII} = - \cC_2 ^{IIIIJ} 
\no \\
\cC_2 ^{IIIJJ} & = & - \cC_2 ^{JJIII} = - \cC_2 ^{IIIJJ} 
\no \\
\cC_2 ^{IIJIJ} & = & - \cC_2 ^{JIJII} = - \cC_2 ^{IIJIJ} 
\eea
The first equality from the left on each line follows from oddness under reversal, while the second equality on the second, third and fourth lines follows from cyclic symmetry. When $(I,J)=(1,2)$ or $(2,1)$ the first line gives 2 identities while each one of the remaining lines gives 10 identities totaling~32. Every line implies that the corresponding components vanish, which proves the vanishing of $\tilde \cC_2$ in item 8.

\newpage

%%%%%%%%%%%%%%%%%%%%%%%%%%%%%%%%%%%%%%%%%%%
%%%%%%%%%%%%%%%%%%%%%%%%%%%%%%%%%%%%%%%%%%%
\section{Kinematic rearrangement of $\mF_8+\mF_9+\mF_{10}$}
\setcounter{equation}{0}
\label{sec:I}
%%%%%%%%%%%%%%%%%%%%%%%%%%%%%%%%%%%%%%%%%%%
%%%%%%%%%%%%%%%%%%%%%%%%%%%%%%%%%%%%%%%%%%%

The purpose of this appendix is to prove formula (\ref{7.kS}) which is the key to express the entire kinematic dependence of the five-point amplitude in terms of $t_8$ tensors. To begin, we obtain the following decomposition of the tensors $S^{\mu \nu}_{ij}$ into the parts $M_{ij}^{\mu \nu}, N_{ij}^{\mu \nu}$ and $Q_{ij}^{\mu \nu} $ which arise from the decompositions of $\mF_8$, $\mF_9$, and $\mF_{10}$, respectively,
\bea
S_{ij}^{\mu \nu} = M_{ij}^{\mu \nu} + N_{ij}^{\mu \nu} + Q_{ij}^{\mu \nu} 
\eea 
The individual contributions, properly symmetrized in $i,j$,  may be read off from matching (\ref{7.F890})
with (\ref{matchwith}), 
\bea
\label{7.MNQ2}
M_{ij}^{\mu \nu} & = & 
- { 1 \over 4}  \sum_{k,\ell,m \not= i,j} 
\Big [ 2 (f_k f_\ell f_m)^{\mu \nu} (f_i f_j) - 2 (f_k f_j f_m)^{\mu \nu} (f_i f_\ell)   - 2 (f_k f_i f_m)^{\mu \nu} (f_j f_\ell) 
  \\ && \hskip 0.8in
+ \big \{  (f_i f_j f_k)^{\mu \nu} - (f_i f_k f_j)^{\mu \nu} + (f_k f_j f_i)^{\mu \nu} 
\no  \\ && \hskip 1in
+ (f_j f_i f_k)^{\mu \nu} - (f_j f_k f_i)^{\mu \nu} + (f_k f_i f_j)^{\mu \nu}\big \} (f_\ell f_m) \Big ]
\no \\
N_{ij}^{\mu \nu} & = &   
 \sum _{k,\ell,m \not= i,j} 
\Big [ (f_i f_j f_k f_\ell f_m)^{\mu \nu} - (f_i f_m f_k f_\ell f_j)^{\mu \nu} - (f_\ell f_j f_k f_i f_m)^{\mu \nu} + (f_\ell f_m f_k f_i f_j)^{\mu \nu} 
\no  \\ && \hskip 0.6in
+ (f_j f_i f_k f_\ell f_m)^{\mu \nu} - (f_j f_m f_k f_\ell f_i)^{\mu \nu} - (f_\ell f_i f_k f_j f_m)^{\mu \nu} + (f_\ell f_m f_k f_j f_i)^{\mu \nu} \Big ]
\no \\
Q_{ij}^{\mu \nu} & = &
 {1 \over 6} \sum _{k,\ell,m \not= i,j} f_k ^{\mu \nu}
\Big ( 2 (f_i f_\ell f_j f_m) - 2 (f_i f_\ell f_m f_j) +(f_i f _j) (f_\ell f_m) - (f_i f_m) (f_j f_\ell) \Big )
\no
\eea
The sums are over mutually distinct $k,\ell,m$ which are different from $i$ and $j$, say six permutations of $3,4,5$ if $(i,j)=(1,2)$. By construction, the tensors $M_{ij}^{\mu \nu}, N^{\mu \nu}_{ij}, Q^{\mu \nu}_{ij}$ are anti-symmetric in $\mu$ and $\nu$ and symmetric in $i$ and $j$. The relations in (\ref{7.Sa}) originate from the corresponding relations of each individual tensor, and in particular guarantee that $\mF_8, \mF_9, \mF_{10}$ are individually independent of the arbitrary point $z_0$.

%%%%%%%%%%%%%%%%%%%%%%%%%%%%%%%%%%%%%%%%%%%
%%%%%%%%%%%%%%%%%%%%%%%%%%%%%%%%%%%%%%%%%%%
\subsection{Kinematics: notation and identities}
\label{sec:G}
%%%%%%%%%%%%%%%%%%%%%%%%%%%%%%%%%%%%%%%%%%%
%%%%%%%%%%%%%%%%%%%%%%%%%%%%%%%%%%%%%%%%%%%

We start by deriving auxiliary identities that will later on facilitate the simplification of 
the tensor contractions $k_i^\mu M_{ij}^{\mu \nu}$ and $k_i^\mu N_{ij}^{\mu \nu} $.
In addition to the conditions $k_i^2 = \ep_i \cdot k_i=0$, we have the linearized Bianchi identity, 
\bea
k_i ^\mu f_i ^{\nu \rho} + k_i ^\nu f_i ^{\rho \mu } + k_i ^\rho f_i ^{\mu \nu} =0
\eea
Throughout we use the notations,
\bea
(f_1 f_2 \cdots f_n) & = & f_1^{\mu_1 \mu_2} f_2^{\mu_2 \mu_3} \cdots f_n ^{\mu _n \mu_1}
\no \\
(f_1 f_2 \cdots f_n)^{\mu \nu}  & = & f_1^{\mu \mu_2} f_2^{\mu_2 \mu_3} \cdots f_n ^{\mu _n \nu}
\eea
The Bianchi identity implies the following identities,
\bea
k_i^\mu \, (f_1 \cdots f_i \cdots f_n) & = & 
(f_i \cdots f_n f_1 \cdots f_{i-1})^{\mu \nu} \, k_i ^\nu 
 + k_i ^\nu \, (f_{i+1} \cdots f_n f_1 \cdots f_i ) ^{\nu \mu}
\eea
as well as,
\bea
 k^\mu _i (f_1  \cdots f_i \cdots f_n)^{\nu \rho} =
 (f_1 \cdots f_{i-1} )^{\nu \sigma} k_i ^\sigma (f_i \cdots f_n)^{\mu \rho} 
 +  (f_1 \cdots f_i )^{\nu \mu} k_i ^\sigma (f_{i+1} \cdots f_n)^{\sigma \rho}
 \qquad
 \label{tobecontracted}
\eea
 for $1 < i < n$, and the following identities for $i=1,n$, 
 \bea
 k^\mu _1 (f_1 f_2 \cdots f_n)^{\nu \rho}  & = & k_1^\nu (f_1 f_2 \cdots f_n)^{\mu \rho} 
+ f_1 ^{\nu \mu} k_1 ^\a (f_2 f_3 \cdots f_n)^{\a \rho}
\no \\
 k^\mu _n (f_1 f_2 \cdots f_n)^{\nu \rho}  & = & 
  (f_1 f_2 \cdots f_{n-1})^{\nu \a} \, k_n ^\a \, f_n ^{\mu \rho} 
 +  (f_1 f_2 \cdots f_n)^{\nu \mu} \, k_n^\rho
 \eea
Further identities that seem to be useful may be obtained by contracting $\mu ,\nu$ in (\ref{tobecontracted}),
\bea
k_i ^\mu (f_1  \cdots f_i \cdots f_n)^{\mu \rho} & = &
 (-)^{i-1} k_i ^\mu (f_{i-1} \cdots f_1 f_i \cdots f_n)^{\mu \rho} 
 +  (f_1 \cdots f_i )k_i ^\sigma (f_{i+1} \cdots f_n)^{\sigma \rho}
 \qquad
 \eea
with the special cases for $i=1,2,n$ given by, 
\bea
k_1 ^\mu (f_1 f_2 \cdots f_n)^{\mu \rho}  & = & 0
\no \\
 k^\mu _n (f_1 f_2 \cdots f_n)^{\mu \rho}  & = & 
 - (f_n f_1 f_2 \cdots f_{n-1})^{\rho \a} \, k_n ^\a  +  (f_1 f_2 \cdots f_n) \, k_n^\rho
 \no \\
  k^\mu _2 (f_1 f_2 \cdots f_n)^{\mu \rho}  & = & \half (f_1 f_2) \, k_2^\mu \, (f_3 f_4 \cdots f_n)^{\mu \rho}
 \eea
 We single out the special cases for $n =3$,
 \bea
 \label{7.kin.3}
 k_i ^\mu \, (f_i f_j f_k) ^{\mu \nu} & = & 0
 \no \\
 k_j ^\mu \, (f_i f_j f_k) ^{\mu \nu} & = & \thalf (f_i f_j) k^\mu _j f^{\mu \nu}_k
 \no \\
 k_k ^\mu \, (f_i f_j f_k)^{\mu \nu}  & = & (f_i f_j f_k) k^\nu _k - (f_k f_i f_j)^{\nu \a} k_k ^\a
 \eea
 and for $n=5$,
 \bea
 \label{7.kin.5}
 k^\mu _i (f_i f_j f_k f_\ell f_m)^{\mu \nu} & = & 0
 \no \\
 k^\mu _j (f_i f_j f_k f_\ell f_m)^{\mu \nu} & = & 
 \thalf (f_i f_j) \, k_j ^\mu ( f_k f_\ell f_m)^{\mu \nu}
 \no \\
 k^\mu _k (f_i f_j f_k f_\ell f_m)^{\mu \nu} - k^\mu _k (f_jf_if_kf_\ell f_m)^{\mu \nu} & = & 
 (f_i f_j f_k) \, k^\mu _k (f_\ell f_m)^{\mu \nu} 
 \no \\
 k^\mu _\ell (f_i f_j f_k f_\ell f_m)^{\mu \nu} +  k_\ell ^\mu (f_k f_j f_i f_\ell f_m)^{\mu \nu}   & = & 
 (f_i f_j f_k f_\ell) \, k_\ell ^\mu f_m ^{\mu \nu}
 \no \\
 k^\mu _m (f_i f_j f_k f_\ell f_m)^{\mu \nu} -  k_m^\mu (f_\ell f_k f_j f_i f_m)^{\mu \nu}     & = & 
k_m^\nu (f_i f_j f_k f_\ell f_m) 
 \eea

\subsection{Inner products with $k_i^\mu$}

To prove the relation (\ref{7.kS}), we shall evaluate the inner products $k_i ^\mu M_{ij} ^{\mu \nu}$, $k_i^\mu N^{\mu \nu} _{ij}$ and $k_i^\mu Q^{\mu \nu}_{ij}$ for all $i \not = j$ and then obtain their sum.  Straightforward application of the kinematic identities in (\ref{7.kin.3}) gives the following result,
\bea
\label{7.kM}
k_i^\mu M_{ij}^{\mu \nu} & = & 
 \sum_{k,\ell,m \not= i,j}  \Big [ 
 \thalf k^\mu _i (f_k f_j f_m)^{\mu \nu} (f_i f_\ell)   
- \thalf  k^\mu _i (f_k f_\ell f_m)^{\mu \nu} (f_i f_j) 
+ \tfrac{1}{4}   (f_i f_k) (f_j f_\ell) k^\mu _i f_m^{\mu \nu}
\no  \\ && \hskip 0.5in
- \tfrac{1}{4} \big \{  
(f_kf_jf_i) k^\nu _i + \thalf (f_i f_j) k^\mu _i f_k^{\mu \nu} + \thalf (f_i f_k) k_i^\mu f_j^{\mu \nu} 
\big \} (f_\ell f_m) \Big ]
\quad
\eea
Similarly, we obtain, 
\bea
k_i^\mu N_{ij}^{\mu \nu}
& = &
 \sum _{k,\ell,m \not= i,j} 
\Big [  k_i^\mu (f_j f_i f_k f_\ell f_m)^{\mu \nu} 
- k_i^\mu (f_\ell f_i f_k f_j f_m)^{\mu \nu} 
- k_i^\mu (f_\ell f_j f_k f_i f_m)^{\mu \nu}
\no  \\ && \hskip 0.65in
 + k_i^\mu (f_\ell f_m f_k f_i f_j)^{\mu \nu} 
 - k_i^\mu (f_j f_m f_k f_\ell f_i)^{\mu \nu}  
 + k_i^\mu (f_\ell f_m f_k f_j f_i)^{\mu \nu} \Big ]
\eea
The evaluation of the first and second terms proceeds directly using the first identity in (\ref{7.kin.5}). 
To evaluate the remaining terms we use the last two lines of (\ref{7.kin.5}) to obtain, 
\bea
k_i^\mu (f_\ell f_j f_k f_i f_m)^{\mu \nu} + k^\mu _i (f_k f_j f_\ell f_i f_m)^{\mu \nu} & = & (f_\ell f_j f_k f_i) k_i^\mu f_m ^{\mu \nu}
\no \\
k_i^\mu (f_\ell f_m f_k f_i f_j)^{\mu \nu} + k^\mu _i (f_k f_m f_\ell f_i f_j)^{\mu \nu} & = & (f_\ell f_m f_k f_i) k_i^\mu f_j ^{\mu \nu}
\no \\
k_i^\mu (f_\ell f_m f_k f_j f_i)^{\mu \nu} - k^\mu _i (f_j f_k f_m f_\ell f_i)^{\mu \nu} & = & (f_\ell f_m f_k f_j f_i ) k_i^\nu 
\eea
After summing over all permutations of $k,\ell,m$, the terms on the left side precisely correspond to the combinations occurring in the summand of $k_i^\mu N_{ij}^{\mu \nu}$, and we have effectively,
\bea
k_i^\mu (f_\ell f_j f_k f_i f_m)^{\mu \nu} & \to & \thalf (f_\ell f_j f_k f_i) k_i^\mu f_m ^{\mu \nu}
\no \\
k_i^\mu (f_\ell f_m f_k f_i f_j)^{\mu \nu} & \to & \thalf (f_\ell f_m f_k f_i) k_i^\mu f_j ^{\mu \nu}
\no \\
k_i^\mu (f_\ell f_m f_k f_j f_i)^{\mu \nu} - k^\mu _i (f_j f_m f_k f_\ell f_i)^{\mu \nu} & \to & (f_\ell f_m f_k f_j f_i ) k_i^\nu 
\eea
Putting all together, we obtain, 
\bea
\label{7.kN}
k_i^\mu N_{ij}^{\mu \nu}
& = &
 \sum _{k,\ell,m \not= i,j} 
\Big [
- \thalf (f_\ell f_j f_k f_i) \, k^\mu _i f_m^{\mu \nu} 
+ \thalf (f_\ell f_m f_k f_i) \, k_i^\mu f_j^{\mu \nu} 
+ k^\nu _i (f_\ell f_m f_k f_j f_i)
\no \\ && \hskip 0.7in
+ \thalf (f_i f_j) \, k^\mu _i (f_k f_\ell f_m)^{\mu \nu} 
- \thalf (f_i f_\ell) \, k^\mu _i (f_k f_j f_m)^{\mu \nu}
\Big ]
\eea
Evaluating $k_i^\mu Q^{\mu \nu}_{ij}$ does not require any use of kinematic identities, and we find, 
\bea
\label{7.kQ}
k_i^\mu Q_{ij}^{\mu \nu} =
 {1 \over 6} \sum _{k,\ell,m \not= i,j} k_i ^\mu f_k ^{\mu \nu}  
\Big [ 2 (f_i f_\ell f_j f_m) - 2 (f_i f_\ell f_m f_j) +(f_i f _j) (f_\ell f_m) - (f_i f_m) (f_j f_\ell) \Big ]
\quad
\eea
Adding up the contributions to $k_i ^\mu S_{ij}^{\mu \nu}$ from (\ref{7.kM}), (\ref{7.kN}) and (\ref{7.kQ}), we see that the two terms $(f_a f_b f_c)^{\mu \nu}$ on the second line of (\ref{7.kN}) cancel the first two terms in (\ref{7.kM}). 
Assembling the remaining contributions, we find, 
\bea
k_i ^\mu S_{ij}^{\mu \nu} & = &
 \sum_{k,\ell,m \not= i,j}  k_i ^\mu f_k^{\mu \nu} \Big [ 
  \tfrac{1}{12}   (f_i f_m) (f_j f_\ell)  + \tfrac{1}{24} (f_i f_j) (f_\ell f_m)
  - \tfrac{1}{6}  (f_i f_\ell f_j f_m)  -  \tfrac{1}{3}   (f_i f_\ell f_m f_j)  \Big ]
  \no \\ && 
+   \sum_{k,\ell,m \not= i,j}  k_i^\mu f_j^{\mu \nu}  \Big [  \thalf (f_\ell f_m f_k f_i) 
- \tfrac{1}{8} (f_i f_k) (f_\ell f_m) \Big ]
\no  \\ &&
+  \sum_{k,\ell,m \not= i,j} k_i^\nu \Big [ 
- \tfrac{1}{2} \big ([f_i, f_j] f_k f_\ell f_m \big ) + \tfrac{1}{8}    \big ([f_i, f_j] f_k \big )   (f_\ell f_m)  \Big ]
\label{kdotSij}
\eea
The sums $k, \ell, m$ are over all permutations of the three distinct values in $\{1,2,3,4,5\} \setminus \{i,j\}$. In the last two lines, we immediately recognize the structure of $t_8$ which is defined only as a sum over cyclic permutations of the last three indices. By writing out the six permutations of $(k,\ell,m)$ in the first line of (\ref{kdotSij}), one can also identify $-\frac{1}{3}t_8(f_i,f_j,f_\ell,f_m)$ along with $k_i^\mu f_k^{\mu \nu}$. In terms of the notation $\mt_i$ and $\mt_{ij}$ in (\ref{newsumm.5}) and (\ref{7.def.T}), we recognize the formula (\ref{7.kS}), whose proof was the purpose of this appendix, and is now complete.

\newpage

%%%%%%%%%%%%%%%%%%%%%%%%%%%%%%%%%%%%%%%%%%%
%%%%%%%%%%%%%%%%%%%%%%%%%%%%%%%%%%%%%%%%%%%
\section{Vanishing of $\mF_C' +\mF_E''$}
\setcounter{equation}{0}
\label{sec:J}
%%%%%%%%%%%%%%%%%%%%%%%%%%%%%%%%%%%%%%%%%%%
%%%%%%%%%%%%%%%%%%%%%%%%%%%%%%%%%%%%%%%%%%%

The $q_\alpha$-dependent combinations $\mF_C' $ and $ \mF_E''$ in (\ref{fcprime}) and (\ref{mfepp})
have the same structure in their worldsheet data $W^I_{a;b}$ and $G^I_{i,j,q_\alpha,k}$. Accordingly,
their kinematic data consistently combines to the following quantity in their sum,
\bea
R^{\rm inv} _{1;2,3} 
= \mt_1 f_1^{\mu \nu} k_2^\mu k_3^\nu + \half k_1\cdot k_2 \, \mt_{13}  - \half k_1\cdot k_3 \, \mt_{12} 
\eea
and permutations thereof. This combination may be related to the quantity $R_{1;2|3,4,5}$ that 
provides an effective description of the bosonic components in pure-spinor calculations \cite{DHoker:2020tcq}
as reviewed in section \ref{sec:int.3a},
\bea
- i R^{\rm inv} _{1;2,3} & = & k_1 \cdot k_2 \, R_{1;3|2,4,5} - k_1 \cdot k_3 \, R_{1;2|3,4,5}
\no \\
R_{1;2|3,4,5} & = & i (\ep_1 \cdot k_2) \mt_1 - { i \over 2} \mt_{12}
\eea
and permutations thereof. Contrarily to the individual $R_{1;2|3,4,5}$, the combination
$R^{{\rm inv}}_{1;2,3}$ is manifestly gauge-invariant, obeys  the symmetry 
$R^{\rm inv} _{1;2,3} = - R^{\rm inv} _{1;3,2}$, and the cyclic identity,
\bea
R^{\rm inv} _{1;2,3} + R^{\rm inv} _{1;2,4}  + R^{\rm inv} _{1;2,5}  =0
\eea
The independence of $q_\a$ of $\mF$ and the vanishing of the combination $\mF_C' +\mF_E''$ hinges on the following remarkable identity, which was proven using {\tt Mathematica}, 
\bea
\label{8.oli}
R^{\rm inv} _{1;2,4}+ R^{\rm inv} _{4;2,1}+ R^{\rm inv} _{3;2,5}+ R^{\rm inv} _{5;2,3}=0
\eea
We note that this identity is manifestly invariant under  swapping $1,4$, independently swapping $3,5$, and independently swapping the pairs $(1,4), (3,5)$. Because, in this presentation, the second index is always 2, it follows that the $\mt_i f_i$ terms factor out a momentum $k_2$. The  terms of the form $\mt_{ij}$ may be similarly rearranged, so that the identity is equivalent to,
\bea
&&
k_2^\mu  \Big [  (k_1-k_4)^\mu \mt_{14} + (k_3-k_5)^\mu \mt_{35} 
+ \thalf (k_1+k_4)^\mu(\mt_{12} +\mt_{42}) + \thalf (k_3+k_5)^\mu(\mt_{32} +\mt_{52})
\no \\ && \qquad
+ 2 \mt_1 f_1^{\mu \nu} k_4^\nu + 2 \mt_4 f_4^{\mu \nu} k_1^\nu + 2 \mt_3 f_3^{\mu \nu} k_5^\nu + 2 \mt_5 f_5^{\mu \nu} k_3^\nu  
 \Big ] =0
\eea
We have properly symmetrized under swapping the pairs $(1,4), (3,5)$ at the cost of introducing the factors of $\thalf$ above. One may speculate whether there exists a vector-valued identity that is first order in momenta $k_i$ and quadri-linear in $f_{j}$ from which this relation follows.

\subsection{The vanishing of $\mF_C' +\mF_E''$}

The combination  $\mF_C' +\mF_E''$ may be expressed exclusively in terms of $R^{\rm inv}$,
\bea
\mF_C' +\mF_E''  & = &
-8 \sum_I \Big (  W_{1;3}^I +W_{1;5}^I -W_3^I -W_5^I -W_{1;4}^I +W_4^I  \Big )
\Big [ R^{\rm inv}_{1;2,3} G^I _{1,3,q_\a,5}
+R^{\rm inv}_{1;2,4} G^I _{1,4,q_\a,5} \Big ]
\no \\ &&
-8  \sum_I \Big (  W_{1;3}^I +W_{1;5}^I -W_3^I -W_5^I  \Big ) 
\Big [ R^{\rm inv}_{1;3,2}  G^I _{1,2,q_\a,5}
+ R^{\rm inv}_{1;3,4}  G^I _{1,4,q_\a,5} \Big ]
\no \\ &&
-8  \sum_I \Big ( W_{1;3}^I  - W_3^I  \Big )
\Big [ R^{\rm inv}_{1;4,2} G^I _{1,2,q_\a,5}
%\no \\ && \hskip 2.7in
+ R^{\rm inv}_{1;4,3} G^I _{1,3,q_\a,5} \Big ]
+ {\rm cycl}(1,2,3,4,5)
\eea
and the five-forms $W_{a;b}, W_a$ in (\ref{basforms}), where the instruction to add cyclic permutations applies to all terms on the right side. We use $R^{\rm inv}_{1;a,b}=-R^{\rm inv}_{1;b,a}$ and the identity (\ref{8.oli}) to express the $R^{\rm inv}$ in terms of two independent terms $R_{1;2,3}^{\rm inv}$ and $R_{1;3,4}^{\rm inv}$ and their cyclic permutations.  The remaining combinations are given by (\ref{8.oli}),
\bea
R^{\rm inv} _{1;2,4} = R_{3;4,5}^{\rm inv} + R_{4;1,2}^{\rm inv} - R_{5;2,3}^{\rm inv} - R_{3;5,1}^{\rm inv}
\eea
and cyclic permutations thereof.  The first term on the right side is in the cyclic orbit of $R^{\rm inv}_{1;2,3}$, while the second, third and fourth terms are in the cyclic  orbit of $R^{\rm inv}_{1;3,4}$. Cyclicly permuting all terms to the same representative in a given cyclic orbit, we find, 
\bea
\mF_C' +\mF_E'' = -8 Z_1 R^{\rm inv}_{1;2,3} -8  Z_2 R^{\rm inv}_{1;3,4} + {\rm cycl}(1,2,3,4,5)
\eea
The coefficients $Z_1,Z_2$ are obtained by collecting all the contributions in the same orbit,
\bea
Z_1 & = &  
\big (W_{4;4}   -W_{1;4} \big ) G_{1,3,q_\a,5} + 
\big( W_{3;3}+W_{5;5} -W_{1;3}-W_{1;5} \big)  G_{1,2,q_\a, 3}
\no \\ &&
+ \big ( W_{4;1} - W_{1;1} \big ) G_{4,2,q_\a,5}
+ \big (  W_{4;3}-W_{4;2}-W_{3;3}+W_{2;2} \big ) G_{4,2,q_\a,3}
\no \\
Z_2 & = &
 \big ( W_{1;1}-W_{2;2}-W_{3;1}+W_{3;2}  \big ) G_{3,1,q_\a,2} 
 + \big (   W_{3;5} -W_{5;5}\big ) G_{3,1,q_\a,4} 
\no \\ &&
+ \big ( W_{1;1} -W_{5;5}  -W_{2;1} + W_{2;5}\big ) G_{2,5,q_\a,1} 
+ \big ( W_{2;4} - W_{4;4} \big ) G_{2,3,q_\a,5} 
\no \\ && 
+ \big ( W_{3;3}  -W_{2;2} -W_{4;3} + W_{4;2}\big ) G_{4,2,q_\a,3} 
+ \big (   W_{1;1} - W_{4;1} \big ) G_{4,2,q_\a,5} 
\no \\ &&
+ \big (   W_{1;5} - W_{5;5} \big ) G_{1,4,q_\a,5} 
+ \big ( W_{3;3} - W_{1;3} \big ) G_{1,3,q_\a,4} 
\label{z1andz2}
\eea
where we suppress the superscripts of $W^I_{a;b}$ and $G^I_{i,j,q_\alpha,k}$ as
well as the associated $\sum_I$ for ease of notation throughout the remainder of this appendix.

\subsection{The vanishing of $Z_1$ and $Z_2$}

The combinations $Z_1$ and $Z_2$ in (\ref{z1andz2}) are,
\begin{enumerate}
\itemsep=-0.03in
\item differential $(1,0)$-forms in $z_i$ for $i=1,2,3,4,5$ and scalars in $q_\a$;
\item free of singularities as $z_i \to q_\a$ so that they are holomorphic in $q_\a$ and therefore independent of $q_\a$;
\item  free of singularities as $z_i \to z_j$ for all $j \not = i$, and are thus holomorphic in $z_i$;
\item zero $Z_1=Z_2=0$.
\end{enumerate}
To show item 2, namely that  $Z_1$ and $Z_2$ are free of singularities as $z_i \to q_\a$, we use the fact that  such singularities are of the form $(z_i{-}q_\alpha)^{-1}$ and can arise from $W_{i;j}$ with $j \not= i$ as well as from the parts proportional to $g_{i,q_\a}$ in~$G$. Furthermore,  we shall use the fact that $G_{i,a,q_\a, b} \to 0 $ as $z_i \to q_\a$, as well as the following identities,
\bea
\label{8.VW}
W_{1;2}^I+W_{1;5}^I &=& W_2^I+W_5^I
\no \\
\Wt_1 +\Wt_3 & = & \varpi(2) \Delta(1,3) \Delta (4,5)
\eea
and permutations thereof.
We begin by establishing the cancellation of all singularities in $Z_1$ as $z_i \to q_\a$. The absence of poles
as $z_1 \to q_\a$ is obvious since any $W_{1;j}$ with $j\neq 1$ is accompanied by a factor
of $G_{1,a,q_\a, b}$ with the corresponding zero. For $z_2 \to q_\a$,  none of the $W_{a;b}$ in $Z_1$ has a singularity, so the leftover source of poles is $g_{2,q_\a}$. After some simplifications, using the first equation in (\ref{8.VW}) this part is given by,
\bea
g_{2,q_\a} \big ( W_{4;1}+W_{4;3}-W_{4;2} -W_{1;1}-W_{1;3}+W_{1;2} \big )
\eea
Using now also the second identity in (\ref{8.VW}), we see that the entire expression is proportional to $g_{2,q_\a}\varpi(2)$ and thus regular as $z_2 \to q_\a$.  For $z_3 \to q_\a$, the pre-factors are regular and  the terms proportional to $g_{3,q_\a}$ simplify using the first equation in (\ref{8.VW}) to give,
\bea
g_{3,q_\a} \big ( 
W_{1;3}-W_{1;2}-W_{1;4}+W_{4;4} -W_{4;3}+W_{4;2}   \big )
\eea
Using a shift forward by 1 of the second equation of (\ref{8.VW}), we see that the entire combination is proportional to $g_{3,q_\a}\varpi(3)$ and thus regular as $z_3 \to q_\a$. For $z_4 \to q_\a$ each individual term has a regular limit
by the zeros of $G_{4,a,q_\a, b}$. For $z_5 \to q_\a$, the pre-factors are regular and  the terms proportional to $g_{5,q_\a}$ simplify using the first equation in (\ref{8.VW}) to give,
\bea
g_{5,q_\a} \big (   W_{1;1} +W_{1;4}-W_{4;4} -W_{4;1}  \big )
\eea
This combination is proportional to $g_{5,q_\a}\varpi(5) $, using the second equation of (\ref{8.VW}) shifted forward by 3, and thus also regular as $z_5 \to q_\a$. This concludes the proof of item 2 for $Z_1$. 

\sm

Item 2 is established in exactly the same way for $Z_2$, so we shall be brief here. For example, as $z_1 \to q_\a$, the potentially singular terms in $Z_2$ are proportional to $g_{1,q_\a}$, 
\bea
g_{1,q_\a} \big (  W_{3;2}  -W_{3;1} + W_{3;5}  +W_{2;1} -W_{2;2} - W_{2;5} \big )
\eea
Using a shift backward by 1 of the second equation of (\ref{8.VW}), we see that the entire combination is proportional to $g_{1,q_\a}\varpi(1)$ and thus regular as $z_1 \to q_\a$. The other limits are computed analogously and all vanish as well so that $Z_2$ is also holomorphic in $q_\a$, thereby completing the proof of item 2. 

\sm

Item 3 is proven by letting $z_ i \to z_j$ for all $1\leq i < j \leq 5$, in which limit all pre-factors $W_{a;b}$ in $Z_1$ and $Z_2$ are regular, so that all singularities arise from the $G$-functions. For $Z_1$, we begin by considering the limits $z_1 \to z_2$ and $z_1 \to z_3$ which are proportional to, 
\bea
&& g_{1,2} \big ( W_{3;3}+W_{5;5}-W_{1;3}-W_{1;5} \big)
\no \\
&& g_{1,3} \big ( W_{1;3}-W_{3;3}+W_{1;5}-W_{5;5}-W_{1;4}+W_{4;4}  \big)
\eea
respectively. The coefficient of $g_{1,2}$ is proportional to $\Delta(1,2)$ which cancels the pole as $z_1 \to z_2$. For the coefficient of $g_{1,3}$ in turn, the first two terms $W_{1;3}-W_{3;3}$ and the remainder separately vanish as $z_1 \rightarrow z_3$. All the constituents of $Z_1$ are manifestly regular as $z_1 \to z_4$. For $z_1 \to z_5$ the relevant terms of $Z_1$ are given by $g_{1,5} ( W_{4;4} - W_{1;4} )$ which is manifestly proportional to $\Delta (5,1)$ and is thus regular in the limit. The remaining limits may easily be established along analogous lines, which concludes the proof that the limits $z_i \to z_j$ of $Z_1$ are all regular.  Item 3 may be established in exactly the same way for $Z_2$.
\sm

To prove item 4 we use the arguments used in appendix \ref{appE.1} for the 
forms $Z_1, Z_2$ which are independent of $q_\a$, and holomorphic in $z_1, \cdots , z_5$.

\end{document}